\documentclass[phd,bottom,nosig,all]{usbthesis}

\usepackage{url}
\usepackage{hyperref} 

\usepackage{graphicx}
\usepackage{color}
\usepackage[intlimits]{amsmath}
\usepackage{amssymb}
\usepackage{setspace} 
\usepackage{epic} 
\usepackage{eepic} 

\usepackage{array} 
\usepackage{longtable} 
\usepackage{tabularx} 
\usepackage{booktabs}
\usepackage{lscape}
\usepackage{multirow}
\usepackage{afterpage}

\usepackage[figuresright]{rotating}
\usepackage[font=small,labelfont=bf,up,textfont=rm,up]{caption}
\usepackage{subfig}
\usepackage{footmisc} 
\usepackage{xspace}
\usepackage{feynmf}
\usepackage[numbers,comma,sort&compress]{natbib}
\usepackage{hypernat} 
\usepackage[section,above,below]{placeins} 

\newcommand{\ee}{\ensuremath{e^+e^-}\xspace}
\newcommand{\epair}{\ensuremath{e^+e^-} pair\xspace}
\newcommand{\epairs}{\ensuremath{e^+e^-} pairs\xspace}
\newcommand{\mumu}{\ensuremath{\mu^+\mu^-}\xspace}
\newcommand{\pion}{\ensuremath{\pi^0}\xspace}
\newcommand{\eexp}[1]{\ensuremath{{\rm e}^{#1}}\xspace}

\newcommand{\eq}[1]{{Eq.~\eqref{#1}}\xspace}
\newcommand{\fig}[1]{{Fig.~\ref{#1}}\xspace}
\newcommand{\tab}[1]{Tab.~\ref{#1}\xspace}

\newcommand{\eg}{\emph{e.\,g.}\xspace}%
\newcommand{\ie}{\emph{i.\,e.}\xspace}%

\newcommand{\pp}{{\ensuremath{p+p}}\xspace}
\newcommand{\dAu}{\ensuremath{d+\rm{Au}}\xspace}
\newcommand{\CuCu}{\ensuremath{\rm{Cu}+\rm{Cu}}\xspace}
\newcommand{\AuAu}{\ensuremath{\rm{Au}+\rm{Au}}\xspace}
\newcommand{\PbPb}{\ensuremath{\rm{Pb}+\rm{Pb}}\xspace}

\newcommand{\sqrts}{\ensuremath{\sqrt{s}}\xspace}
\newcommand{\sqrtsnn}{\ensuremath{\sqrt{s_{NN}}}\xspace}

\newcommand{\kt}{\ensuremath{k_T}\xspace}
\newcommand{\mt}{\ensuremath{m_T}\xspace}
\newcommand{\pt}{\ensuremath{p_T}\xspace}

\newcommand{\mevc}{MeV/\ensuremath{c}\xspace}
\newcommand{\gevc}{GeV/\ensuremath{c}\xspace}
\newcommand{\kevcc}{keV/\ensuremath{c^2}\xspace}
\newcommand{\mevcc}{MeV/\ensuremath{c^2}\xspace}
\newcommand{\gevcc}{GeV/\ensuremath{c^2}\xspace}

\newcommand{\mee}{\ensuremath{m_{ee}}\xspace}

\newcommand{\exodus}{{\sc Exodus}\xspace}
\newcommand{\pythia}{{\sc Pythia}\xspace}

\providecommand{\renewoperator}[3]{\renewcommand*{#1}{\mathop{#2}#3}}
\renewoperator{\Re}{\mathrm{Re}}{\nolimits}
\renewoperator{\Im}{\mathrm{Im}}{\nolimits}

\author{Torsten Dahms}
\title{Dilepton spectra in $\boldsymbol{p+p}$ and $\boldsymbol{\rm Au + Au}$ collisions at RHIC}

\month{December} \year{2008}

\program{Physics}

\director{Axel Drees}{Professor, Department of Physics and Astronomy}
\chairman{Edward Shuryak}{Professor, Department of Physics and Astronomy}
\fstmember{Thomas C. Weinacht}{Professor, Department of Physics and Astronomy}
\outmember{Yasuyuki Akiba}{Vice Chief Scientist}{Radiation Laboratory,
  RIKEN Nishina Research Center, Wako, Japan}
\dean{Lawrence Martin}

\begin{document}
\setlength{\unitlength}{1mm}

\singlespacing
\pagenumbering{roman}

\maketitle
\makeapproval

\begin{abstract}
    Recent experimental results show that a strongly coupled quark-gluon
plasma (sQGP) is created in heavy ion collisions at the Relativistic
Heavy Ion Collider (RHIC) at Brookhaven National Laboratory
(BNL). Electromagnetic radiation, \ie, photons and lepton pairs, are
penetrating probes that allow investigating the full time evolution
and dynamics of the matter produced, as they do not undergo strong
interaction in the final state.

This work presents measurements of electron-positron pairs from \pp
collisions at \sqrts = 200~GeV collected during the 2005 RHIC run and
compares them to results from \AuAu collisions at \sqrtsnn = 200~GeV
taken in 2004 with the PHENIX detector. The invariant mass
distribution of \ee pairs in \pp is consistent with the expected
contributions from Dalitz decays of light hadrons, dielectron decays
of vector mesons and correlated charm production, which have been
measured in the same experiment. The charm and bottom cross section
extracted from the measured dielectron yield are
$\sigma_{c\overline{c}} = 544 \pm 39 ({\rm stat.}) \pm 142 ({\rm
  syst.})  \pm 200 ({\rm model})~\mu{\rm b}$ and
$\sigma_{b\overline{b}} = 3.9 \pm 2.4 ({\rm stat.}) ^{+3}_{-2} ({\rm
  syst.})~\mu{\rm b}$, respectively. The dielectron continuum
measurement in \pp provides a crucial baseline for the modification of
the dielectron continuum observed in \AuAu.

In min. bias \AuAu collisions the yield of dielectrons in the low mass
region ($150 < \mee < 750$~\mevcc) is enhanced by a factor of $4.0 \pm
0.3 {\rm (stat.)} \pm 1.5 {\rm (syst.)} \pm 0.8 {\rm (model)}$
compared to the known hadronic sources. The centrality dependence of
this enhancement suggests emission from in-medium scattering
processes. The excess dominates the yield in the transverse momentum
region below 1~\gevc and shows significantly lower $\langle\pt\rangle$
than the expected sources. The low \pt enhancement is currently not
understood by any theoretical model of heavy ion collisions. The
enhancement extends to larger transverse momenta ($\pt > 1$~\gevc)
where it is also observed in \pp and explained by virtual direct
photons. The \pp measurement serves as an important test to pQCD
calculations of direct photon production from hard scattering
processes in this momentum range. An excess with an inverse slope of
$T_{\rm eff} = 221 \pm 23 {\rm (stat.)} \pm 18{\rm (syst.)}$~MeV is
observed in central \AuAu collisions above the binary scaled direct
photon yield in \pp. This can be qualitatively explained by
hydrodynamical models including thermal photon radiation with initial
temperatures of $300 \leq T_{\rm init} \leq 600$~MeV and formation
times of $0.12 \leq \tau_0 \leq 0.6$~fm/$c$.

\end{abstract}

\begin{dedication}
  To my family.
\end{dedication}

\tableofcontents
\clearpage
\listoffigures
\clearpage
\listoftables

\begin{acknowledgements}
    I would like to gratefully acknowledge my advisor Axel Drees for his
support and guidance over the past years. He introduced me to this
exciting field of physics and the PHENIX experiment when I came to
Stony Brook as an exchange student from W\"{u}rzburg. Axel was always
helpful and encouraging and has been an excellent advisor and a good
friend.

I would like to extend my thanks to Alberica Toia with whom I had the
pleasure to work closely for over three years. Working with her was
inspirational and fun. In her I found another strong supporter of my
work.

It has been a wonderful experience to work within the PHENIX
collaboration. In particular, I am grateful to Yasuyuki Akiba at BNL
for his help and inspiring discussions. I thank Barbara Jacak, Tom
Hemmick, Ralf Averbeck and everybody else in the Relativistic Heavy
Ion Group at Stony Brook. I also want to thank all my fellow graduate
students and friends that provided welcome distraction from work and
made life in Stony Brook enjoyable.

Last, I would like to thank my parents for their unconditional love,
support, encouragement, and trust. It is impossible to put into words
what I owe to them.

\end{acknowledgements}

\pagestyle{thesis}


\newpage
\pagenumbering{arabic}

\chapter{Introduction}
\label{cha:introduction}
Relativistic heavy ion collisions, \ie, collisions in which the
projectile energies are much larger than their rest masses, have been
an excellent tool to study nuclear matter under extreme conditions and
explore the phase diagram of strongly interacting matter for new
states of matter. At the Relativistic Heavy Ion Collider (RHIC) at
Brookhaven National Laboratory (BNL) gold ions are brought to
collisions at center of mass energies of up to \sqrtsnn = 200
GeV. Before RHIC, the Alternating Gradient Synchrotron (AGS) at BNL
and the Super Proton Synchrotron (SPS) have been colliding heavy ions
at energies up to \sqrtsnn $\approx 5$~GeV (AGS) and \sqrtsnn $\approx
17$~GeV and therefore studying nuclear matter at lower temperatures
and higher net baryon densities. With even lower energies the heavy
ion synchrotron ``SchwerIonen Synchrotron'' (SIS) provides heavy ion
collisions at beam energies up to 2~$A$GeV.  The Large Hadron Collider
(LHC) will achieve even higher temperatures in lead on lead collisions
at \sqrtsnn = 5.5 TeV.

At energy densities of $\varepsilon \approx 1$~GeV/fm$^3$ a phase
transition from hadronic matter to deconfined quarks and gluons, the
so called quark-gluon plasma (QGP), is
predicted~\cite{Shuryak:1978ij}. Calculations of lattice quantum
chromodynamics (lQCD) predict that such an energy density is reached
at a temperature of $T \approx 170~{\rm MeV}\footnote{Natural units
  $\hbar = c = k_B = 1$ are employed throughout this thesis except
  where noted otherwise} \approx 10^{12}$~K~\cite{Karsch:2001cy}. This
state may have existed in the early universe a few tens of
microseconds after the big-bang and may still exist in the core of
neutron stars at densities of 10$^{18}$~kg/m$^3$ = 0.6~fm$^{-3}$ which
exceeds normal nuclear matter density by more than a factor of
four. The same calculations also show that, at vanishing net baryon
density, with the transition into the deconfined phase, chiral
symmetry, which is spontaneously broken in vacuum due to non-vanishing
effective quark masses, is restored~\cite{Karsch:2001cy}.

\begin{figure}
  \centering
  \includegraphics[width=\textwidth]{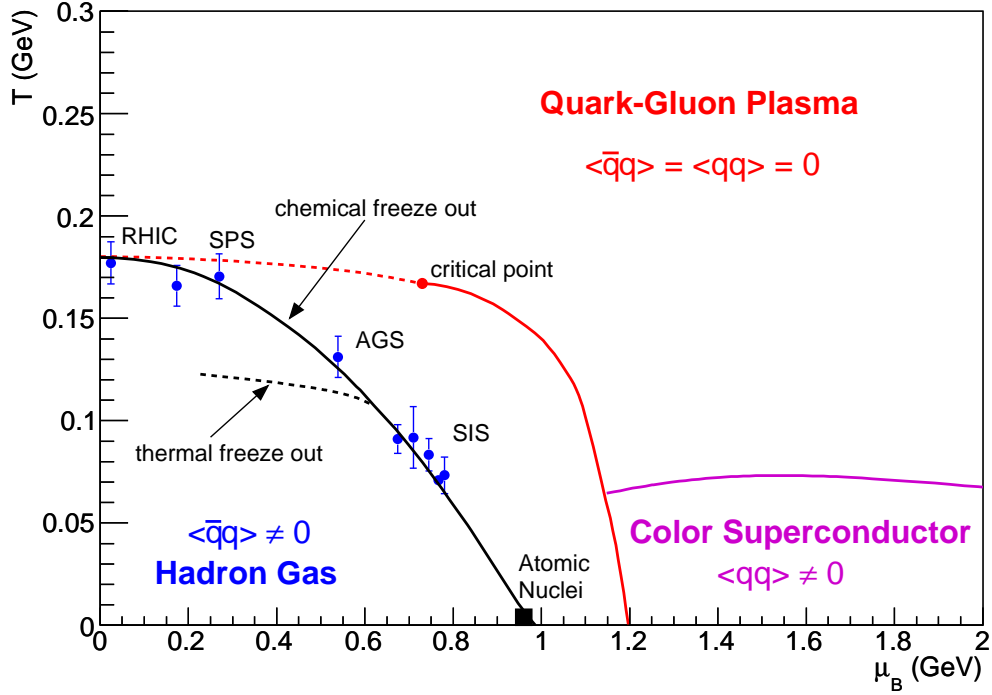}
  \caption[QCD Phase Diagram]{Schematic phase diagram of QCD matter as
    function of temperature $T$ and baryonic chemical potential
    $\mu_B$. The measured chemical freeze out points for SIS, AGS,
    SPS, and RHIC energies are shown as
    points~\cite{BraunMunzinger:2003zd}. The {\em dashed} line denotes
    the thermal freeze out. The existence and exact location of a
    critical point are unknown.}
  \label{fig:qcd_phase_diagram}
\end{figure}

In \fig{fig:qcd_phase_diagram} a schematic phase diagram of QCD
matter~\cite{Rapp:1999ej,BraunMunzinger:1998cg,BraunMunzinger:2003zd}
is shown as function of temperature $T$ and baryon chemical potential
$\mu_B$. Below the phase boundary quarks and gluons are confined to
hadrons which form a interacting hadron gas. Above the phase
transition, matter of asymptotically free quarks and gluons with
partonic degrees of freedom is created. The exact details such as the
order of the phase transition are unknown. Lattice QCD calculations
indicate a cross over at vanishing net baryon density, while at large
net baryon densities a first order phase transition is expected. If
such a scenario is realized, a critical point, below which the
transition becomes a cross over, would terminate the first order phase
transition.

Also shown are for different collision energies, the points at which
the fireball created in the collision, reaches its chemical freeze
out, \ie the end of inelastic collisions and therefore the production
of new particles has stopped. The thermal freeze out refers to the end
of elastic collision, \ie the end of momentum transfer between
particles, as the mean free path of the particles has become larger
than the size of the system. The experimental points of chemical
freeze out are calculated based on the final multiplicities of
particle species~\cite{BraunMunzinger:1995bp,BraunMunzinger:2001ip}.

Experimental results from all four RHIC experiments (BRAHMS, PHENIX,
PHOBOS, and STAR) give clear evidence that a new state of matter has
been created in \AuAu collisions at \sqrtsnn =
200~GeV~\cite{Arsene:2004fa,Adcox:2004mh,Back:2004je,Adams:2005dq}.
These results include that a very high energy density $\langle
\varepsilon \rangle \approx 15$~GeV/fm$^3$ is achieved, as indicated
by the large energy loss of light
hadrons~\cite{adcox:022301,adler:202301}, and heavy
quarks~\cite{adler:032301,adare:172301,adare:252002}. Furthermore, a
large elliptic flow is developed on a partonic level including heavy
quarks which is evidence for an early thermalization of the
medium\cite{adler:182301,adare:162301,adler:032301,afanasiev:052301}.

Colorless, electromagnetic probes such as photons and dileptons
(electron-positron or muon pairs) are particular useful probes to
study the the quark-gluon plasma, as they, once emitted, do not
undergo strong final state interactions with the hadronic
medium. Dileptons are created via various processes during all stages
of the collision. There is an approximate time ordering in the
invariant mass of the lepton pair; the earlier produced the larger its
mass. In the earliest stage, Drell-Yan annihilation between incoming
$q\overline{q}$ pairs creates dileptons dominating the invariant mass
region well above 3~\gevcc. As the formed hot and dense medium rapidly
thermalizes, a significant contribution of dileptons from
$q\overline{q}$ annihilation is expected, but in contrast to the DY
pairs, these pairs should follow a thermal distribution in the
intermediate mass region (IMR) $1< m_{ll} < 3$~\gevcc. The major
background source for the thermal radiation in the IMR comes from
semi-leptonic decays of open charm. Pairs of $c\overline{c}$ quarks
created in the initial hard scattering independently hadronize to
$D\overline{D}$ mesons, which consist of one charm and one light
quark, therefore often referred to as ``open charm'' mesons. These
mesons inherit the strong initial back-to-back correlation of the
charm quarks. Their individual weak decay $D \rightarrow K l \nu_l$
leads to a continuum of lepton pairs dominating the IMR. Medium
modifications of charm quarks, which leads to a significant flow and
suppression at large transverse momenta of single electrons from open
charm decays~\cite{adare:172301}, may also be reflected in the
invariant mass distribution of dileptons from open charm decays. After
hadronization the main contribution is expected from annihilation of
pions and other hadrons. Of particular interest are the two body
annihilation of pions through the light vector mesons $\rho$,
$\omega$, and $\phi$, as they decay directly into lepton pairs, whose
invariant mass reflects the mass of the vector meson at the time of
the decay. With the short life time of the $\rho$ of $\tau_{\rho} =
1.3$~fm and a life time of the hadronic medium of $\approx 10$~fm,
this channel provides an excellent probe to the effects of chiral
symmetry restoration~\cite{Pisarski:1981mq}. Once the chemical freeze
out is reached, the hadronic resonances and Dalitz decays of light
pseudo-scalar mesons as \pion, $\eta$ contribute to the low mass
region (LMR) below 1~\gevcc.

\begin{figure}
  \centering
  \includegraphics[width=\textwidth]{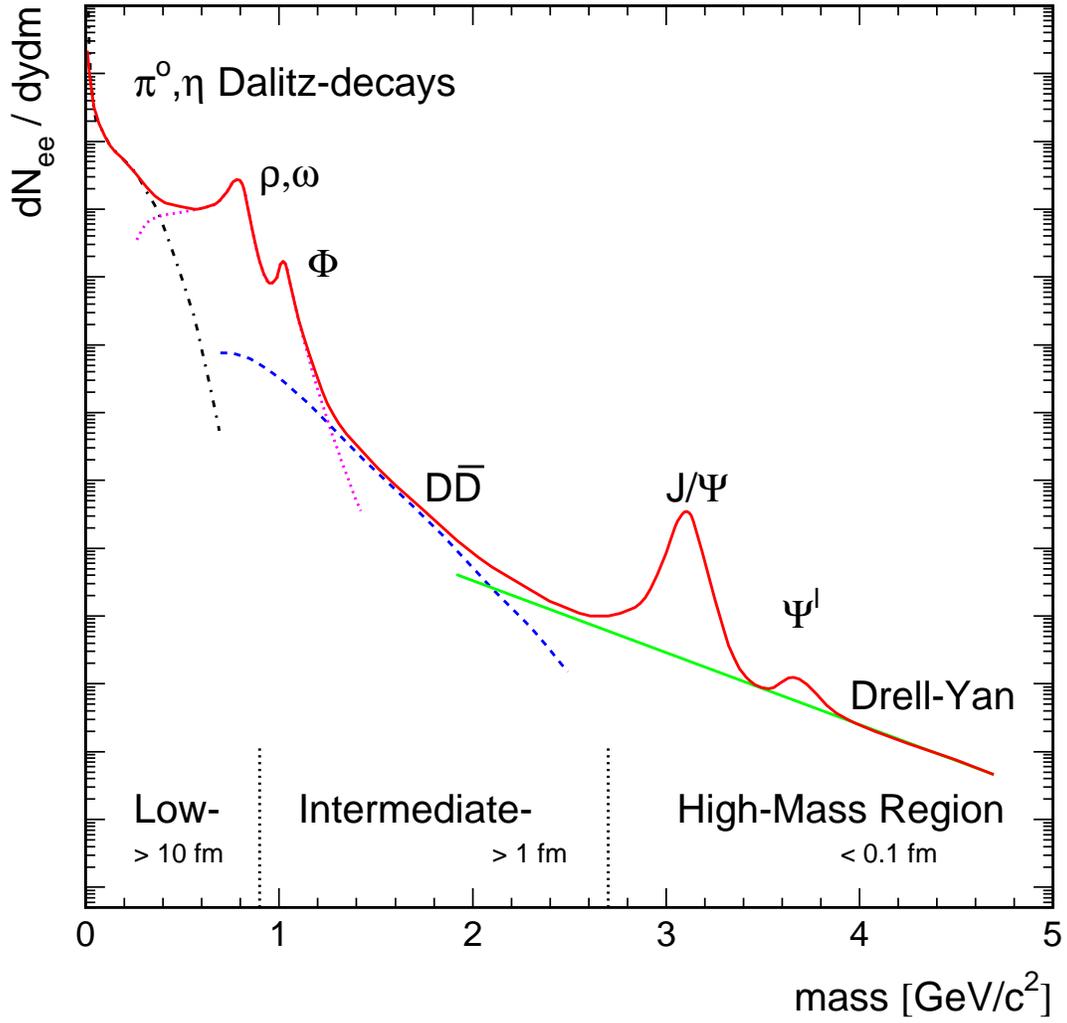}
  \caption[Expected sources of dilepton production in heavy ion
  collisions]{Expected sources of dilepton production as function of
    invariant mass in ultra-relativistic heavy ion collisions.}
  \label{fig:dilepton_scheme}
\end{figure}

\begin{figure}
  \centering
  \includegraphics[width=\textwidth]{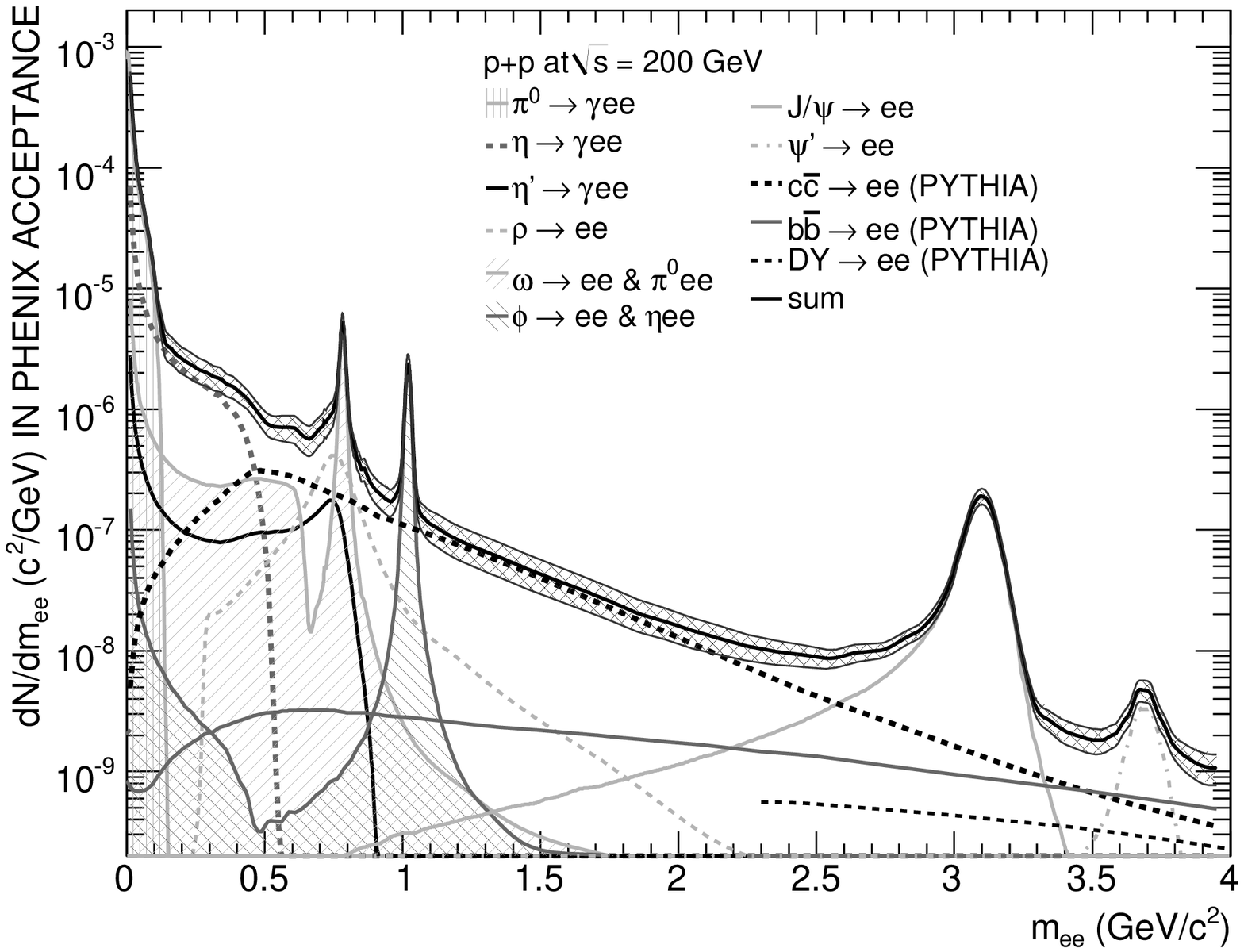}
  \caption[Expected sources of dilepton production in \pp
  collisions]{Expected sources of dilepton production as function of
    invariant mass in \pp collisions at \sqrts = 200 GeV.}
  \label{fig:cocktail}
\end{figure}

A schematic view of all contributions is shown in
\fig{fig:dilepton_scheme} and is supplemented with a realistic
``cocktail'' of hadron decays as expected to be measured with PHENIX
in \pp collisions shown in \fig{fig:cocktail}. This cocktail includes
the effects of detector resolution and acceptance.

This thesis is about the analysis of the dielectron continuum in the
low and intermediate mass region in \pp collisions at \sqrts = 200 GeV
recorded in 2005 with the PHENIX detector and the comparison to the
dielectron continuum measured in \AuAu collisions taken in 2004. As
part of this effort the cross sections of the $\omega$ and $\phi$ in
\pp collisions mesons have been measured. The \ee pairs in the
intermediate mass region allowed the extraction of the total charm
cross section in \pp~\cite{adare:2008asa}, which is in very good
agreement with previous measurements via single
electrons~\cite{adare:252002}. The high mass region (HMR) $\mee >
3$~GeV gave access to the first measurement of the bottom cross
section in \pp collisions at \sqrts = 200~GeV.

In the continuum of the low mass region a contribution from internal
conversions of direct virtual photons has been extracted and provided
an important reference for the measurement of direct virtual photons
in \AuAu collisions~\cite{Adare:2008fq}. Furthermore, the measurement
of the full dielectron continuum over a wide range in mass ($0 < \mee
< 8$~\gevcc) and \pt ($0< \pt < 5$\gevc) in \pp provides an important
baseline for the interpretation of the \AuAu
result~\cite{afanasiev:2007xw}.

This thesis is structured in the following way: In
Chapter~\ref{cha:introduction} an introduction into the physics of the
dielectron continuum is given. After a brief discussion of the
theoretical background in Section~\ref{sec:qcd} and the different
dilepton sources~\ref{sec:dileptons} an overview over recent results
at other dilepton experiments is given in
Section~\ref{sec:exp-results}. In Chapter~\ref{cha:phenix-detector} a
description of the PHENIX detector is given. The analysis details are
presented in Chapter~\ref{cha:analysis}. The various results of this
analysis are shown and discussed in Chapter~\ref{cha:results}. The
thesis ends with a summary and an outlook for further analyses of the
dielectron continuum in Chapter~\ref{cha:summary}.

\section{Quantum Chromodynamics}
\label{sec:qcd}

\begin{figure}
  \centering
  \begin{fmffile}{qeda}
    \begin{fmfgraph*}(35,20)
      \fmfpen{thin}
      \fmftop{i2,t1,t2,o2}
      \fmfbottom{i1,b1,b2,o1}
      \fmfset{thin}{0.5pt}
      \fmf{plain_arrow, label=$t$, label.side=right}{b1,b2}
      \fmfset{thin}{1pt}
      \fmf{fermion}{i1,v1,o1}
      \fmf{fermion}{o2,v2,i2}
      \fmf{photon, tension=0.6, label=$\gamma$}{v1,v2}
      \fmflabel{$e^-$}{i1}
      \fmflabel{$e^+$}{i2}
      \fmflabel{$e^-$}{o1}
      \fmflabel{$e^+$}{o2}
    \end{fmfgraph*}
  \end{fmffile}
  \hspace{10\unitlength}
  \begin{fmffile}{qedc}
    \begin{fmfgraph*}(35,20)
      \fmfpen{thin}
      \fmfleft{i1,i2}
      \fmfright{o1,o2}
      \fmf{fermion}{i1,v1,v2,o2}
      \fmf{photon}{o1,v1}
      \fmf{photon}{v2,i2}
      \fmflabel{$e^-$}{i1}
      \fmflabel{$e^-$}{o2}
      \fmflabel{$\gamma$}{o1}
      \fmflabel{$\gamma$}{i2}
    \end{fmfgraph*}
  \end{fmffile}
  \\\vspace{10\unitlength}
  \caption[Feynman diagram of the electromagnetic
  interaction]{Examples of Feynman diagrams for the interaction
    between two charged particles, and the Compton scattering of a
    photon on an electron. Time is advancing from the left to the
    right.}
  \label{fig:qed}
\end{figure}
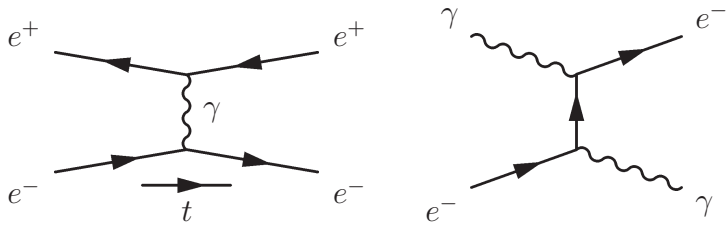

The theory of quantum electrodynamics (QED) is probably the best
understood and tested theory in the field of physics. On very short
time scales, which can be estimated by the uncertainty principle
($\Delta E\Delta t\approx\hbar$), electrons are allowed to emit and
absorb a virtual photon with non-zero rest mass. The electromagnetic
interaction between two charged particles is described by the exchange
of such virtual photons. As the photon is charge neutral it does not
interact with itself. The strength of the electromagnetic interaction
is given by the coupling constant $g_{e} = \sqrt{4 \pi \alpha}$ which
depends on the fine structure constant $\alpha = e^2/(\hbar c 4 \pi
\varepsilon_0) \approx 1/137$. QED predicts a logarithmic increase of
the interaction strength with increasing momentum transfers. The
interaction between two charged particles, annihilation or scattering,
can be represented by the Feynman diagram shown in \fig{fig:qed}. Also
shown are the two diagrams for the Compton scattering of a photon from
an electron.

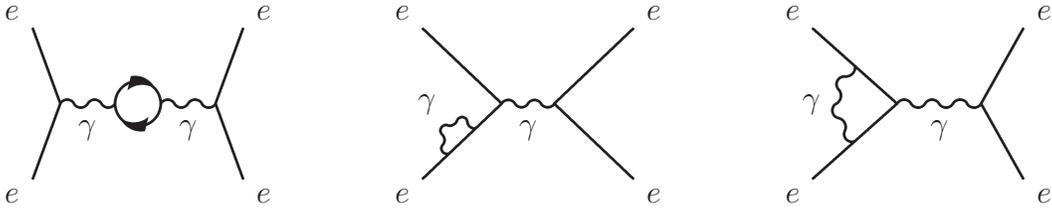
\begin{figure}
  \centering
  \begin{fmffile}{qede}
    \begin{fmfgraph*}(35,20)
      \fmfpen{thin}
      \fmfleft{i1,i2}
      \fmfright{o1,o2}
      \fmf{plain}{i1,v1,i2}
      \fmf{plain}{o2,v4,o1}
      \fmf{photon, label=$\gamma$}{v1,v2}
      \fmf{photon, label=$\gamma$}{v3,v4}
      \fmf{fermion,left, tension=0.6}{v2,v3,v2}
      \fmflabel{$e$}{i1}
      \fmflabel{$e$}{i2}
      \fmflabel{$e$}{o1}
      \fmflabel{$e$}{o2}
    \end{fmfgraph*}
  \end{fmffile}
  \hspace{10\unitlength}
  \begin{fmffile}{qedf}
    \begin{fmfgraph*}(35,20)
      \fmfpen{thin}
      \fmfleft{i1,i2}
      \fmfright{o1,o2}
      \fmf{phantom}{i1,v1,v2,v3,v4,v5,i2}
      \fmf{phantom}{o1,v6,v7,v8,v9,v10,o2}
      \fmf{photon, label=$\gamma$}{v3,v8}
      \fmffreeze
      \fmf{photon, left=1.2, tension=0.1, label=$\gamma$}{v1,v2}
      \fmf{plain}{i1,v1,v2,v3,v4,v5,i2}
      \fmf{plain}{o1,,v6,v7,v8,v9,v10,o2}
      \fmflabel{$e$}{i1}
      \fmflabel{$e$}{i2}
      \fmflabel{$e$}{o1}
      \fmflabel{$e$}{o2}
    \end{fmfgraph*}
  \end{fmffile}
  \hspace{10\unitlength}
  \begin{fmffile}{qedg}
    \begin{fmfgraph*}(35,20)
      \fmfpen{thin}
      \fmfleft{i1,i2}
      \fmfright{o1,o2}
      \fmf{phantom}{i1,v1,v2,v3,i2}
      \fmf{photon, label=$\gamma$}{v2,v4}
      \fmf{plain}{o2,v4,o1}
      \fmffreeze
      \fmf{photon, left=0.5, tension=0.3, label=$\gamma$}{v1,v3}
      \fmf{plain}{i1,v1,v2,v3,i2}
      \fmflabel{$e$}{i1}
      \fmflabel{$e$}{i2}
      \fmflabel{$e$}{o1}
      \fmflabel{$e$}{o2}
    \end{fmfgraph*}
  \end{fmffile}
  \\\vspace{10\unitlength}
  \caption{Higher order processes of the electromagnetic interaction}
  \label{fig:qed_corr}
\end{figure}

Higher order processes are responsible for changes in the strength of
the electromagnetic interaction. As shown in \fig{fig:qed_corr} the
virtual photon exchanged between two charges can create an \ee pair
which subsequently annihilate each other. This virtual \ee pair acts
as an electric dipole, effectively screening the two interacting
charges. Such vacuum polarization at short distances (large momenta)
reduces the effective charge of the electron as well as the fine
structure constant:
\begin{equation}\label{eq:alpha_em}
  \alpha(|q^2|) = \left(\frac{1}{\alpha(0)} - \frac{1}{3\pi}\ln\frac{|q^2|}{m^2}\right)^{-1}
\end{equation}
$|q^2|$ is the square of the momentum transfer, $m$ the mass of the
charged particle and \eq{eq:alpha_em} an approximation for large
$|q^2|/m$.  Other higher order processes are responsible for the
electron's self energy and its anomalous magnetic moment $(g_s - 2)
\neq 1$ and are shown in \fig{fig:qed_corr}.

\begin{figure}
  \centering
  \begin{fmffile}{qcda}
    \begin{fmfgraph*}(35,20)
      \fmfset{curly_len}{2mm}
      \fmfpen{thin}
      \fmfleft{i1,i2}
      \fmfright{o1,o2}
      \fmf{fermion}{i1,v1,i2}
      \fmf{fermion}{o2,v2,o1}
      \fmf{gluon, label=$g$}{v1,v2}
      \fmflabel{$q$}{i1}
      \fmflabel{$\overline{q}$}{i2}
      \fmflabel{$q$}{o1}
      \fmflabel{$\overline{q}$}{o2}
    \end{fmfgraph*}
  \end{fmffile}
  \begin{fmffile}{qcdc}
    \begin{fmfgraph*}(35,20)
      \fmfset{curly_len}{2mm}
      \fmfpen{thin}
      \fmfleft{i1,i2}
      \fmfright{o1,o2}
      \fmf{plain}{i1,v1,v2,o2}
      \fmf{gluon}{o1,v2}
      \fmf{gluon}{v1,i2}
      \fmflabel{$q$}{i1}
      \fmflabel{$q$}{o2}
      \fmflabel{$g$}{o1}
      \fmflabel{$g$}{i2}
    \end{fmfgraph*}
  \end{fmffile}
  \\\vspace{10\unitlength}
  \caption[Feynman diagram for the strong interaction between
  quarks]{Exemplary Feynman diagram for the strong interaction between
    quarks}
  \label{fig:qcd}
\end{figure}
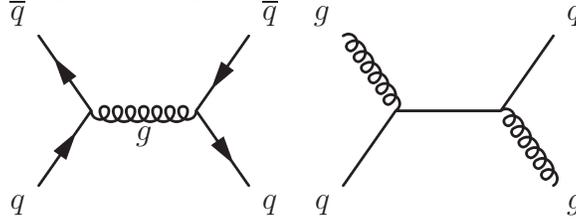

In an analog way, in the Standard Model quantum chromodynamics (QCD)
describes the strong interaction between quarks via the exchange of
color charges\footnote{The necessity of color charges arises
  experimentally from the existence of baryons with three quarks of
  identical flavor, \eg the $\Delta^{++}$ consists of three $u$
  quarks. The Pauli exclusion principle demands an extra quantum number
  to allow such a configuration, as any two fermions must not occupy
  the same state but must at least differ in one quantum number. The
  introduction of {\em color} charges ``red'', ``blue'' and ``green''
  lifts the degeneracy of the three quarks.} carried by
gluons. \fig{fig:qcd} shows the the lowest order Feynman diagrams for
quark-antiquark annihilation and quark-gluon scattering.

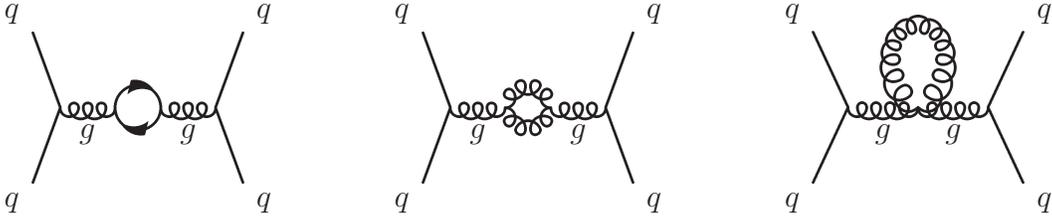
\begin{figure}
  \centering
  \begin{fmffile}{qcdd}
    \begin{fmfgraph*}(35,20)
      \fmfset{curly_len}{2mm}
      \fmfpen{thin}
      \fmfleft{i1,i2}
      \fmfright{o1,o2}
      \fmf{plain}{i1,v1,i2}
      \fmf{plain}{o2,v4,o1}
      \fmf{gluon, label=$g$}{v1,v2}
      \fmf{gluon, label=$g$}{v3,v4}
      \fmf{fermion, left, tension=0.6}{v2,v3,v2}
      \fmflabel{$q$}{i1}
      \fmflabel{$q$}{i2}
      \fmflabel{$q$}{o1}
      \fmflabel{$q$}{o2}
    \end{fmfgraph*}
  \end{fmffile}
  \hspace{10\unitlength}
  \begin{fmffile}{qcde}
    \begin{fmfgraph*}(35,20)
      \fmfset{curly_len}{2mm}
      \fmfpen{thin}
      \fmfleft{i1,i2}
      \fmfright{o1,o2}
      \fmf{plain}{i1,v1,i2}
      \fmf{plain}{o2,v4,o1}
      \fmf{gluon, label=$g$}{v1,v2}
      \fmf{gluon, label=$g$}{v3,v4}
      \fmf{gluon, left, tension=0.6}{v2,v3,v2}
      \fmflabel{$q$}{i1}
      \fmflabel{$q$}{i2}
      \fmflabel{$q$}{o1}
      \fmflabel{$q$}{o2}
    \end{fmfgraph*}
  \end{fmffile}
  \hspace{10\unitlength}
  \begin{fmffile}{qcdf}
    \begin{fmfgraph*}(35,20)
      \fmfset{curly_len}{2mm}
      \fmfpen{thin}
      \fmfleft{i1,i2}
      \fmfright{o1,o2}
      \fmf{plain}{i1,v1,i2}
      \fmf{plain}{o2,v3,o1}
      \fmf{gluon, label=$g$}{v1,v2}
      \fmf{gluon, label=$g$}{v2,v3}
      \fmf{gluon, right, tension=0.6}{v2,v2}
      \fmflabel{$q$}{i1}
      \fmflabel{$q$}{i2}
      \fmflabel{$q$}{o1}
      \fmflabel{$q$}{o2}
    \end{fmfgraph*}
  \end{fmffile}
  \\\vspace{10\unitlength}
  \caption{Higher order processes of the strong interaction}
  \label{fig:qcd_corr}
\end{figure}

As gluons themselves carry a non-zero color charge, they can interact
with themselves. This allows gluons not only to split into virtual
quark-antiquark pairs, but also to split in pairs of gluon as shown
in \fig{fig:qcd_corr}. This leads to an important difference from the
electromagnetic interaction. The coupling strength of the strong
interaction $\alpha_s$ increases with increasing distance of two
quarks.

The momentum transfer dependence of the coupling strength can be
written in analogy to \eq{eq:alpha_em} as:
\begin{equation}\label{eq:alpha_s}
  \alpha_{\rm s}(|q^2|) = \left(\frac{1}{\alpha_{\rm s}(\mu^2)} + \frac{1}{12\pi}(11 N_c - 2 N_f)\ln\frac{|q^2|}{\mu^2}\right)^{-1}
\end{equation}
for $(|q^2| \gg \mu^2)$ with the number of colors $N_c=3$ and the
number of flavors $N_f=6$ (in the Standard Model). As for large
distances or small momentum transfers the the coupling increases,
known as anti-screening, $\alpha_s$ cannot be expanded around $|q^2| =
0$, quite in contrast to $\alpha$.  With increasing momentum transfer
the coupling decreases leading to quasi-free quarks and gluons, known
as {\em asymptotic
  freedom}~\cite{PhysRevLett.30.1343,PhysRevLett.30.1346,PhysRevD.8.3633,PhysRept.14.129}. This
allows an expansion of $\alpha_{\rm s}$ around large momentum
transfers and therefore a perturbative treatment as long as
$\alpha_{\rm s} \leq 1$. However, for small momentum transfers on the
order of $q \simeq 1$~\gevc the perturbative treatment breaks down and
both the effective coupling and the relevant degrees of freedom change
rapidly with scale --- at large distances the degrees of freedom are
colorless objects of two (mesons) or three (baryons) confined
quarks. This regime of ``strong QCD'' poses the biggest challenge to
the theory of the strong interaction.

The QCD Lagrangian which describes the dynamics of the strong
interaction can be expressed as:
\begin{equation}\label{eq:qcd_lagrangian}
  {\cal L}_{\rm QCD} = \overline{\psi} (i \hbar \gamma^{\mu} {\bf D}_{\mu} -
  {\bf M}) \psi_ - \frac{1}{4}{\bf G}_{\mu \nu}^i {\bf G}^{\mu\nu\,i}
\end{equation}
where $\psi$ is a vector of Dirac spinors\footnote{Each quark can
  carry one of three colors, red, blue or green. Therefore the vector
  of Dirac spinors reads: $\psi = \begin{pmatrix} \psi_r\\ \psi_b\\
    \psi_g\end{pmatrix}$.}  representing the wave function of a
spin-1/2 quark-fields. $\gamma^{\mu}$ are the Dirac matrices and ${\bf
  M} = {\rm diag}(m_u, m_d, m_s, \ldots, m_t)$ is composed of the bare
quark masses $m_u \simeq 3$, $m_d \simeq 7$, $m_s \simeq 100$~\mevcc
and $m_c \simeq 1.25$, $m_b \simeq 4.1$, $m_t \simeq 175$~\gevcc.
${\bf D}_{\mu} = \partial_{\mu} + i g \frac{\lambda^i}{2} {\bf
  A}_{\mu}^i$ is the covariant derivative with the Gell-Mann matrices
$\lambda^i$ and the spin-1 gauge fields ${\bf A}_{\mu}^i$ with color
index $i = 1,\ldots8$. This term describes the interaction of quarks
with gluons, to which the eight gauge fields correspond to. The field
strength tensor
\begin{equation}\label{eq:gluonic_lagrangian}
  {\bf G}_{\mu \nu}^i = \partial_{\mu} {\bf A}_{\nu}^i - \partial_{\nu}
  {\bf A}_{\mu}^i + i g f_{ijk} {\bf A}_{\mu}^j {\bf A}_{\nu}^k
\end{equation}
describes the interaction of gluons with other gluons, where $g =
\sqrt{4 \pi \alpha_{\rm s}}$ is the coupling constant of the strong
interaction and $f_{ijk}$'s denote the structure constants of the
$SU(3)$ group~\cite{Itzykson:1980rh}.

\subsection{Chiral Symmetry}
\label{sec:chiral_symmetry}

The Lagrangian is invariant under local $SU(3)$ gauge transformations,
\ie, it is invariant under arbitrary rotations in color space. In
addition, the Lagrangian exhibits a global symmetry $U(1)$ that
corresponds to the baryon number conservation. In the limit of
vanishing quark masses (which for momentum transfers of $q \simeq
1$~\gevc is a good approximation for the light quarks $u$ and $d$, and
to a lesser extent also for the strange quark) the Lagrangian reveals
another symmetry under global vector and axial vector transformations
in the $SU(3)$ flavor space which are defined as:
\begin{equation}\label{eq:va_rotation}
  \psi \rightarrow \eexp{-i \alpha_V^i \frac{\lambda^i}{2}} \psi\quad\mbox{and}\quad  \psi \rightarrow \eexp{-i \alpha_A^i \frac{\lambda^i}{2}} \gamma_5 \psi.
\end{equation}
The conserved Noether currents associated with these symmetries are
\begin{equation}\label{eq:va_currents}
  j_{V,i}^{\mu} = \overline{\psi} \gamma^{\mu} \frac{\lambda^i}{2} \psi\quad\mbox{and}\quad  j_{A,i}^{\mu} = \overline{\psi} \gamma^{\mu} \frac{\lambda^i}{2} \gamma_5 \psi.
\end{equation}

The quark-spinors in \eq{eq:qcd_lagrangian} can be decomposed into a
left- and a right-handed component
\begin{equation}\label{eq:lr_spinors}
  \psi_{L,R} = \frac{1}{2} (1 \mp \gamma_5) \psi
\end{equation}
which transform under the rotations defined in \eq{eq:va_rotation} as:
\begin{align}\label{eq:lr_rotation}
  \psi_L &\rightarrow \eexp{-i \alpha_L^i \frac{\lambda^i}{2}}  \psi_L\quad\mbox{and}\quad  \psi_R \rightarrow \psi_R\\
  \psi_R &\rightarrow \eexp{-i \alpha_R^i \frac{\lambda^i}{2}}  \psi_R\quad\mbox{and}\quad  \psi_L \rightarrow \psi_L,
\end{align}
which constitutes a global $SU(3)_L \otimes SU(3)_R$ chiral symmetry in
flavor space. This symmetry conserves the handedness, \ie, the
projection of the spin on the momentum direction, of a
quark. Therefore the two kinds of quarks, left- and right-handed, do
not mix dynamically.

\subsubsection{Spontaneous Breaking of Chiral Symmetry}
\label{sec:chiral_ssb}

The observation of the mass splitting of chiral partners, \eg, the
$\rho$ and $a_1$ mesons have a mass difference of $\simeq 550$~\mevcc
($m_{\rho} = 776$~\mevcc and $m_{a_1} = 1230$~\mevcc~\cite{pdg}),
implies a spontaneous breaking of chiral symmetry due to a
non-vanishing vacuum expectation value of the quark condensate
$\langle \overline{\psi} \psi \rangle \neq 0$. While the vector
current $j_V = j_L + j_R$ is still conserved the axial-vector symmetry
($j_A = j_L - j_R$) is spontaneously broken, which means that the
axial-vector charge $Q_A^k = \int d^3x\, \psi^{\dag}
\frac{\lambda_k}{2} \gamma_5 \psi$ still commutes with the Hamiltonian
but the ground state has a non-zero expectation value: $Q_A^k
|0\rangle \neq 0$.  An often used analogy is a ferromagnet below the
Curie temperature, in which the ground state of aligned spins breaks
the rotational symmetry. The choice of a particular ground state leads
to a spontaneous breaking of chiral symmetry which according to the
Goldstone theorem~\cite{Goldstone:1961eq} results in the appearance of
eight massless Goldstone bosons ($\pi^{\pm}$, $\pi^0$, $K^{\pm}$,
$K^0$, $\overline{K^0}$, and $\eta$)

\begin{figure}
  \centering
  \includegraphics[width=0.9\textwidth]{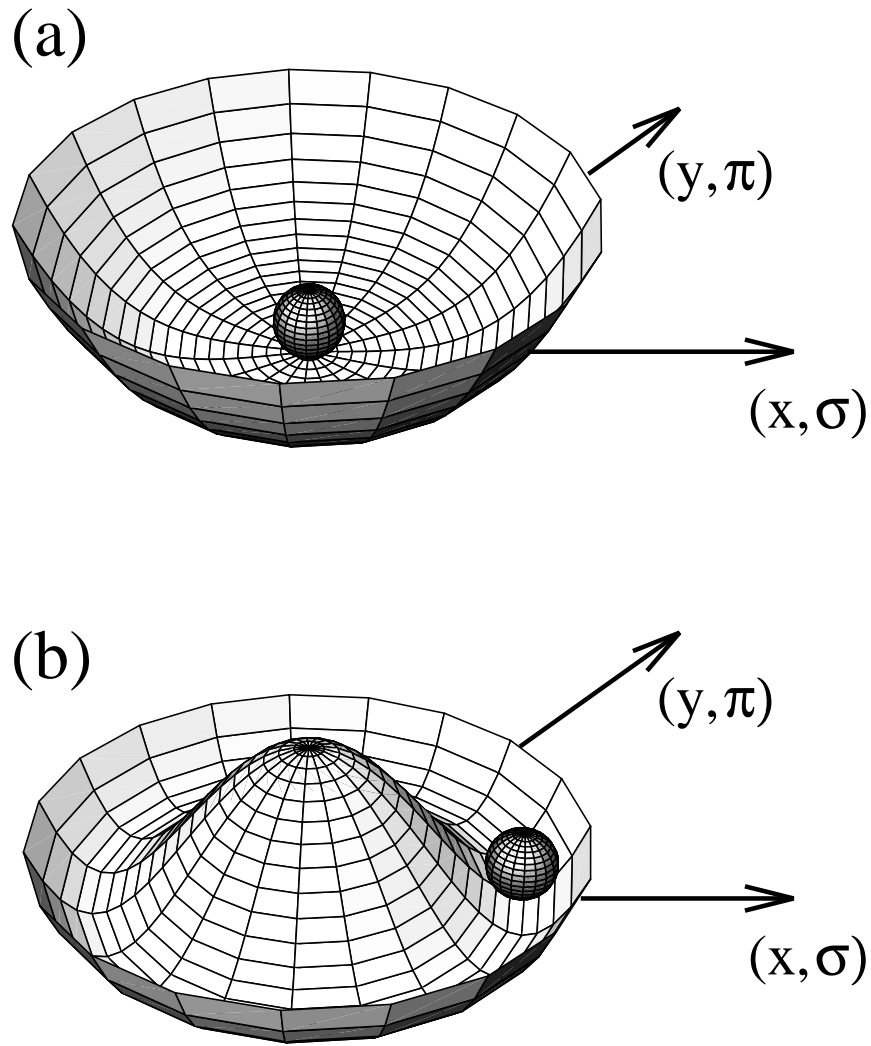}
  \caption[Mexican Hat Potential]{Mexican hat potential as defined in
    \eq{eq:mexican_hat}. The case $\mu^2 > 0$ is shown in (a) has a
    symmetric ground state at $\sigma = \pi = 0$. For $\mu^2 < 0$ an
    infinite number of degenerate ground states lie on the circle
    around $\sigma^2 + \pi^2 = \mu^2/\nu^2$, while $\sigma = \pi = 0$
    is a local maximum. The choice of a ground state breaks the
    symmetry spontaneously.}
  \label{fig:mexican_hat}
\end{figure}

Such a system can be visualized by the following potential
\begin{equation}\label{eq:mexican_hat}
  V = -\frac{1}{2}\mu^2(\sigma^{\dag}\sigma + \pi^{\dag}\pi)
  + \frac{1}{4}\nu^2(\sigma^{\dag}\sigma + \pi^{\dag}\pi)^2
\end{equation}
which is shown for $\mu^2>0$ and $\mu^2<0$ in
\fig{fig:mexican_hat}~\cite{Koch:1997ei}. For $\mu^2>0$ the potential
has a rotational symmetric ground state at $\sigma=\pi=0$. For
$\mu^2<0$, the state $\sigma=\pi=0$ is a local maximum and therefore
unstable. An infinite number of degenerate ground states lie on the
circle $\sigma^2 + \pi^2 = \mu^2/\nu^2$, which are not rotational
symmetric. Therefore, picking a ground state spontaneously breaks the
symmetry. However, effects of the symmetry are still present, as
excitations around the symmetry axis, \ie axial-vector rotations, do
not cost energy. With the identification of the fields $\vec{\pi}
\equiv i \overline{\psi}\vec{\lambda}\gamma_5\psi$ and $\sigma \equiv
\overline{\psi}\psi$ these rotations correspond to the massless
Goldstone bosons. In contrast, radial excitations along the $\sigma$
field do cost energy and correspond to massive particles.

The strength of the symmetry breaking is measured by the vacuum
expectation value of such Goldstone bosons which is, \eg, for pions:
\begin{equation}\label{eq:pion_strength}
  \langle0| j_{A,k}^{\mu}(x)|\pi_j(p)\rangle = i \delta_{jk} f_{\pi}
  p^{\mu} \eexp{-ipx}.
\end{equation}
with the measured pion decay constant $f_{\pi} = 93$~MeV, which is the
order parameter of the chiral symmetry. A second order parameter is
the vacuum expectation value or quark condensate
\begin{equation}\label{eq:chir_cond}
  \langle \overline{\psi}\psi \rangle = \langle 0| \overline{\psi}_L\psi_R +
  \overline{\psi}_R\psi_L | 0 \rangle = \langle 0| \overline{u}u + \overline{d}d|0 \rangle.
\end{equation}
The relation between quark condensate $\langle \overline{\psi}\psi\rangle$
and $f_{\pi}$ is given by the Gell-Mann-Oakes-Renner relation
(GOR)~\cite{PhysRev.175.2195}:
\begin{align}
  m_{\pi}^2 f_{\pi}^2 &= -\overline{m} \langle \overline{\psi} \psi \rangle\notag\\
\label{eq:gor}  &= -2 \overline{m} \langle \overline{q}q \rangle
\end{align}
where $\overline{m} = \frac{1}{2} (m_u + m_d)$ and $\langle \overline{q}q
\rangle = \langle \overline{u}u \rangle = \langle \overline{d}d \rangle$.

\subsubsection{Explicit Breaking of Chiral Symmetry}
\label{sec:chiral_esb}

The small but non-vanishing mass of the pseudoscalar mesons is
explained by the explicit breaking of the chiral symmetry due the
finite quark masses. 

\begin{figure}
  \centering
  \subfloat[Symmetric Potential]{\label{fig:ssb_potential}\includegraphics[width=0.44\textwidth]{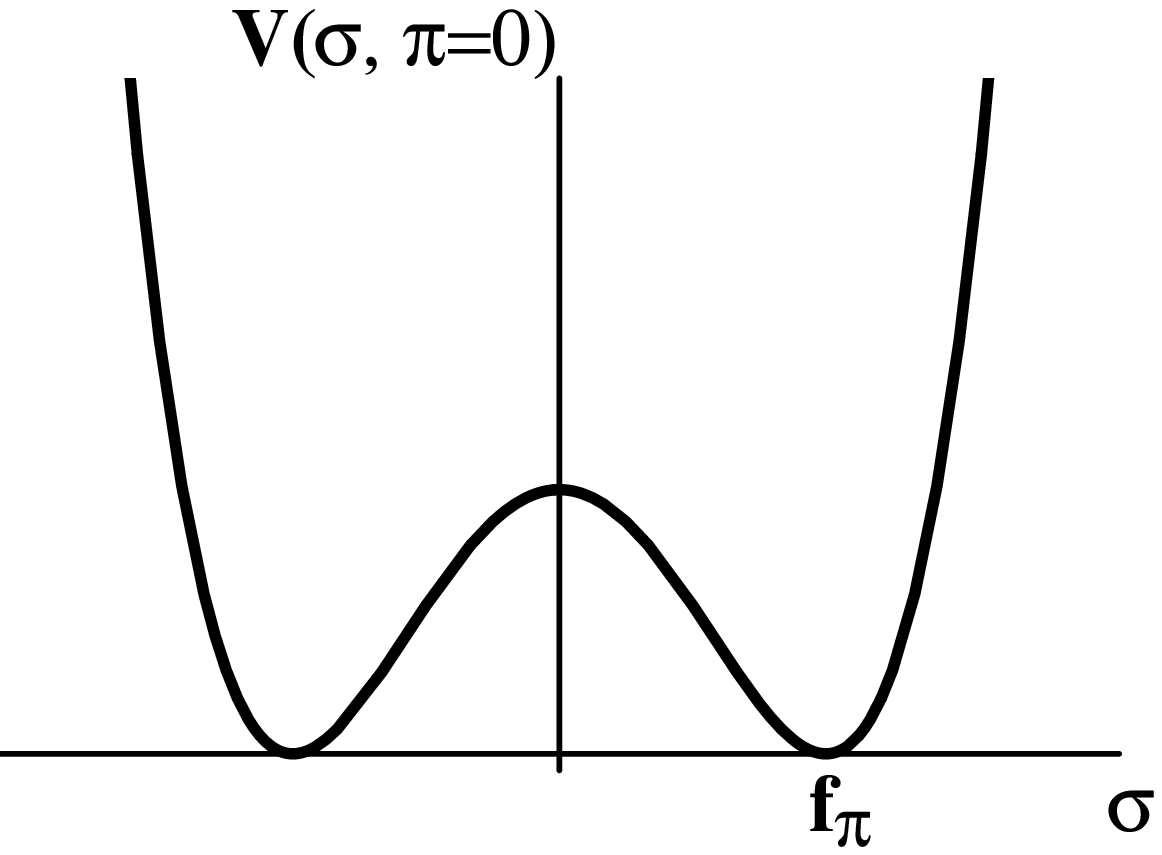}}
  \subfloat[Tilted Potential]{\label{fig:esb_potential}\includegraphics[width=0.44\textwidth]{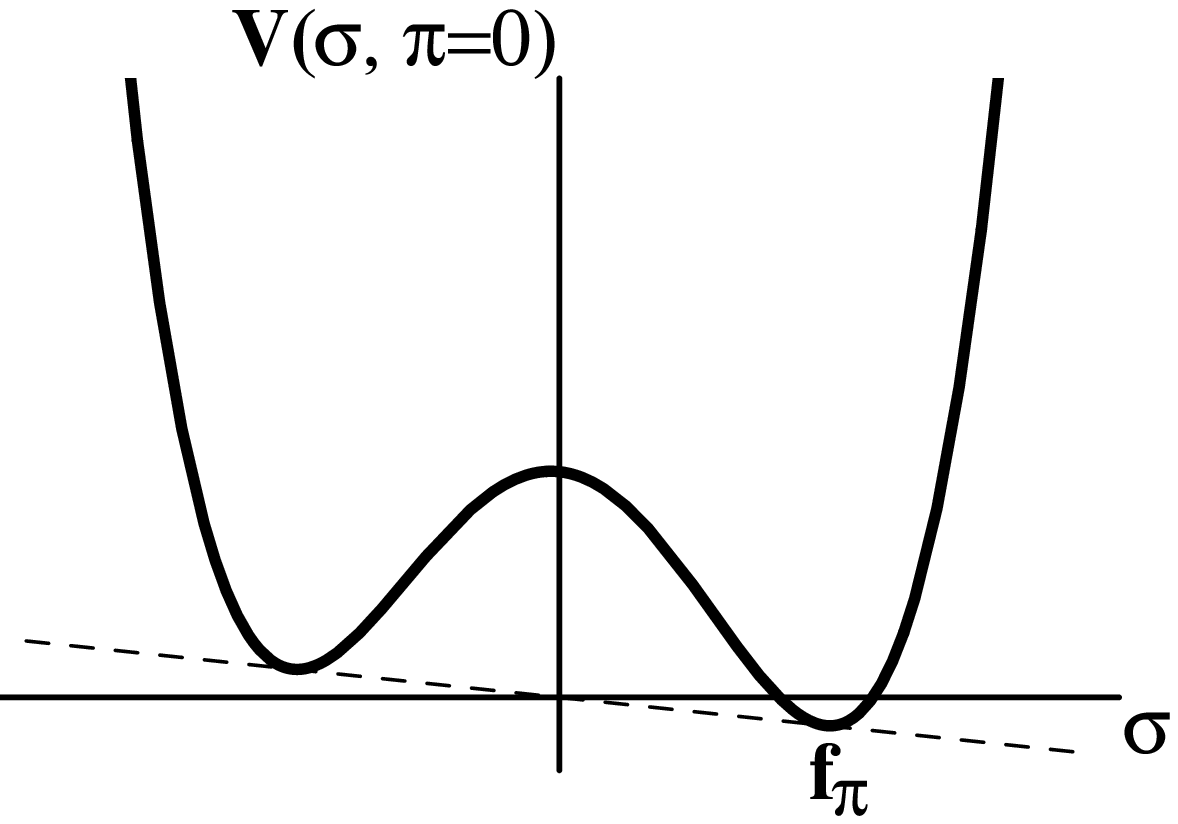}}
  \caption[Mexican Hat Potentials at $\pi=0$]{Mexican Hat Potential at
    $\pi = 0$. In case of \subref{fig:esb_potential} the potential is
    tilted along the $\sigma$ field breaking explicitly the chiral
    symmetry}
  \label{fig:sb_potential}
\end{figure}

The finite quark masses create a contribution $-\overline{m} \langle
\overline{\psi}\psi \rangle$ in the QCD Lagrangian in
\eq{eq:qcd_lagrangian}. This leads to a Mexican hat potential which is
slightly tilted towards the positive $\sigma$ direction which breaks
the symmetry of the potential:
\begin{equation}\label{eq:mexican_hat_tilt}
  V = -\frac{1}{2}\mu^2(\sigma^{\dag}\sigma + \pi^{\dag}\pi)
  + \frac{1}{4}\nu^2(\sigma^{\dag}\sigma + \pi^{\dag}\pi)^2 -
  f_{\pi} m_{\pi}^2 \sigma
\end{equation}
which is illustrated in \fig{fig:sb_potential}. With the choice of
$\mu^2/\nu^2 = f_{\pi}^2$ the tilting leads to a minimum at $\sigma =
f_{\pi}$ and axial-vector currents are no longer conserved, \ie the
pseudoscalar mesons, which were the Goldstone bosons of the
spontaneous symmetry breaking, acquire a finite mass which is related
to the current quark mass by \eq{eq:gor}. With $\overline{m} \approx
5$~\mevcc the vacuum value for the quark condensate becomes $\langle
\overline{q}q \rangle \approx -(254~{\rm MeV})^3$.

\subsubsection{In-Medium Quark Condensate}
\label{sec:inmedium_quarkcond}

So far only vacuum properties of hadronic matter have been
discussed. Of particular interest are the dynamics in the presence of
a hot and dense medium as it is expected to be created in relativistic
heavy-ion collisions. At high pressure or temperature hadronic matter
is expected to undergo a phase transition to deconfined quarks and
gluons accompanied by the melting of the quark condensate. Already
before reaching deconfinement, chiral symmetry is partially restored.

The expected modifications can be derived in the limit of low
temperatures $T$ and density starting from the grand canonical
partition function for a hadron gas in volume $V$ in contact with a
heat bath
\begin{equation}\label{eq:partition_function}
  {\cal Z}(V,T,\mu_q) = {\rm Tr}(\eexp{-({\bf H} - \mu_q {\bf N})/T})
\end{equation}
with the Hamiltonian of the system ${\bf H}$, the quark chemical
potential $\mu_q$, and the quark number generator ${\bf N}$.

\begin{figure}
  \centering
  \includegraphics[width=0.9\textwidth]{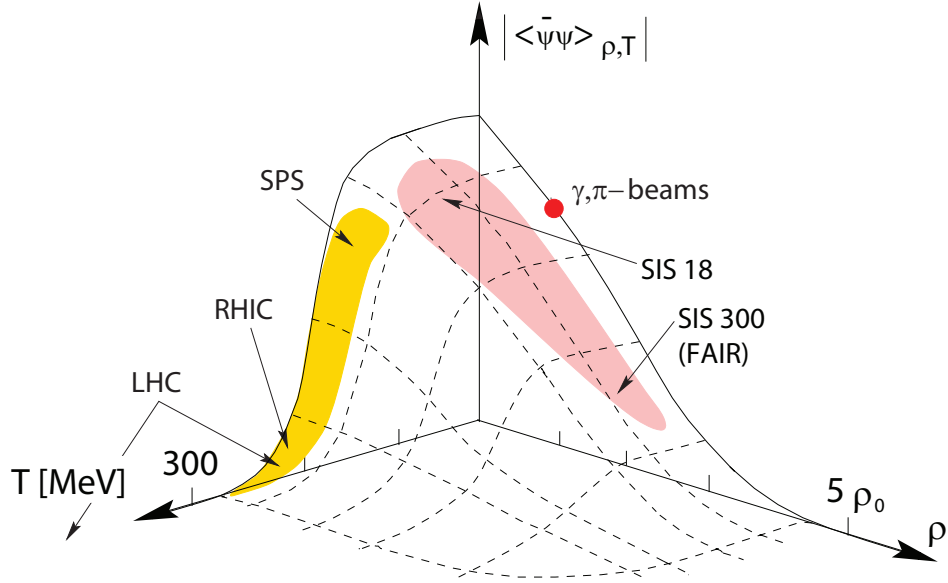}
  \caption[Expectation Value of the Chiral Condensate]{The expectation
    value of the chiral condensate as function of temperature $T$ and
    nuclear matter density $\rho$ as calculated with the
    Nambu-Jona-Lasinio model~\cite{Klimt:1990ws}. Figure based
    on~\cite{hering}.}
  \label{fig:qbarq}
\end{figure}

The expectation value of the quark condensate is given by the thermal
average
\begin{equation}\label{eq:thermal_quark_cond}
  \langle\!\langle \overline{q}q \rangle\!\rangle = {\cal Z}^{-1} \sum_n \langle
  n|\overline{\psi}\psi|n\rangle \eexp{-(E_n-\mu_q)/T},
\end{equation}
where the sum is carried out over all eigenstates of ${\bf H}$ with
the corresponding eigenvalues $E_n$. For a non-interacting hadron gas
the leading order result to \eq{eq:thermal_quark_cond} with respect to
the vacuum expectation value for low densities is:
\begin{align}\label{eq:qbarq_temperature}
  \frac{\langle\!\langle \overline{q}q \rangle\!\rangle}{\langle \overline{q}q
    \rangle} &\simeq 1 - \frac{\Sigma_{\pi}
    \rho_{\pi}^2(T)}{f_{\pi}^2m_{\pi}^2}\\
  &= 1 - \frac{(N_f^2-1)}{N_f}\frac{T^2}{12f_{\pi}^2} +
  \frac{(N_f^2-1)}{2N_f} \left(\frac{T^2}{12f_{\pi}^2}\right)^2\notag\\
  &\quad - N_f(N_f^2-1)\left(\frac{T^2}{12f_{\pi}^2}\right)^3
  \ln\left(\frac{\Lambda_q}{T}\right) + {\cal O}(T^8).
\end{align}
The scale $\Lambda_q = 470\pm110$~MeV is fixed from pion scattering
data. To first order the quark condensate at low densities decreases
quadratically with the temperature. Analog, the result for finite
nuclear densities and low temperatures is a linear decrease with
density:
\begin{equation}\label{eq:qbarq_density}
  \frac{\langle\!\langle \overline{q}q \rangle\!\rangle}{\langle \overline{q}q
    \rangle} \simeq 1 - \frac{\Sigma_{N}
    \rho_{N}^2(\mu_N)}{f_{\pi}^2m_{\pi}^2}
\end{equation}
with baryon density $\mu_N = 3 \mu_q$. At nuclear matter densities
($\rho_0 = 0.17~{\rm fm}^{-3}$) the quark condensate has dropped by
35\% with a experimentally determined nucleon $\sigma$-term of
$\Sigma_N \approx 45$~MeV.  Based on these calculations one can derive
a prediction of the scalar quark condensate $\langle \overline{q}q
\rangle$ as function of temperature and density, which is shown in
\fig{fig:qbarq}\cite{Klimt:1990ws}.

Based on connections between hadron masses and the quark condensate
various scenarios for the change of meson masses with decreasing
quark condensates are proposed. Brown and Rho predict in
~\cite{PhysRevLett.66.2720} a dropping of the $\rho$ mass by 15--20\%
at normal nuclear matter densities and a universal Brown-Rho (BR)
scaling of the in-medium vector meson masses with the in-medium pion
decay constant $f_{\pi}^{\ast}$:
\begin{equation}\label{eq:br_scaling}
  \frac{\langle\!\langle \overline{\psi}\psi \rangle\!\rangle}{\langle
    \overline{\psi}\psi \rangle} =
  \left(\frac{f_{\pi}^{\ast}}{f_{\pi}}\right)^3\quad \mbox{and}\quad \frac{f_{\pi}^{\ast}}{f_{\pi}} =
  \frac{m_{\sigma}^{\ast}}{m_{\sigma}} =
  \frac{m_{N}^{\ast}}{m_{N}} =
  \frac{m_{\rho}^{\ast}}{m_{\rho}} = \frac{m_{\omega}^{\ast}}{m_{\omega}}
\end{equation}
Other models explain medium modification of vector mesons by
interactions with surrounding hadrons in the hot and dense medium. For
an excellent review see, \eg, Ref.~\cite{Rapp:1999ej}.

\section{Dileptons}
\label{sec:dileptons}

Dileptons are produced during all stages of the collisions and carry a
variety of signals. In this Section the various sources are
discussed.

\subsection{Drell-Yan}
\label{sec:drell_yan}

\begin{figure}
  \centering
  \begin{fmffile}{dya}
    \begin{fmfgraph*}(70,40)
      \fmfset{curly_len}{2mm}
      \fmfpen{thin}
      \fmfleft{i1,i2}
      \fmfright{o1,o2,o3,o4}
      \fmf{fermion, tension=2}{i1,v1,o1}
      \fmf{fermion, tension=2}{i2,v2,o4}
      \fmf{fermion, label=$q$}{v1,v3}
      \fmf{fermion, label=$\overline{q}$}{v3,v2}
      \fmf{boson, label=$\gamma^{\ast}$}{v3,v4}
      \fmf{fermion}{o3,v4,o2}
      \fmfblob{0.15w}{v1}
      \fmfblob{0.15w}{v2}
      \fmflabel{$A$}{i1}
      \fmflabel{$B$}{i2}
      \fmflabel{$l^+$}{o3}
      \fmflabel{$l^-$}{o2}
      \fmffreeze
      \fmfi{plain}{vpath (__i1,__v1) shifted (thick*(0,2))}
      \fmfi{plain}{vpath (__i1,__v1) shifted (thick*(1,-2))}

      \fmfi{plain}{vpath (__i2,__v2) shifted (thick*(0,2))}
      \fmfi{plain}{vpath (__i2,__v2) shifted (thick*(1,-2))}

      \fmfi{plain}{vpath (__v2,__o4) shifted (thick*(1,-2))}
      \fmfi{plain}{vpath (__v1,__o1) shifted (thick*(0,2))}

    \end{fmfgraph*}
  \end{fmffile}
  \\\vspace{10\unitlength}
  \caption[Drell-Yan process]{Feynman diagram for the production of a
    dilepton pair via the Drell-Yan process.}
  \label{fig:DY}
\end{figure}
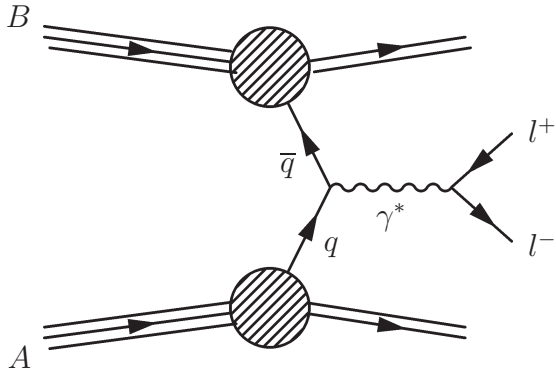

In a collision of two nuclei, the Drell-Yan process is the
annihilation of a quark in one nucleus with a sea antiquark from the
other nucleus into a virtual photon which subsequently converts into a
lepton pair. This process is shown in \fig{fig:DY} for two nuclei $A$
and $B$ and has particular important at large invariant masses
($m_{ll} > 3$~\gevcc). At high energies the invariant mass of the
lepton pair $m_{ll}$ is given by the product of the quark momenta $x_1
\sqrt{s}$ and $x_2 \sqrt{2}$:
\begin{equation}\label{eq:mass_DY}
  m_{ll}^2 = x_1 x_2 s
\end{equation}
with $\sqrt{s}$ the center of mass energy of the incoming nuclei. The
cross section of Drell-Yan dilepton production can be calculated to
leading order using parton distribution functions obtained from deep
inelastic lepton-nucleon collisions. An additional factor $K \approx
2$ accounts for higher order corrections in $\alpha_{\rm s}$ is
necessary to describe experimental dilepton data.

One can show that in collisions of two equal nuclei with mass number
$A$ (\eg, for gold: $A_{\rm Au} = 197$), the contribution of dileptons
from the Drell-Yan process scales with respect to the nucleon-nucleon
case as $A^{4/3}$~\cite{Wong:1995jf}. In a thermalized partonic medium
the same process, \ie, $q \overline{q} \rightarrow \ee$, is possible
between quarks and antiquarks with thermal momentum
distributions. Theoretical models predict such a contribution to be
dominant in the intermediate mass region.

\subsection{Open Charm and Bottom}
\label{sec:open_charm}

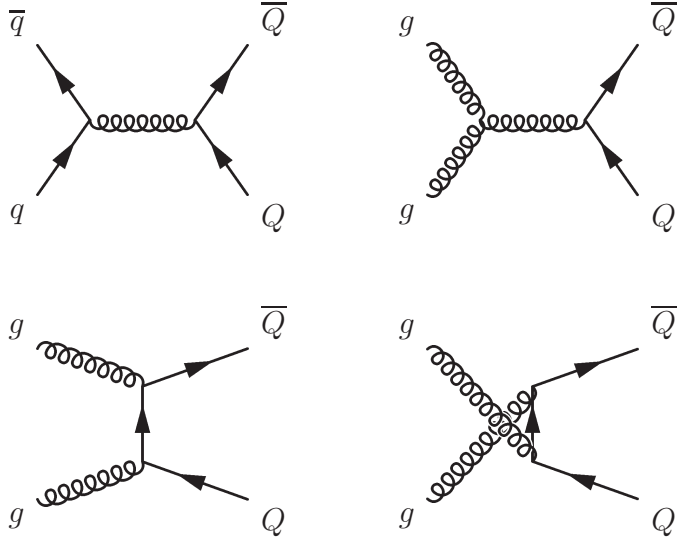
\begin{figure}
  \centering
  \begin{fmffile}{cca}
    \begin{fmfgraph*}(35,20)
      \fmfset{curly_len}{2mm}
      \fmfpen{thin}
      \fmfleft{i1,i2}
      \fmfright{o1,o2}
      \fmf{fermion}{i1,v1,i2}
      \fmf{fermion}{o1,v2,o2}
      \fmf{gluon}{v1,v2}
      \fmflabel{$q$}{i1}
      \fmflabel{$\overline{q}$}{i2}
      \fmflabel{$Q$}{o1}
      \fmflabel{$\overline{Q}$}{o2}
    \end{fmfgraph*}
  \end{fmffile}
  \hspace{10\unitlength}
  \begin{fmffile}{ccb}
    \begin{fmfgraph*}(35,20)
      \fmfset{curly_len}{2mm}
      \fmfpen{thin}
      \fmfleft{i1,i2}
      \fmfright{o1,o2}
      \fmf{gluon}{i1,v1,i2}
      \fmf{fermion}{o1,v2,o2}
      \fmf{gluon}{v1,v2}
      \fmflabel{$g$}{i1}
      \fmflabel{$g$}{i2}
      \fmflabel{$Q$}{o1}
      \fmflabel{$\overline{Q}$}{o2}
    \end{fmfgraph*}
  \end{fmffile}
  \\\vspace{20\unitlength}
  \begin{fmffile}{ccc}
    \begin{fmfgraph*}(35,20)
      \fmfset{curly_len}{2mm}
      \fmfpen{thin}
      \fmfleft{i1,i2}
      \fmfright{o1,o2}
      \fmf{gluon}{i1,v1}
      \fmf{fermion}{o1,v1,v2,o2}
      \fmf{gluon}{v2,i2}
      \fmflabel{$g$}{i1}
      \fmflabel{$g$}{i2}
      \fmflabel{$Q$}{o1}
      \fmflabel{$\overline{Q}$}{o2}
    \end{fmfgraph*}
  \end{fmffile}
  \hspace{10\unitlength}
  \begin{fmffile}{ccd}
    \begin{fmfgraph*}(35,20)
      \fmfset{curly_len}{2mm}
      \fmfpen{thin}
      \fmfleft{i1,i2}
      \fmfright{o1,o2}
      \fmf{phantom}{i1,v1}
      \fmf{phantom}{i2,v2}
      \fmf{gluon,tension=0}{i1,v2}
      \fmf{gluon,tension=0,rubout}{v1,i2}
      \fmf{fermion}{o1,v1,v2,o2}
      \fmflabel{$g$}{i1}
      \fmflabel{$g$}{i2}
      \fmflabel{$Q$}{o1}
      \fmflabel{$\overline{Q}$}{o2}
    \end{fmfgraph*}
  \end{fmffile}
  \\\vspace{10\unitlength}
  \caption[LO processes of heavy quark production]{Leading
    order processes of heavy quark production.}
  \label{fig:lo}
\end{figure}

Another contribution to the dilepton continuum are semi-leptonic
decays of charm and bottom mesons. Heavy quark pairs ($Q\overline{Q} =
c\overline{c}$ or $b\overline{b}$, respectively) are produced in
nucleon-nucleon collisions in inelastic hard-scattering processes
between constituent quarks of the two nucleons ($q\overline{q}
\rightarrow g^{\ast} \rightarrow Q\overline{Q}$). The lowest order
diagrams for heavy quark production are shown in \fig{fig:lo}. In
addition to the annihilation of two light quarks into a virtual gluon,
heavy quarks can be produced by interactions of a gluon of one nucleon
with a gluon in the other nucleon ($gg \rightarrow g^{\ast}
\rightarrow Q\overline{Q}$).

Open charm and bottom mesons are formed out of a heavy quark and a
light antiquark ($\overline{u}$, $\overline{d}$, or $\overline{s}$) or
a heavy antiquark and a light quark, \eg $D^+ = c\overline{d}$, $B^+ =
u\overline{b}$. After production of a heavy quark pair, the quarks
undergo fragmentation and can form a $D^+D^-$ pair. Due to the large
mass of the $c$ quark, the $D^+D^-$ pair preserves most of the
initial correlation of the $Q\overline{Q}$ pair. The charmed (or
bottom) mesons then can undergo weak decays into, \eg, $D^+
\rightarrow \overline{K}^0 l^{+} \nu_l$. The total branching ratio of
semi-leptonic decays of a $D$ meson is on the order of $10\%$.  The
semi-leptonic decays of both the $D^+$ and the $D^-$ results in the
creation of a dilepton pair.

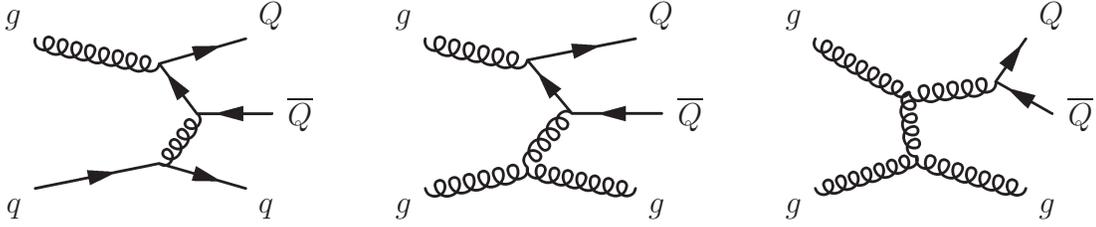
\begin{figure}
  \centering
  \begin{fmffile}{cce}
    \begin{fmfgraph*}(35,20)
      \fmfset{curly_len}{2mm}
      \fmfpen{thin}
      \fmfleft{i1,i2}
      \fmfright{o1,o2,o3}
      \fmf{fermion}{i1,v1,o1}
      \fmf{gluon}{i2,v3}
      \fmf{gluon}{v1,v2}
      \fmf{fermion}{o2,v2,v3,o3}
      \fmflabel{$q$}{i1}
      \fmflabel{$g$}{i2}
      \fmflabel{$q$}{o1}
      \fmflabel{$\overline{Q}$}{o2}
      \fmflabel{$Q$}{o3}
    \end{fmfgraph*}
  \end{fmffile}
  \hspace{10\unitlength}
  \begin{fmffile}{ccf}
    \begin{fmfgraph*}(35,20)
      \fmfset{curly_len}{2mm}
      \fmfpen{thin}
      \fmfleft{i1,i2}
      \fmfright{o1,o2,o3}
      \fmf{gluon, tension=1.5}{i1,v1}
      \fmf{gluon, tension=1.5}{i2,v3}
      \fmf{gluon}{v2,v1,o1}
      \fmf{fermion}{o2,v2,v3,o3}
      \fmflabel{$g$}{i1}
      \fmflabel{$g$}{i2}
      \fmflabel{$g$}{o1}
      \fmflabel{$\overline{Q}$}{o2}
      \fmflabel{$Q$}{o3}
    \end{fmfgraph*}
  \end{fmffile}
  \hspace{10\unitlength}
  \begin{fmffile}{ccg}
    \begin{fmfgraph*}(35,20)
      \fmfset{curly_len}{2mm}
      \fmfpen{thin}
      \fmfleft{i1,i2}
      \fmfright{o1,o2,o3}
      \fmf{gluon}{i1,v1,o1}
      \fmf{gluon}{i2,v2,v1}
      \fmf{gluon}{v2,v3}
      \fmf{fermion}{o2,v3,o3}
      \fmflabel{$g$}{i1}
      \fmflabel{$g$}{i2}
      \fmflabel{$g$}{o1}
      \fmflabel{$\overline{Q}$}{o2}
      \fmflabel{$Q$}{o3}
    \end{fmfgraph*}
  \end{fmffile}
  \\\vspace{10\unitlength}
  \caption[Examples of NLO processes of $Q\overline{Q}$
  production]{Examples of next-to-leading order processes contributing
    to the production of heavy quark pairs.}
  \label{fig:nlo}
\end{figure}

Perturbative QCD calculations to leading order are not able to fully
describe charm production in nucleon-nucleon collisions. Similar to
the Drell-Yan process, a $K$ factor has to be employed to correct for
the difference to the measured charm cross section. Higher order
calculations such as next-to-leading order (NLO), for which
exemplary processes are shown in \fig{fig:nlo}, and
fixed-order-plus-next-to-leading order (FONLL) are in agreement with
the $D$ meson cross section measured by CDF in $p\overline{p}$ collisions
at \sqrts = 1.96 TeV~\cite{PhysRevLett.91.241804} as well as with
measurements of single electrons~\cite{adare:252002} and
muons~\cite{adler:092002} from semi-leptonic charm decays performed by
PHENIX in \pp collisions at \sqrts = 200~GeV. However, the theoretical
uncertainties are large and the data prefer larger cross sections
within these uncertainties.

\begin{figure}
  \centering
  \includegraphics[width=0.9\textwidth]{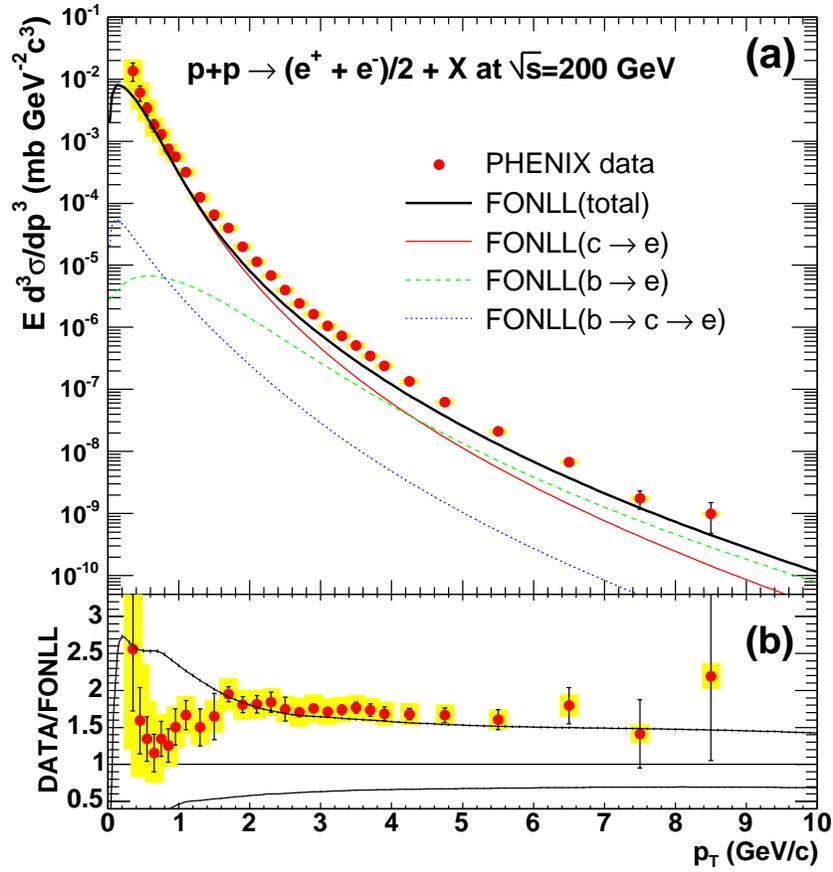}
  \caption[Differential cross section of electrons from heavy flavor
  decays in \pp collisions at \sqrts = 200GeV]{(a) Shown is the
    differential cross section of electrons from heavy flavor decays
    in \pp collisions at \sqrts = 200GeV~\cite{adare:252002}. The data
    are compared to the central value of a FONLL
    calculation~\cite{cacciari:122001}. (b) Ratio of data to FONLL
    calculation. The upper (lower) curve show the upper (lower)
    theoretical limit of the FONLL calculation.}
  \label{fig:ppg065_spectra}
\end{figure}

Shown in \fig{fig:ppg065_spectra} is the invariant differential cross
sections of electrons from heavy flavor decays~\cite{adare:252002} in
comparison to a FONLL calculation~\cite{cacciari:122001}. The shape of
the FONLL is in good agreement with the data, overall the data are a
factor of $\approx 1.7$ higher than the calculation, as visible in the
bottom panel of the same figure, which shows the ratio of data to
FONLL.

In the absence of medium modifications, the production of particles by
inelastic hard scattering processes in heavy ion collision is given by
the production cross section in \pp collisions scaled by the nuclear
overlap factor $T_{AB}$, which is the integral over the product of the
thickness functions of the two colliding nuclei $A$ and $B$ in the
geometric overlap region (or $T_{AA}$ in case of two identical nuclei
$A$). There is a simple relation between the nuclear overlap factor,
the inelastic \pp cross section and the average number of binary
nucleon-nucleon collisions occurring in a nucleus-nucleus collisions:
$T_{AA} = \langle N_{\rm coll} \rangle/\sigma_{pp}$.

In order to quantify the deviation of the yield measured in a
nucleus-nucleus collision from the binary scaled \pp
expectation the nuclear modification factor $R_{AA}$ is defined as:
\begin{equation}\label{eq:raa}
  R_{AA} = \frac{dN_{AA}}{T_{AA}\, d\sigma_{pp}} =
  \frac{dN_{AA}}{\langle N_{\rm coll}\rangle\, dN_{pp}}
\end{equation}
were $dN_{AA}$ ($dN_{pp}$) is the differential particle yield measured
in the nucleus-nucleus (\pp) collision, and $d\sigma_{pp} =
\sigma_{pp}\, dN_{pp}$.

\begin{figure}
  \centering
  \includegraphics[width=0.9\textwidth]{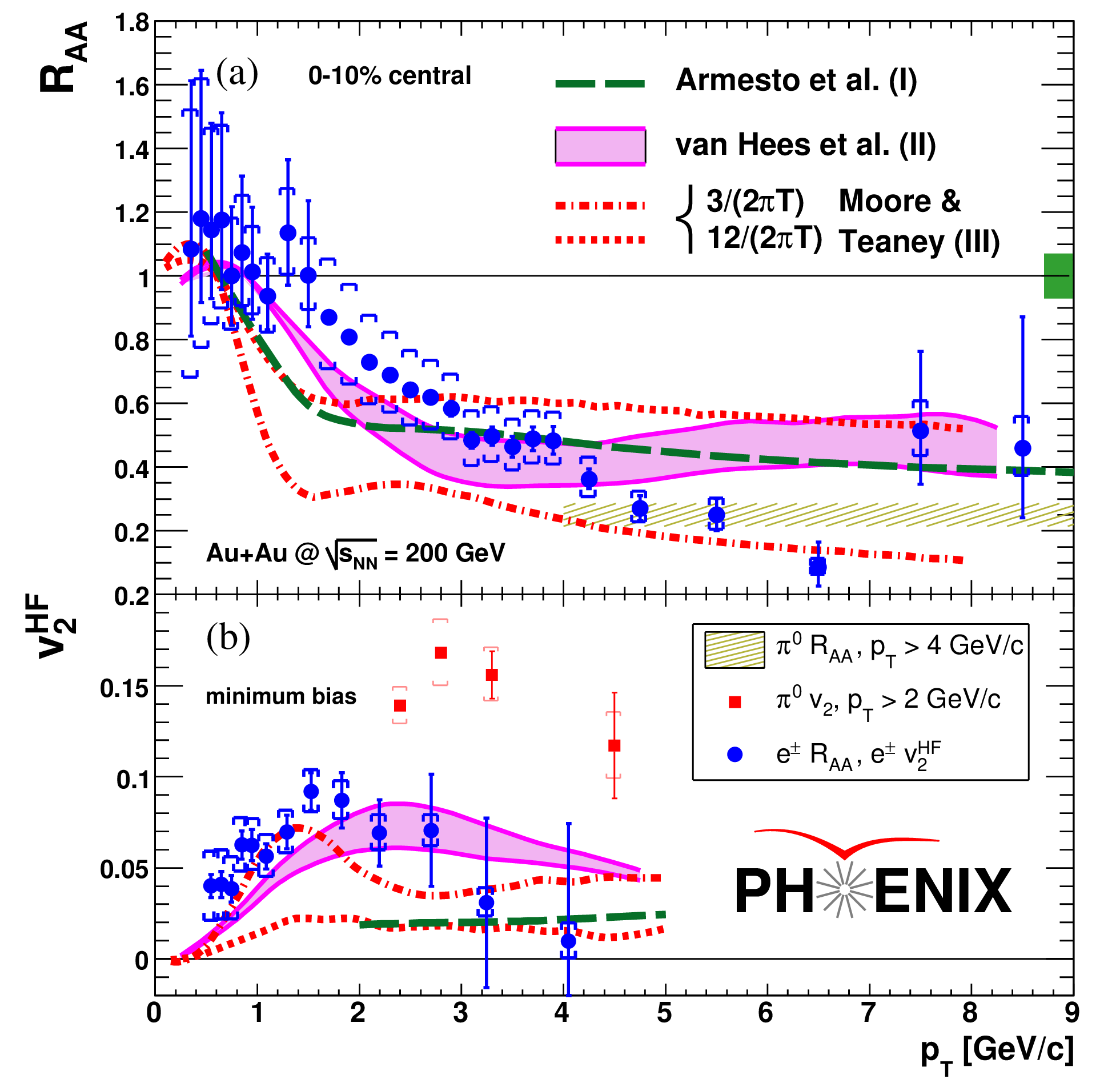}
  \caption[$R_{AA}$ and $v_2$ of single electrons from heavy flavor
  decays]{(a) Shown is the nuclear modification factor $R_{AA}$ of
    electrons from heavy flavor decays. (b) Elliptic flow of electrons
    form heavy flavor decays. A comparison to various models of medium
    effects is shown~\cite{adare:172301}.}
  \label{fig:ppg066_raa_v2}
\end{figure}

In \AuAu collisions at \sqrtsnn = 200 GeV the single electron spectra
from heavy flavor decays are strongly modified with respect to the
expected yield form a hard probe~\cite{adare:172301}. Their nuclear
modification factor is shown in \fig{fig:ppg066_raa_v2}. At high \pt
single electrons from heavy flavor decays are suppressed with respect
the binary scaled yield of electrons in \pp collisions. Furthermore,
they exhibit significant elliptic flow, measured as the second Fourier
coefficient $v_2$ in the expansion of the azimuthal particle
distribution, shown in the bottom panel of
\fig{fig:ppg066_raa_v2}. Both these effects indicate that charm quarks
are subject to significant energy loss and may thermalize in the
medium created in \AuAu collisions at \sqrtsnn = 200 GeV. This should
leave an imprint also in the \ee pair spectrum in the IMR, where
dielectrons from open charm decays are the dominant contribution.

\subsection{Direct Photons}
\label{sec:direct_photons}

Direct photons are another important electromagnetic probe of the
medium created in heavy ion collisions. As discussed below, direct
photons are not only created as massless real photons, but also as
virtual photons with non-zero mass, which internally convert into a
dilepton pair.

\begin{figure}
  \centering
  \begin{fmffile}{dirga}
    \begin{fmfgraph*}(35,20)
      \fmfset{curly_len}{2mm}
      \fmfpen{thin}
      \fmfleft{i1,i2}
      \fmfright{o1,o2}
      \fmf{fermion}{i1,v1,v2,i2}
      \fmf{gluon}{v1,o1}
      \fmf{photon}{v2,o2}
      \fmflabel{$q$}{i1}
      \fmflabel{$\overline{q}$}{i2}
      \fmflabel{$g$}{o1}
      \fmflabel{$\gamma$}{o2}
    \end{fmfgraph*}
  \end{fmffile}
  \hspace{10\unitlength}
  \begin{fmffile}{dirgb}
    \begin{fmfgraph*}(35,20)
      \fmfset{curly_len}{2mm}
      \fmfpen{thin}
      \fmfleft{i1,i2}
      \fmfright{o1,o2}
      \fmf{fermion}{i1,v1,v2,i2}
      \fmf{gluon}{v2,o2}
      \fmf{photon}{v1,o1}
      \fmflabel{$q$}{i1}
      \fmflabel{$\overline{q}$}{i2}
      \fmflabel{$g$}{o2}
      \fmflabel{$\gamma$}{o1}
    \end{fmfgraph*}
  \end{fmffile}
  \\\vspace{20\unitlength}
  \begin{fmffile}{dirgc}
    \begin{fmfgraph*}(35,20)
      \fmfset{curly_len}{2mm}
      \fmfpen{thin}
      \fmfleft{i1,i2}
      \fmfright{o1,o2}
      \fmf{fermion}{i1,v1,v2,o2}
      \fmf{gluon}{i2,v2}
      \fmf{photon}{v1,o1}
      \fmflabel{$q$}{i1}
      \fmflabel{$q$}{o2}
      \fmflabel{$g$}{i2}
      \fmflabel{$\gamma$}{o1}
    \end{fmfgraph*}
  \end{fmffile}
  \hspace{10\unitlength}
  \begin{fmffile}{dirgd}
    \begin{fmfgraph*}(35,20)
      \fmfset{curly_len}{2mm}
      \fmfpen{thin}
      \fmfleft{i1,i2}
      \fmfright{o1,o2}
      \fmf{fermion}{i1,v1,v2,o2}
      \fmf{gluon}{i2,v1}
      \fmf{photon}{v2,o1}
      \fmflabel{$q$}{i1}
      \fmflabel{$q$}{o2}
      \fmflabel{$g$}{i2}
      \fmflabel{$\gamma$}{o1}
    \end{fmfgraph*}
  \end{fmffile}
  \\\vspace{10\unitlength}
  \caption[Feynman diagrams for LO direct photon production]{Feynman
    diagrams for direct photon production to leading order. The top
    row represents the annihilation process, the bottom two graphs the
    Compton process.}
  \label{fig:direct_photon}
\end{figure}
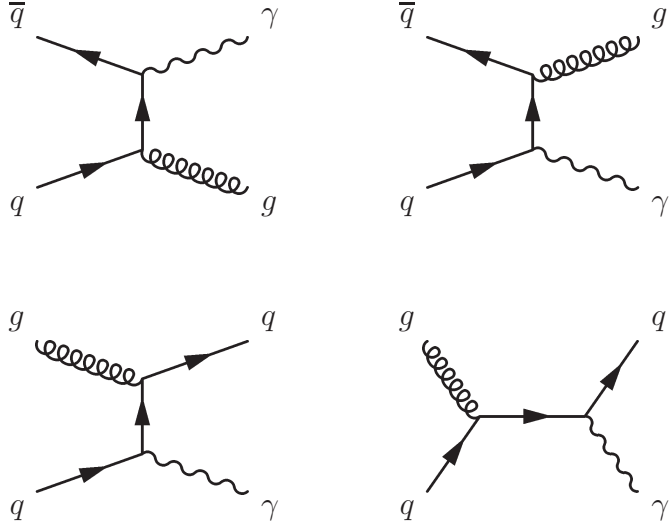

Direct photons are produced by inelastic scattering processes between
the incoming partons. The lowest order processes as shown in
\fig{fig:direct_photon} are quark-antiquark annihilation into a gluon
and a photon and quark-gluon Compton scattering producing a quark and a
photon. 

\begin{figure}
  \centering
  \includegraphics[width=0.9\textwidth]{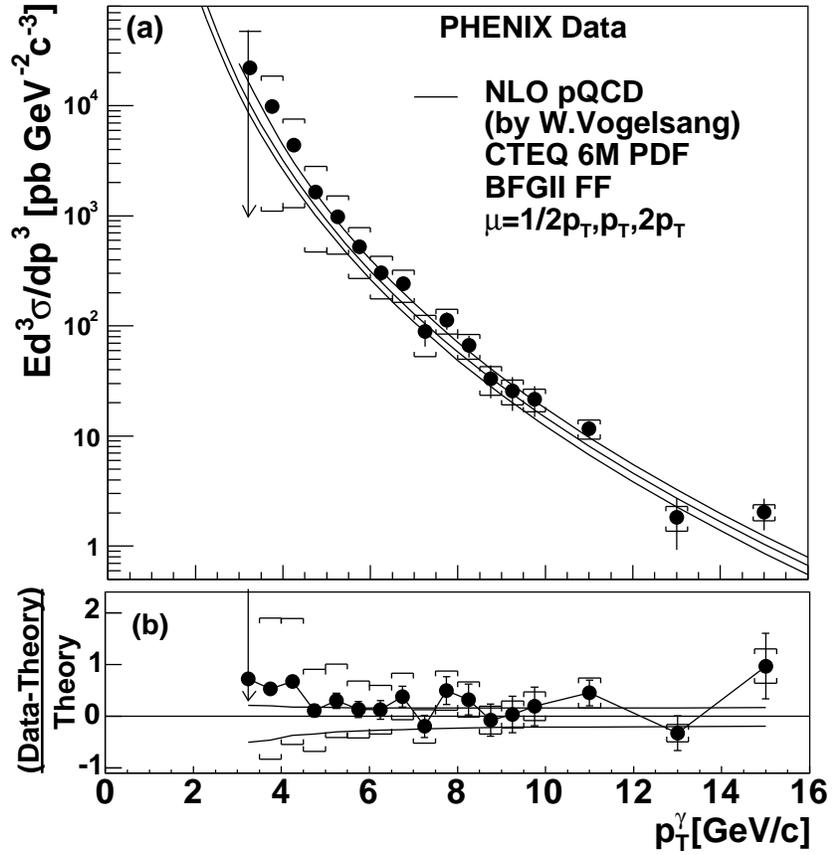}
  \caption[Direct photons in \pp collisions at \sqrts = 200
  GeV]{Direct photon section measured in \pp collisions at \sqrts =
    200GeV~\cite{adler:012002}. The data are compared with a NLO pQCD
    calculation~\cite{PhysRevD.48.3136,PhysRevD.50.1901,Aurenche:1983ws,Aurenche:1987fs}. The
    bottom panel shows the ratio of data to theory.}
  \label{fig:directpp}
\end{figure}

The production of direct photons in \pp collisions at \sqrts = 200 GeV
has been measured by PHENIX~\cite{adler:012002}. The measured direct
photon cross section is in excellent agreement with NLO pQCD
calculations~\cite{PhysRevD.48.3136,PhysRevD.50.1901,Aurenche:1983ws,Aurenche:1987fs}
as shown in \fig{fig:directpp}. The three curves which are shown
correspond to three different choices of the momentum scale $\mu =
0.5\pt,\, 1.0\pt,\, 2.0\pt$. There are actually three different scales
involved: the renormalization scale $\mu_R$, the factorization scale
$\mu_F$, and the fragmentation scale $\mu_F'$. The fragmentation scale
is included, because the calculation includes direct photons from
parton fragmentation into photons. All three scales are set to a
common value $\mu = \mu_R = \mu_F = \mu_F'$.

\begin{figure}
  \centering
  \includegraphics[width=0.9\textwidth]{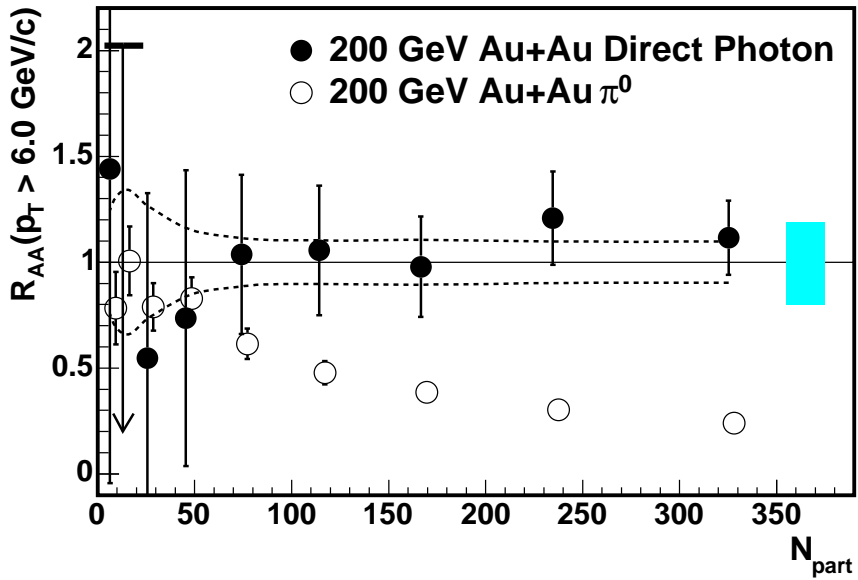}
  \caption[$R_{AA}$ of direct photons]{Nuclear modification factor of
    direct photons and \pion's with $\pt > 6.0$~\gevc as function of
    $N_{\rm part}$~\cite{adler:232301}.}
  \label{fig:directg_raa}
\end{figure}

In \AuAu collisions, direct photons have been measured at high \pt in
various centrality bins~\cite{adler:232301}. The nuclear modification
factor $R_{AA}$ for direct photons with $\pt > 6$~\gevc is shown in
\fig{fig:directg_raa} as function of the participating nucleons $N_{\rm
  part}$. The production of direct photons is consistent for all
centralities $T_{AA}$ scaled direct photons cross section measured in
\pp, \ie, $R_{AA} = 1$, \ie, the assumption that direct photons with
transverse momenta above $\approx 6$~\gevc are produced by inelastic
scattering processes between the incoming partons and escape the
medium unperturbed by final state interactions. This observation is
supporting evidence that the observed \pion
suppression~\cite{adler:072301}, also shown in \fig{fig:directg_raa},
is indeed a final state effect due to energy loss of the partons in
the hot and dense medium. At very high \pt a deviation from one is
observed which may, while still under investigation and despite large
uncertainties, be attributed to isospin differences between the proton
and a gold nucleon~\cite{David:2006sr}.

As direct photons carry the information about the momentum
distribution of the partons involved in their production, the momentum
distribution of direct ``thermal'' photons produced by partons in a
thermalized medium directly reflects the temperature of the medium.

\begin{figure}
  \centering
  \includegraphics[width=0.8\textwidth,angle=-90]{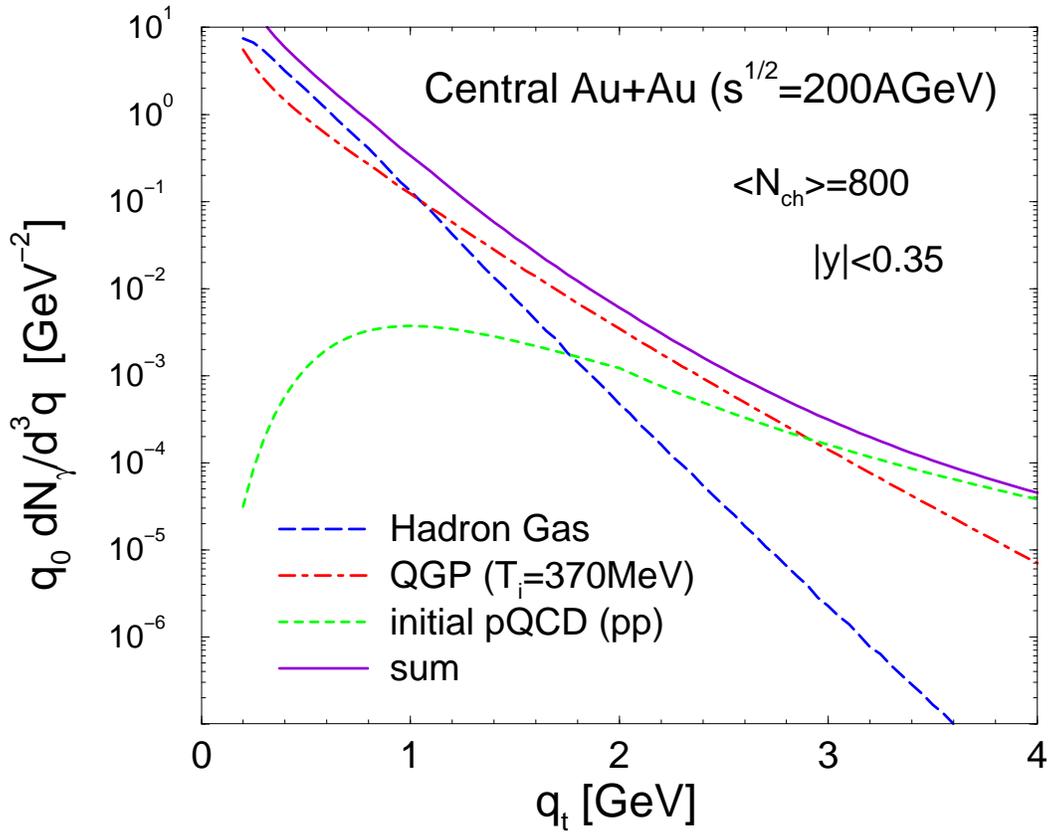}
  \caption[Prediction of thermal photons at RHIC]{Direct photon
    emission from central \AuAu collisions at \sqrtsnn = 200 GeV as
    function of transverse momentum. The {\em short dashed} line shows
    the predicted contribution of photons form the initial hard
    scattering processes, the {\em dashed-dotted} line thermal
    radiation from the QGP and the {\em long-dashed} line emission
    from the hadron gas.\cite{PhysRevC.69.014903}.}
  \label{fig:thermalg}
\end{figure}

In \AuAu collisions at RHIC energies, thermal photons are predicted to
be the dominant source of direct photons in $1 < \pt <
3$~\gevc~\cite{PhysRevC.69.014903}. This is illustrated in
\fig{fig:thermalg} which shows the predicted contributions to the
total direct photon spectrum; from the initial hard scattering
dominating the direct photon yield for $\pt>3$~\gevc, and the
relatively soft thermal emission of from the hadron gas being the main
contributor at $\pt<1$~\gevc, leaving a window for QGP radiation at
$1<\pt<3$~\gevc. As initial conditions a formation time of $\tau_0 =
0.33$~fm and a initial temperature $T = 370$~MeV were used.

However, the inclusive photon yield is dominated by a large background
of hadron decays such as $\pion \rightarrow \gamma \gamma$ for $\pt
\lesssim 5$~\gevc, limiting the measurement of direct photons at low
\pt with the Electromagnetic Calorimeter (EMCal) in PHENIX, which is
based on a statistical subtraction of the decay photon background. The
uncertainties in the knowledge of the decay background can be reduced
by avoiding its explicit measurement, but rather tagging of the decay
photons~\cite{Gong:2007hr}. Furthermore, to circumvent the limitations
due to the energy resolution at low photon energies, the excellent
capabilities of the PHENIX detector to measure electrons have been
utilized by measuring photons via their external conversion into \ee
pairs, a method discussed in Appendix~\ref{cha:bpconv}.

In contrast to massless real photons, virtual photons bring an
additional observable, their invariant mass, which as discussed in the
following brings the advantage of measuring thermal photons in a
better signal to background region than real photons. In general any
source of real photons, \eg the ones shown in \fig{fig:direct_photon},
can also create a virtual photon which is emitted as \ee pair. The
relation between photon production and the associated \ee pair
production can be written as~\cite{PhysRev.98.1355}:
\begin{equation}\label{eq:kroll-wada}
  \frac{d^2n_{ee}}{d\mee} = \frac{2 \alpha}{3 \pi} \frac{1}{\mee} \sqrt{1 -
    \frac{4 m_e^2}{\mee^2}} \left(1+\frac{2m_e^2}{\mee^2}\right) S dn_{\gamma}
\end{equation}
with \mee being the invariant mass of the \ee pair and
$m_e=511$~\kevcc the mass of the electron. The process dependent
factor $S$ aproaches 1 as $m \rightarrow 0$ or $m \ll \pt$. For \pion
and $\eta$ decays, $S$ is given by $S = |F(\mee^2)|^2
(1-\mee^2/m_h^2)^3$~\cite{landsberg} where $m_h$ is the hadron mass
and $F(\mee^2)$ the form factor. For \ee pair masses approaching
$m_h$, the factor $S$ goes to zero. While the measurement of real
thermal photons suffers from a large background of hadron decays (80\%
of the background comes from \pion decays), measuring virtual photons
allows to select a mass range $\mee > m_{\pion} = 135$~\mevcc, in
which the signal to background ratio is improved by a factor five,
thus allowing a 10\% signal of direct photons to be observed as a 50\%
excess of \ee pairs.

For quark-gluon Compton scattering the exact relation between the
photon production process and the \ee pair process can be calculated
as~\cite{Akiba:2008AN695}:
\begin{equation}\label{eq:qg_compton}
  \frac{d^3n_{ee}}{d\mee\,dt} = \frac{2 \alpha}{3 \pi} \frac{1}{\mee} \sqrt{1 -
    \frac{4 m_e^2}{\mee^2}} \left(1+\frac{2m_e^2}{\mee^2}\right) (1 +
  \frac{2 u}{t^2 + s^2} \mee^2) \frac{dn_{\gamma}}{dt}
\end{equation}
in which the Mandelstam variables $s$, $t$, and $u$ are defined as:
\begin{subequations}\label{eq:mandelstam}
  \begin{align}
    s &= (p_q + p_g)^2\\
    t &= (p_q - p_{\gamma})^2\\
    u &= (p_q - p_q')^2
  \end{align}
\end{subequations}
with the four-momentum vectors of the incoming quark $p_q$, incoming
gluon $p_g$, the outgoing real or virtual photon $p_g$, and the
outgoing quark $p_q'$. Comparing \eq{eq:qg_compton} with
\eq{eq:kroll-wada}, the process specific $S$ factor for the
quark-gluon Compton process can be identified as:
\begin{equation}\label{eq:qg_sfactor}
  \begin{split}
    S &= 1 + \frac{2 u}{t^2 + s^2} \mee^2\\
    &= 1 - \frac{2x}{(x + \sqrt{1+x^2}) (3x^2+1+2x\,\sqrt{1+x^2})}
  \end{split}
\end{equation}
with $x = \mee/\pt$ which for $\pt \gg \mee$ simplifies to $S \approx
1$.

\subsection{Medium Modifications of Vector Mesons}
\label{sec:medium_modifications}

At later stages of the collision, after the medium has expanded and
cooled down below the critical temperature, quarks and gluons are
confined to hadrons in a hadron gas. The dominant dilepton production
during this stage is expected from pion and kaon annihilation and
scattering between other hadrons. The cross section for $\pi \pi$
scattering is dynamically enhanced through the formation of light
vector mesons $\rho$, $\omega$, and $\phi$. As these vector mesons
carry the same quantum numbers as a photon, they couple according to
the vector dominance model (VDM) directly to a lepton pair: $\pi^+
\pi^- \rightarrow \rho \rightarrow \gamma^{\ast} \rightarrow l^+
l^-$~\cite{Sakurai:1969}.
\begin{figure}
  \centering
  \begin{fmffile}{vdma}
    \begin{fmfgraph*}(70,30)
      \fmfset{curly_len}{2mm}
      \fmfpen{thin}
      \fmfleft{i1,i2}
      \fmfright{o1,o2}
      \fmf{fermion}{i1,v1,i2}
      \fmf{fermion}{o2,v3,o1}
      \fmf{dbl_plain, label=$\rho$}{v1,v2}
      \fmf{photon, label=$\gamma^{\ast}$}{v2,v3}
      \fmflabel{$\pi^-$}{i1}
      \fmflabel{$\pi^+$}{i2}
      \fmflabel{$l^-$}{o1}
      \fmflabel{$l^+$}{o2}
    \end{fmfgraph*}
  \end{fmffile}
  \\\vspace{10\unitlength}
  \caption[Vector Dominance]{Vector dominance model in dilepton
    production via $\pi^+ \pi^-$ annihilation.}
  \label{fig:vdm}
\end{figure}
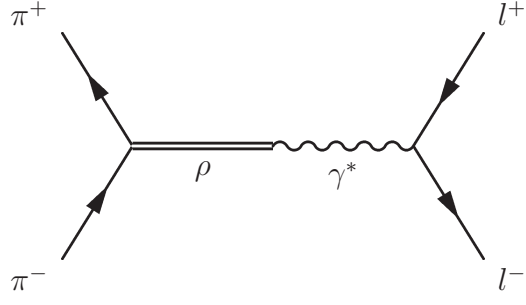
Thus the invariant mass of the lepton pair directly reflects the mass
distribution of the vector meson at the time of its decay. As the
lifetime of the $\rho$ meson ($\tau \approx 1.3$~fm) is shorter than
the lifetime of the hadron gas $\approx 10$~fm, most of the $\rho$
mesons will decay in-medium imprinting any medium effect on the lepton
pair. The invariant mass distribution of \ee pairs from $\rho$ decays
is given by the $\rho$ spectral function (modulo a thermal Bose factor
$f_B$).  Schematically $\rho$ spectral function as function of its
energy $\omega$ and 3-momentum $\vec{q}$, defined as the imaginary
part of the propagator, can be presented in the form of:
\begin{equation}\label{eq:rho_spectralf}
  D_{\rho}(\omega,\vec{q}) = \frac{\Im
    \Sigma_{\rho}(\omega,\vec{q})}{|\omega^2 -q^2 - m_{\rho}^2 + \Re
    \Sigma_{\rho}(\omega,\vec{q})|^2
    + |\Im \Sigma_{\rho}(\omega,\vec{q})|^2},
\end{equation}
with the $\rho$ the pole mass $m_{\rho}$. The $\rho$ self-energy
$\Sigma_{\rho}$ includes the summation of all scattering amplitudes
which includes in-medium interactions with baryons and mesons which
broaden the spectral function.

\begin{figure}
  \centering
  \subfloat[]{\label{fig:rapp_rho}\includegraphics[width=0.44\textwidth]{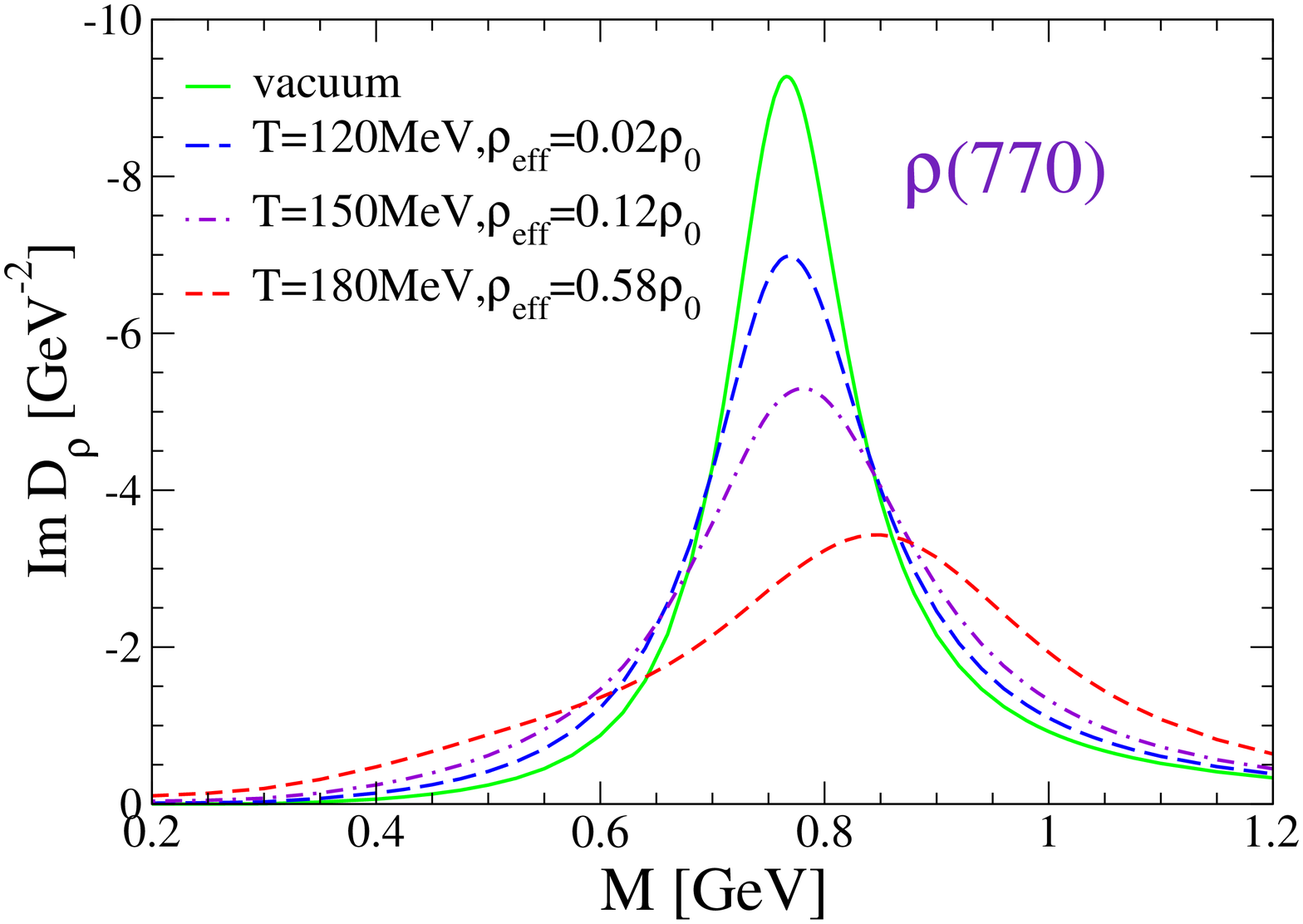}}
  \subfloat[]{\label{fig:rapp_om_phi}\includegraphics[width=0.44\textwidth]{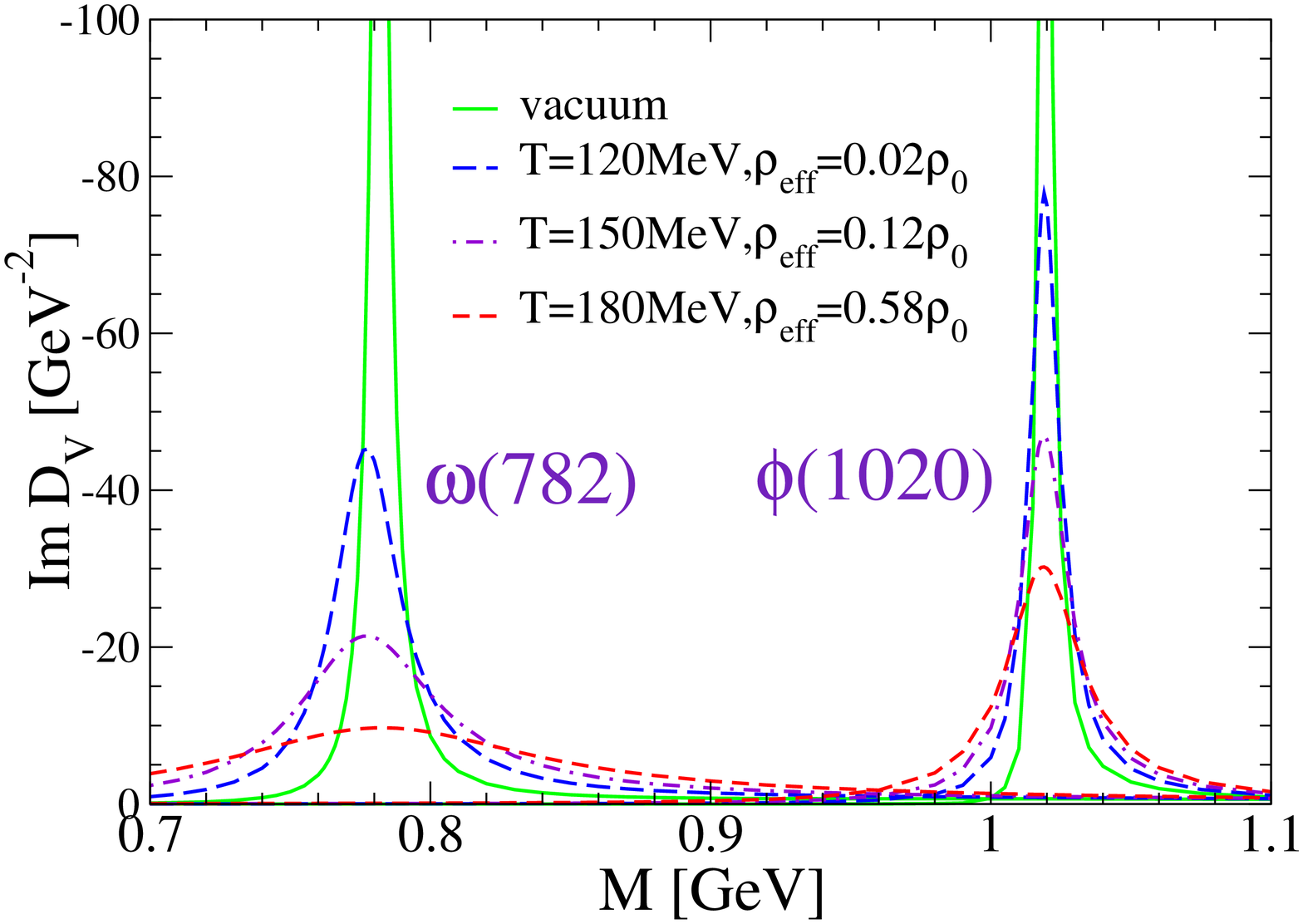}}
  \caption[Medium Modifications to Light Vector Mesons]{In-medium
    spectral function of the vector mesons for different temperatures
    and densities~\cite{Rapp:2002mm,PhysRevC.63.054907}.}
  \label{fig:rapp_inmedium}
\end{figure}

As an example, a prediction of the spectral functions of $\rho$,
$\omega$, and $\phi$ at RHIC energies by
R. Rapp~\cite{Rapp:2002mm,PhysRevC.63.054907} is shown in
\fig{fig:rapp_inmedium}. Recent reviews of the role of the $\rho$
meson in dilepton emission can be found, \eg, in
Refs.~\cite{Rapp:1999ej,Alam:1999sc,Gale:2003iz,Harada:2003jx,Brown:2003ee}.
In this calculation the $\rho$ and $\omega$ spectral function shows
the strong broadening towards higher temperatures and densities, \ie,
towards the phase boundary. The slight upward shift in mass is due to
repulsive parts in the self energy, \eg, from baryonic particle-hole
excitations. The $\phi$ does not show such strong modifications, but
at the highest temperatures the width is significantly increased by a
factor of $\approx 7$.

This explains the special role of the light vector mesons, and their
in-medium modifications, in dilepton measurements. In contrast,
heavier vector mesons such as the $J/\psi$ or the $\Upsilon$ have a
substantially longer lifetime and decay predominantly after freeze
out. Also they carry information of modifications to heavy quarks,
\eg, suppression, but it is in the magnitude of their signal rather
than their spectral shape, see \eg Ref.~\cite{adare:232301}.

After the thermal freeze out temperature is reached, the dominant
sources of lepton pairs are resonance and Dalitz decays of light
mesons with their vacuum properties, such as \pion, $\eta$,
$\eta'$. $\omega$, and $\phi$.

\section{Previous Experimental Results}
\label{sec:exp-results}

A number of experiments have measured the dilepton continuum in heavy
ion collisions under a variety of conditions. The chapter gives a
brief summary of their results, but is by no means meant to be
complete. Section~\ref{sec:hades} summarizes the recent results of the
HADES experiment at the SIS accelerator at
GSI. Section~\ref{sec:ceres} presents the result of the NA45/CERES
collaboration at the SPS accelerator at CERN. The measurement of the
dimuon continuum by the NA60 collaboration is summarized in
Section~\ref{sec:na60}. The section concludes with the direct photon
measurement of the WA98 collaboration in Section~\ref{sec:wa98}.

\subsection{HADES}
\label{sec:hades}

\begin{figure}
  \centering
  \subfloat[C-C at 1~$A$GeV~\cite{Agakishiev:2007ts}]{\label{fig:hades_1agev}\includegraphics[width=0.44\textwidth]{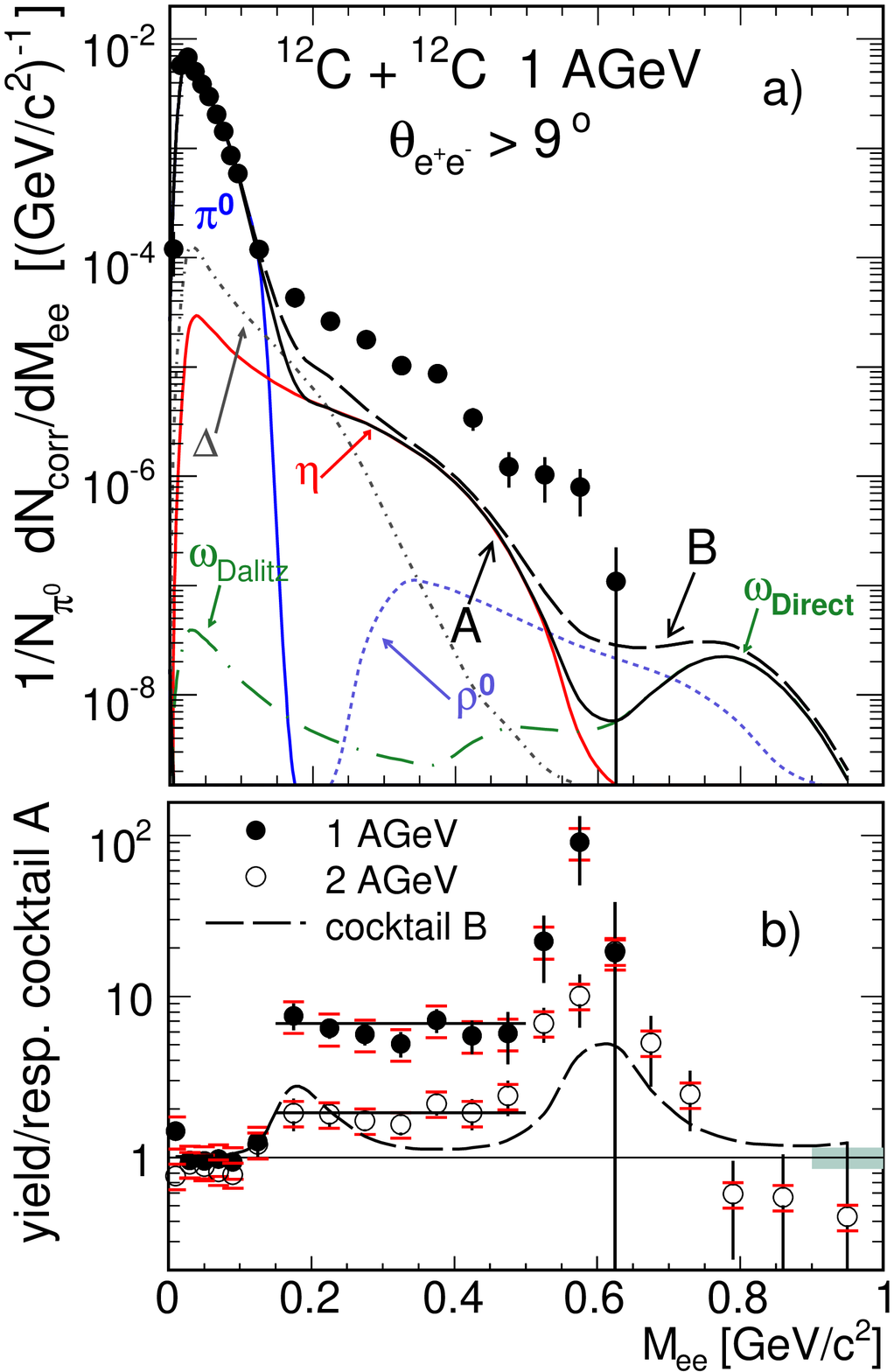}}
  \subfloat[C-C at 2~$A$GeV~\cite{agakichiev:052302}]{\label{fig:hades_2agev}\includegraphics[width=0.44\textwidth]{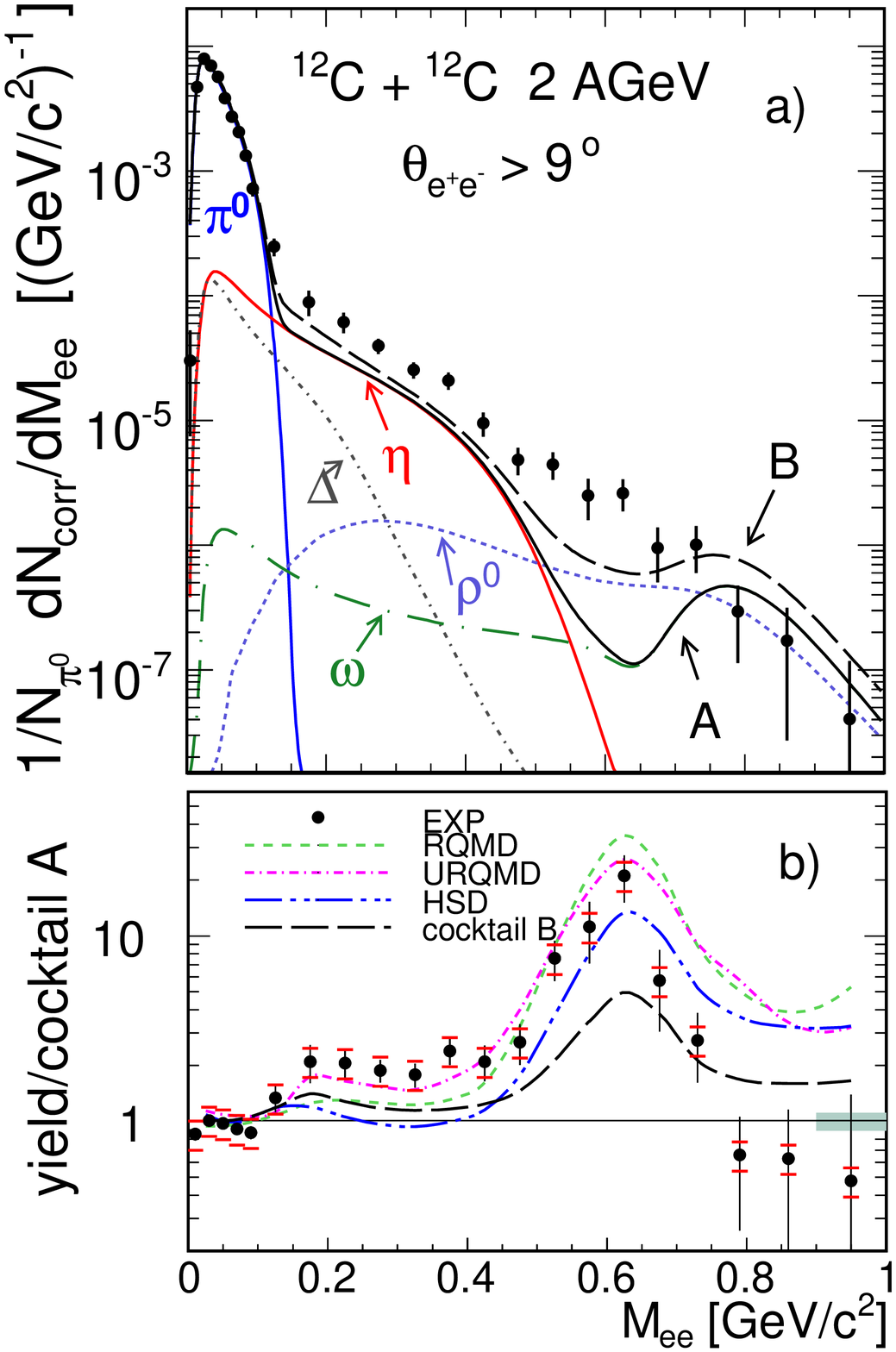}}
  \caption[Dielectron continuum in C-C collisions at 1 and
  2~$A$GeV]{Invariant mass spectrum of \ee pairs measured by HADES in
    C-C collisions at 1~$A$GeV \subref{fig:hades_1agev} and 2~$A$GeV
    \subref{fig:hades_2agev}, respectively. The measured yields are
    compared to a cocktail of \ee pairs from decays of \pion, $\eta$
    and $\omega$ (cocktail A, {\em solid line}). A second cocktail,
    including $\Delta$ and $\rho$ resonance decays is shown as {\em
      dashed line} (cocktail B). The bottom panels show the ratios of
    data and cocktail A.}
  \label{fig:hades}
\end{figure}

The High Acceptance DiElectron Spectrometer (HADES) is a fixed target
experiment at SIS accelerator at GSI. It has measured the dielectron
continuum up to \mee = 1~\gevcc in low energy collisions of light
ions, such as C-C at 1 and
2~$A$GeV~\cite{Agakishiev:2007ts,agakichiev:052302} as shown in
\fig{fig:hades}. The yield of \ee pairs is compared to a cocktail of
decays of $\pi^0 \rightarrow \gamma \ee$, $\eta \rightarrow \gamma
\ee$, $\omega \rightarrow \pi^0 \ee$, and $\omega \rightarrow
\ee$. While the mass region below 150~\mevcc is well explained by the
cocktail, an enhanced yield of \ee pairs in the mass region of
0.15--0.50~\gevcc above the cocktail is observed. Also including in
addition $\rho$ and $\Delta$ resonance decays does not fully explain
the measured yield.  HADES also compared the yield of \ee pairs to the
results of the DLS experiment~\cite{PhysRevLett.79.1229} and finds
them in good agreement~\cite{Agakishiev:2007ts,Pachmayer:2008yn}
challenging current theoretical descriptions of the dielectron
production at these energies. The excess yield scales as the \pion
yield with increasing beam energy, rather than the $\eta$
yield~\cite{Agakishiev:2007ts}. From preliminary
results~\cite{Frohlich:2007gu} from elementary reactions ($p-p$,
$n-p$) it has been shown that a isospin dependent Bremsstrahlung
contribution, increased by a factor of $\approx 4$ with respect to
early theoretical calculations, can almost entirely explain the excess
observed in C-C collisions at the same
energy~\cite{Kaptari:2005qz,Bratkovskaya:2007jk}. This clearly shows
the importance of a baseline measurement in elementary collisions at
the same energy as the heavy-ion collisions.

\subsection{NA45/CERES}
\label{sec:ceres}

The NA45 experiment\footnote{named after its location in the North
  Area of the SPS, in contrast to the West Area} better known as
ChErenkov Ring Electron Spectrometer (CERES) was a fixed target
experiment at the Super Proton Synchrotron (SPS) at CERN. It has
measured the dielectron continuum in heavy-ion reactions at kinetic
beam energies of
40--200~$A$GeV~\cite{PhysRevLett.75.1272,PhysRevLett.91.042301,Agakichiev:2005ai}.

\begin{figure}
  \centering
  \subfloat[$p$-Be at 450~$A$GeV]{\label{fig:ceres_p-be}\includegraphics[width=0.44\textwidth]{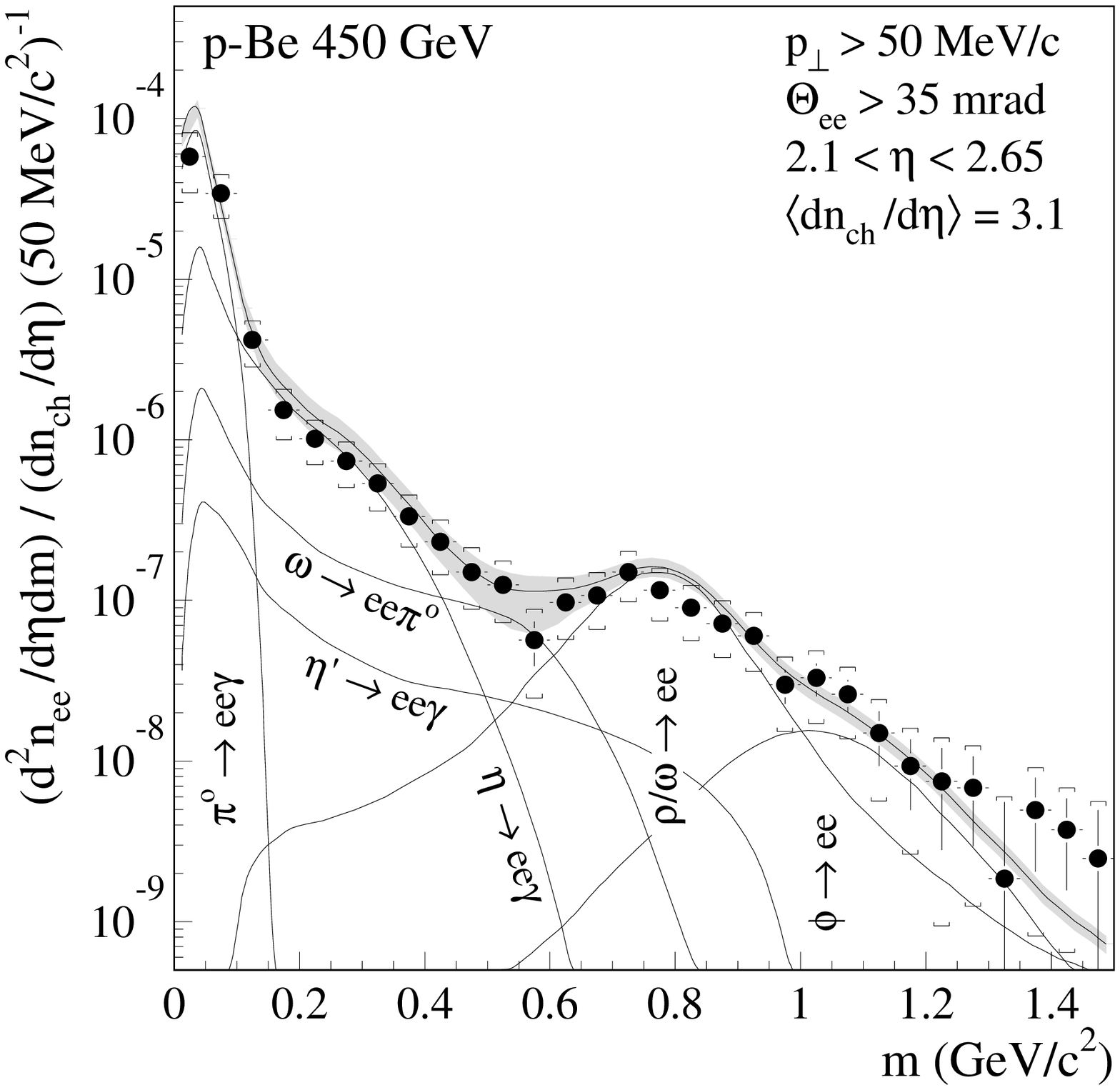}}
  \subfloat[S-Au at 200~$A$GeV]{\label{fig:ceres_s-au_200agev}\includegraphics[width=0.44\textwidth]{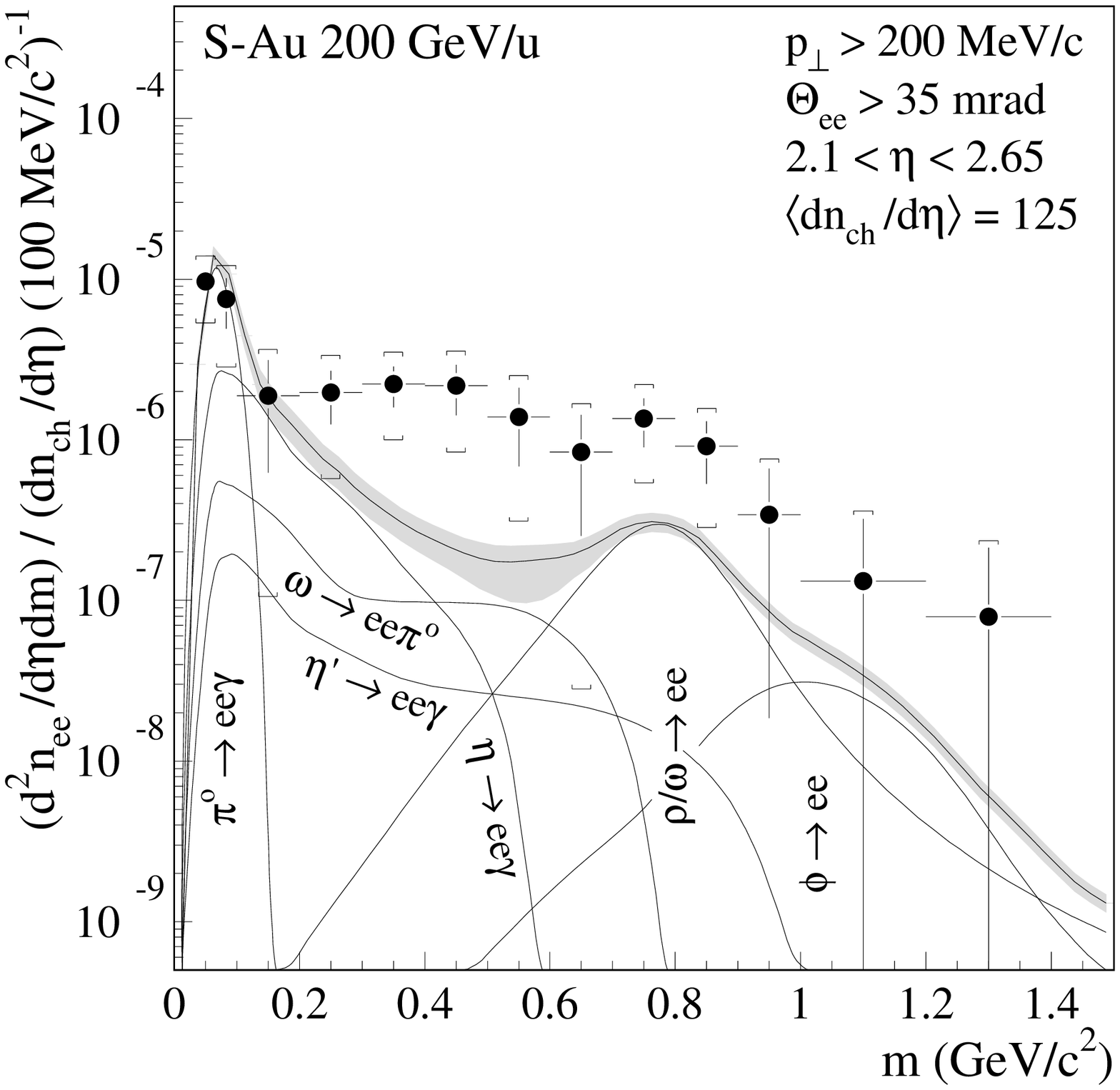}}
  \caption[Dielectron continuum in $p$-Be and S-Au collisions at SPS
  energies]{Invariant mass spectrum of \ee pairs measured by CERES in
    $p$-Be collisions at 450~$A$GeV~\subref{fig:ceres_p-be} and S-Au
    collisions at 200~$A$GeV~\subref{fig:ceres_s-au_200agev},
    respectively. The measured yields are compared to a cocktail of
    \ee pairs from hadron decays; the various sources are shown
    individually. While the cocktail agrees with the measured yield of
    \ee pairs in $p$-Be collisions, an enhancement above the cocktail
    is observed in Pb-Au collisions~\cite{PhysRevLett.75.1272}.}
  \label{fig:ceres}
\end{figure}

While the \ee pair yield in proton induced collisions such as $p$-Be
(as shown in \fig{fig:ceres_p-be}) and $p$-Au at 450~$A$GeV is fully
reproduced by a cocktail of hadron decays, an enhanced yield for $\mee
\geq 200$~\mevcc has been observed in S-Au collisions at 200 $A$GeV
which is shown in
\fig{fig:ceres_s-au_200agev}\cite{PhysRevLett.75.1272}. CERES has also
measured a low mass enhancement of \ee pairs in Pb-Au collisions at 40
and 158 $A$GeV. At the latter energy the \pt and centrality dependence
of the enhancement have been studied, which was found to be localized
at low \pt and to increase stronger than linear with the charged
particle density~\cite{Agakichiev:2005ai} suggesting binary
annihilation processes such as $\pi^+ \pi^- \rightarrow \rho
\rightarrow \gamma^{\ast} \rightarrow \ee$. Their results are compared
to model calculations of three scenarios, (i) free $\pi\pi$
annihilation in vacuum, (ii) assuming Brown-Rho scaling of the in
medium $\rho$ meson~\cite{Brown:2001nh}, and (iii) a modified
in-medium $\rho$ spectral function~\cite{Rapp:1999ej}. The models have
employed a thermal fireball model to include the full time evolution
of the system which includes the experimentally determined freeze out
conditions $(T, \rho_B)_{\rm fo} = (115~{\rm MeV}, 0.33\rho_0)$ and
initial conditions of $(T, \rho_B)_{\rm init} = (190~{\rm MeV},
2.55\rho_0)$, as well as a finite pion chemical potential. For the
dropping mass scenario a in-medium mass of:
\begin{equation}
  m_{\rho}^{\ast} = m_{\rho} \left(1 -C \frac{\rho_B}{\rho_0}\right) \left(1 - \frac{T}{T_c^{\chi}}\right)^{\alpha}
\end{equation}
with $C$ = 0.15, $T_c^{\chi}$ = 200 MeV and $\alpha$ = 0.3 has been
assumed. Clearly, the in vacuum $\pi\pi$ annihilation fails to
describe the data, it gives too much yield at the $\omega$ resonance
peak and too little in the mass range $0.2 < \mee < 0.7$~\mevcc, which
leads to the conclusion that in-medium modifications must play a
role. Within the experimental uncertainties both the dropping mass as
the in-medium collisional broadening scenarios give a reasonable
description of the enhanced dielectron yield.
\begin{figure}
  \centering
  \includegraphics[width=1.0\textwidth]{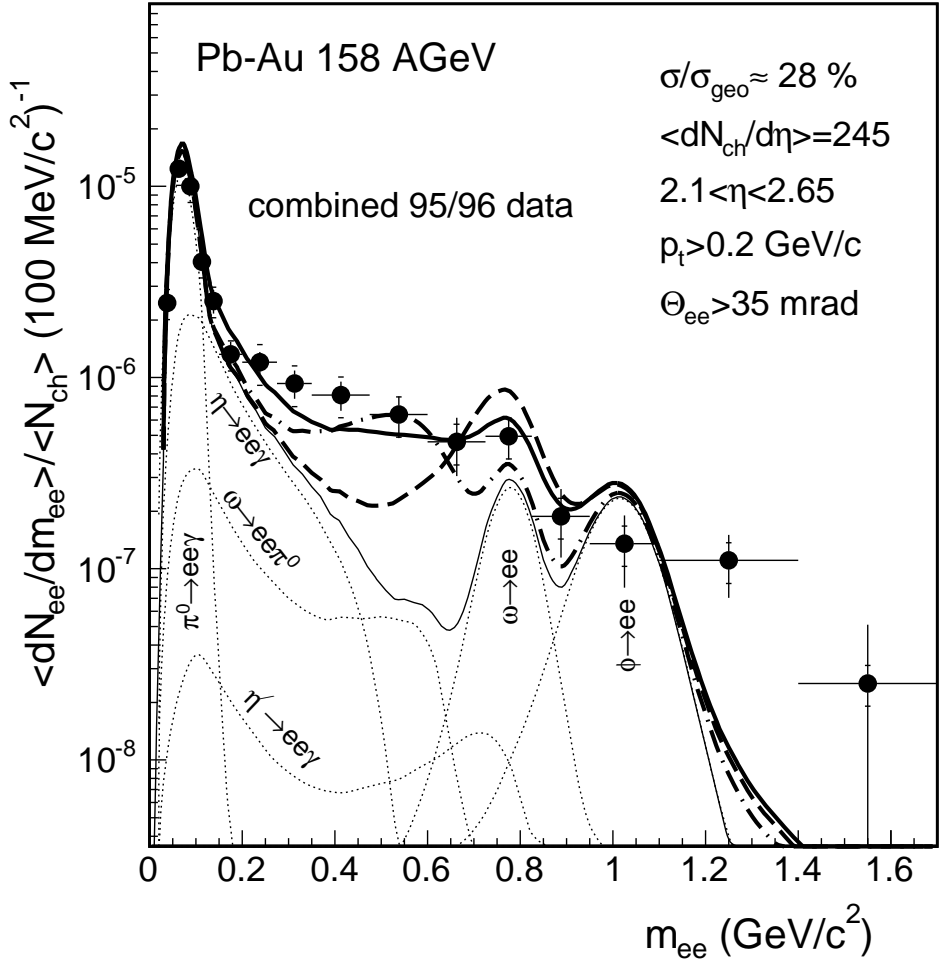}
  \caption[Dielectron continuum in Pb-Au collisions at
  158~$A$GeV]{Invariant mass spectrum of \ee pairs measured by CERES
    in Pb-Au collisions at 158~$A$GeV. The measured yield is compared
    to a cocktail of \ee pairs from free hadron decays and (i) without
    $\rho$ decays ({\em thin line}), (ii) a vacuum $\rho$ spectral
    function ({\em thick dashed line})~\cite{Brown:2001nh}, (iii) a
    dropping in-medium $\rho$ mass ({\em thick dashed-dotted
      line})~\cite{Rapp:1999ej}, (iv) a medium-modified $\rho$
    spectral function ({\em thick solid
      line})~\cite{Agakichiev:2005ai}.}
  \label{fig:ceres_pbau}
\end{figure}

\subsection{NA60}
\label{sec:na60}

The NA60 experiment at SPS was created out of the NA50
experiment\footnote{itself an upgrade of the NA38
  experiment~\cite{Baglin:1990iv} to study Pb-Pb
  collisions}~\cite{Abreu:1997ji} upgraded with a silicon-vertex
tracker. NA50 had measured \mumu pairs, rather than \ee pairs
eliminating the major background source of \pion Dalitz decays due to
the larger muon mass ($m_{\mu} = 105.66$~\mevcc). For Pb-Pb collisions
at 158 $A$GeV, they reported an enhanced dimuon yield in the
intermediate mass region of $1.15 < m_{\mu\mu} < 2.5$6~\gevcc of up to
a factor of $\approx 1.65$ above the expected sources from open charm
decays and Drell-Yan~\cite{Abreu:2000nj}. They were able to rule out
an enhanced Drell-Yan production, as the mass shape of the enhancement
was much steeper then expected for Drell-Yan and concluded the
enhancement was consistent with a charm production enhanced with
respect to the one observed in $p$-A collisions, but did not excluded
the contribution of thermal radiation.

\begin{figure}
  \centering
  \subfloat[]{\label{fig:na60_imr_mass}\includegraphics[width=0.44\textwidth]{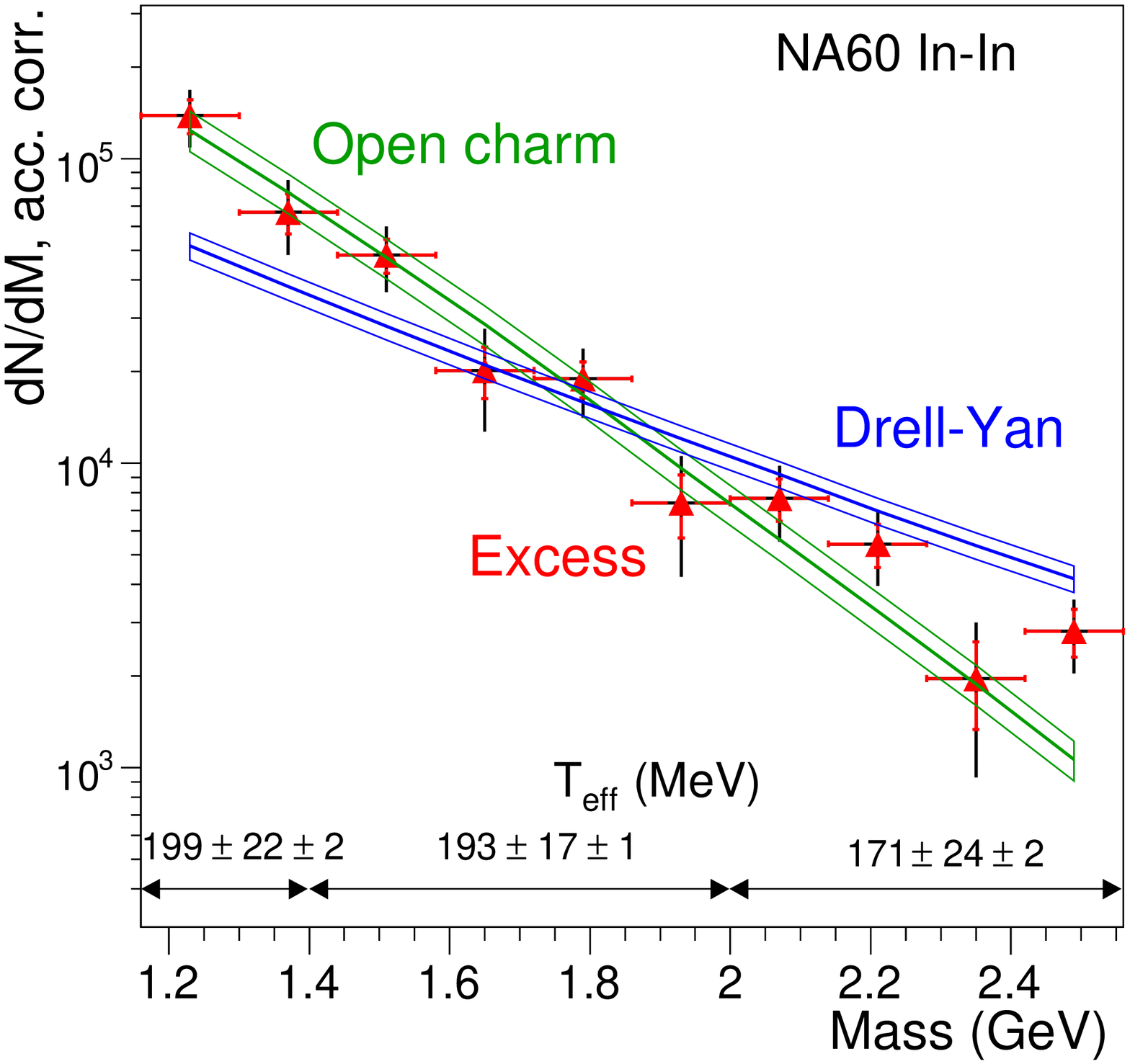}}
  \subfloat[]{\label{fig:na60_imr_offset}\includegraphics[width=0.44\textwidth]{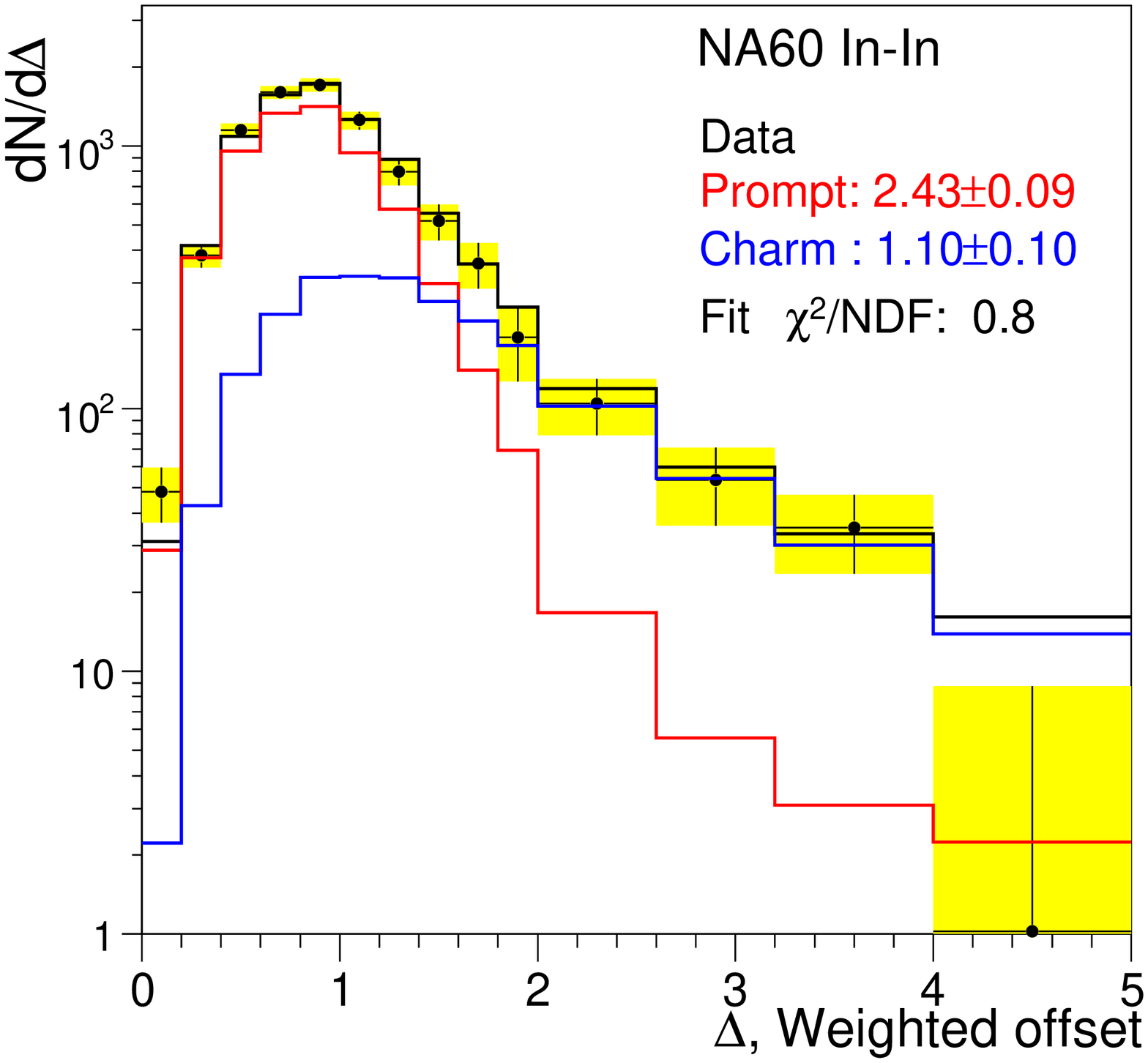}}
  \caption[IMR dimuon continuum in In-In collisions at
  158~$A$GeV]{Shown in \subref{fig:na60_imr_mass} is the mass
    distribution of the acceptance corrected excess yield compared to
    the expected shapes from open charm and Drell-Yan decays. A fit of
    prompt and charm decays to the weighted offset distribution of
    \mumu pairs in the intermediate mass region is shown in
    \subref{fig:na60_imr_offset}~\cite{Damjanovic:2008ta}.}
  \label{fig:na60_imr}
\end{figure}

The upgrade with a silicon-vertex tracker allowed NA60 to determine
the distance between the muon tracks and the collision vertex. Due to
the long life-time of $D$ mesons, their decays vertex would be offset
from the collision vertex in contrast to a prompt source of \mumu
pairs~\cite{0954-3899-34-8-S149}. \fig{fig:na60_imr_mass} shows the
invariant mass spectrum of excess \mumu pairs compared to the expected
shapes from open charm decays and
Drell-Yan~\cite{Damjanovic:2008ta}. The shape is compatible with
semi-leptonic decays of charmed mesons. \fig{fig:na60_imr_offset}
shows the weighted offset distribution of all \mumu pairs. The weight
in the \mumu pair offset from the vertex considers the momentum
dependent resolution of the vertex measurement of a muon. The
distribution of \mumu pairs is fitted with the expected shapes from
open charm and prompt decays. The result is compatible with an open
charm yield as extrapolated from the NA50 $p$-A result and a prompt
excess with an enhancement factor of 2.4 over the Drell-Yan
contribution.

\begin{figure}
  \centering
  \subfloat[]{\label{fig:na60_lmr_all}\includegraphics[width=0.44\textwidth]{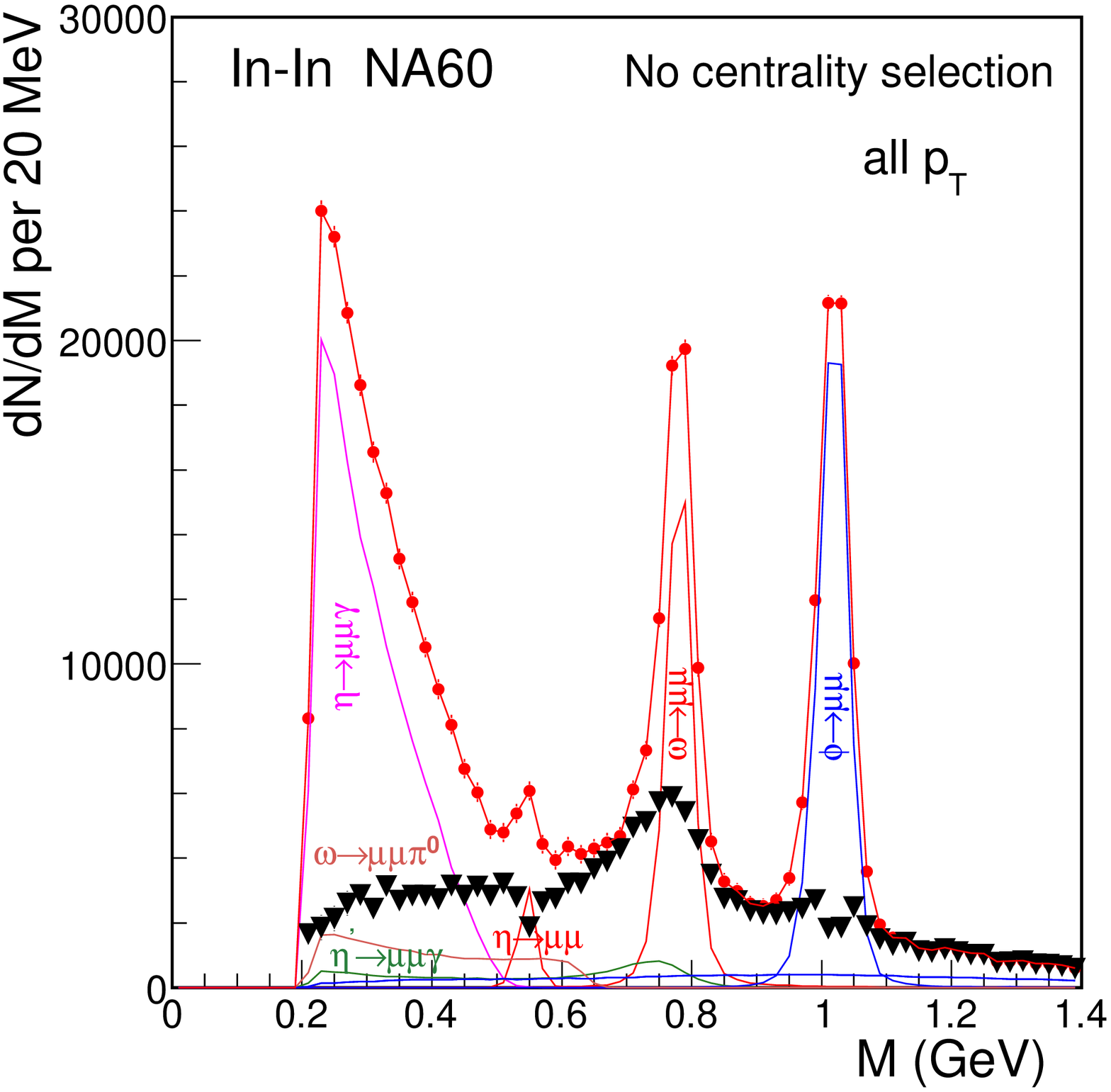}}
  \subfloat[]{\label{fig:na60_lmr_theory}\includegraphics[width=0.44\textwidth]{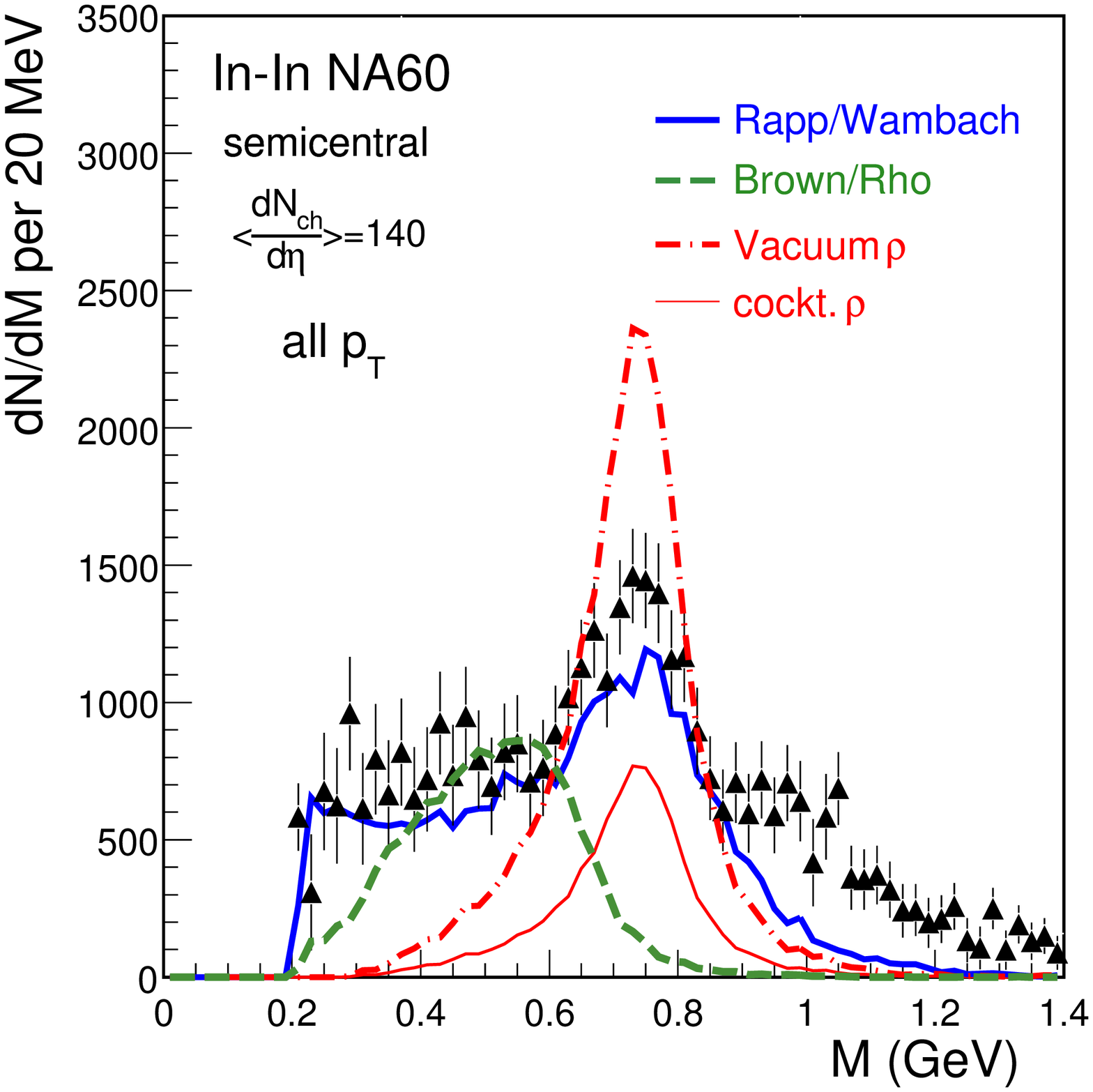}}
  \caption[LMR dimuon continuum in In-In collisions at
  158~$A$GeV]{Shown in \subref{fig:na60_lmr_all} is the invariant mass
    spectrum of \mumu pairs measured by NA60 in In-In collisions at
    158~$A$GeV. The measured yield ({\em open circles}) is compared to
    individual sources of \mumu pairs from hadron decays without any
    $\rho$ contribution. Their contributions are subtracted from the
    data, and the difference ({\em solid triangles}) attributed to
    $\rho$ decays~\cite{arnaldi:022302,Damjanovic:2008ta}. In
    \subref{fig:na60_lmr_theory} the excess yield is compared to a
    number of theoretical predictions in~\cite{Damjanovic:2008ta}.}
  \label{fig:na60_lmr}
\end{figure}

Furthermore, NA60 has observed an enhancement in the low mass dimuon
continuum in In-In collisions at 158 $A$GeV shown in
\fig{fig:na60_lmr_all}~\cite{Damjanovic:2008ta,arnaldi:162302,arnaldi:022302}. The
high precision of the data allowed them to subtract all sources of
hadron decays (besides the contribution of $\rho$ decays) and extract
the enhancement. The two scenarios which were in reasonable agreement
with the CERES data as shown in the previous chapter, an in-medium
broadened $\rho$ spectral function~\cite{Rapp:1999ej} and the dropping
mass~\cite{Brown:2001nh} are compared to the excess yield in
\fig{fig:na60_lmr_theory}. While the broadening scenario is in good
agreement, the Brown-Rho scaling does not explain the observed
enhancement.

\begin{figure}
  \centering 
  \subfloat[]{\label{fig:na60_mt}\includegraphics[height=0.44\textwidth]{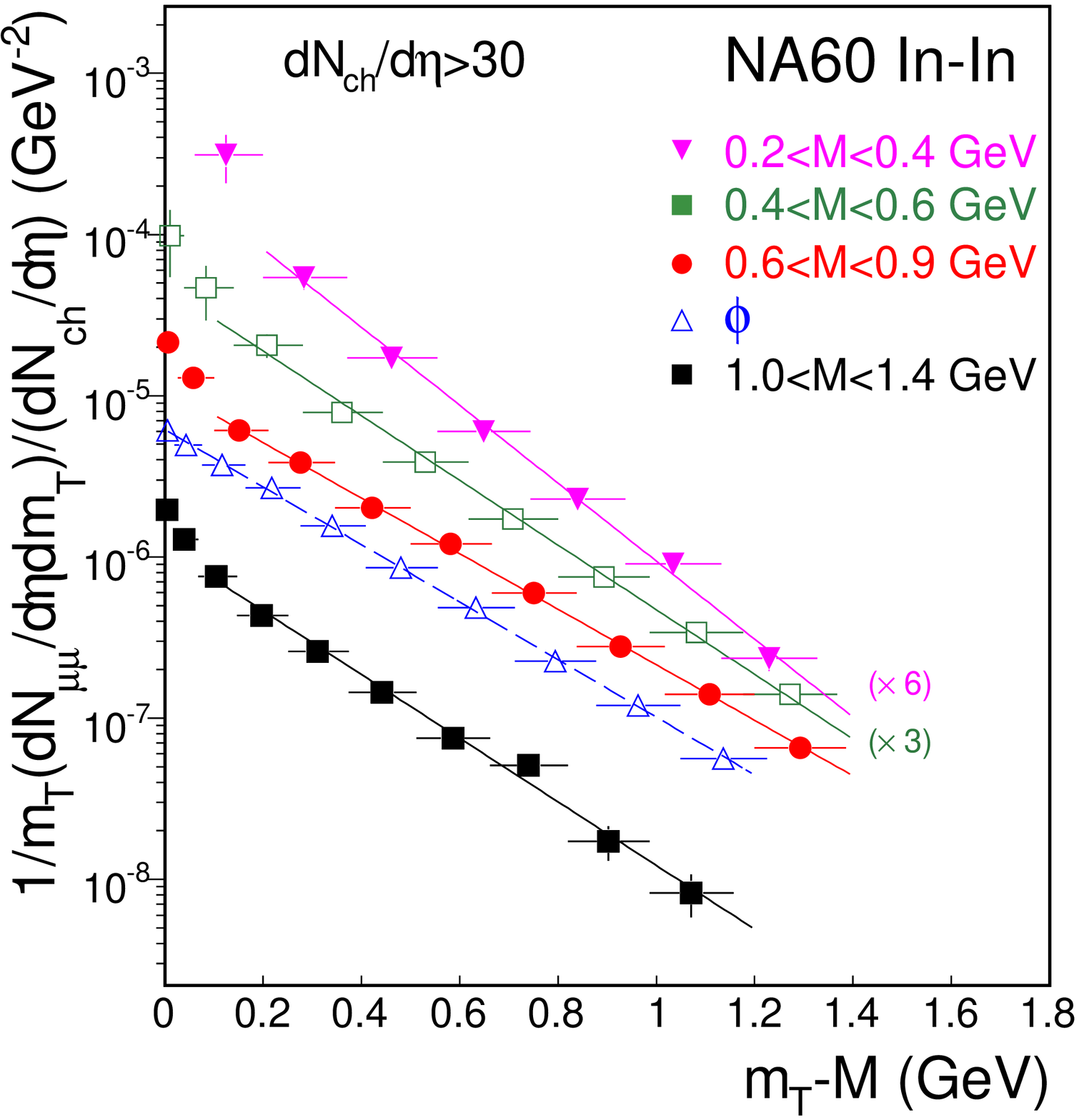}}
  \subfloat[]{\label{fig:na60_teff}\includegraphics[height=0.44\textwidth]{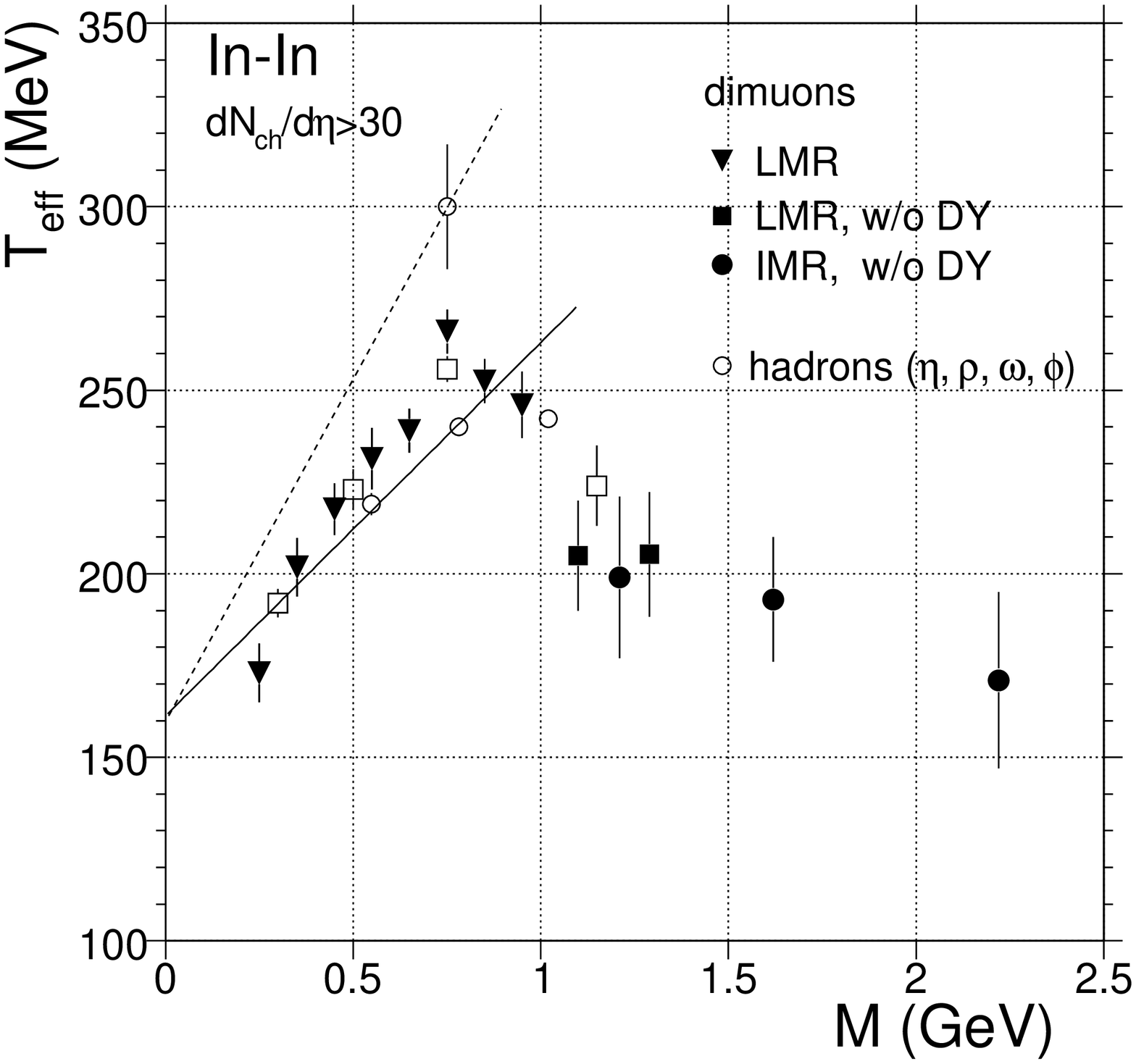}}
  \caption[Dimuon \mt spectra in In-In collisions]{Shown in
    \subref{fig:na60_mt} are the \mt spectra of the excess yield of
    \mumu pairs in four mass windows in comparison to the $\phi$
    summed over all centralities excluding the most peripheral bin. In
    \subref{fig:na60_teff} the inverse slopes $T_{\rm eff}$ are shown
    as function of the invariant mass. The inverse slopes are
    extracted from a fit of the \mt spectra in \subref{fig:na60_mt}
    ({\em open symbols}) and of narrower mass windows ({\em filled
      triangles}). The inverse slopes are compared to the slopes of
    hadrons ({\em open
      circles})~\cite{arnaldi:022302,Damjanovic:2008ta}.}
  \label{fig:na60_results}
\end{figure}

The \mt spectra of the excess yield is shown in \fig{fig:na60_mt} for
four slices in mass and the $\phi$. They are fitted to an exponential
and the inverse slope parameter $T_{\rm eff}$ is shown in
\fig{fig:na60_teff} together with the results of the same fitting
procedure of the excess yield observed in the intermediate mass
region. The inverse slope of the low mass enhancement follows closely
the trend observed for the hadrons $\eta$, $\omega$ and $\phi$ which
is consistent with the expected linear increase due to radial flow
indicated by the solid line $T_{\rm eff} = T_{\rm fo} + m \langle
\beta_T \rangle$. The effective temperature of the vacuum $\rho$ is
much higher, which is interpreted as a later decoupling of the $\rho$
from the medium and therefore obtaining a larger radial flow. Above
the $\phi$ meson the effective temperature suddenly drops and stays
constant with mass. This behaviour is consistent with an thermal
emission from the early partonic phase, before significant radial flow
has developed~\cite{renk:024907}, but hadronic scenarios are not
excluded~\cite{hees:102301}.

At very low \mt ($\mt - M < 0.2$~\gevcc) a steepening of the \mt
spectra shown in \fig{fig:na60_mt} is observed for all four mass
windows of the excess yield. This trend is opposite to the expectation
for radial flow and is not seen for the $\phi$ meson.

\subsection{WA98}
\label{sec:wa98}

\begin{figure}
  \centering
  \includegraphics[width=1.0\textwidth]{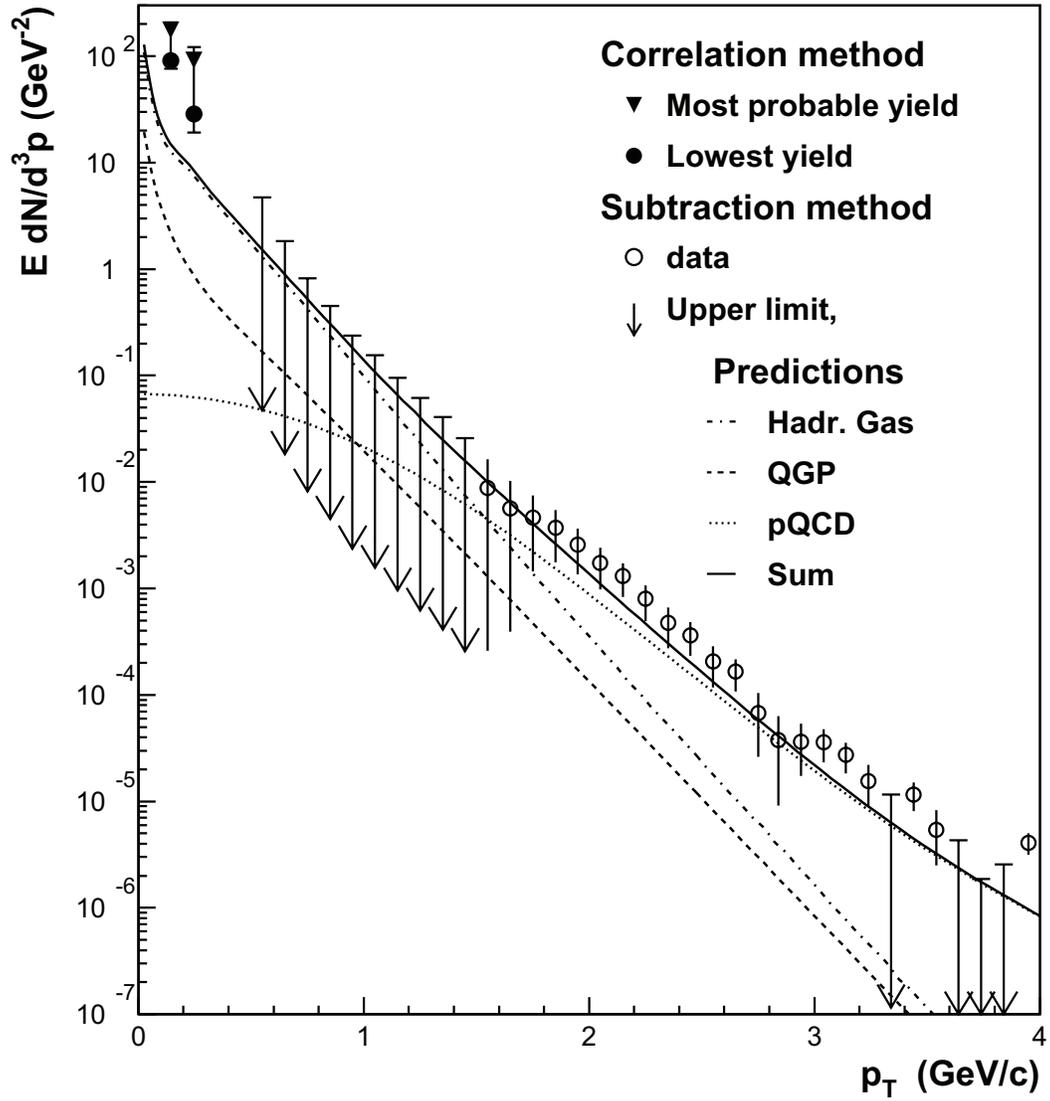}
  \caption[Direct photons in Pb-Pb collisions]{Yield of direct photons
    in Pb-Pb collisions at 158~$A$GeV measured by WA98. The lowest to
    points are extracted from the strength of the two-photon
    correlation~\cite{PhysRevLett.93.022301}, the other points with a
    statistical subtraction method~\cite{PhysRevLett.85.3595}.}
  \label{fig:wa98_directg}
\end{figure}

The WA98 Collaboration has measured direct photons in Pb-Pb collisions
\footnote{with an array of lead glass calorimeter which is now
  installed in PHENIX}~\cite{PhysRevLett.85.3595} based on a
statistical subtraction of decay photons, which show an excess above
the yield expected from the direct photon measurements in proton
induced reaction. The result is shown in \fig{fig:wa98_directg} in
comparison to a prediction of direct
photons~\cite{PhysRevC.69.014903}, which includes a contribution of
thermal emission from a QGP phase; $\approx 30\%$ of the total thermal
photon yield. This prediction is based on the same fireball model
which successfully describes the low mass enhancement observed in the
dilepton continuum in CERES and NA60 with a formation time of $\tau_0
= 1$~fm and an initial temperature of $T_i \simeq 210$~MeV. The two
recently published low \pt data points~\cite{PhysRevLett.93.022301},
extracted via photon HBT (Hanbury Brown and Twiss
Effect~\cite{Hanbury_Brown:1956pf,Hanbury_Brown:1954wr})
interferometry, are not described by current theoretical models which
attribute this yield to the hadronic stage of the
fireball~\cite{PhysRevC.69.014903,srivastava:034905}. But the
inclusion of soft Bremsstrahlung off $\pi \pi$ and $\pi K$ scattering
improves the situation~\cite{Liu:2007zzw}.

\section{RHIC}
\label{sec:rhic}

The Relativistic Heavy Ion Collider (RHIC) at Brookhaven National
Laboratory (BNL) provides heavy-ion collisions at the highest energies
currently available. It provides collisions of \pp, \dAu, \CuCu, and
\AuAu with energies up to \sqrtsnn = 200 GeV, but can also vary the
collision energies to close the gap to SPS energies on the search for
a critical point in the QCD phase diagram. Besides the study of the
quark-gluon plasma, the polarized \pp collisions are of particular
interest for the study of the proton spin structure and allows to
measure the gluon's contribution to the spin of the
proton~\cite{adare:051106}. For the near future RHIC plans to provide
polarized \pp collisions at \sqrts = 500 GeV, energetic enough to
produce $W$ bosons, which will allow to probe $\overline{u}$ and
$\overline{d}$ quarks independently.

The ions are accelerated in two intersecting rings, one clockwise, the
other counter-clockwise and brought to collisions at the six ring
intersections. Four of the intersections have been equipped with
experiments. The BRAHMS experiment was designed to measure charged
hadrons over a wide range of rapidity and transverse momentum. It
completed its data taking program in 2006. PHOBOS was equipped with
subsystems to measure charge particle multiplicities over almost the
entire solid angle, and in addition with two magnetic spectrometers
providing particle identification in a narrow aperture. Phobos
completed its running time at RHIC in 2005. The Solenoid Tracker At
RHIC (STAR) with its large acceptance Time Projection Chamber (TPC)
covering the full azimuth and $|y|<1.5$ is a multipurpose detector
with focus on global event analyses, particle correlations and
particle identification.

Reviews of the results of the first three years of RHIC operation by
all four collaborations can be found in their ``white
papers''~\cite{Arsene:2004fa,Adcox:2004mh,Back:2004je,Adams:2005dq}.

\chapter{The PHENIX Experiment}
\label{cha:phenix-detector}

The PHENIX experiment\footnote{A variety of explanations for the name
  PHENIX exist, from the a posteriori definition as Pioneering
  High-Energy Nuclear Interaction eXperiment to the experiment which
  rose from the ashes of the original proposals for the RHIC
  experiments TALES, SPARC, OASIS, and DIMUON.} is a detector system
which consists of four spectrometer arms and two sets of global
detectors. Of the four spectrometer arms, two, the so called central
arms, are located at mid-rapidity covering each $\eta < 0.35$ in
pseudo-rapidity and $\pi/2$ in azimuth. The two arms at forward
rapidity, the muon arms, cover both $2\pi$ in azimuth. The north arm
covers the pseudo-rapidity range of $1.15 < \eta < 2.44$ while the
south arms extends over the pseudo-rapidity range $-1.15 > \eta >
-2.25$.

\begin{figure}[p]
\includegraphics[width=1.0\textwidth]{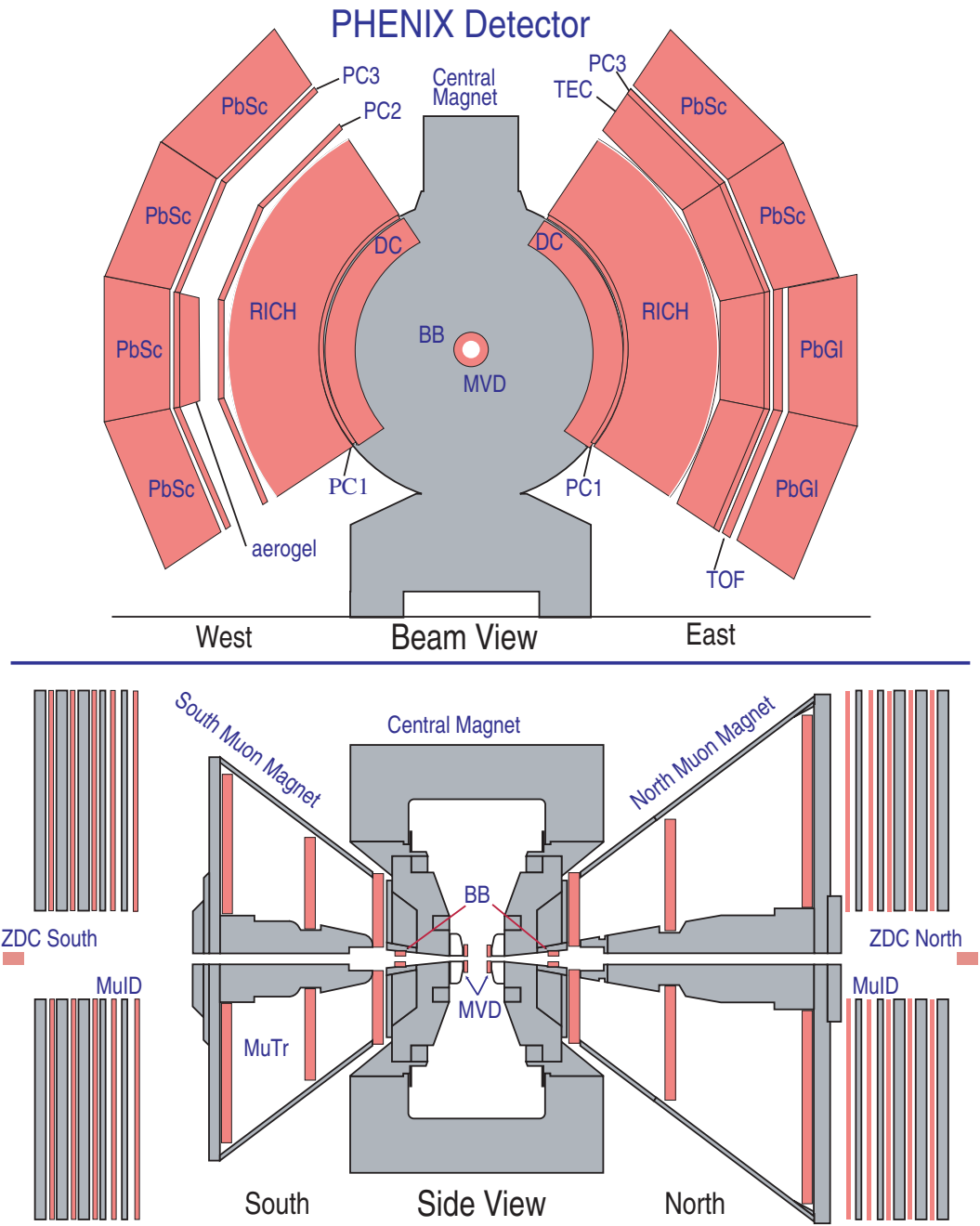}
\caption[The PHENIX Detector]{The PHENIX Detector configuration as of
  2004. The upper figure shows the two central arms in a view
  perpendicular to the beam axis. The lower one shows a cut away view
  along the direction of the beam.}
\label{fig:phenix} 
\end{figure}

The detector setup is shown in Fig.~\ref{fig:phenix}; the upper panel
shows the two central arms in a cut-away view perpendicular to the
beam direction, while a view along the beam direction is displayed in
the lower panel. The latter one also shows the two Muon Arms at
forward rapidity. The azimuthal and rapidity coverage of each
subsystem is summarized in Tab.~\ref{tab:phenix}.

\begin{table}[pt]
\centering
\small
\caption[PHENIX Detector Subsystems]{\label{tab:phenix} Summary of the PHENIX detector subsystems~\cite{Adcox2003}.\\}
\begin{tabularx}{\linewidth}{>{\raggedright\arraybackslash}Xll>{\raggedright\arraybackslash}X}
\toprule
Element & $\Delta\eta$ & $\Delta\phi$ & Purpose and special features\\\midrule
\emph{Magnet} & & &\\
Central (CM) & $\pm 0.35$ & $2\pi$ & Up to 1.15~Tm\\
Muon (MMS) & $-1.1$ to $-2.2$ & $2\pi$ & 0.72~Tm for $\eta = 2$\\
Muon (MMN) & 1.1 to 2.4 & $2\pi$ & 0.72~Tm for $\eta = 2$\\
Beam-beam Counters (BBC) & $\pm$ (3.1 to 3.9) & $2\pi$ & Start timing, fast vertex\\
Zero-degree Calorimeter (ZDC) & $\pm 2$~mrad & $2\pi$ & Minimum bias trigger\\
Drift Chambers (DC) & $\pm 0.35$ & $2 \times \pi/2$ & Good momentum and mass resolution\\
~ & ~ & ~ & $\sigma_m/m = 1\%$ at $m = 1$~GeV\\
Pad Chambers (PC) & $\pm 0.35$ & $2 \times \pi/2$ & Pattern recognition, tracking for non-bend direction\\
Time Expansion Chambers (TEC) & $\pm 0.35$ & $\pi/2$ & Pattern recognition, dE/dx\\
Ring Imaging Cherenkov Counter (RICH) & $\pm 0.35$ & $2 \times \pi/2$ & Electron identification\\
Time of Flight (ToF) & $\pm 0.35$ & $\pi/4$ & Hadron identification, $\sigma < 100$~ps\\
\emph{EMCal} & ~ & ~ & ~\\
Lead-Scintillator (PbSc) & $\pm 0.35$ & $\pi/2 + \pi/4$ & Electron and photon identification and energy measurement\\
Lead-Glass (PbGl) & $\pm 0.35$ & $\pi/4$ & $e^{\pm}/\pi^{\pm}$ separation at $p > 1$~GeV/$c$ by EM shower and $p < 0.35$~GeV/$c$ by ToF\\
~ & ~ & ~ & $\mathrm{K}^{\pm}/\pi^{\pm}$ separation up to 1~GeV/$c$ by ToF\\
\emph{Muon Tracker (MuTr)} & ~ & ~ & Tracking for muons\\
MuTr South & $-1.15$ to $-2.25$ & $2\pi$ & ~\\
MuTr North & 1.15 to 2.44 & $2\pi$ & ~\\
\emph{Muon Identifier (MuID)} & ~ & ~ & Steel absorber and Iarocci tubes for $\mu/$hadron separation\\
MuID South & $-1.15$ to $-2.25$ & $2\pi$ & ~\\
MuID North & 1.15 to 2.44 & $2\pi$ & ~\\
\bottomrule
\end{tabularx}
\end{table}

Two global detectors, the beam-beam counters (BBCs) and the
zero-degree calorimeters (ZDCs), serve as event trigger and are
responsible to measure global event parameters like the collision
time, vertex and centrality. They are described in
Section~\ref{sec:global_detectors}.

The two central arms provide tracking and momentum measurements of
charged particles over a large range in \pt from 0.2~GeV/$c$ to
20~GeV/$c$, as well as particle identification; electrons are
identified via the Ring Imaging Cherenkov Counter (RICH) and the
energy-momentum matching measured in the Electromagnetic Calorimeter
(EMCal) and the Drift Chamber (DC), respectively, photons via their
electromagnetic showers in the EMCal and hadrons via their time of
flight. All detector subsystem that have been used for the
measurements presented in this thesis are described in
Section~\ref{sec:central_detectors}.

For a description of the other central arm detectors, \ie the Time
Expansion Chamber (TEC), the Time of Flight detector (ToF), as well as
the muon arms, in which two detectors (MuTr and MuID) allow tracking
and identification of muons at forward rapidity, one is referred to
Refs.~\cite{Adcox2003a,Aizawa2003,Akikawa2003}.

\section{Global Detectors}
\label{sec:global_detectors}

Global detectors are used to measure the event topology, \ie, the
vertex position, the orientation of the reaction plane, and the
centrality of the collision. Furthermore they measure the time at
which a collision occurs. There are two pairs of detector systems
installed on either side of the interaction point which are
responsible for event selection and characterization in heavy ion
collisions: two sets of beam beam counters cover $3.1 < |\eta| < 3.9$
and two sets of zero degree calorimeters are installed at $\eta
\approx \pm 6.9$.

\subsection{Beam-Beam Counters}
\label{sec:bbc}

\begin{figure*}[t]
  \centering
  \subfloat[]{\label{fig:bbc_a}\includegraphics[width=0.44\textwidth]{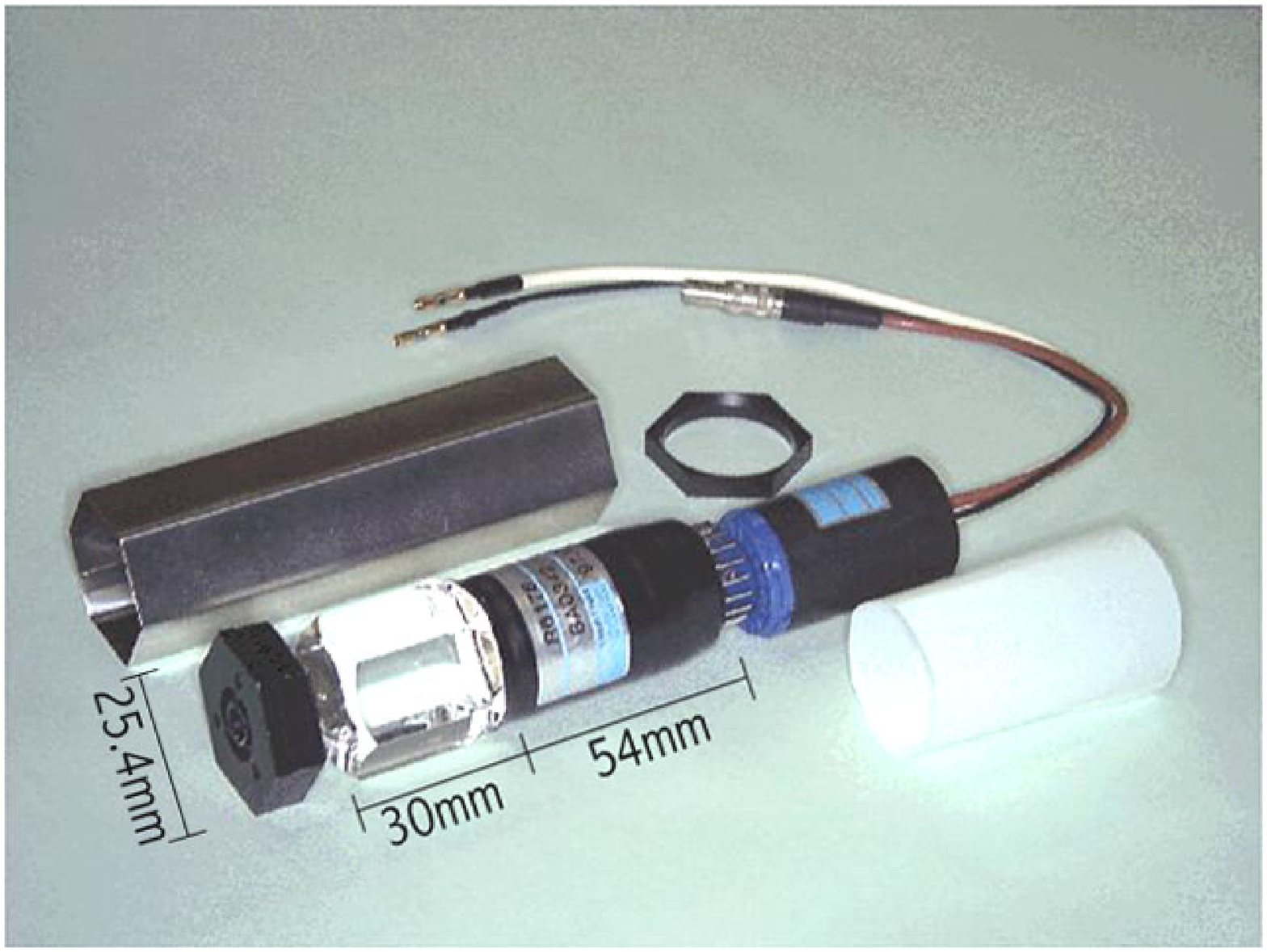}}
  \subfloat[]{\label{fig:bbc_b}\includegraphics[width=0.44\textwidth]{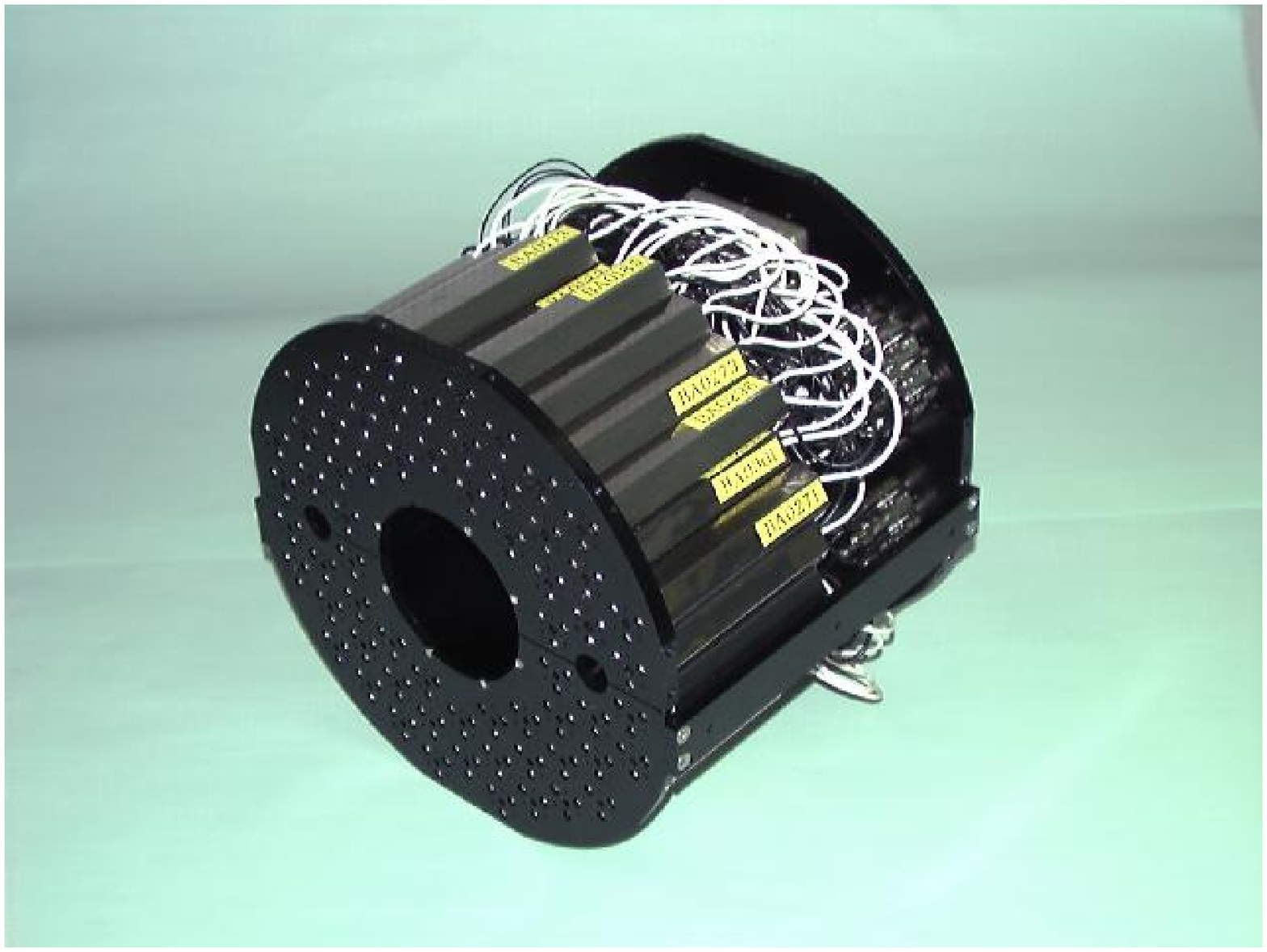}}
  \caption[BBC Components]{Components of the BBC. \subref{fig:bbc_a}
    shows a photomultiplier tube (PMT) with a diameter of 25.4~mm and
    a 30~mm thick Quartz window mounted in front of
    it. \subref{fig:bbc_b} shows 64 PMTs assembled to one unit.}
  \label{fig:bbc}
\end{figure*}

The major tasks of the Beam-Beam Counters
(BBC)~\cite{Allen2003,Ikematsu1998} are to serve as a trigger for
collisions at the interaction point and to provide time and vertex
information of the collision. The BBC comprises of two identical sets
of 64 hexagonal shaped Cherenkov counters as shown in \fig{fig:bbc},
which are installed around the beam pipe at a distance of $\pm 144$~cm
on the north and south side of the interaction point.

Measuring the time difference between the BBC North and the BBC South,
allows for the determination of the collision time as well as
collision vertex:
\begin{subequations}\label{eq:bbc}
  \begin{align}
    t_{0} &= \frac{1}{2} (t_{\rm BBCS} + t_{\rm BBCN})\\
    z_{\rm vertex} &= \frac{c}{2} (t_{\rm BBCS} - t_{\rm BBCN})
  \end{align}
\end{subequations}
where $t_{\rm BBCS}$ and $t_{\rm BBCN}$ are the average arrival time
of particles in the BBC South and BBC North, respectively. The time
resolution of the BBC is $52 \pm 4$~ps (rms). This corresponds to a
vertex position resolution of $1.1$~cm.

\subsection{Zero Degree Calorimeter}
\label{sec:zdc}

\begin{figure}[t]
  \centering
  \includegraphics[width=1.0\textwidth]{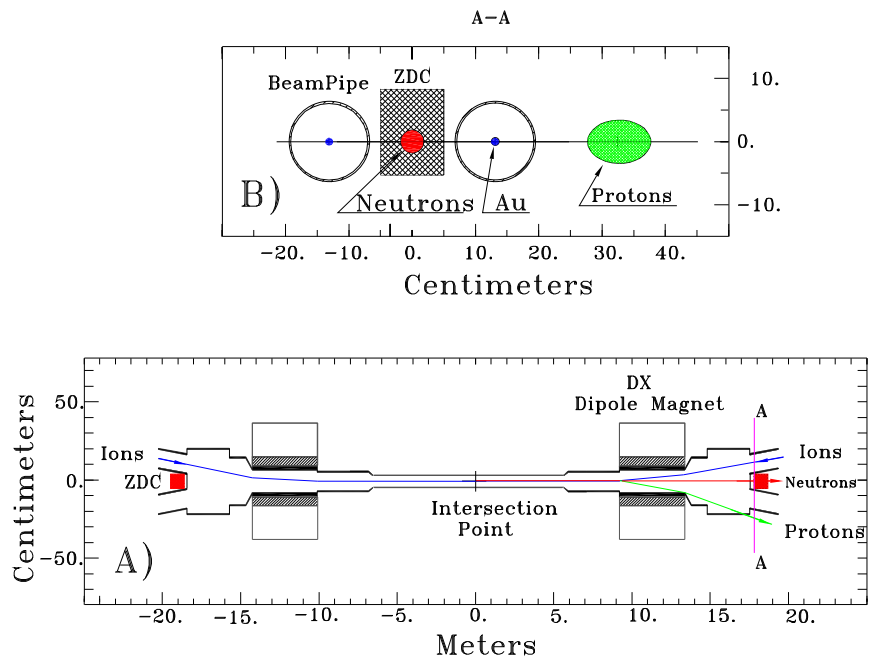}
  \caption[ZDC Layout]{Top: Cut-away view of the ZDC in the plane
    perpendicular to the beam axis indicating deflection of protons
    and neutrons downstream of the ``DX'' Dipole magnet. Bottom: Plan
    view of the collision region~\cite{Adler2001}.}
  \label{fig:zdc_layout}
\end{figure}

All four RHIC experiments are equipped with a pair of Zero Degree
Calorimeters (ZDC)~\cite{Adler2001} located at a distance of 18 m
downstream of each interaction point behind the first accelerator
``DX'' dipole magnet as shown in~\fig{fig:zdc_layout}. Their task is
to measure the energy of spectator neutrons, which did not participate
in the collision and therefore carry still a large fraction of the
beam momentum. While charged particles like spectator protons are
deflected by the ``DX'' dipole magnet in front, neutrons hit the ZDC
and create a hadronic shower. Neutral particles created within the
heavy ion collision moving in forward direction have typically a much
smaller energy. The ZDC consist of Cherenkov sampling hadronic
calorimeter made of a tungsten alloy with a conical coverage of 21
mrad around the beam direction.  The energy resolution of the ZDC is
$\sigma_E/E = 85\%/\sqrt{E} \oplus 9.1\%$.

\begin{figure}[t]
  \centering
  \includegraphics[width=1.0\textwidth]{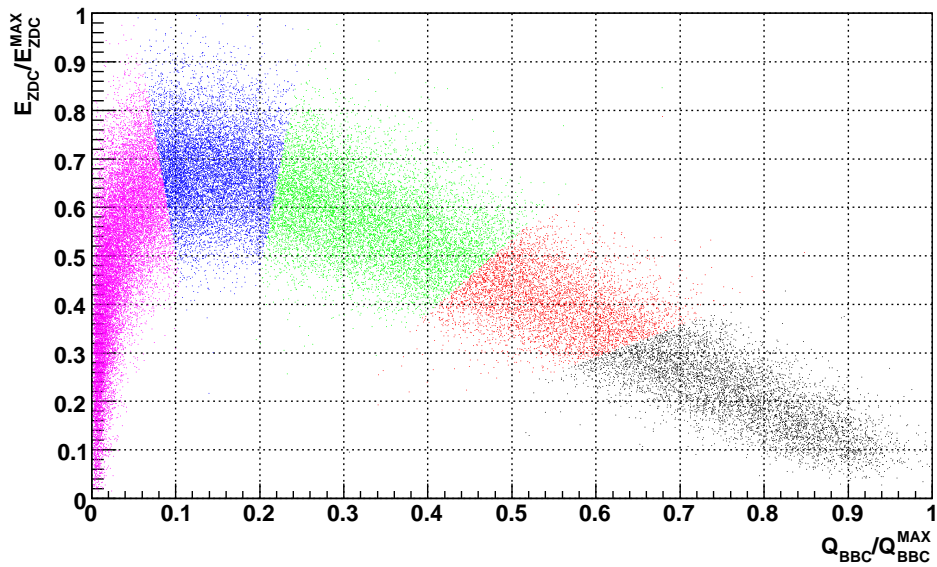}
  \caption[Centrality]{Correlation between the total energy deposited
    in the ZDC and the total charge measured in the BBC for \AuAu
    collisions at \sqrtsnn = 200~GeV. The colored regions show the
    definition of centrality classes based on this correlation (black:
    0--10\%, red: 10--20\%, green: 20--40\%, blue: 40--60\%, and
    magenta: 60--92\%). Their boundaries are perpendicular to the
    centroid of the distribution.}
  \label{fig:bbc_zdc_centrality}
\end{figure}

The total energy deposited by spectator neutrons can be used in
anti-correlation with the total charge deposited in the BBC to
determine the centrality of the collision as shown
in~\fig{fig:bbc_zdc_centrality}. The centrality is a measure for how
much the two colliding ions overlap. In addition to the centrality
determination the ZDC also serves as part of the minimum bias trigger
in heavy ion collisions and provides timing information, but with a
resolution of $\approx 200$ ps it is less accurate than the BBC.

\section{Central Arm Detectors}
\label{sec:central_detectors}

The two Central Arms consist out of several subsystems for charged
particle tracking, momentum measurement and particle
identification. Each arm covers $|\eta| < 0.35$ in pseudo-rapidity and
$|\phi| < \pi/2$ in azimuth. The central arm coordinate system is
chosen with its origin at the nominal interaction point such that the
$\hat{z}$-axis is aligned with the beam direction pointing north, the
$\hat{x}$-axis pointing west and the $\hat{y}$-axis upwards
perpendicular to the two other axis. In this coordinate system the
west arm covers $-\frac{3}{16}\pi < \phi < \frac{5}{16}\pi$ and the
east arm $\frac{11}{16}\pi < \phi < \frac{19}{16}\pi$.

\subsection{Central Magnet}
\label{sec:central_magnet}

\begin{figure}[t]
  \centering
  \includegraphics[width=1.0\textwidth]{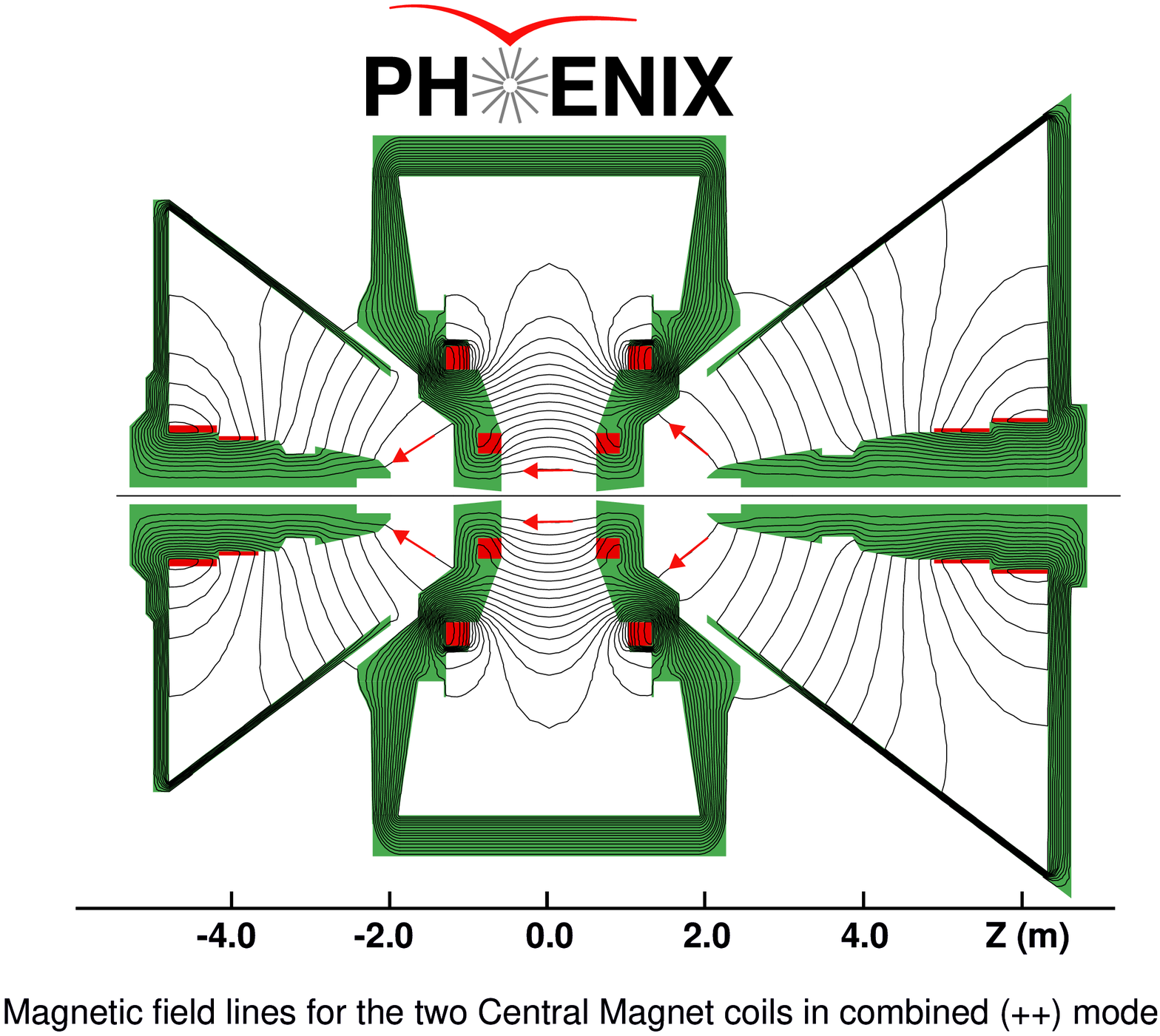}
  \caption[Central Magnet]{The PHENIX Central and Muon Magnets and
    their field lines shown in a cut-away view for adding ($++$)
    configuration~\cite{Aronson2003}. Also shown are the two magnets
    in the Muon Tracker arms on either side of the Central Magnet.}
  \label{fig:cm}
\end{figure}

The transverse momentum of charged particles is determined by their
bending in the magnetic field provided by the Central Magnet
(CM)~\cite{Aronson2003}. It consists out of an inner and an outer pair
of concentric Helmholtz coils inside a steel yoke which provide a
axially-symmetric magnetic field around the interaction point that is
parallel to the beam direction as shown in~\fig{fig:cm}. They cover a
polar angle range of $70^{\circ} < \theta < 110^{\circ}$ which
corresponds to a pseudo-rapidity range of $|\eta| < 0.35$.

The two pairs of Helmholtz coils can be run with the same ($++$) or
opposite ($+-$) polarity. In the $++$ configuration a total field
integral of $\int B dl = 1.5$~Tm is achieved over the first 2 m from
the interaction point, while the magnetic field in the region of the
tracking devices ($R > 2$~m) is nearly zero in order to allow a
tracking model which assumes straight tracks and to minimize the
smearing of Cherenkov rings in the Ring Imaging Cherenkov Counter.

The $+-$ configuration leads to a cancellation of the field in the
first 50~cm around the vertex to approximately zero field integral,
which is used in combination with the Hadron-Blind Detector
(HBD)~\cite{Kozlov:2003zr,Fraenkel:2005wx}, which was installed for
the first time in 2007 inside this field free region and will be
briefly discussed in Chapter~\ref{cha:summary}.

\subsection{Drift Chambers}
\label{sec:dc}

A charged particle which traverses a gas-filled detector randomly
ionizes the gas. The electrons from the primary ionization process are
drifted in an electrical field towards an anode (sense) wire after a
time proportional to the distance of the the track to the wire. Hits
in subsequent anodes can be reconstructed to a track as described
Section~\ref{sec:tracking}.

Two Drift Chambers (DC) are installed in both Central Arms as the main
tracking device for charged particles in PHENIX. They each consist out
of a multiwire gas chamber located at a distance of 2.02~m to 2.40~m
from the interaction point outside the magnetic field of the Central
Magnet. The DC reconstructs the trajectory of charged particles in the
$r-\phi$ plane in order to determine their transverse momentum \pt.

\begin{figure}[t]
  \centering
  \includegraphics[width=0.45\textwidth]{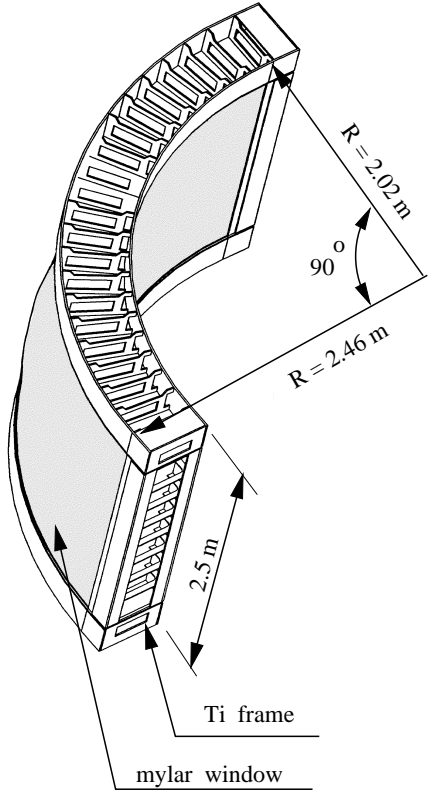}
  \caption[Drift Chamber Frame]{The PHENIX Drift Chamber
    Frame~\cite{Adcox2003a}.}
  \label{fig:dc_frame}
\end{figure}

Both chambers extend over 2 m along the beam direction corresponding
to $\Delta \eta = \pm 0.35$ in pseudo-rapidity; while the Drift
Chamber installed in the Central Arm West covers $-\frac{3}{16}\pi <
\phi < \frac{5}{16}\pi$ in azimuth, the one in the East Arm covers
$\frac{11}{16}\pi < \phi < \frac{19}{16}\pi$. Their active volume is
confined by Mylar windows and supported by a cylindrical shaped
titanium frame as shown in~\fig{fig:dc_frame}. The detectors are
filled with a gas mixture of 50\% Argon and 50\% Ethane.

\begin{figure}
  \centering
  \includegraphics[width=1.0\textwidth]{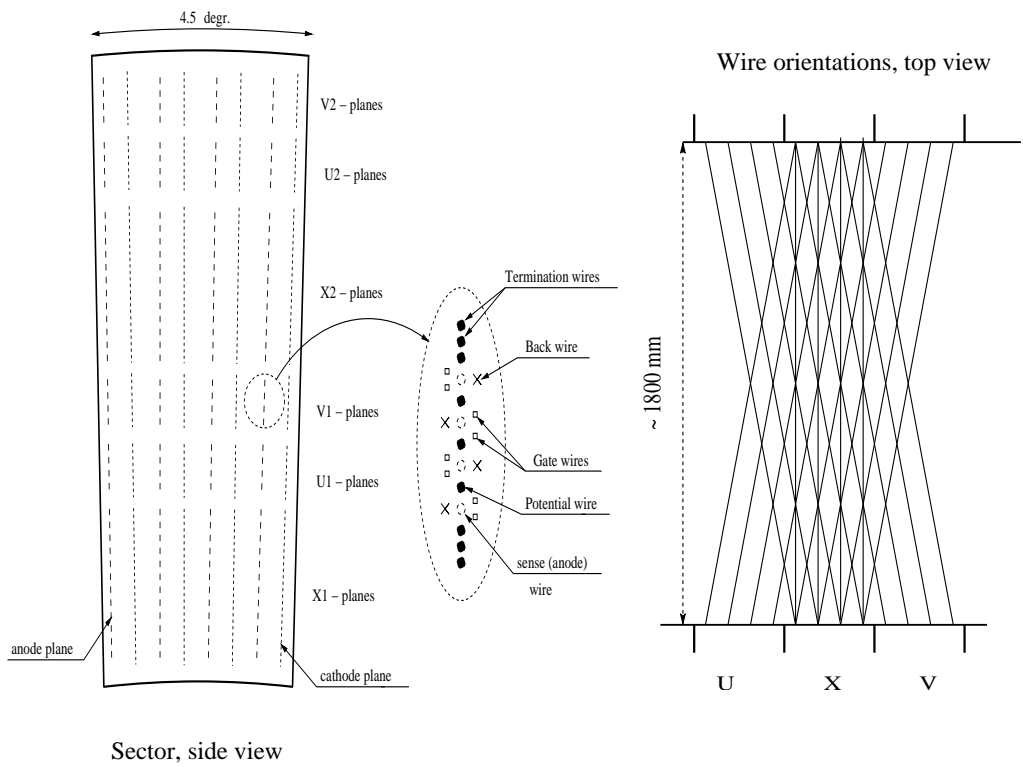}
  \caption[Drift Chamber Wire Layout]{Left: Cut-away view of the wire
    layout within one keystone of the Drift Chamber. Right: Plan view
    of the stereo wire orientation~\cite{Adcox2003a}.}
  \label{fig:dc_wirelayout}
\end{figure}

A design goal of the drift chamber was to measure the mass of the
$\phi$ meson in the $\phi \rightarrow e^+ e^-$ decay channel with a
resolution better than its natural width of 4.4~\mevcc. In conjunction
with the necessity to perform in a high particle multiplicity
environment (as many as two hundred tracks in the central \AuAu
collisions) this imposes the following requirements on the DC:
\begin{itemize}
\item Single wire resolution better than 150~$\mu$m in $r-\phi$.
\item Single wire two track separation better than 1.5 mm.
\item Single wire efficiency better than 99\%.
\item Spacial resolution in $\hat{z}$-direction better than 2 mm.
\end{itemize}
Each Drift Chamber consists of 20 identical sectors covering
4.5$^{\circ}$. As illustrated in~\fig{fig:dc_wirelayout}, each sector
contains six different types of wire modules, X1, U1, V1, X2, U2, and
V2 stacked in radial direction. Every module contains, alternating in
azimuth direction, four anode (sense) and four cathode planes. The X1
and X2 wires run parallel to the beam direction to track particles in
the $r-\phi$ plane. Behind each X wire module two smaller U and V
modules follow whose wires have a small stereo angle of 6$^{\circ}$
with respect to the X wires in order to measure the
$\hat{z}$-coordinate of the track.

\begin{figure}
  \centering
  \includegraphics[width=1.0\textwidth]{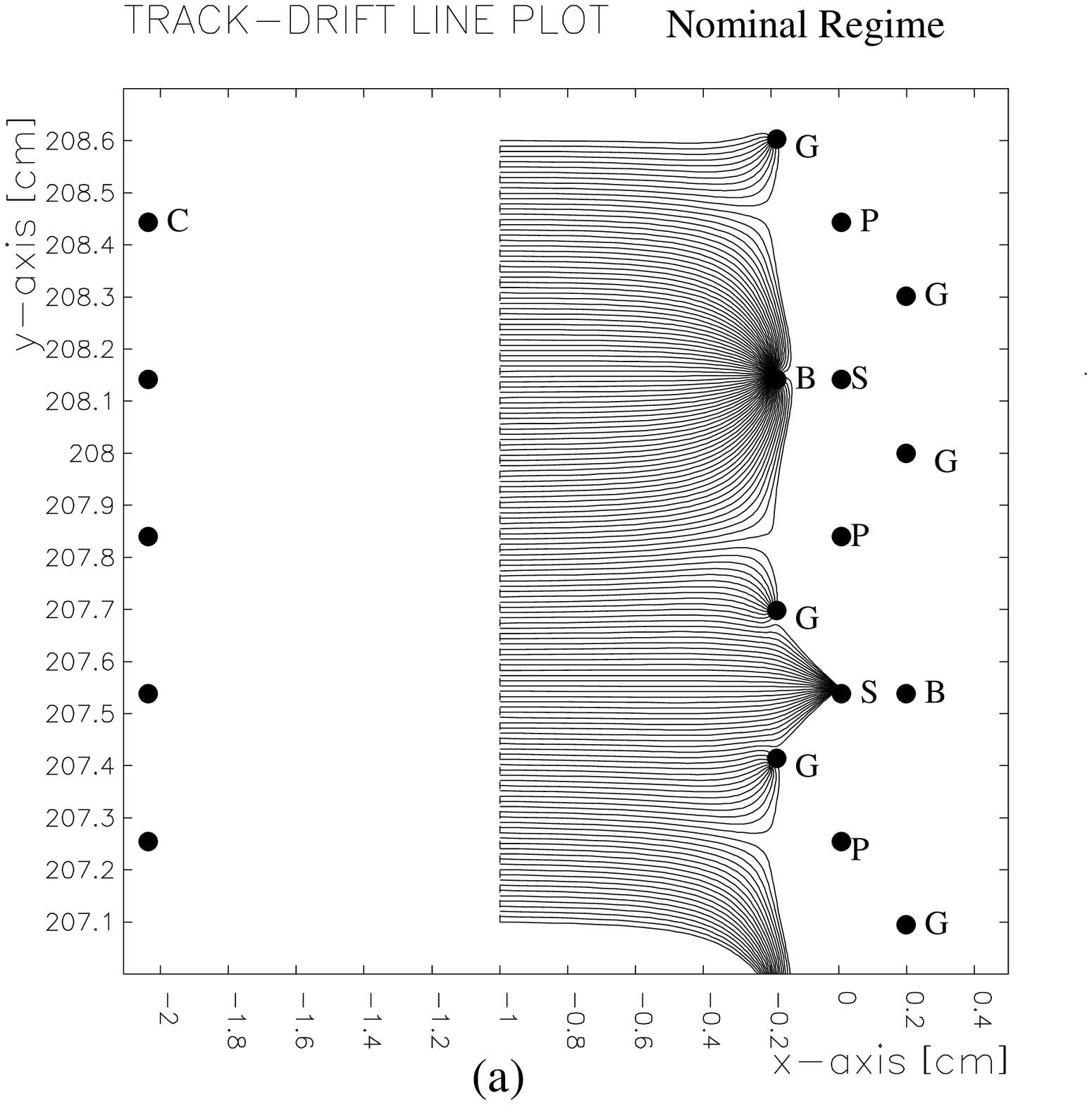}
  \caption[Drift Chamber Drift Lines]{Calculation of the drift lines
    for the nominal electrostatic field configuration. Different wires
    are marked by letters: Back (B), Gate (G), Potential (P), and
    Sense (S) wires~\cite{Adcox2003a}.}
  \label{fig:dc_fieldlines}
\end{figure}

Each X module contains twelve sense wires separated by Potential (P)
wires and surrounded by Back (B) and Gate (G) wires as shown in the
left panel of~\fig{fig:dc_wirelayout} to shape the electrical field
lines such that every sense wire is alternating sensitive to drift
charges from only one side therefore limiting the left-right ambiguity
to a region of $\pm 2$~mm. A calculation of the nominal drift field
configuration due to this wire layout is shown
in~\fig{fig:dc_fieldlines}. The layout of the U,V-stereo modules is
identical, but they contain only four sense wires. The stereo wires
start in one sector and end on the other side of the Drift Chamber in
the neighboring sector, as illustrated in the right panel
of~\fig{fig:dc_wirelayout}.

For the pattern recognition to work with up to 500 tracks, each sense
wire is separated in two halves at the center. Each half is read out
independently. To electrically isolate the two halves, they are
connected by a 100 $\mu$m thick Kapton strip. In total the Drift
Chamber contains 6500 wires and therefore 13000 read out channels.

\subsection{Pad Chambers}
\label{sec:pc}
The Pad Chambers are multiwire proportional chambers with a cathode
pad readout that determine space points along the straight trajectory
of charged particles to determine the polar angle $\theta$ which
allows to reconstruct the $\hat{z}$-component of the momentum vector.

The Central Arms are equipped with three layers of Pad Chambers in the
West Arm and two layers in the East Arm, respectively. The first layer
of Pad Chambers (PC1) is installed just behind the Drift Chambers,
while the third layer (PC3) is situated right in front of the
Electromagnetic Calorimeter. The second layer of Pad Chambers (PC2) is
only present in the West Arm following the Ring Imaging Cherenkov
Counter.

\begin{figure}[t]
  \centering
  \includegraphics[width=1.0\textwidth]{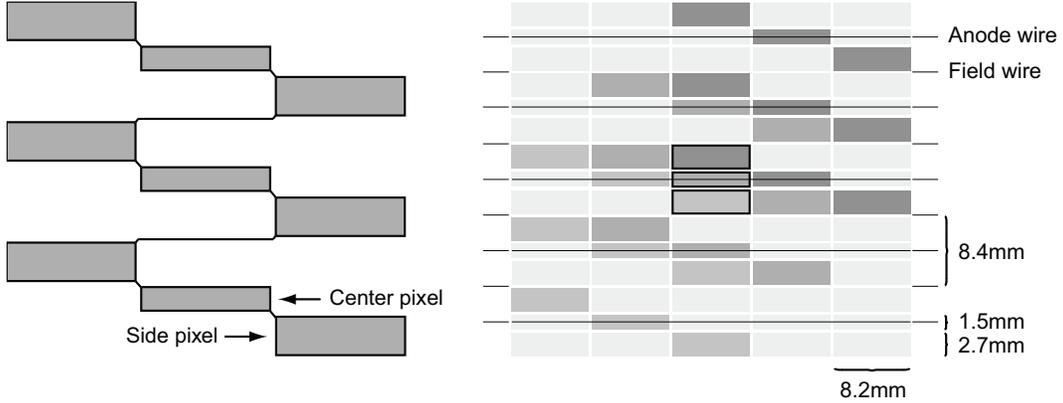}
  \caption[Pad Chamber Pixel Design]{Left: the pad and pixel
      geometry. Right: Interleaved pad design.~\cite{Adcox2003a}.}
  \label{fig:pc_paddesign}
\end{figure}

Each PC contains a single layer of wires within a gas volume that is
confined by two cathode planes. One cathode plane is solid copper,
while the the other one is segmented into a fine array of pixels as
shown in~\fig{fig:pc_paddesign}. The basic unit is a pad formed by
nine non-neighboring pixels, which are read out by a common
channel. Three pixels within a pad form a cell. For a valid hit, three
neighboring pads must sense the avalanche. The interleaved design
allows a fine position resolution of 1.7~mm in $\hat{z}$ direction and
2.5~mm in $\hat{x}$ and $\hat{y}$ and reduces at the same time the
number of electronic channels by a factor of nine.

Associating hits in PC1 with tracks reconstructed in the DC is
essential to determine the three dimensional momentum of a particle.

\subsection{Ring Imaging Cherenkov Counter}
\label{sec:rich}
A charged particle travelling in a medium with a velocity $\beta c$
that is greater than the speed of light in this medium, $c_n = c/n$
for a medium with refractive index $n$, emits Cherenkov radiation
under angle $\cos \theta_C = 1/(n\beta)$.

In each of the two central arms a Ring Imaging Cherenkov Counter
(RICH) is installed between the inner and outer tracking detectors
following the first layer of Pad Chambers~\cite{Aizawa2003}. Its main
purpose is the separation of electrons from the large background of
charged pions produced in heavy-ion collisions. In combination with
the Electromagnetic Calorimeter it also provides information for an
electron trigger in \pp collisions. Behind the entrance window with an
area of 8.9 m$^2$ a volume of 40 m$^3$ is filled with CO$_2$ as
radiator gas, which has a refractive index of $n - 1 = 410 \times
10^{-6}$~\cite{pdg}, corresponding to a threshold velocity $\beta_t =
1/n = 0.99590168$ and a $\gamma$-factor of $\gamma_t =
1/\sqrt{1-\beta_t^2} = 34.932$. This leads to a Cherenkov threshold of
$p_t = m_{\pi} \gamma_t \beta_t = 4.87$ \gevc for charged pions
($m_{\pi} = 139.570$~\mevcc), while electrons ($m_e = 0.511$~\mevcc)
exceed the Cherenkov threshold already with a momentum of $p_t =
0.018$~\gevc. Below the pion threshold the RICH has a hadron rejection
of $10^4$ to 1.
\begin{figure}[t]
  \centering
  \includegraphics[width=1.0\textwidth]{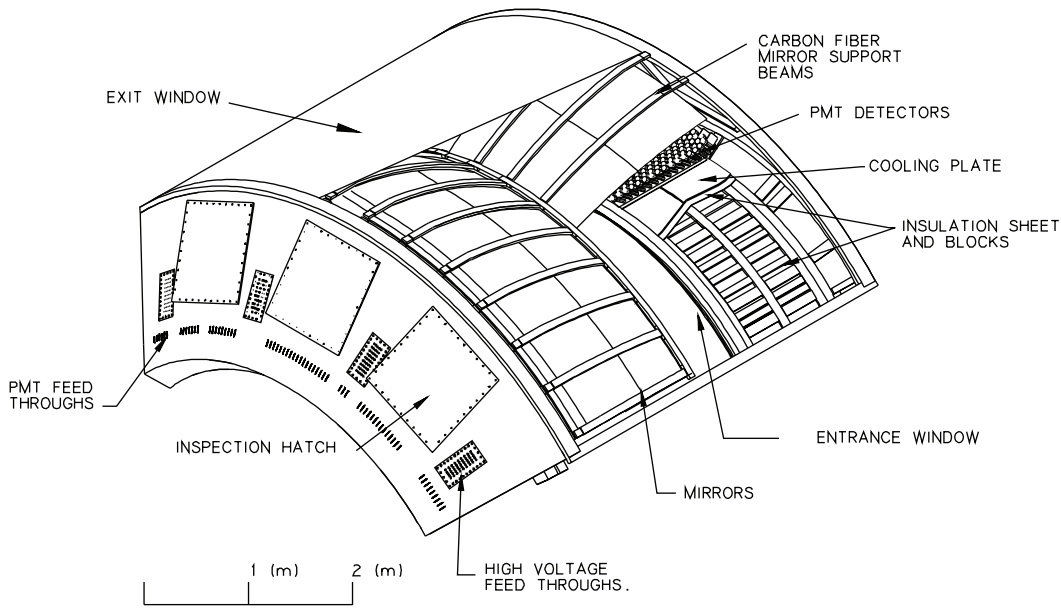}
  \caption[RICH Detector]{A cut-away view of one arm of the PHENIX RICH detector~\cite{Aizawa2003}.}
  \label{fig:rich}
\end{figure}

A cut-away view of the RICH detector is shown in \fig{fig:rich}. The
Cherenkov light is focused by two intersecting spherical mirrors with
a total area of 20~m$^2$ onto two arrays of 1280 photo-multiplier
tubes (PMT) each which are located on either side of the entrance
window. An average number of 10 photons per $\beta \approx 1$ particle
are emitted under the angle of $\theta_C \approx 9$~mrad. They are
focused to a ring on the PMT array with an asymptotic radius of
$\approx 11.8$~cm. The glass in front of the photo tube absorbs light
with wave lengths below 200~nm. The mirror reflectivity is 83\% at
this wave length and rises to 90\% at 250~nm.

In \pp collisions the RICH also serves as Level-1 trigger on rare
events with electrons. The trigger is comprised of 64 non overlapping
trigger tiles in each PMT array. Each trigger tile consists of
$4(\phi)\times5(z)$ PMTs, an area which approximately corresponds to
the size of a Cherenkov ring of a $\beta \approx 1$ particle.

\subsection{Electromagnetic Calorimeter}
\label{sec:emcal}
High-energy electrons and photons lose energy in matter predominantly
via Bremsstrahlung and \ee pair production, respectively. The amount
of energy they lose is defined by the radiation length $X_0$, which is
both (a) the mean length of traversed matter after which an electron
has lost all but $1/e$ of its energy and (b) $7/9$ of the mean free
path for \epair production by a photon.

The Electromagnetic Calorimeter (EMCal) measures the energy and
position photons and electrons. Furthermore, it serves as trigger on
rare events with high momentum photons. It comprises of eight sectors
covering each 22.5$^{\circ}$ in azimuth and $\Delta \eta = \pm 0.35$
in pseudo-rapidity. All four sectors of the West Arm and the two top
sectors in the East Arm are shashlik type lead-scintillator (PbSc)
sampling calorimeter. The two bottom sectors are lead-glass (PbGl)
Cherenkov calorimeters, which had been used previously in the CERN
experiment WA98 at the SPS.

\begin{figure}[t]
  \centering
  \includegraphics[width=1.0\textwidth]{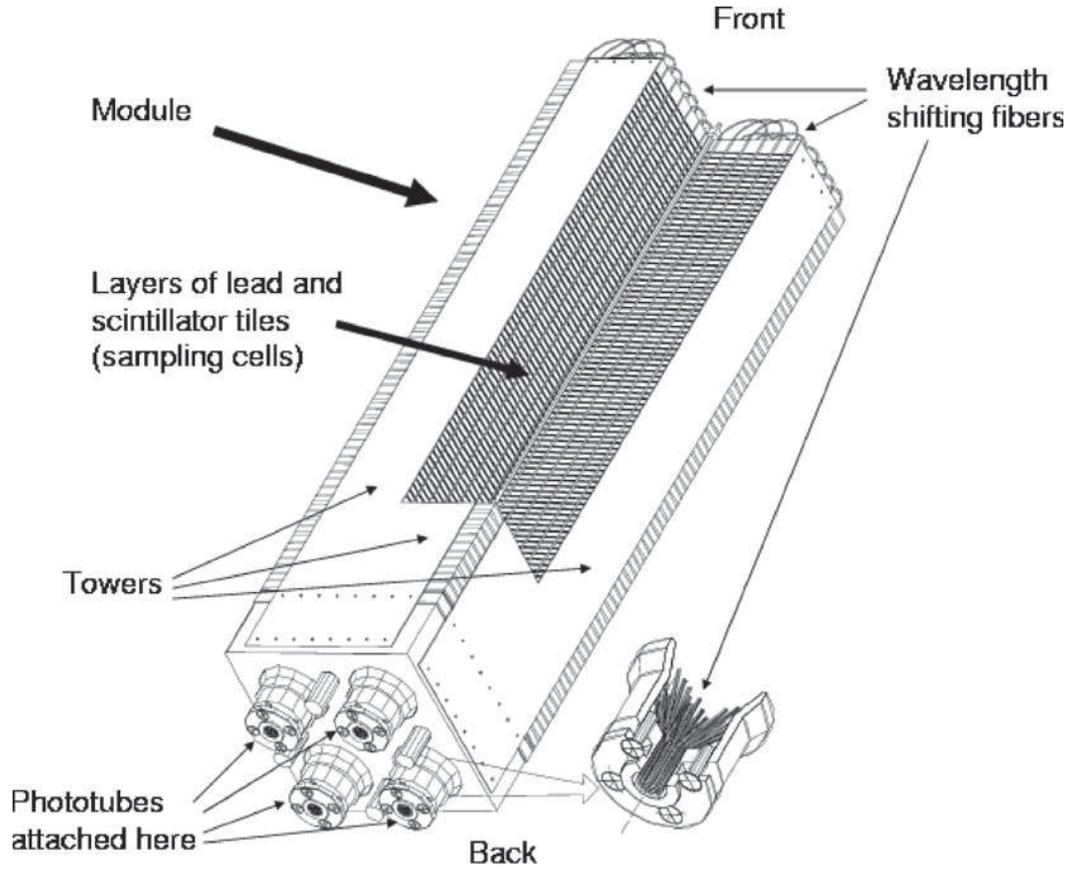}
  \caption[PbSc Module]{View of a PbSc module showing the layers of Pb
    and scintillator, the wavelength shifting fibers and the
    phototubes attached to the back.}
  \label{fig:pbsc_module}
\end{figure}

The PbSc calorimeter contains a total of 15,552 individual towers
which are made of 66 sampling cells with alternating layers of 1.5~mm
Pb and 4~mm scintillator
(1.5\%PT/0.01\%POPOP)~\cite{Aphecetche2003}. A module as shown
in~\fig{fig:pbsc_module} comprises of four optically isolated towers
which are read out individually. Each tower has measures $5.535 \times
5.535$~cm$^2$ across and has a length of 37.5 cm, which corresponds to
18~$X_0$. 36 modules are held by a common support structure called
super module. 18 super modules form a sector. The energy resolution of
the PbSc Calorimeter is
\begin{equation}\label{eq:pbsc_res}
  \frac{\sigma_E}{E} = \frac{8.1\%}{\sqrt{E}} \oplus 2.1\% .
\end{equation}

\begin{figure}[t]
  \centering
  \includegraphics[width=1.0\textwidth]{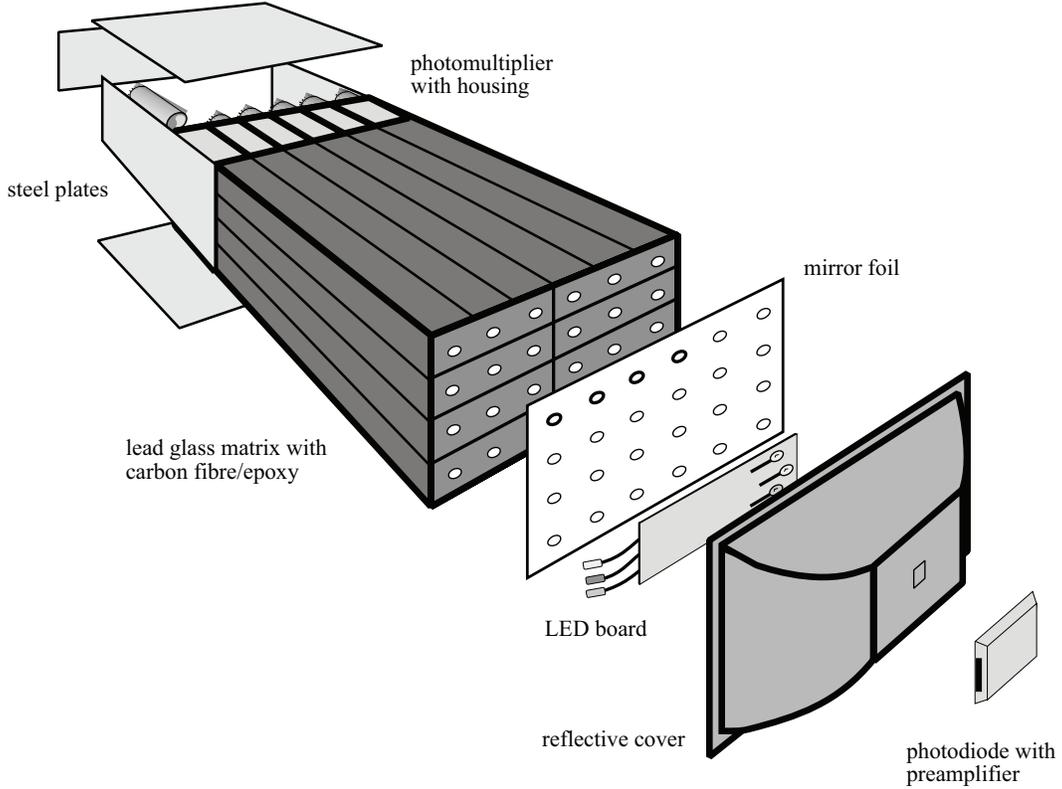}
  \caption[PbGl Super Module]{View of a PbGl super module.}
  \label{fig:pbgl_module}
\end{figure}

Each sector of the PbGl calorimeter comprises of 192 super modules
(SM) which contain each 24 modules as shown
in~\fig{fig:pbgl_module}. Each module measures $4 \times 4$~cm$^2$
across, is 40 cm long (14.3 $X_0$), and read out with a
photomultiplier at its end. The energy resolution of the PbGl
Calorimeter is
\begin{equation}\label{eq:pbgl_res}
  \frac{\sigma_E}{E} = \frac{5.9\%}{\sqrt{E}} \oplus 0.76\%.
\end{equation}

With a thickness of 18~$X_0$ in the PbSc and 14.3~$X_0$ in the PbGl,
respectively, electrons and photons will deposit their energy within
the calorimeter as electromagnetic shower of subsequent Bremsstrahlung
and \ee pair creation. In the PbSc the electrons within the
electromagnetic shower created in the Pb-layer produces scintillation
light in the scintillator layers. The scintillation light is guided by
wavelength shifting fibers to the phototubes located at the back of
each tower. In the PbGl, which has a refractive index of $n = 1.648$,
the electromagnetic showers is detected by Cherenkov light radiated by
electrons in the shower. The Cherenkov light is read out at the end of
the Calorimeter by photomultiplier tubes.

\begin{figure}[t]
  \centering
  \includegraphics[width=1.0\textwidth]{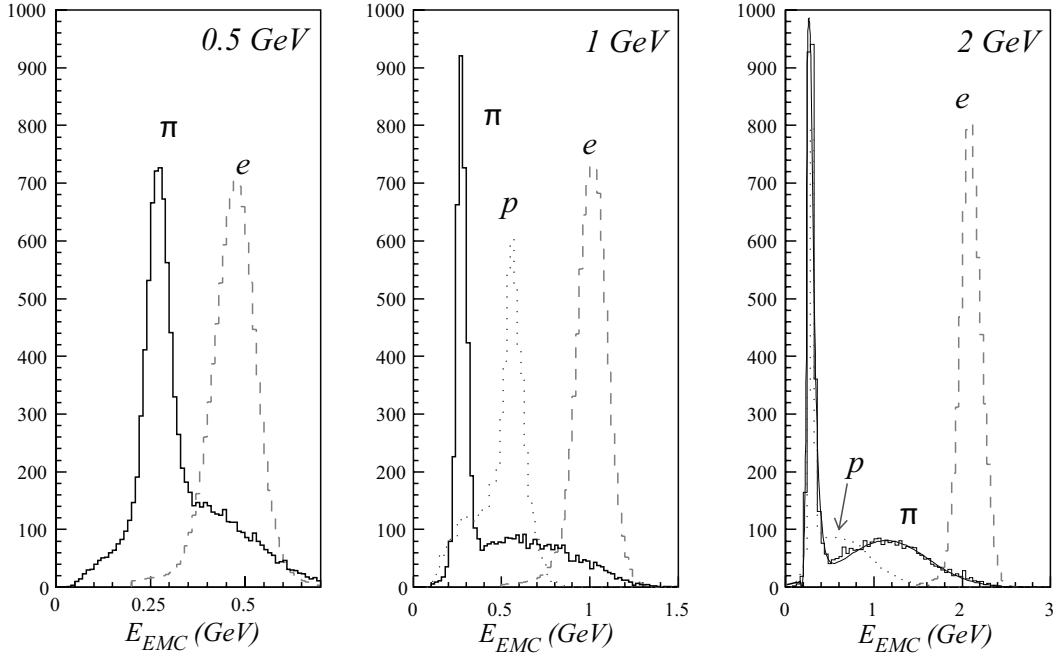}
  \caption[Energy spectrum of mip's]{Energy spectrum measured with the
    PbSc calorimeter, when exposed to pions, protons and electrons for
    incident energies of 0.5, 1, and 2~GeV~\cite{Aphecetche2003}. The
    y-axis shows counts in arbitrary units.}
  \label{fig:mips}
\end{figure}

In contrast to electrons and photons, the energy loss of hadrons in
matter occurs primarily through ionization and atomic excitations. For
typical hadron energies ($0.1 \leq E \leq 10$~GeV) the energy
deposited in matter is nearly independent of the particle's energy,
therefore these particles are called minimum ionizing particles, or
mip's. Furthermore, PbSc has a nuclear interaction length $\lambda_I =
0.85$ and PbGl $\lambda_I = 1.05$, respectively. Therefore, only few
hadrons will interact strongly and deposit a significant fraction of
their energy. This leads to mip peaks in the energy spectrum as shown
in~\fig{fig:mips} for charged pions and protons along with electrons
for comparison.

The EMCal serves as Level-1 trigger for events with high momentum
photons, triggering when the energy deposited in an area of $4\times4$
overlapping towers surpasses a defined threshold. In addition the
energy in a area of $2\times2$ overlapping towers can be used in
coincidence with the RICH trigger to trigger on events with electron
candidates.

\subsection{Charged Particle Acceptance}
\label{sec:acceptance}

\begin{figure}
  \centering
  \includegraphics[width=0.9\textwidth]{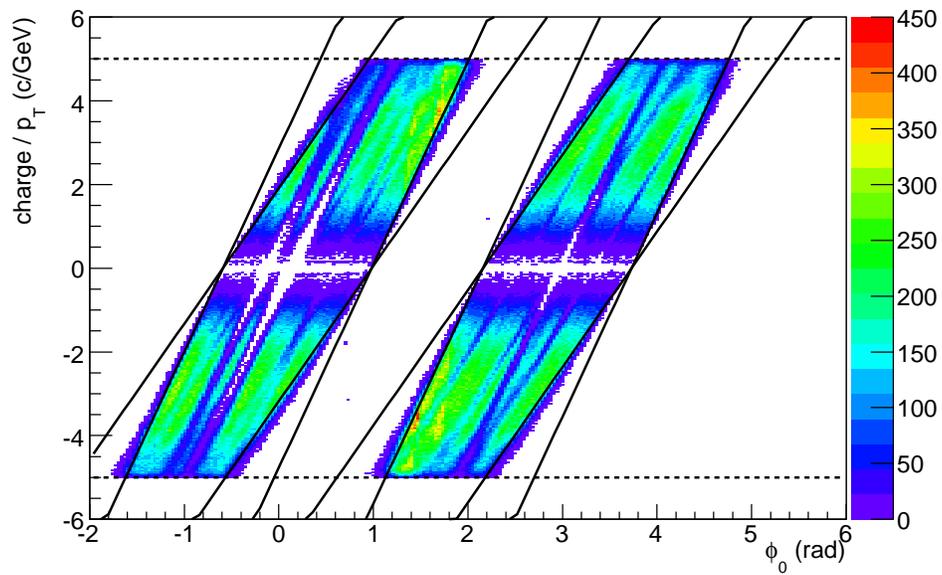}
  \caption[Single electron acceptance parameterization]{Single
    electron acceptance. The diagonal line represent the acceptance
    limits due to RICH and DC as defined in \eq{eq:track_acc}. The
    dashed lines indicate the low \pt cut off at 200~\mevc.}
  \label{fig:single_acc}
\end{figure}

The central arm acceptance of charged tracks depends on their charge
sign $q$, their transverse momentum and their azimuthal angle. The
magnetic field of the central magnets will bend a particle emitted
under $\phi_0$ and momentum \pt; its azimuthal angle measured at a
distance $r$ from the vertex increases inverse proportional to
\pt. Therefore the relation between azimuthal angle at the vertex and
at the DC (RICH) is given by:
\begin{subequations}\label{eq:phi_acc}
  \begin{align}
    \phi_{\rm DC} &= \phi_0 + q\frac{k_{\rm DC}}{\pt}\label{eq:dc_phi}\\
    \phi_{\rm RICH} &= \phi_0 + q\frac{k_{\rm RICH}}{\pt}\label{eq:rich_phi}
  \end{align}
\end{subequations}
An electron is accepted if its azimuthal angle is within the coverage
of DC and RICH in one of the two central arms, \ie:
\begin{equation}\label{eq:track_acc}
  \phi_{\rm min} < \phi_{\rm DC} \leq \phi_{\rm max}\quad \&\&\notag\quad \phi_{\rm min} < \phi_{\rm RICH} \leq \phi_{\rm max}
\end{equation}
\fig{fig:single_acc} shows the distribution of single electrons
measured in \pp collisions in $q/\pt$ vs. $\phi_0$. The west arm
coverage extends around $\phi \approx 0$ and the the east arm around
$\phi \approx \pi$. The shapes can be described by the low \pt cut of
0.2~\gevc limits the distribution within $|q/\pt| < 5~c$/GeV shown as
horizontal dashed black lines and the conditions defined in
\eq{eq:track_acc}, with $k_{\rm DC} = 0.206$ rad \gevc, $k_{\rm RICH}
= 0.309$ rad \gevc, $\phi_{\rm min} = -\frac{3}{16}\pi$~rad and
$\phi_{\rm max} = \frac{5}{16}\pi$~rad for the west arm and $\phi_{\rm
  min} = \frac{11}{16}\pi$~rad, and $\phi_{\rm max} =
\frac{19}{16}\pi$~rad for the east arm, respectively, whose boundaries
are shown as solid black lines.

\chapter{Analysis}
\label{cha:analysis}

In this chapter the analysis of the dielectron continuum in \pp as
well as \AuAu collisions at a \sqrtsnn = 200GeV is presented. The
analysis procedure of both data sets has large overlap so they are
discussed in parallel, but emphasizing differences where they are
present.

The major steps of the analysis are outlined briefly here. The
analysis begins with the event selection in
Section~\ref{sec:global_evt}, which is discussed for \pp in
Section~\ref{sec:global_evt_pp} and in
Section~\ref{sec:global_evt_auau} for \AuAu based on trigger and
vertex information. This is followed by a single electron analysis,
including the tracking discussed in Section~\ref{sec:tracking} and
electron identification in Section~\ref{sec:eid}.

Single electrons are paired to electron-positron pairs as explained in
Section~\ref{sec:pair_ana}. The subtraction of combinatorial and
correlated background is discussed in the
Sections~\ref{sec:event_mixing}--\ref{sec:like_subtraction}. The
resulting invariant mass and \pt spectra of \ee pairs are corrected
for reconstruction and electron identification efficiencies as well as
trigger efficiencies as presented in Section~\ref{sec:eff_corr}. The
\pt spectra of \ee pairs require an additional correction for the
detector acceptance which is discussed in Section~\ref{sec:acc_corr}.

Section~\ref{sec:exodus} presents a calculation {\em(Cocktail)} of the
expected \ee pair yield from hadronic decays. The results will be
compared to the experimental data in Chapter~\ref{cha:results}.

\section[Data Set and Event Selection]{Data Set and Event Selection}
\label{sec:global_evt}

\subsection[\pp Collisions]{$\boldsymbol{p+p}$ Collisions}
\label{sec:global_evt_pp}

The analysis is based on the data set of \pp collisions at \sqrts =
200 GeV which was collected during the run period in 2005.

Two data samples are used for the measurement of the dielectron
continuum: a reference sample of events which were selected by the
minimum bias trigger (MB) and a data set recorded with the single
electron trigger (ERT). The MB trigger for \pp collisions requires
that the BBC has registered hits in at least two photo tubes and an
online determined collision vertex within 30~cm:
\begin{equation}\label{eq:pp_mb}
  {\rm MB} \equiv ({\rm BBC} \geq 2) \cap (|z_{\rm vertex}| < 30~{\rm cm})
\end{equation}

The MB trigger cross section is $\sigma_{\rm BBC} = 23.0 \pm 2.2$~mb
corresponding to $54.5 \pm 6$\% of the inelastic \pp cross section
$\sigma_{pp} = 42.2$~mb. Simulations, and data collected without
requiring the BBC trigger, indicate that the triggered events include
$79 \pm 2$\% of events with particles in the central arm
acceptance. This number coincides with the fraction of non-diffractive
events triggered by the BBC from which it is concluded that for
non-diffractive collisions the BBC trigger can have only little bias
towards events with particles produced in the central arms.

\begin{figure}[h]
  \centering
  \subfloat[]{\label{fig:zvtx_events}\includegraphics[width=0.44\textwidth]{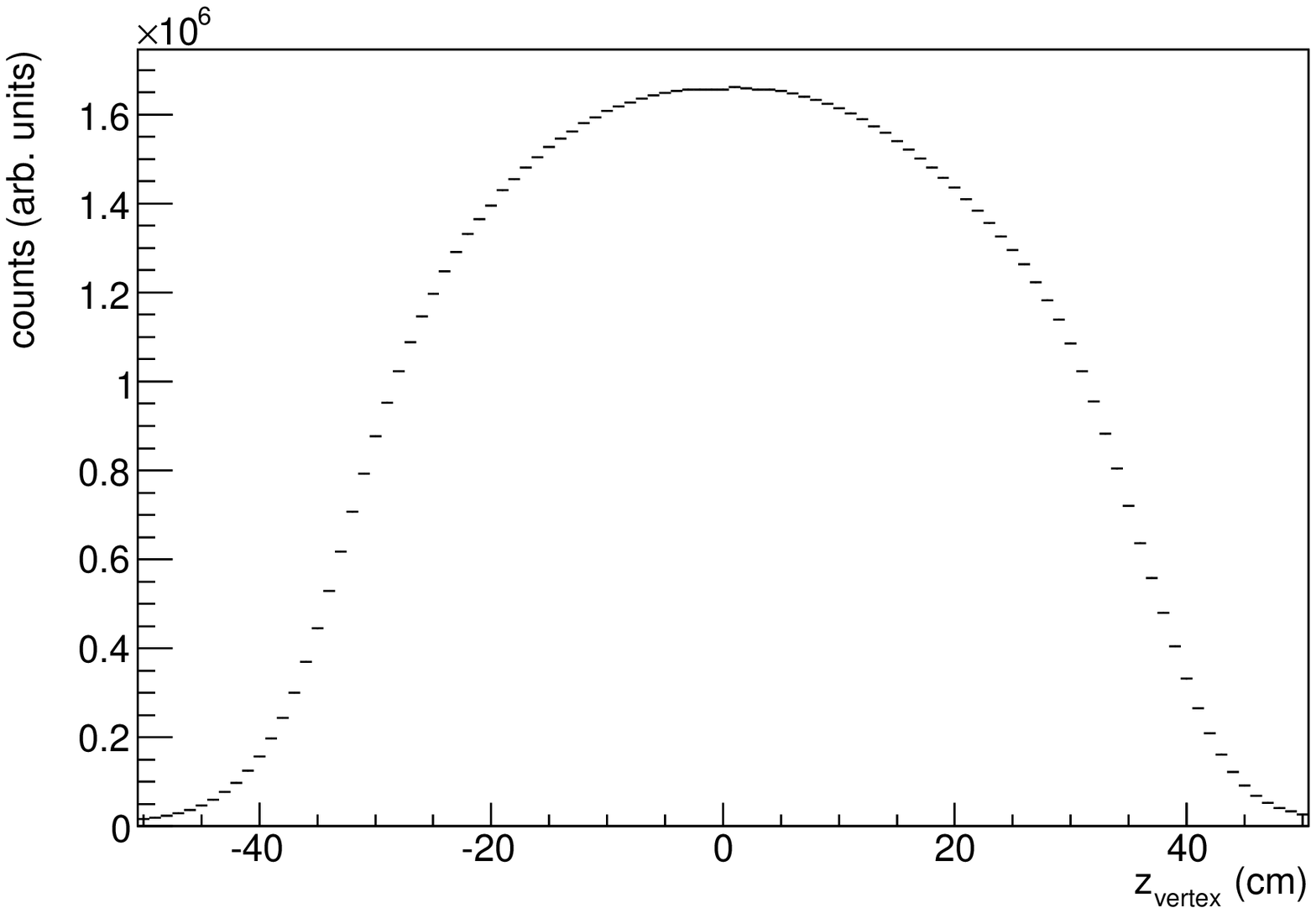}}
  \subfloat[]{\label{fig:zvtx_tracks}\includegraphics[width=0.44\textwidth]{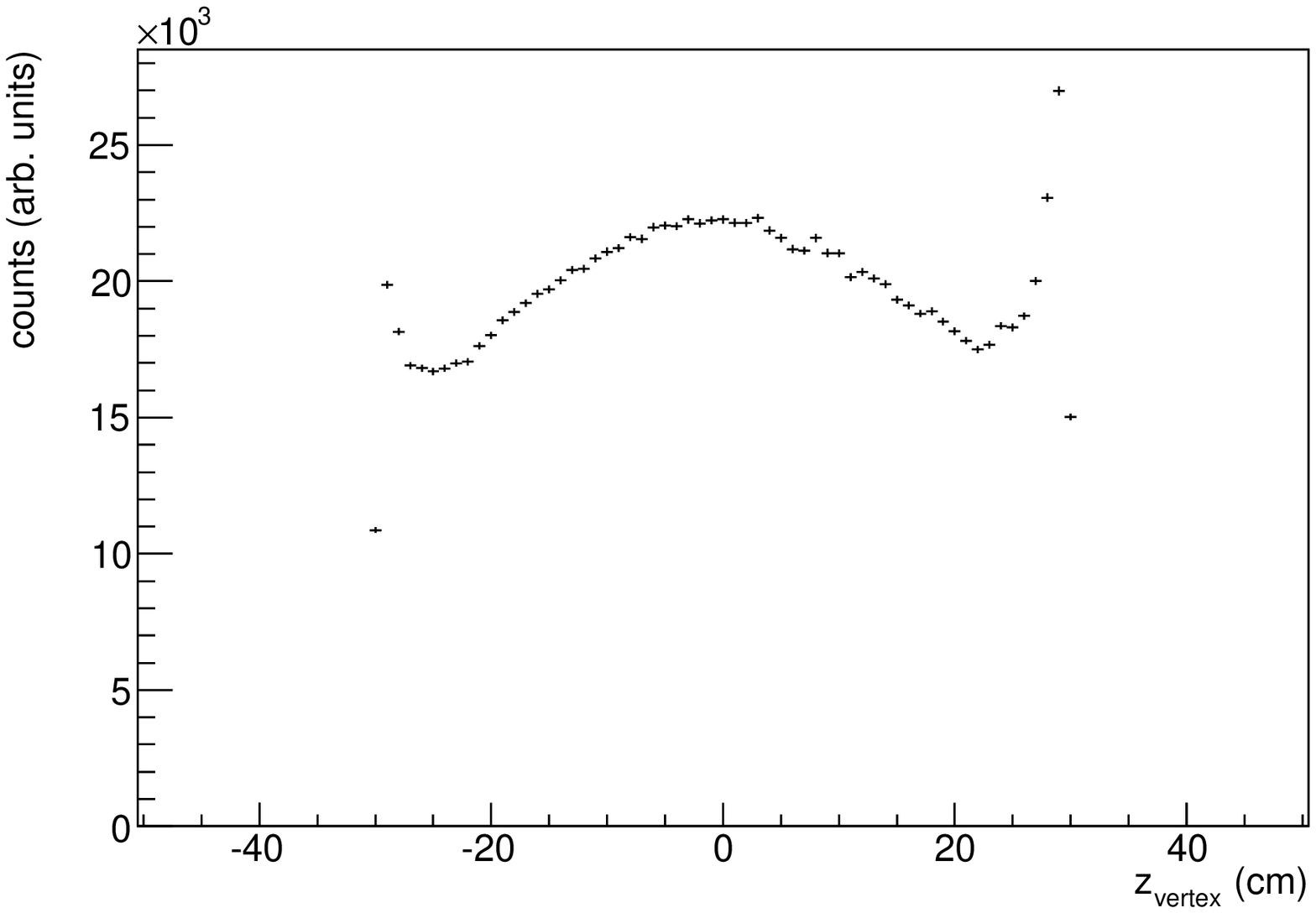}}
  \caption[Distribution of the Collision Vertex]{$z_{\rm vertex}$
    distribution of events \subref{fig:zvtx_events} and of electrons
    \subref{fig:zvtx_tracks} in \pp collisions at \sqrts = 200 GeV.}
  \label{fig:zvtx}
\end{figure}

Events with a collision vertex far from the origin have a higher
chance to create particles that hit the nose cones of the central
magnet. This creates additional particles, \eg, from photon
conversions which are reconstructed in the central arm detectors that
do not originate from the actual collision. The $z_{\rm vertex}$
distribution of events is shown in~\fig{fig:zvtx_events}. While this
distribution is centered around $z_{\rm vertex} = 0$~cm and has a full
width at half maximum of $\approx 30$~cm, the number of electrons in
the central arm peaks strongly for $|z_{\rm vertex}| > 25$~cm, as
shown in~\fig{fig:zvtx_tracks}. Therefore, an vertex cut of $z_{\rm
  vertex} < 25$~cm, with the vertex position determined offline by the
BBC with better accuracy, is applied to avoid such contamination.
Although with $\sim 4$~kHz the rate at which PHENIX can record data
exceeds the capability of any other RHIC experiments, it would not be
useful to only record events based on the MB trigger if one wants to
study rare events, containing, \eg, electrons. Therefore, PHENIX has a
variety of Level-1 triggers to select particularly interesting events,
\eg, events with a high \pt particle in the central arm or an electron
candidate. They are called ERT (EMCal RICH triggers), as they are
based on the energy measurement in the EMCal, which in case of the
electron trigger is matched to hits in the RICH.

The high \pt photon trigger is based on the energy measured within a
$4\times4$ neighboring EMCal towers. If this energy exceeds a set
threshold, the trigger is fired. During the \pp run in 2005 there were
three triggers with different energy thresholds:
\begin{itemize}
\item ERTLL1\_4x4a: $E > 2.1$~GeV
\item ERTLL1\_4x4b: $E > 2.8$~GeV
\item ERTLL1\_4x4c: $E > 1.4$~GeV
\end{itemize}

In addition to the high \pt photon trigger an electron trigger
(ERTLL1\_E) was active. This trigger requires a minimum deposited
energy of 400~MeV in an overlapping tile of 2x2 EMCal towers matched
to a hit in the RICH. The RICH hit is required within a trigger tile
of 4x5 photo tubes. The location of the RICH trigger depends on the
momentum of the trigger particle and is determined from a look-up
table, assuming an electron to be the trigger particle, \ie the
momentum being equal to the energy deposited within the 2x2 EMCal
towers. Only events which are triggered in coincidence with the MB
trigger (ERTLL1\_E\&BBCLL1) are considered for the analysis so that a
cross section for MB collisions can be extracted. The efficiency of
the ERT trigger is discussed in Section~\ref{sec:ert_eff}. When in the
following the term ``ERT trigger'' is used, it refers to the
ERTLL1\_E\&BBCLL1 trigger unless stated otherwise.

The downscale factors on the MB trigger changes frequently to adjust
for changing beam conditions. These downscale factors are recorded in
a database and have to be considered when determining the total
luminosity recorded.

\begin{figure*}[t]
  \centering
  \subfloat[$N_{\mathrm{ERT}}^{\mathrm{MB}}/N_{\mathrm{MB}}^{\mathrm{ERT}}$ as function of run \#]{\label{fig:ratiorbr}\includegraphics[width=0.44\textwidth]{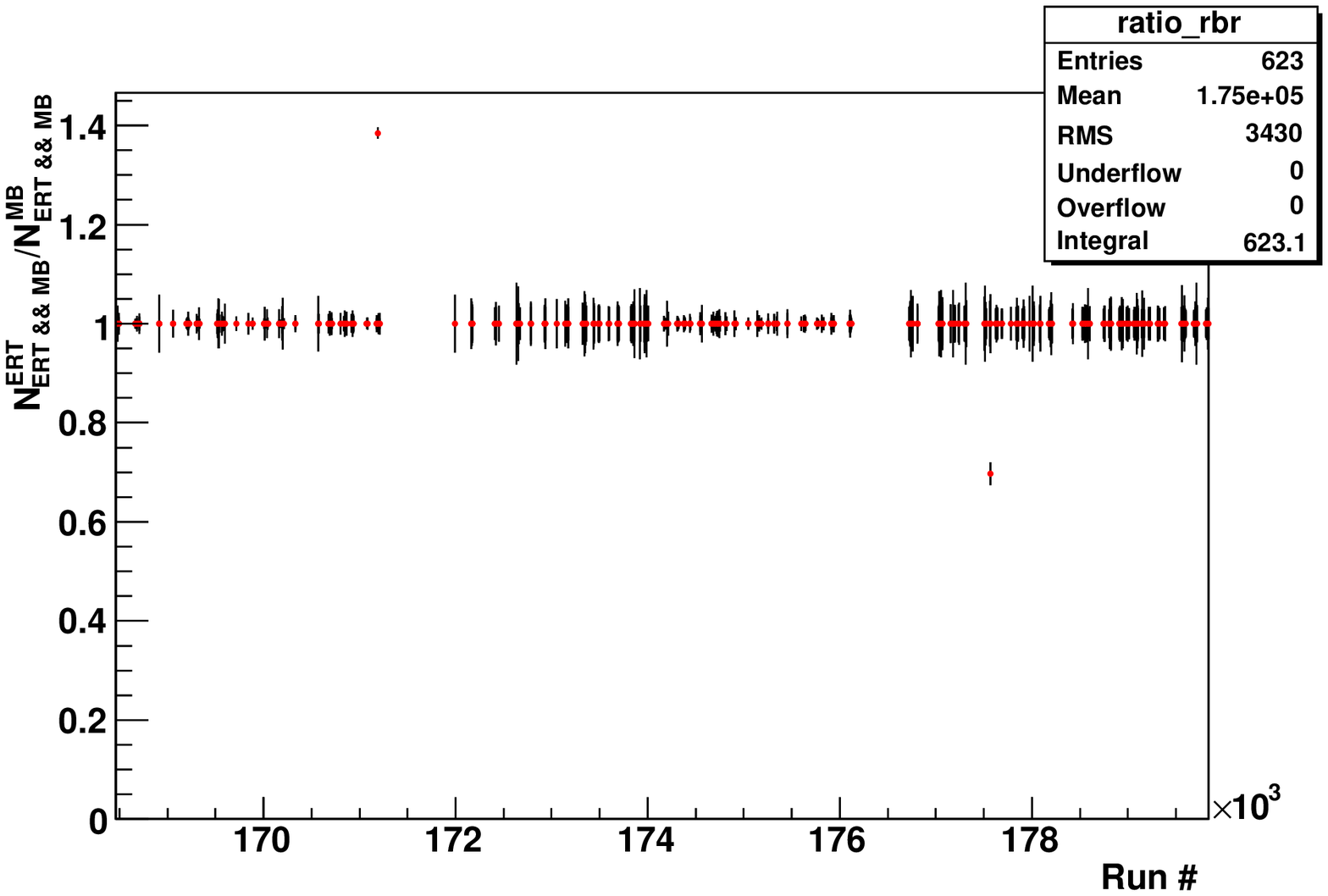}}
  \subfloat[$N_{\mathrm{ERT}}^{\mathrm{MB}}/N_{\mathrm{MB}}^{\mathrm{ERT}}$]{\label{fig:ratioint}\includegraphics[width=0.44\textwidth]{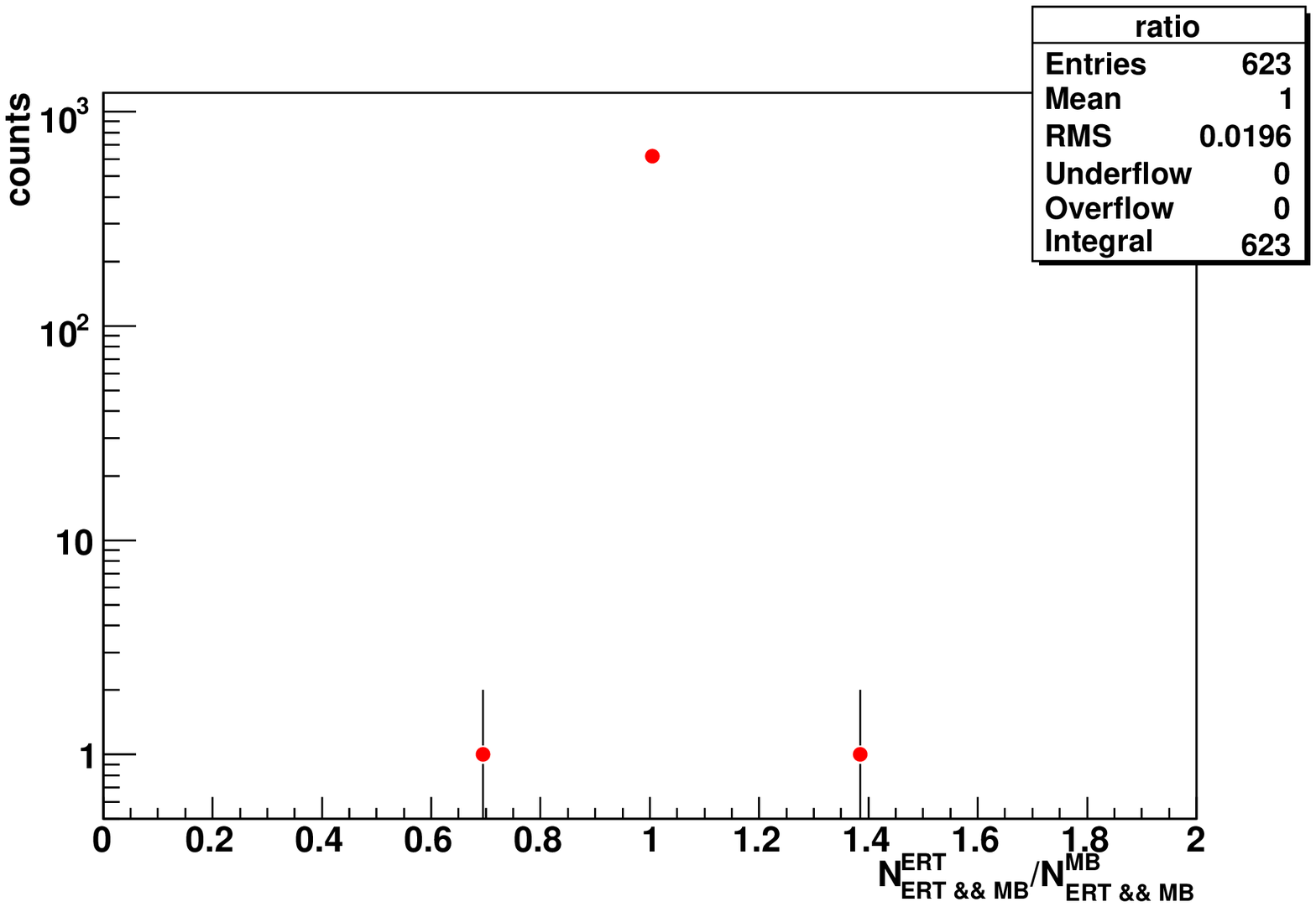}}
  \label{fig:ratio}
  \caption{Ratio of triggered events in the ERT and MB samples}
\end{figure*}

This analysis uses a MB data sample of 1.5 billion events and an ERT
triggered sample of 270 million events. As the ERT trigger requires
the coincidence of the MB trigger one can calculate the number of
sampled MB events that the ERT data set of a given run corresponds to
by as:
\begin{equation}\label{eq:nevt}
  N_{\rm MB}^{\rm sampled} = N_{\rm MB} \cdot f_{\rm scale-down} \cdot N_{\rm ERT}^{\rm MB}/N_{\rm MB}^{\rm ERT}
\end{equation}
Where $N_{\rm MB}$ is the number of events recorded with the MB
trigger in this run and $f_{\rm scale-down}$ is the scale-down factor
for the MB trigger.  The correction factor $N_{\rm ERT}^{\rm
  MB}/N_{\rm MB}^{\rm ERT}$ is necessary in occasions when during the
data reconstruction a file segment of either the MB sample or the ERT
sample is lost. In this case the number of ERT triggered events in the
MB sample ($N_{\rm MB}^{\rm ERT}$) does not equal the number of MB
triggered events in the ERT sample ($N_{\rm ERT}^{\rm MB}$). This
ratio is shown in \fig{fig:ratiorbr} for all run numbers. Runs for
which this ratio is $>2$ or $<0.5$ are rejected (5 runs), all others
which are not equal to 1 are corrected (2 runs).

The total number of MB events sampled by the ERT trigger is the sum
of~\eq{eq:nevt} over all runs:
\begin{equation}
  N_{\rm MB}^{\rm sampled} = 5.18558\times 10^{10}\nonumber
\end{equation}

\subsection[\AuAu Collisions]{$\boldsymbol{\rm Au + Au}$ Collisions}
\label{sec:global_evt_auau}

The analysis of the dielectron continuum in \AuAu at \sqrtsnn = 200
GeV is performed on the data set recorded during the RHIC running time
in 2004. Collisions were selected based on a minimum bias trigger,
requiring at least two hits in each of the BBC and a coincident hit in
one of the ZDCs. In addition the collision vertex had to be within
38~cm:
\begin{equation}\label{eq:auau_mb}
  {\rm MB} \equiv ({\rm BBCN} \geq 2) \cap ({\rm BBCS} \geq 2) \cap ({\rm ZDCS} \geq 1 \cup {\rm ZDCN} \geq 1) \cap (|z_{\rm vertex}| < 38~{\rm cm})
\end{equation}
$92_{-3.0}^{+2.5}$\% of the inelastic \AuAu cross section are selected
by this trigger. For the same reason as in \pp a vertex cut is applied
at $z_{\rm vertex} < 25$~cm in the offline analysis.

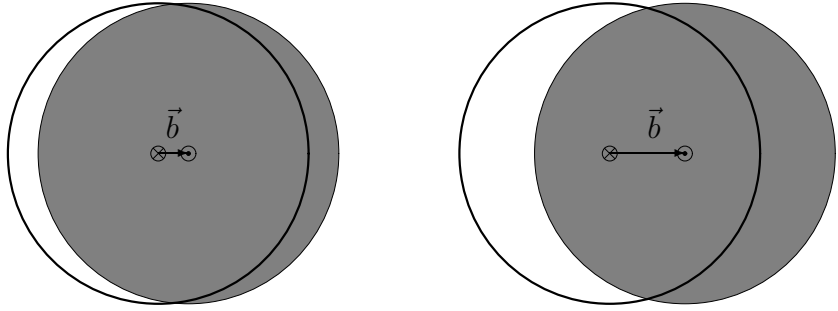
\begin{figure}
  \centering
  \begin{picture}(120, 60)
    \put(34, 30){\shade\circle{40}}
    \put(30, 30){\thicklines\circle{40}}
    \put(34,30){\circle{2}}
    \put(34,30){\circle*{0.5}}
    \put(30,30){\circle{2}}
    \put(29.3,29.3){\line(1,1){1.41}}
    \put(29.3,30.7){\line(1,-1){1.41}}
    \put(30,30){\vector(1,0){4}}
    \put(31,32){$\vec{b}$} 
    \put(100, 30){\shade\circle{40}}
    \put(90, 30){\thicklines\circle{40}}
    \put(100,30){\circle{2}}
    \put(100,30){\circle*{0.5}}
    \put(90,30){\circle{2}}
    \put(89.3,29.3){\line(1,1){1.41}}
    \put(89.3,30.7){\line(1,-1){1.41}}
    \put(90,30){\vector(1,0){10}}
    \put(95,32){$\vec{b}$} 
  \end{picture}
  \caption{Overlap of two gold nuclei for different impact parameters $\vec{b}$.}
  \label{fig:auau_overlap}
\end{figure}

In contrast to protons, gold ions cannot be treated as point-like
particles. They consist of 197 nucleons, 79 protons an 119
neutrons. When accelerated to an energy of 200~GeV per nucleon they
are Lorentz contracted in longitudinal directions and look much like
pan cakes with a radius of $\approx 7$~fm in the lab frame. Not every
collision of two gold ions is head on, but their centers are shifted
by a distance called the impact parameter $\vec{b}$ as illustrated in
\fig{fig:auau_overlap}.

The number of nucleons participating in a collision $N_{\rm part}$,
depends on the impact parameter. Every nucleon that participates in a
collision can undergo a number of binary collisions $N_{\rm coll}$
with other nucleons. Both numbers can be calculated with a Glauber
Monte Carlo simulation, which is a simple geometric model of the
nuclei in which the nucleons are distributed following a Wood-Saxon
potential. For the centrality classes used in this analysis the
results of such a Glauber Monte Carlo
simulations~\cite{Reygers:2003AN169} are tabulated
in~\tab{tab:glauber}.
\begin{table}
  \centering
  \caption[Glauber Monte Carlo Simulation]{\label{tab:glauber}Average
    values of $N_{\rm part}$, $N_{\rm coll}$, and $\vec{b}$ for
    different centrality classes of \AuAu and \pp collisions at \sqrtsnn = 200
    GeV~\cite{Reygers:2003AN169}. Quoted errors are systematic
    uncertainties. Also listed are the number of events and the number
    of signal \ee pairs for each centrality class.\\}
  \begin{tabular}{rr@{.}l@{ $\pm$ }r@{.}lr@{ $\pm$ }lr@{.}l@{
        $\pm$ }r@{.}lr@{.}l@{ $\times$ }lr@{.}l@{ $\times$ }l}
    \toprule
    Centrality & \multicolumn{4}{c}{$\langle N_{\rm part} \rangle$} &
    \multicolumn{2}{c}{$\langle N_{\rm coll} \rangle$} &
    \multicolumn{4}{c}{$\langle \vec{b}\, \rangle$} & \multicolumn{3}{c}{$N_{\rm evt}$} & \multicolumn{3}{c}{Signal Pairs}\\\midrule
    \multicolumn{17}{c}{\AuAu}\\\midrule
    0--10\%  & 325&2 & 3&3 & 955 & 94  & 3&2  & 0&2       &  8&6 & $10^7$  & 9&2 & $10^4$ \\
    10--20\% & 234&6 & 4&7 & 603 & 59  & 5&7  & 0&3       &  8&6 & $10^7$  & 6&6 & $10^4$ \\
    20--40\% & 140&4 & 4&9 & 297 & 31  & 8&1  & 0&3       &  1&7 & $10^8$  & 8&1 & $10^4$ \\
    40--60\% & 59&95 & 3&6 &  91 & 12  & 10&4 & 0&4       &  1&7 & $10^8$  & 3&3 & $10^4$ \\
    60--92\% &  14&5 & 2&5 &14.5 & 4.0 & 13&0 & 0&5       &  2&9 & $10^8$  & 1&1 & $10^4$ \\
    0--92\%  & 109&1 & 4&1 & 258 & 25  & 9&5  & 0&4       &  8&1 & $10^8$  & 28&3& $10^4$\\\midrule
    \multicolumn{17}{c}{\pp}\\\midrule
    MB~ & \multicolumn{4}{c}{2} & \multicolumn{2}{c}{1} & \multicolumn{4}{c}{0} &  1&5 & $10^9$ & 1&4 & $10^4$ \\
    ERT & \multicolumn{4}{c}{2} & \multicolumn{2}{c}{1} & \multicolumn{4}{c}{0} &  2&7 & $10^8$ & 22&8 & $10^4$\\
    \bottomrule
  \end{tabular}
\end{table}

\section{Charged Particle Tracking}
\label{sec:tracking}

\begin{figure}
  \centering
  \includegraphics[height=0.4\textheight]{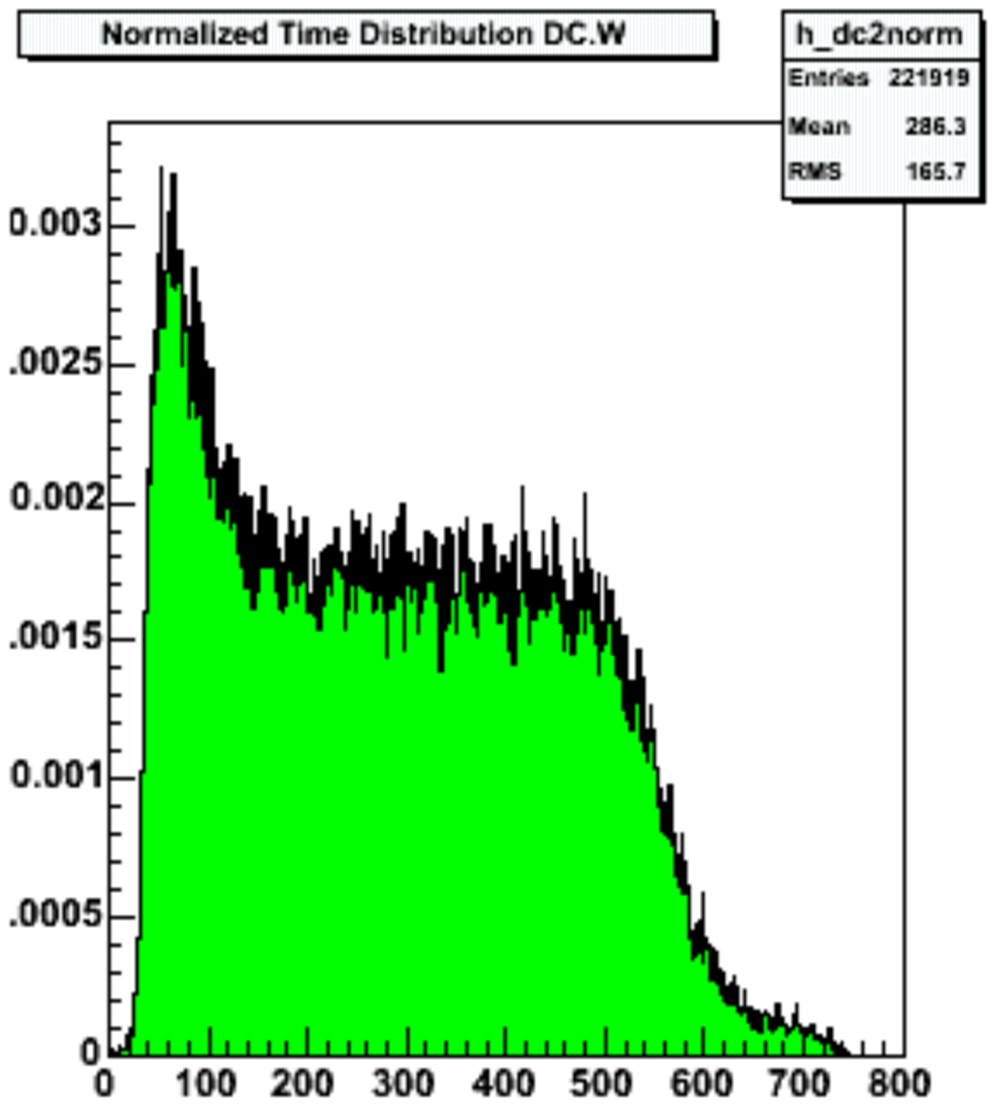}
  \caption[Timing Distribution of DC Hits]{Example of a time
    distribution of hits in the Drift Chamber West arm summed over all
    wires (collected over $\approx 15,000$ \pp events). The x-axis
    displays time in units of ns.}
  \label{fig:dc_timing_dist}
  ~\\
  \includegraphics[height=0.4\textheight]{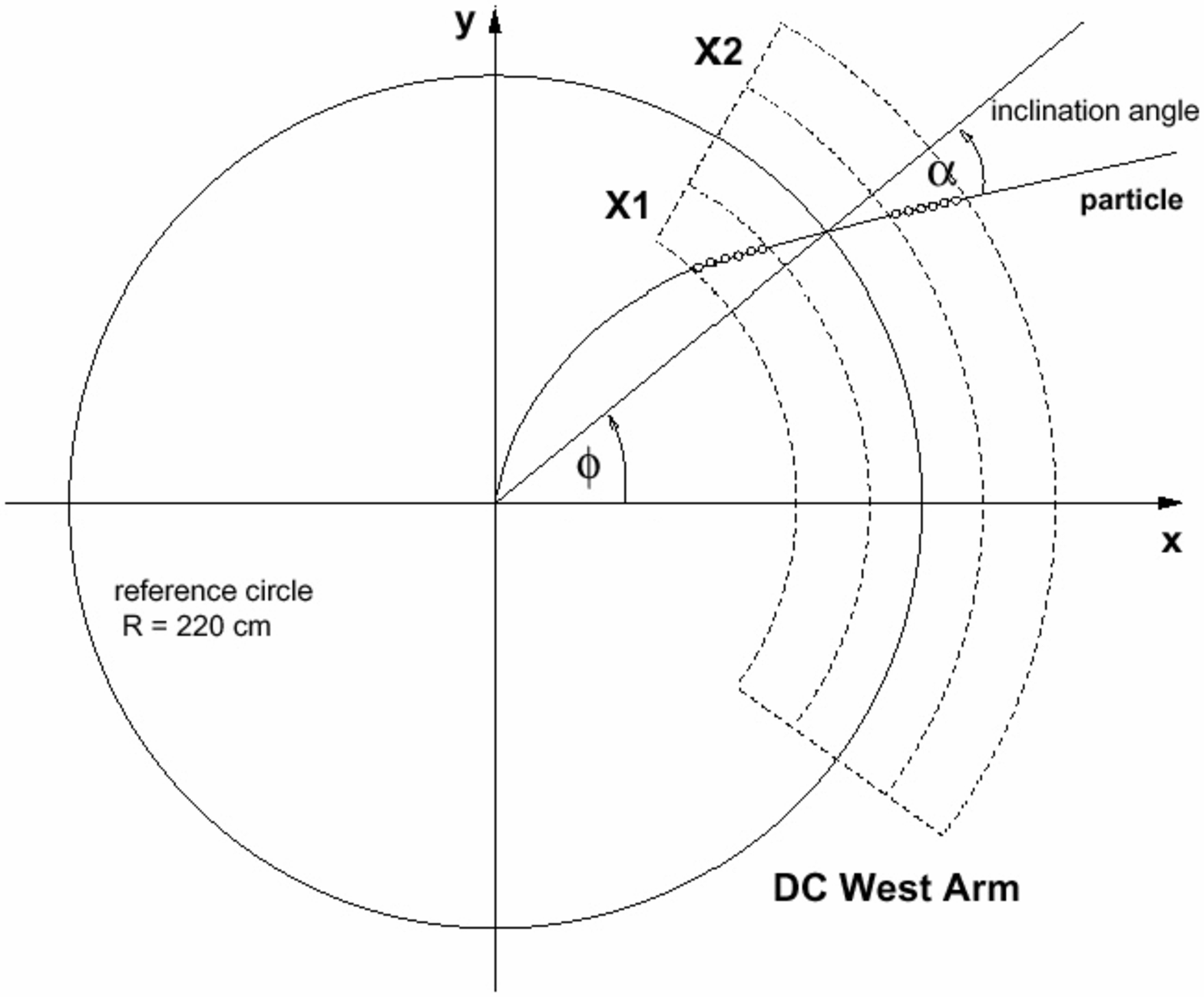}
  \caption[Definition of Track Reconstruction Coordinates]{Definition
    of the coordinates $\phi$ and $\alpha$ used in the Drift Chamber
    track reconstruction.}
  \label{fig:dc_tracking}
\end{figure}

With the help of the Drift and Pad Chambers charged particles are
tracked down to momenta of 200~\mevc with a resolution of
\begin{equation}\label{eq:dc_res}
  \sigma_p/p = 0.7\% \oplus 1\%p/(\gevc).
\end{equation}
A track is reconstructed in the $r-\phi$ plane with six measurements
in the X1 and another six the X2 plane of DC over a radial distance of
20~cm. The drift time after which a hit is registered by the anode
wire is measured with respect to the collision time $t_0$ which is
measured by the BBC. The working gas is chosen such that the drift
velocity is uniform in the active region. This allows to convert the
time after which a hit is measured by a particular wire into a
distance at which the track passed this wire:
\begin{equation}\label{eq:drifttime}
  x(t) = v_{\rm drift} \cdot t.
\end{equation}
The drift velocity is obtained by measuring the time distribution of
all hits in a DC arm. A typical distribution is shown
in~\fig{fig:dc_timing_dist}. The distribution has a characteristic
peak which is about twice as high as the following plateau. This is
due to tracks within a distance of 2~mm from the anode wire, the
region in which the back wires do not shield the field and the
left-right ambiguity remains and therefore twice as many hits are
registered as outside that region. The time interval between leading
edge and trailing edge corresponds to a drift distance of 2~cm, which
is the separation between the cathode and anode wires. The drift
velocity is calibrated based on these time distributions for every
run. The expected value for the DC working gas is: $v_{\rm drift} =
50~\mu\rm{m}/\rm{ns}$. Further calibrations are necessary to consider
effects due to the signal width dependence of the arrival time, edge
effects in the drift field, geometrical shifts between wires and the
global alignment with respect to the collision vertex.

Once the hit times are converted into distances a track candidate is
reconstructed with a combinatorial Hough transformation over all
possible hit combinations~\cite{johnson:1998}. This is done for in a
2-dimensional space of the coordinates $\phi$ and $\alpha$. The angle
$\phi$ is the azimuthal angle of a track candidate at the DC reference
radius of 223~cm, as shown in~\fig{fig:dc_tracking}. The inclination
$\alpha$ of the track candidate with respect to an infinite momentum
(\ie, straight) track at angle $\phi$ is inverse proportional to the
momentum and its sign depends on the charge of the particle. Any
combination of hits that have a local maximum and surpass the
threshold criterion are considered a track candidate. After this Hough
transformation further steps are performed to remove background
tracks. First, hits are tested on their association to a track. A fit
is performed that weights hits according to their distance from the
track, such that the weight goes to zero for hits far away from the
track. Then it is required that each hit can only belong to one
track. Any track with less than eight hits is rejected.

In addition the hits are measured in the UV wires as well as PC1 hits
are matched to track candidates to reconstruct the polar angle of the
track candidate. If there are more than one PC1 hit that can be
associated to a track, the hit associated with the most UV hits is
used. The following bits are used to determine the track quality:
\begin{verbatim}
0 (1)   X1 used
1 (2)   X2 used
2 (4)   UV found
3 (8)   UV unique
4 (16)  PC1 found
5 (32)  PC1 unique

Valid patterns include:

49,50,51  1 1 0 0 x x PC1 found/unique, no UV
61,62,63  1 1 1 1 x x PC1 found/unique, UV found/unique

17,18,19  0 1 0 0 x x PC1 found/ambiguous, no UVs
21,22,23  0 1 0 1 x x PC1 found/ambiguous, UV found but tied
29,30,31  0 1 1 1 x x PC1 found/ambiguous, UV found w/ one best choice
\end{verbatim}

The highest quality tracks can have is 63, \ie, it is reconstructed
based on hits in the X1 and X2 planes, has a unique PC1 and UV hit. In
addition to these, tracks with quality 31, \ie, requiring hits in X1
and X2 plane and a unique UV hit, but only a ambiguous PC1 hit, as
well as with quality 51, \ie, demanding hits in X1 and X2 plane and a
unique PC1 hit, but without a matching UV hit, are considered in this
analysis.

The measurement of $\alpha$ allows to reconstruct the transverse
momentum of charged particles. There is a close relationship between
$\alpha$ and the magnetic field integral along the trajectory of a
particle:
\begin{equation}
  \alpha \simeq \frac{K}{\pt}
\end{equation}
where $K=206$~mrad~\gevc is the total magnetic field integral. With
the measurements of $\alpha$ and $\phi$ at the DC reference radius,
one can determine the initial azimuthal angle at the collision vertex
$\phi_0 = \phi - \alpha + \Delta\phi$, with $\Delta\phi = 0.3~{\rm
  GeV}/c~\frac{\int B\,dl}{\pt}$.

\section{Electron Identification}
\label{sec:eid}
Electrons emit an average number of 10 Cherenkov photons in the RHIC
under an angle of $cos(\theta) = 1/(n \beta)$, which are focused to
rings in the PMT plane with an asymptotic radius of $\approx 5.4$~cm.
\begin{figure}
  \centering
  \begin{picture}(75, 50)
    \put(0,0){\line(1,0){75}}
    \put(0,0){\line(0,1){50}}
    \put(0,50){\line(1,0){75}}
    \put(75,0){\line(0,1){50}}
    \Thicklines
    \texture{bbbbbbbb 0 0 0 bbbbbbbb 0 0 0
             bbbbbbbb 0 0 0 bbbbbbbb 0 0 0
             bbbbbbbb 0 0 0 bbbbbbbb 0 0 0
             bbbbbbbb 0 0 0 bbbbbbbb 0 0 0}
    \put(37.5,25){\shade\circle{43.4}} 
    \put(37.5,25){\whiten\circle{17.6}} 
    \put(37.5,25){\line(-4,1){8.5}}
    \put(31,26.8){{\small $R_{\rm min}$}}
    \put(37.5,25){\line(2,1){19.4}}
    \put(48,27){{\small $R_{\rm max}$}}
    \put(36.8,24.3){\line(1,1){1.41}}
    \put(36.8,25.7){\line(1,-1){1.41}}
    \put(34.5,21){{\small $P_{\rm cross}$}}

    \thinlines
    \multiput(15,12.2)(15,0){4}{\circle{15}}
    \multiput(7.5,25)(15,0){5}{\circle{15}}
    \multiput(15,37.8)(15,0){4}{\circle{15}}
    \put(3,23.5){{\small PMT}}
    \put(18,23.5){{\small PMT}}
    \put(10.5,11.2){{\small PMT}}
  \end{picture}
  \caption[RICH ring mask]{Ring mask on the RICH PMT array. $P_{\rm
      cross}$ is the coordinate of the track projection around which
    the ring mask (shaded) is applied. The ring mask is limited by
    $R_{\rm min} < r < R_{\rm max}$}
  \label{fig:rich_pmt}
\end{figure}
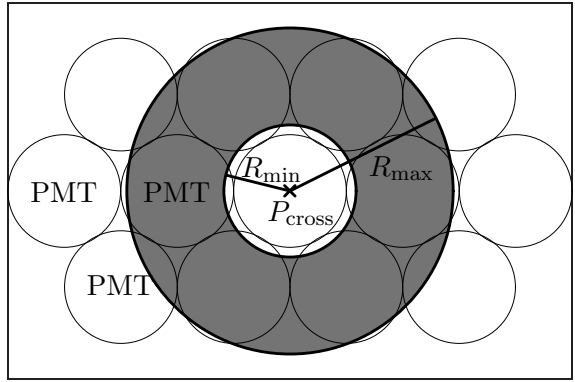

Based on the tracking information from Drift and Pad Chambers, a
projection of the track onto the PMT plane is calculated and stored in
coordinates of the PMT plane: $P_{\rm cross} = (z_{\rm cross},
\phi_{\rm cross})$. As illustrated in \fig{fig:rich_pmt}, around this
point the number of PMT hits ${\rm n}_0$ with a distance between
$R_{\rm min} = 3.3$~cm and $R_{\rm max} = 8.4$~cm are counted. Based
on the pulse height measured in a PMT the number of photo electrons
(${\rm npe}_i$) is determined. The ring center is reconstructed with
${\rm n}_0$, the location $R_i$ of the hit PMT, and the number of
photo-electrons in this area (${\rm npe}_0 = \sum_i {\rm npe}_i$) as:
\begin{equation}
  R_{\rm center} = \frac{\sum_i {\rm npe}_i \cdot R_i}{\sum_i {\rm npe}_i}
\end{equation}
with coordinates $R_{\rm center} = (z_{\rm center}, \phi_{\rm
  center})$. The distance to the track projection ({\em displacement})
is calculated as:
\begin{equation}
  {\rm disp} = \sqrt{(z_{\rm cross} - z_{\rm center})^2 - (\phi_{\rm cross} - \phi_{\rm center})^2}.
\end{equation}
In addition the quality of the measured ring shape is expressed in terms
of the difference to the expected ring shape:
\begin{equation}
  \chi^2/{\rm npe}_0 = \frac{\sum_i (R_i - R_0)^2 \cdot {\rm npe}_i}{\sum_i {\rm npe}_i}
\end{equation}
with $R_0 = 5.9$~cm.

The distance between the track projection onto the EMCal and the
position of the associated EMCal cluster is calculated in azimuthal
and $\hat{z}$ direction, emcdphi\_e and emcdz\_e, respectively. These
matching variables in units of radians and cm are converted into units
of one standard deviation $\sigma$ by fitting the distribution to a
Gaussian and stored as {\rm emcsdphi\_e} and {\rm emcsdz\_e}. As the
cluster position in the EMCal is particle species dependent due to the
different shower shapes, this calibration is done for electron
candidates. A circular 5~$\sigma$ cut on the distance is applied in
\pp while in \AuAu a 3~$\sigma$ matching is required.

In comparison to their momentum of $p > 200$~\mevc, electrons have a
negligible mass ($m_e = 511$~\kevcc). Therefore the energy they
deposit in the EMCal must match the momentum ($E = \sqrt{p^2 + m_e^2}
\simeq p$). In contrast, hadrons only deposit a fraction of their
energy in the EMCal which leads to a measured energies which are
smaller than their momenta.

\begin{figure}
  \centering
  \subfloat[\pp]{\label{fig:pp_eop_mb}\includegraphics[width=0.44\textwidth]{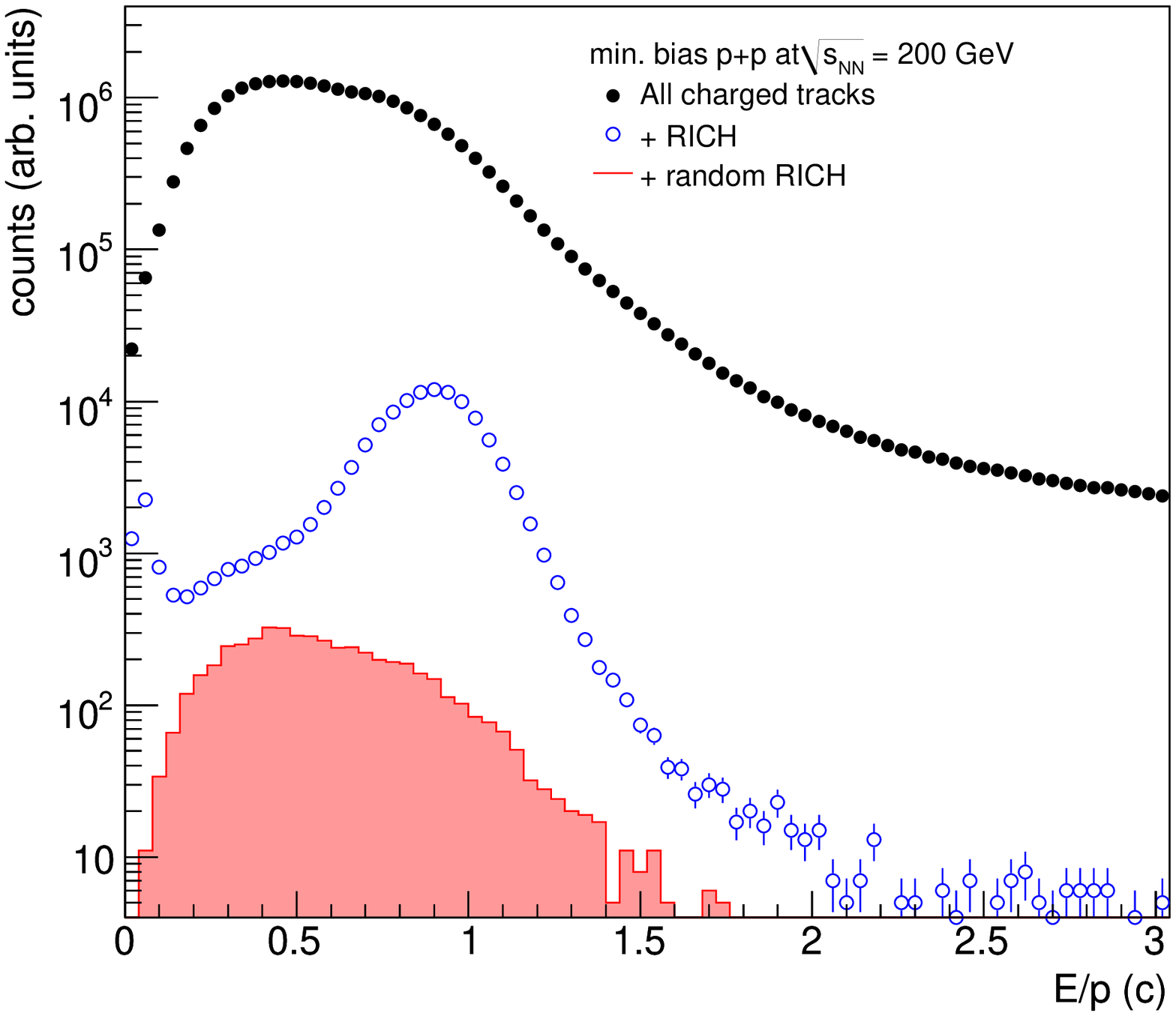}}
  \subfloat[\AuAu]{\label{fig:au_eop_mb}\includegraphics[width=0.44\textwidth]{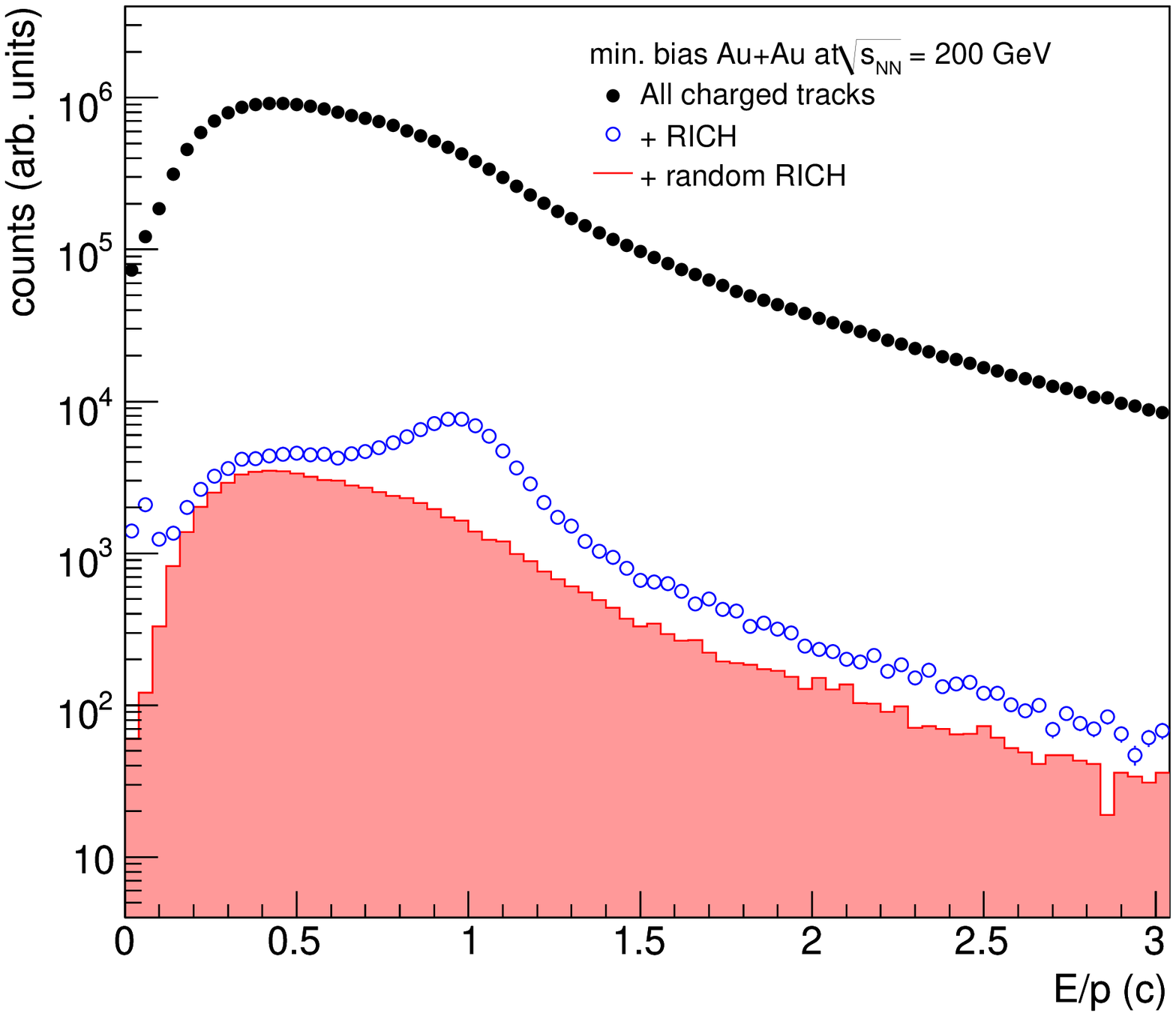}}
  \caption[Energy-Momentum matching]{$E/p$ distribution in min. bias
    \pp~\subref{fig:pp_eop_mb} and min. bias
    \AuAu~\subref{fig:au_eop_mb} for charged tracks (black), tracks
    after applying the RICH cuts (blue) and contribution of hadrons
    randomly associated to hits in the RICH (red).}
  \label{fig:eop}
\end{figure}

\fig{fig:pp_eop_mb} shows the $E/p$ distribution of charged tracks in
min. bias \pp collisions in comparison with electron candidates, \ie,
charged tracks fulfilling all but the $E/p$ eID cut. While the
distribution of all charged tracks shows no clear peak due to
electrons at $E/p = 1$, requiring the eID cut greatly improves the
signal to background ratio. The width of the peak of $\approx 14\%$ is
dominated by the energy resolution of the EMCal given in
Eqs.~(\ref{eq:pbsc_res}, \ref{eq:pbgl_res}).

The reconstructed energy does not match the momentum for all
electrons, as in some cases their shower overlaps with a photon shower
which leads to a larger energy and causes the tail at
$E/p>1$. Electrons from off-vertex decays or late conversions have a
misreconstructed momentum, as the tracking algorithm assumes all
tracks to originate from the collision vertex. Off-vertex decays
traverse less magnetic field integral and are therefore bent less,
which leads to a larger reconstructed momentum and $E/p < 1$. In the
momentum range of $p \lesssim 4$~\gevc, relevant for this analysis,
the momentum resolution of the drift chamber \eq{eq:dc_res} is better
than the energy resolution of the EMCal Eqs.~(\ref{eq:pbsc_res},
\ref{eq:pbgl_res}). Therefore, in the \ee pair analysis the invariant
mass and \pt of the \ee pair is calculated using the momentum
information rather than the energy \eq{eq:invmass}. As a consequence,
electrons with a mismeasured energy can be kept, while electrons with
misreconstructed momenta are removed with a cut on $E/p > 0.5$ in
\pp. In \AuAu the momentum dependence of the $E/p$ distribution is
corrected and the matching is stored in units of a standard deviation
${\rm dep} = (E/p -1)/\sigma_{E/p}$. This parameterization is
determined by a Gaussian fit of the $E/p$ distribution of electron
candidates as function of \pt. A value of ${\rm dep} > -2$ is required
for tracks to pass the electron identification cut.

In addition, some background remains underneath the peak which is
attributed to random associations of RICH hits to hadrons. This
contribution can be estimated by matching tracks in the north side of
the drift chamber to hits in the south side of the RICH and vice
versa. This is done by swapping the RICH sides during the track
reconstruction. In contrast to a single electron analysis, this hadron
contamination is unreducable in the pair analysis. \fig{fig:au_eop_mb}
shows the same distributions for min. bias \AuAu collisions, in which
the hadron contamination is much larger. In the region $0.7<E/p<1.2$
the contribution to the sample of electron candidates is about 1.6\%
in \pp and 16\% in \AuAu collisions All electron identification cuts
are summarized in~\tab{tab:eid}.

\begin{table}
\centering
\caption[Electron Identification Cuts]{\label{tab:eid}Electron
  identification cuts for \pp and \AuAu.\\}
\begin{tabular}{ccc}
\toprule
eID cut & \pp & \AuAu\\\midrule
quality & \multicolumn{2}{c}{$63 || 31 || 51$} \\
${\rm n}_0$   & $\geq 1$ & $\geq 2$\\
${\rm disp}$  & $< 10$~cm   & $< 5$~cm\\
$\chi^2/{\rm npe}_0$ & $< 15$~cm$^2$ & $< 10$~cm$^2$ \\
$E/p$   & $> 0.5$  & $-$\\
${\rm dep}$   & $-$      & $> -2~\sigma$\\
$\sqrt{{\rm emcsdphi\_e}^2 + {\rm emcsdz\_e}^2}$  & $< 5~\sigma$ & $< 3~\sigma$ \\
\bottomrule
\end{tabular}
\end{table}

\section{Pair Analysis}
\label{sec:pair_ana}

In an event the source of any electron or positron is unknown and
therefore all electrons and positrons are combined to {\em foreground}
pairs, like sign $FG_{++}$, $FG_{--}$ and unlike sign $FG_{+-}$. In
the \pp data set it is checked that at least one of the tracks in each
pair is associated to a hit of the ERT trigger. From the single track
information, the invariant mass \mee and the transverse momentum \pt
of the \ee pair are calculated as:
\begin{align}
  m_{ee}^2 &= (p_+ + p_-)^2\notag\\
  &= (E_+ + E_-)^2 - (\vec{p}_++\vec{p}_-)^2\label{eq:invmass}\\
  \pt^2 & = (p_{x,+} + p_{x,-})^2 + (p_{y,+} + p_{y,-})^2\label{eq:pairpt}
\end{align}
with $E_{\pm} = \sqrt{\vec{p}_{\pm}^{\,2} + m_e^2}$ , $m_e =
511$~\kevcc, and the 3-momentum vector $\vec{p}_{\pm}$ as measured
with the drift chamber.

The {\em foreground} pairs can be generally separated into {\em
  physical} and {\em unphysical} pairs. With {\em physical} pairs it
is referred to an \ee pair which originates from the same parent,
which can be either a hadron, \eg, $\omega \rightarrow \ee$ or a
photon. The photon can either be virtual and convert internally (such
a virtual photon appears for instance in the \pion Dalitz decay:
$\pi^0 \rightarrow \gamma \gamma^{\ast} \rightarrow \gamma \ee$), or
real and convert in detector material (\eg, the two photon decay of a
\pion: $\pi^0 \rightarrow \gamma \gamma \rightarrow \gamma \ee$). The
latter type of photon conversions, also referred to as external
conversion, are for reasons that will be explained in
Section~\ref{sec:photon_conversions} not reconstructed with their real
mass $m=0$ but contaminate the invariant mass spectrum of \ee pairs up
to $\mee\approx0.3$~GeV and need to be removed from the sample of {\em
  physical} pairs in order to extract the physics signal of hadron
decays and internal photon conversions. By definition, {\em physical}
pairs can only exist as unlike sign pairs. However, the chance to
reconstruct both the electron and the positron of such pairs is very
much reduced by the limited azimuthal acceptance and the low momentum
cut off at $p < 200$~\mevc.

The other, much larger contribution to the {\em foreground}, are {\em
  unphysical} pairs, \ie, electrons and positrons that are not from
the same hadron, but just a result of combining all electrons in the
event. {\em Unphysical} pairs can either be of unlike- or like-sign
charge combinations. Most of these pairs are completely {\em
  uncorrelated} combinations of electrons, which can be described with
a combinatorial background generated with an event mixing
technique. However, some fraction of the {\em unphysical} pairs is
{\em correlated}, either due to correlations early on in the decay
history, \ie, the pair does not share the same parent, but the same
``grand parent'' particle, in which case they are referred to as {\em
  cross} pairs or because they were produced within the same jet,
therefore denoted {\em jet} pairs. Another source of {\em correlated}
pairs are {\em ghost} pairs, which appear when tracks share the same
detector information.

This classification scheme has ignored one very important source of
\ee pairs, \ie, \ee pairs from semi-leptonic open charm and bottom
decays correlated through flavor conservation, discussed in
Section~\ref{sec:open_charm}. They do not originate from the same
parent, however, they are considered {\em physical} pairs.

In this Section all the details of the various background subtractions
are discussed. Beforehand, a schematic overview over the various
background sources is given as orientation. Noted in parenthesis are
the Section number in which the particular background is discussed and
a keyword referring to the method with which this background is
removed.
\\

All Pairs:
\begin{itemize}
\item physical pairs
  \begin{itemize}
  \item hadron and open charm decays, internal conversions (Signal)
  \item external photon conversions ($\phi_V$ cut, Section~\ref{sec:photon_conversions})
  \end{itemize}
\item unphysical pairs
  \begin{itemize}
  \item uncorrelated (Mixed Events, Section~\ref{sec:event_mixing})
  \item correlated
    \begin{itemize}
    \item[$\circ$] Detector Ghosts (Pair Cuts, Section~\ref{sec:pair_cuts})
    \item[$\circ$] Jet and Cross Pairs (Simulations, Section~\ref{sec:correlated_background})
    \end{itemize}
  \end{itemize}
\end{itemize}

\subsection{Pair Cuts}
\label{sec:pair_cuts}
\begin{figure}
  \centering
  \begin{picture}(120, 80)
    \put(60,5){\line(1,0){50}}  
    \put(60,15){\line(1,0){50}} 
    \put(115,8){DC} 
    \put(60,16){\line(1,0){50}} 
    \put(115,15){PC1} 
    \put(60,74){\line(1,0){50}} 
    \put(115,73){PC3} 
    \put(60,78){\line(1,0){50}} 
    \put(115,77){EMCal} 
    \put(60,18){\thicklines\arc{100}{-1.57}{0}} 
    \put(115,30){RICH mirror} 
    \put(18,18){\line(1,0){40}}  
    \multiput(18,15)(1,0){40}{\line(1,4){0.75}}  
    \put(28,11){PMT plane} 
    \put(90,1){\Thicklines\vector(-1,4){20}}  
    \put(66,70){$e^-$} 

    \thicklines
    \dashline{2}(84,25)(65,67.5)  
    \put(70,40){{\small$\gamma_{\rm ch}$}} 
    \dashline{2}(65,67.5)(29.7,18)  

    \dashline{2}(81.5,35)(79.5,64.1)  
    \dashline{2}(79.5,64.1)(24.7,18)  

    \thinlines
    \put(94,1){\vector(-1,4){20}}  

    \dottedline{2}(90,25)(87.57,59.71)  
    \dottedline{2}(87.57,59.71)(24.7,18)  
  \end{picture}
  \caption[RICH ring overlap]{Overlapping rings in the RICH PMT plane
    due to two parallel tracks.}
  \label{fig:rich_ghost}
\end{figure}
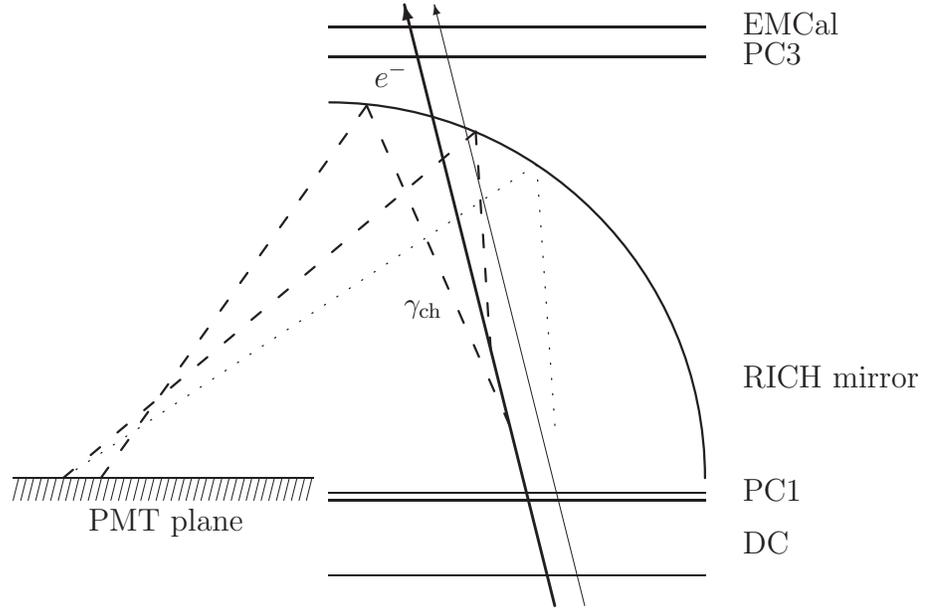

In order to reproduce the shape of the combinatorial background of
uncorrelated \ee pairs with mixed events, it needs to be assured that
any detector correlations are removed from the sample of \ee pairs in
real events. Such correlations can occur, when two electron candidates
share detector information, \eg have overlapping rings in the RICH,
share the same hit in PC1 or have overlapping cluster in the
EMCal. For instance, two electrons which traverse the RICH along
parallel trajectories would emit Cherenkov photons which are focused
on the same ring, as any two parallel photons are focused by the
spherical mirror onto the same focal point. Therefore, a RICH hit due
to an electron would be matched also to any charged track crossing the
RICH parallel to the electron and therefore fulfill all eID cuts on
RICH variables, even if the charged track itself does not emit
Cherenkov photons. This effect is illustrated in~\fig{fig:rich_ghost}.

While an uncorrelated hadron contamination would be reproduced by
mixed event, such a {\em ghost} pair is highly correlated in its
geometry and therefore in the invariant mass spectrum. They have
typical small opening angle and are therefore reconstructed with a
small invariant mass. As like-sign tracks are bent by the magnetic
field bents in the same direction, in contrast to unlike-sign pars
which are bent in opposite directions, they have even smaller masses than
unlike-sign pairs. 

\begin{figure}
  \centering
  \subfloat[]{\label{fig:dcenter_phi}\includegraphics[width=0.44\textwidth]{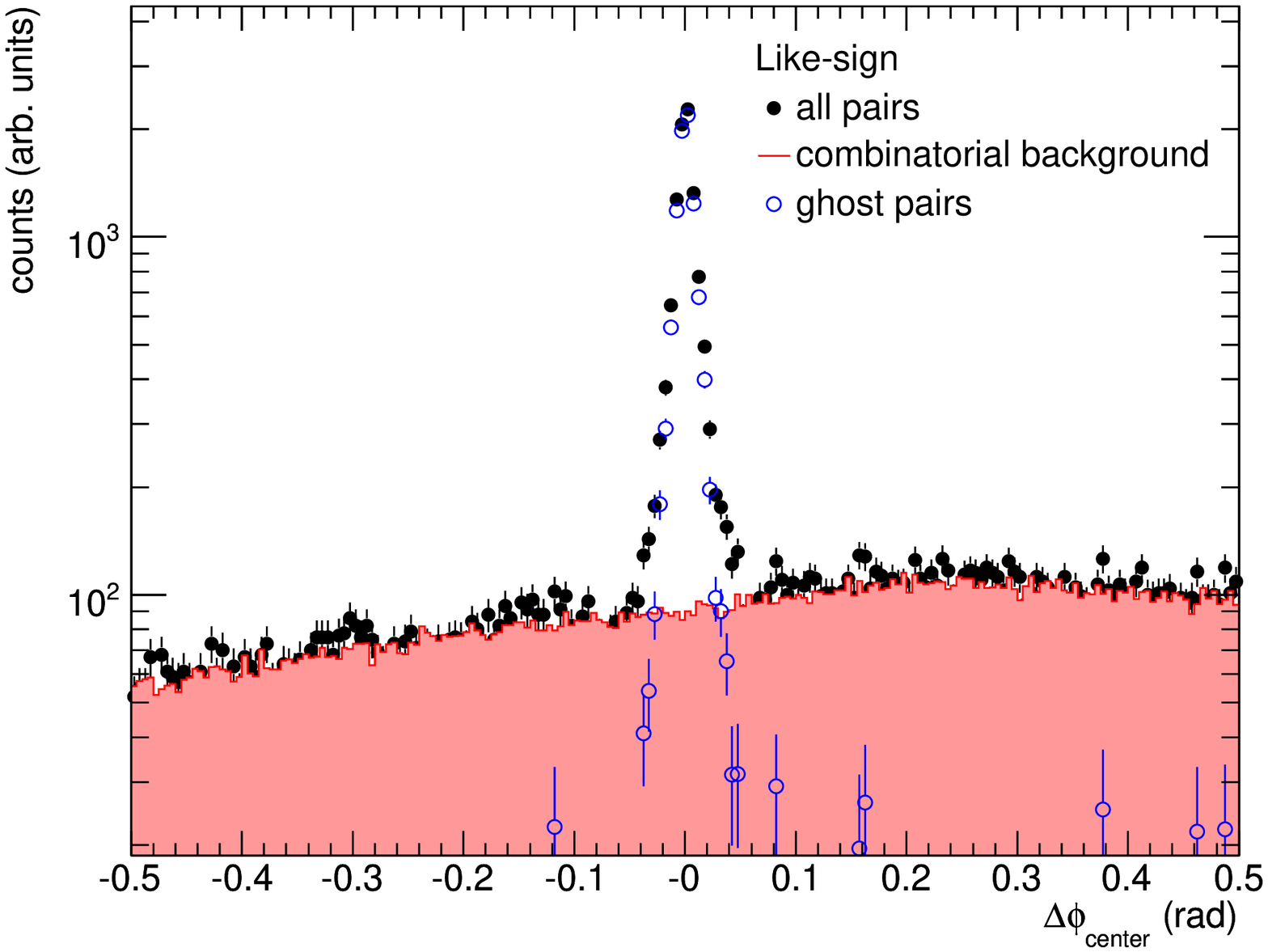}}
  \subfloat[]{\label{fig:dcenter_zed}\includegraphics[width=0.44\textwidth]{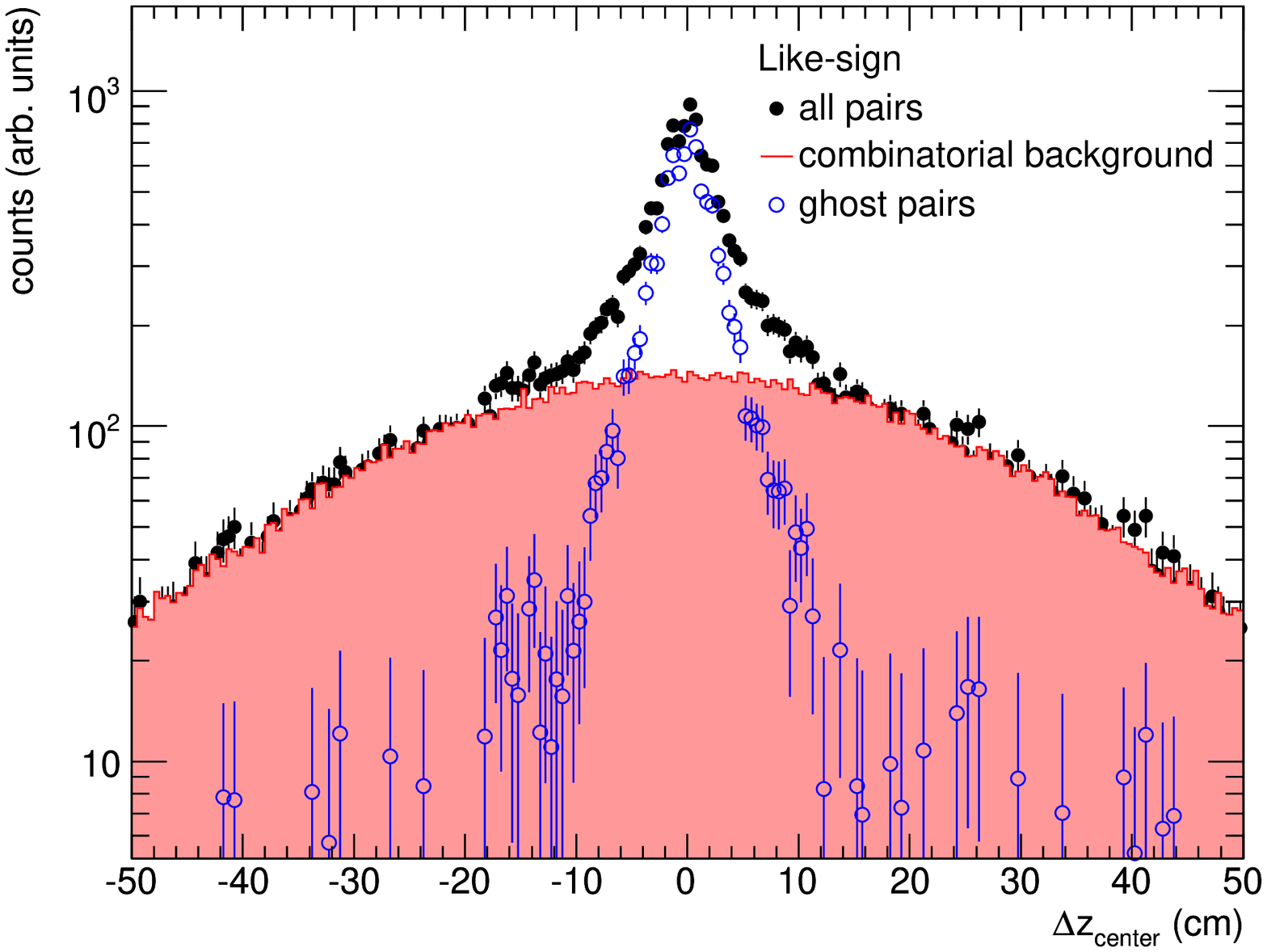}}
  \caption[RICH ring distance]{Distance of the ring centers associated
    to two tracks in $\Delta \phi_{\rm center}$~\subref{fig:dcenter_phi} and $\Delta
    z_{\rm center}$~\subref{fig:dcenter_zed}.}
  \label{fig:dcenter}
\end{figure}

Naturally, these {\em ghost} pairs do not exist in mixed events and
therefore have to be rejected in real events. They are rejected with
an event cut which rejects events (real and mixed) if there are two
tracks which overlap in RICH, EMCal, or PC1. On overlaps in the RICH a
cut is applied based on the distance between the ring centers
associated to the two electron candidates of a pair. The distance
between two tracks $i$ and $j$ is calculated as:
\begin{equation}\label{eq:dcenter}
\Delta_{\rm center} = \sqrt{(z_{\rm center}^i - z_{\rm
    center}^j)^2/\sigma_{z_{\rm center}}^2 + (\phi_{\rm center}^i - \phi_{\rm center}^j)^2/\sigma_{\phi_{\rm center}}^2}
\end{equation}
with $\sigma_{z_{\rm center}} = 3.6$~cm and $\sigma_{\phi_{\rm
    center}} = 10$~mrad being the rms of the distributions of ring
distances (without any pair cut) in $\hat{z}$ and $\hat{\phi}$
directions, respectively, as shown in~\fig{fig:dcenter}. A cut of
$\Delta_{\rm center} > 10$ (equivalent to 36~cm) is applied, which
corresponds to twice the expected ring diameter ($\sim$16.8~cm). One
may consider to cut only on one ring diameter as this would be enough
to remove false positive ring association to charged hadrons, but in
addition one observes a reduction in the reconstruction efficiency of
two electrons which have close rings, as their rings may overlap and
distort their shape, which increases, \eg, $\chi^2/{\rm npe}_0$. This effect
only vanishes once the two ring are clearly separated by more than one
ring diameter, \ie their centers by more than two ring diameters. In
the EMCal events are rejected if two electron candidates are
associated to two clusters within a $3\times3$ tower region (which is
the average size of an electromagnetic shower). In the PC1 the cut is
applied at $\Delta z \leq 0.5$~cm and $\Delta \phi \leq 20$~mrad,
which is the size of a cell. The pair cut on RICH variables has the
largest effect and the combined cuts remove $\sim 0.08$\% of all
events in \AuAu collisions and a fraction of pairs which varies from
4\% in the most central to 2\% in the most peripheral collisions. In
\pp this cut removes only 0.1\% of all pairs.

\begin{figure}
  \centering
  \subfloat[like-sign pairs]{\label{fig:like_mass_ghost}\includegraphics[width=0.44\textwidth]{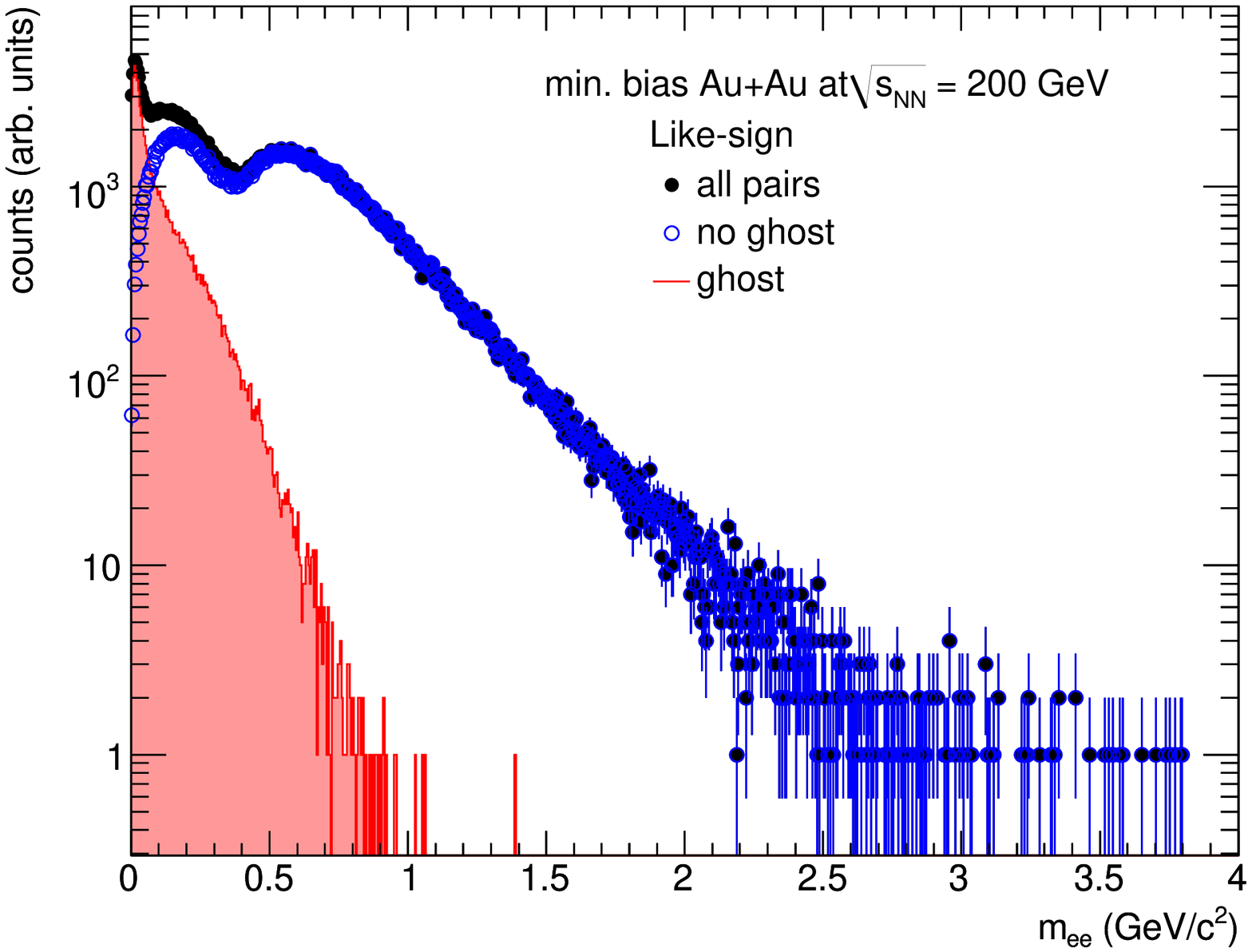}}
  \subfloat[unlike-sign pairs]{\label{fig:unlike_mass_ghost}\includegraphics[width=0.44\textwidth]{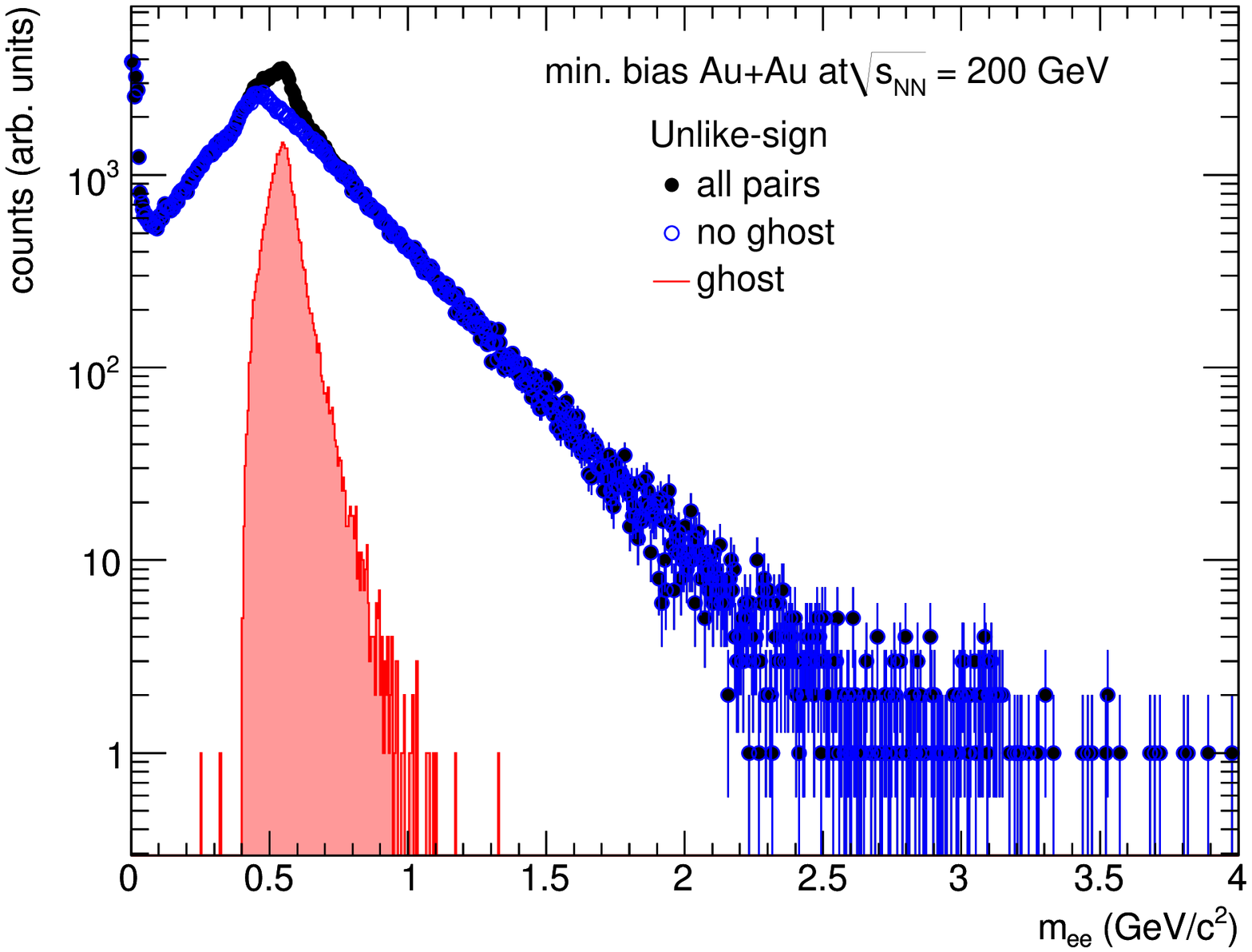}}
  \caption[Ghost Pairs]{Invariant mass distribution of like-sign
    \subref{fig:like_mass_ghost} and unlike-sign
    pairs\subref{fig:unlike_mass_ghost}. All pairs are shown in black,
    ghost pairs in blue and all others in red.}
  \label{fig:ghost}
\end{figure}

The effect of ghost pairs on the \ee pair distribution can be studied
with the invariant mass distribution of like-sign pairs, \ie, $e^+e^+$
and $e^-e^-$ pairs, which do not contain any physics signal. The real
and mixed event distributions of like-sign pairs in MB \AuAu
collisions is shown in black and red, respectively,
in~\fig{fig:like_mass_ghost}. The mixed events clearly deviate in
shape from the real event distribution. In the invariant mass
distribution of \ee pair the ghosts are also clearly visible at
$m\approx500$~\mevcc as shown in~\fig{fig:unlike_mass_ghost}, where
the mixed events deviate from the real events.

\subsection{Photon Conversions}
\label{sec:photon_conversions}
\begin{figure}
  \centering
  \begin{picture}(70, 70)
    \put(20,40){\circle*{1}}
    \put(20,40){\thicklines\circle{30}} 
    \put(10,20){\vector(1,3){2}}
    \put(2,15){beam pipe}
    \thinlines
    \spline(20,40)(22.5,42.8)(25,44)(28,43.5)(35,40)
    \spline(20,40)(22.5,37.2)(25,36)(28,36.5)(35,40)
    \dottedline(20,40)(30,55)
    \dottedline(20,40)(30,25)
    \dashline{1}(20,40)(35,40)  
    \put(34,40){\vector(1,0){1}}
    \put(25,38){$\gamma$}
    \color{red}
    \thicklines
    \spline(35,40)(40,37)(50,30)(60,20) 
    \put(55,25){\vector(1,-1){7}}
    \put(58,26){$e^+$}
    \color{blue}
    \spline(35,40)(40,43)(50,50)(60,60) 
    \put(55,55){\vector(1,1){7}}
    \put(58,53){$e^-$}
    \color{black}
    \dashline{1}(35,40)(50,45)
    \dashline{1}(35,40)(50,35)
    \put(5,60){\circle{4}}
    \put(5,60){\circle*{0.5}}
    \put(7,62){$\vec{B}$}
  \end{picture}
  \caption[Opening angle of conversion pairs]{Misreconstruction of the
    opening of a conversion pair. Shown are the photon and the
    conversion pair which is created in the beam pipe. The real
    opening angle between the original trajectories (dashed lines) is
    much smaller than the reconstructed opening angle, artificially
    created by tracking the particles back to the collision vertex
    (dotted lines).}
  \label{fig:conv_opening}
\end{figure}
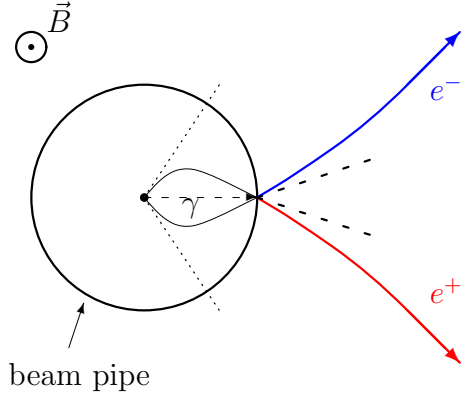

Photon conversions contribute to the final spectrum of \ee pairs and
should be rejected. Electrons from conversions that are created off
vertex in detector material, \eg, the beam pipe made of Beryllium (0.3
$X_0$), pass through less magnetic field integral than the PHENIX
tracking algorithm assumes, which leads to a misreconstructed
momentum. In addition the opening angle of the \ee pair is
misreconstructed, as it is tracked back to the collision vertex,
rather than to the conversion vertex, as illustrated in
\fig{fig:conv_opening}. Therefore, conversion pairs have an apparent
mass that increases with the distance from the collision vertex and
therefore with the misreconstructed opening angle.

\begin{figure}
  \centering
  \includegraphics[width=0.9\textwidth]{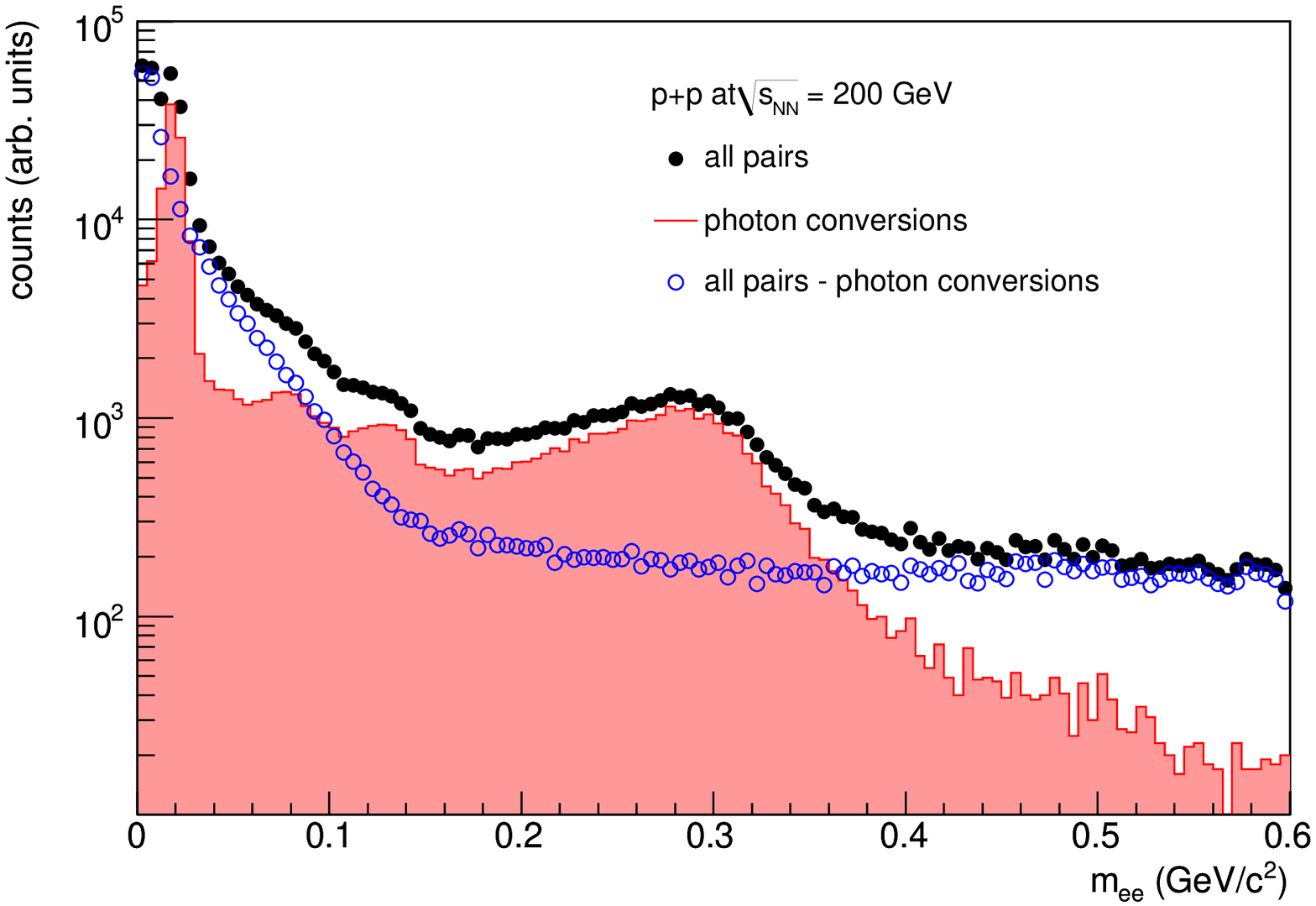}
  \caption[Invariant mass distribution of conversions]{Invariant mass
    distribution of all \ee pairs in real events (black), photon
    conversions (red) and remaining pairs (blue). The peak at
    20~\mevcc is caused by photon conversions in the beam pipe. The
    remaining shape is due to other detector support structures (see
    text).}
  \label{fig:invmass_conversions}
\end{figure}

Photon conversions occur in the beam pipe at a radius of 4~cm which
corresponds to the peak at $\mee \approx 20$~\mevcc in the invariant
mass spectrum shown in \fig{fig:invmass_conversions}, in detector
support structures ($\mee \approx 80$\mevcc and $\mee \approx
125$~\mevcc) as well as in the air before, and much reduced within,
the installed Helium bag which contributes as a continuum out to $\mee
\leq 300$~\mevcc, corresponding to the DC entrance window, beyond
which electrons do not bend anymore because the region is
field-free. Therefore, electrons from photon conversions within the DC
are straight and rejected with the high-\pt cut of 20~\gevc.
As photons are massless, conversion pairs do not have an intrinsic
opening angle, but are aligned perpendicular to the axial magnetic
field. A cut on the orientation of \ee pairs is used to remove those
conversion photons. To measure the orientation of the \ee pair plane
with respect to the $\hat{z}$ direction, \ie, the magnetic field, the
angle $\phi_V$ is defined~\cite{Akiba:2001AN044}:
\begin{subequations}
\begin{align}
  \label{eq:phivdef}
  \vec{u} &= \frac{\vec{p}_1 + \vec{p}_2}{|\vec{p}_1 + \vec{p}_2|}\\
  \vec{v} &= \vec{p}_1 \times \vec{p}_2\\
  \vec{w} &= \vec{u} \times \vec{v}\\
  \vec{u}_a &= \frac{\vec{u} \times \hat{z}}{|\vec{u} \times \hat{z}|}
\end{align}
\end{subequations}
\begin{align}
  \label{eq:phiv}
  \phi_V &= \arccos \left(\frac{\vec{w} \cdot \vec{u}_a}{|\vec{w}| |\vec{u}_a|}\right).
\end{align}
Here $\vec{p}_1$ is the 3-momentum vector of the electron and
$\vec{p}_2$ the 3-momentum vector of the positron. The following cuts
are applied:
\begin{itemize}
\item $m < 0.60~\&\&~\phi_V > 0.1$~rad
\item $m < 0.03~\&\&~\phi_V > 0.25$~rad
\end{itemize}
To remove beam pipe conversions a harder cut has to be applied,
because the multiple scattering within Beryllium smears the
orientation with respect to the magnetic field more than in air. In
addition the resolution of the $\phi_V$ measurement improves with
increasing conversion radius. This cut removes $>98\%$ of all
conversion pairs.

\subsection{Event Mixing}
\label{sec:event_mixing}

As mentioned in the introduction to this Section, most of the \ee
pairs are of uncorrelated origin. While the unlike-sign spectrum
contains both physics signal and background pairs, the like-sign
spectrum does not contain any physics signal and therefore is a direct
measure of the background. It can be shown (see
Appendix~\ref{cha:background_normalization}) that, as long as
electrons and positrons are produced in pairs, the unlike-sign
combinatorial background is the geometrical mean of the like-sign
backgrounds, independent of efficiency and acceptance differences,
\ie:
\begin{equation}\label{eq:twosqrt}
  \langle BG_{+-} \rangle = 2 \sqrt{\langle BG_{++} \rangle \langle BG_{--} \rangle}.
\end{equation}
In experiments with equal acceptance for like- and unlike-sign pairs,
the distribution of like-sign pairs in real events can be directly
related to the unlike-sign background through their geometrical mean
$BG_{+-} = 2 \sqrt{FG_{++} FG_{--}}$ and subtracted from the
unlike-sign pair spectrum to obtain the physics signal. Because of a
different acceptance for like- and unlike-sign pairs a correction
needs to be applied to use this method in PHENIX.

Another possibility, which leads to a higher statistical accuracy, is
the event mixing. Electrons from events with similar topology are
combined to pairs to statistically approximate the combinatorial
background. To ensure the same topology, events are categorized in
pools according to their centrality, z-vertex and reaction plane. In
\AuAu, mixed events were generated in ten centrality, six z-vertex and
one reaction plane pools\footnote{One reaction plane pool means no
  separation by reaction plane. However, it was tested that the
  reaction plane pooling has no effect on the mixed event distribution
  and can be neglected}, while five z-vertex pools were used for \pp.

The ERT trigger used in the \pp analysis biases the single electron
distribution towards higher \pt. Therefore, to generate the correct
combinatorial background shape of \ee pairs, the mixed events must be
generated from the minimum bias sample, but as in real events it is
required that in every pair at least one of the electrons must fulfill
the ERT trigger condition.

The shape of the like-sign background from mixed events can be
compared to the like-sign distribution in real events to determine in
which region and how well the background shape is reproduced in the
event mixing. The absolute normalization of unlike-sign combinatorial
background is given by the geometrical mean of the observed positive
and negative like-sign pairs $2\sqrt{BG_{++} BG_{--}}$ where $BG_{++}$
and $BG_{--}$ are determined by integrating the mixed event
distributions after they have been normalized in a region where the
mixed events reproduce the distribution from real events, which means
it is enforced that:
\begin{equation}
  \frac{\int_A BG_{++} + BG_{--}}{\int_A FG_{++} + FG_{--}} \equiv 1
\end{equation}
where $A$ denotes the normalization region.

\subsubsection{Background Normalization}
\label{sec:background_norm}
\begin{figure}
  \centering
  \includegraphics[width=0.9\textwidth]{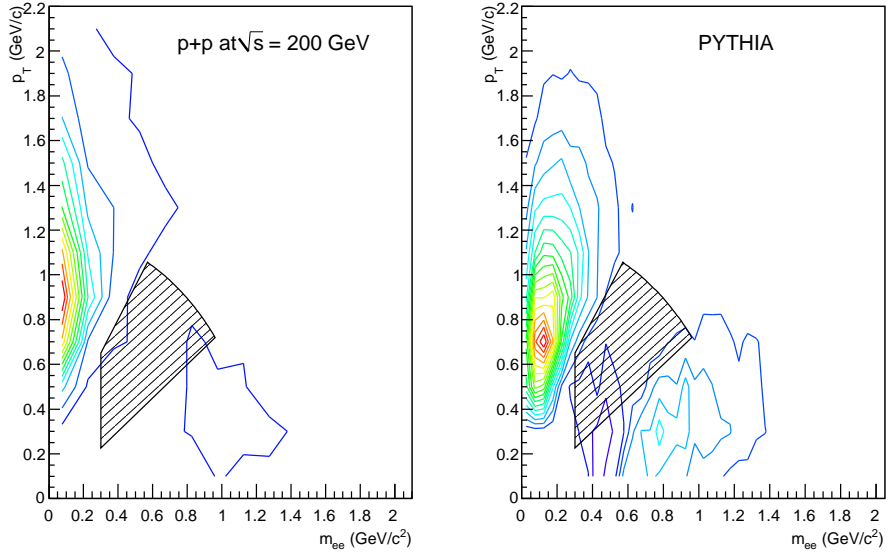}
  \caption[Background subtracted like-sign distribution in \mee and
  \pt]{Mixed event subtracted like-sign spectrum in data~(left) and as
    calculated with \pythia~(right) as function of invariant mass and
    \pt. The dashed region outlines the normalization area as defined
    in \eq{eq:regionA}.}
  \label{fig:mpt_like_sub}
\end{figure}

The like-sign spectra as function of mass and \pt after mixed event
subtraction are shown in the left panel of \fig{fig:mpt_like_sub}. The
figure clearly shows that mixed and real events do not have the same
distribution like-sign pairs, which is an indication of correlated
pairs. However, in the region $\mee \sim \pt$ the distributions are
very similar. Since for a pair with opening angle $\delta$ it is
$m_{ee} = p_1 p_2 (1-\cos \delta)$, the condition $\mee \sim \pt$
roughly corresponds to an azimuthal opening angle of $\Delta\phi \sim
\pi/2$.
\begin{figure}
  \centering
  \includegraphics[width=0.9\textwidth]{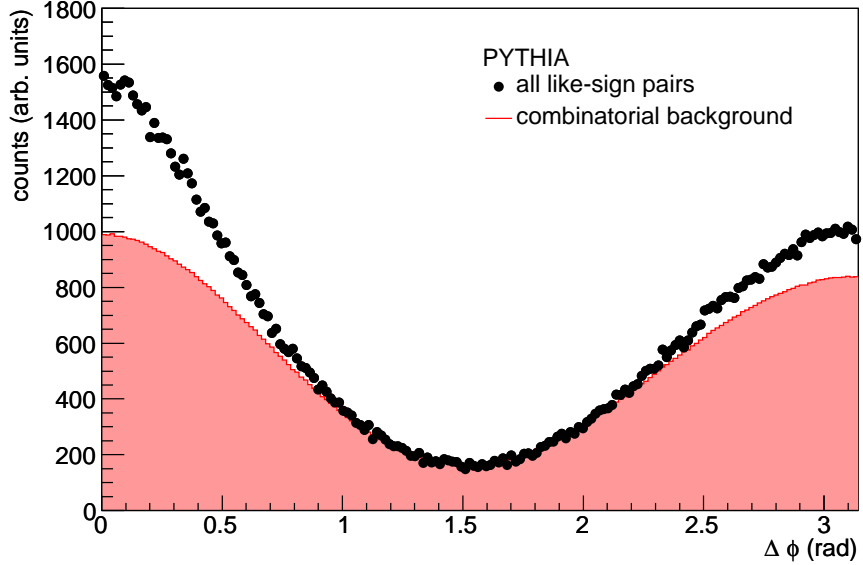}
  \caption[$\Delta\phi$ of \pythia like-sign pairs]{$\Delta\phi$
    distribution of \pythia like-sign pairs in same ({\em black}) and mixed
    ({\em red}) events.}
  \label{fig:py_dphi}
\end{figure}
This was checked with a \pythia simulation of minimum bias events which is
described in Section~\ref{sec:jet_pairs}. The $\Delta\phi$
distribution of like-sign pairs from these simulated events is shown
in \fig{fig:py_dphi}. This distribution is compared to the mixed event
distribution also generated from the \pythia simulation. The mixed
events agree well with the real events around $\Delta\phi \sim
\pi/2$. Thus, the mixed event distribution in \pythia is normalized in
the region defined by:
\begin{subequations}
  \label{eq:regionA}
  \begin{align}
    \mee&>300~{\rm MeV}/c^2&\&\&\\
    \mt&<1.2~{\rm MeV}/c^2&\&\&\\
    \pt/c-1.5 \mee &\leq 200~{\rm MeV}/c^2&\&\&\\
    \pt/c-0.75 \mee &\geq 0~{\rm MeV}/c^2.
  \end{align}
\end{subequations}
which is shown as the dashed contour in \fig{fig:mpt_like_sub}. The
yield remaining after the subtraction such normalized mixed event is
considered correlated and discussed in
Section~\ref{sec:jet_pairs}. The same region is used to normalize the
mixed events to the real events in data. The normalization to this
region has a statistical accuracy of 2.8\%.

\subsection{Correlated Background}
\label{sec:correlated_background}
\begin{figure}
  \centering
  \includegraphics[width=0.9\textwidth]{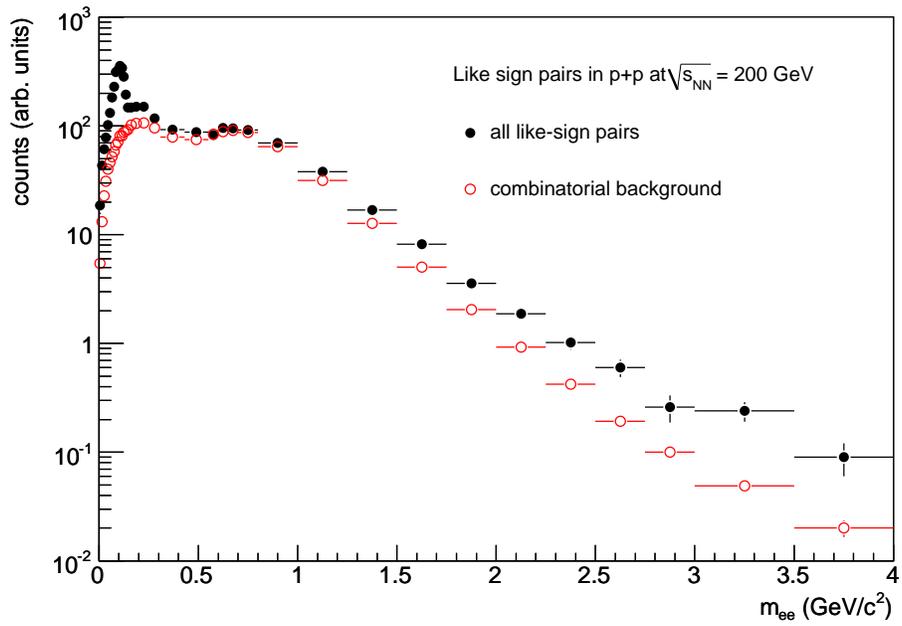}
  \caption[Invariant mass distribution of like-sign pairs]{Invariant
    mass distribution of all like-sign pairs in real events (black)
    and mixed events (red). The normalization is discussed in
    Section~\ref{sec:background_norm}. A clear shape difference is
    observed at small and at large masses, indicating a source of
    correlated background.}
  \label{fig:pp_like_mass}
\end{figure}

\fig{fig:pp_like_mass} shows the invariant mass distribution of
like-sign pair of real and mixed events in \pp collisions. The pairs
from mixed events do not reproduce the shape of like-sign pairs in
real events which is an indication for correlated background whose two
sources are discussed in the following.

\subsubsection{Cross Pairs}
\label{sec:cross_pairs}
The first correlation of like-sign pairs stems from hadron decays with
two \ee pairs in the final state, \eg, for a \pion these are the
double Dalitz decay ($\pi^0 \rightarrow e^+ e^- e^+ e^-$), the Dalitz
decay ($\pi^0 \rightarrow \gamma e^+ e^-$) in which the photon
converts into an \ee pair in detector material, and the two photon
decay ($\pi^0 \rightarrow \gamma \gamma$) in which both photons
convert into two \ee pairs\footnote{All these decays are in principle
  two photon decays, but with two, one or zero virtual photons}:
\begin{align}
  \pi^0 &\rightarrow \gamma_1 \gamma_2 \notag\\
  &\rightarrow e_1^+ e_1^- e_2^+ e_2^-\label{eq:pi0cross}
\end{align}
The \ee pair with the same parent photon is considered physics signal
(\ie, $e_1^+ e_1^-$ and $e_2^+ e_2^-$), and in case of a real photon
conversion in detector material removed with a cut (see
Section~\ref{sec:photon_conversions}). The {\em``cross''} combinations
into two unlike-sign pairs ($e_1^+ e_2^-$ and $e_2^+ e_1^-$) as well
as into two like-sign pairs ($e_1^+ e_2^+$ and $e_1^- e_2^-$) are not
purely combinatorial, but correlated via the \pion mass ($\mee <
m_{\pi^0}$). Therefore, their contribution is present in the real
event spectra of both unlike- and like-sign pairs, but not in mixed
events. \fig{fig:pp_like_mass} shows the invariant mass distribution
of like sign pairs in real (black) and mixed events (red). The peak
visible in the real event distribution of like-sign pairs at a mass of
$\mee \approx 100$~\mevcc is due to cross pairs. The same cross
combinations are possible for $\eta$ decays, with the additional decay
channel of $\eta \rightarrow \pi^0 \pi^0 \pi^0$. Therefore these
correlated pairs extend up to a mass of $m_{\eta} \approx 550$~\mevcc.

For every decay, contributions to unlike-sign signal $U$, unlike-sign
cross $C$ and like-sign cross pairs $L$ can be calculated in terms of
branching ratios ($BR_{\gamma\gamma} = 98.8\%$, $BR_{\rm Dalitz} =
1.2\%$, and $BR_{\rm dbl. Dalitz} = 3.14\times10^{-5}$~\cite{pdg}) and
conversion probability ($P_{\rm conv} = 0.3\%$) relative to the \pion
multiplicity:
\begin{itemize}
\item $\pion \rightarrow \gamma \gamma \rightarrow \ee \ee$:
  \begin{itemize}
    \item $U = 0 $
    \item $C = 2\cdot BR_{\gamma\gamma}\cdot P_{\rm conv}^2 = 2 \cdot 98.8\% \cdot 0.3\%^2 = 1.76\times10^{-5}$
    \item $L = 2\cdot BR_{\gamma\gamma}\cdot P_{\rm conv}^2 = 2 \cdot 98.8\% \cdot 0.3\%^2 = 1.76\times10^{-5}$
  \end{itemize}
\item $\pion \rightarrow \gamma \gamma^{\ast} \rightarrow \ee \ee$:
  \begin{itemize}
    \item $U = BR_{\rm Dalitz} = 1.2\%$
    \item $C = 2\cdot BR_{\rm Dalitz}\cdot P_{\rm conv} = 2 \cdot 1.2\% \cdot 0.3\% = 7.2\times10^{-5}$
    \item $L = 2\cdot BR_{\rm Dalitz}\cdot P_{\rm conv} = 2 \cdot 1.2\% \cdot 0.3\% = 7.2\times10^{-5}$
  \end{itemize}
\item $\pion \rightarrow \gamma \gamma^{\ast} \rightarrow \ee \ee$:
  \begin{itemize}
    \item $U = 2\cdot BR_{\rm dbl.~Dalitz} = 2 \cdot 3.14\times10^{-5} = 6.28\times10^{-5}$
    \item $C = 2\cdot BR_{\rm dbl.~Dalitz} = 2 \cdot 3.14\times10^{-5} = 6.28\times10^{-5}$
    \item $L = 2\cdot BR_{\rm dbl.~Dalitz} = 2 \cdot 3.14\times10^{-5} = 6.28\times10^{-5}$
  \end{itemize}
\end{itemize}
which gives a total of $U = 1.2\times10^{-2}$ unlike-sign signal
pairs, $C = 1.5\times10^{-4}$ unlike-sign cross pairs, and $L =
1.5\times10^{-4} $ like-sign cross pairs per \pion.

The same calculation for the $\eta$ yields a total of $U =
6\times10^{-3}$ unlike-sign signal pairs, $C = 6.7\times10^{-4}$
unlike-sign cross pairs, and $L = 6.7\times10^{-4} $ like-sign cross
pairs per $\eta$. Here the branching ratios ($BR_{\gamma\gamma} =
39.38\%$, $BR_{\rm Dalitz} = 0.6\%$~\cite{pdg}) have been used and the
unknown double Dalitz branching ratio of the has been assumed to be
the square of the single Dalitz: $BR_{\rm dbl.~Dalitz} = BR_{\rm
  Dalitz}^2 = 1.23\times10^{-5}$. The three \pion decay of the $eta$
contribute with an additional $C = L =6.09\times10^{-4}$ to the
unlike- an like-sign cross pairs.

All these contribution are simulated with a fast Monte Carlo (\exodus)
of \pion and $\eta$ double Dalitz decay as well as $\eta \rightarrow
\pi^0 \pi^0 \pi^0$ decays and filtered into the PHENIX acceptance. The
resulting unlike- and like-sign cross pairs are shown
\fig{fig:mass_like_unlike}.

\subsubsection{Jet Pairs}
\label{sec:jet_pairs}
Besides cross pairs another source contributes to correlated
background pairs. This contribution can be reproduced with minimum bias
events from \pythia
simulations~\cite{Sjostrand:2000wi,Sjostrand:2001yu,Sjostrand:1993yb}.
\pythia is a computer program widely used as Monte Carlo simulation of
high-energy elementary particle collisions, \eg $e + p$ or \pp.  The
particle interactions are calculated using leading order matrix
elements and the Lund string fragmentation
model~\cite{Andersson:1983ia}. \pythia 6.319 with {\sc cteq5l} parton
distribution function~\cite{Lai:1999wy} has been used. The following
hard QCD processes have been activated:
\begin{description}
\item[{\sc msub 11}:] $f_i f_j \rightarrow f_i f_j$
\item[{\sc msub 12}:] $f_i \overline{f}_i \rightarrow f_k \overline{f}_k$
\item[{\sc msub 13}:] $f_i \overline{f}_j \rightarrow g g$
\item[{\sc msub 28}:] $f_i g \rightarrow f_i g$
\item[{\sc msub 53}:] $g g \rightarrow f_k \overline{f}_k$
\item[{\sc msub 68}:] $g g \rightarrow g g$
\end{description}
Here $g$ denotes a gluon and $f_i$ a fermion with flavor $i$, \ie,
$u$, $d$, $s$, $b$, $t$, $e^-$, $\nu_e$, $\mu^-$, $\nu_{\mu}$,
$\tau^-$, or $\nu_{\tau}$, and their corresponding antiparticles
$\overline{f}_i$. Flavors appearing in the initial carry indices
$i$ or $j$, while flavors created only in the final state are denoted
$k$.  A Gaussian width of 1.5 GeV for the primordial \kt distribution
is used ({\sc PARP}(91)=1.5), and the minimum parton \pt has been set
to 2 GeV ({\sc CKIN}(3)=2.0).

In this simulation the branching ratio of the \pion Dalitz decay was
set to 100\% to increase the sample of \ee pairs. The resulting
$\Delta\phi$ distribution of like sign pairs from real (black) and
mixed events (red) in \pythia is shown in \fig{fig:py_dphi}. The mixed
event like-sign pairs have been normalized to the like-sign pairs in
real events in the range $70^{\circ} < \Delta\phi < 110^{\circ}$, in
which the distributions of like-sign pairs in real and mixed events
agree well. The deviation in shape at $\Delta\phi \approx 0$ can be
attributed to electrons which originate from two \pion decays which
occurred within the same jet, while the correlation at $\Delta\phi
\approx \pi$ comes from electrons that are decay products of \pion's
in back-to-back jets. \pion's within the same jet are close in $\phi$
which leads to a small opening angle of the \ee pair, which also
causes these pairs to have a rather small mass and at the same time a
large \pt as their momentum vectors add constructively. \ee pairs
correlated via back-to-back jet have large opening angles and
therefore larger invariant masses, while their \pt is smaller than the
one of pairs in the same jet, as their momentum vectors cancel
mostly. These correlations in \mee and \pt are shown in
\fig{fig:mpt_like_sub} for like-sign pairs from \pythia after the
subtraction of mixed events.

\begin{figure}
  \centering
  \includegraphics[width=0.9\textwidth]{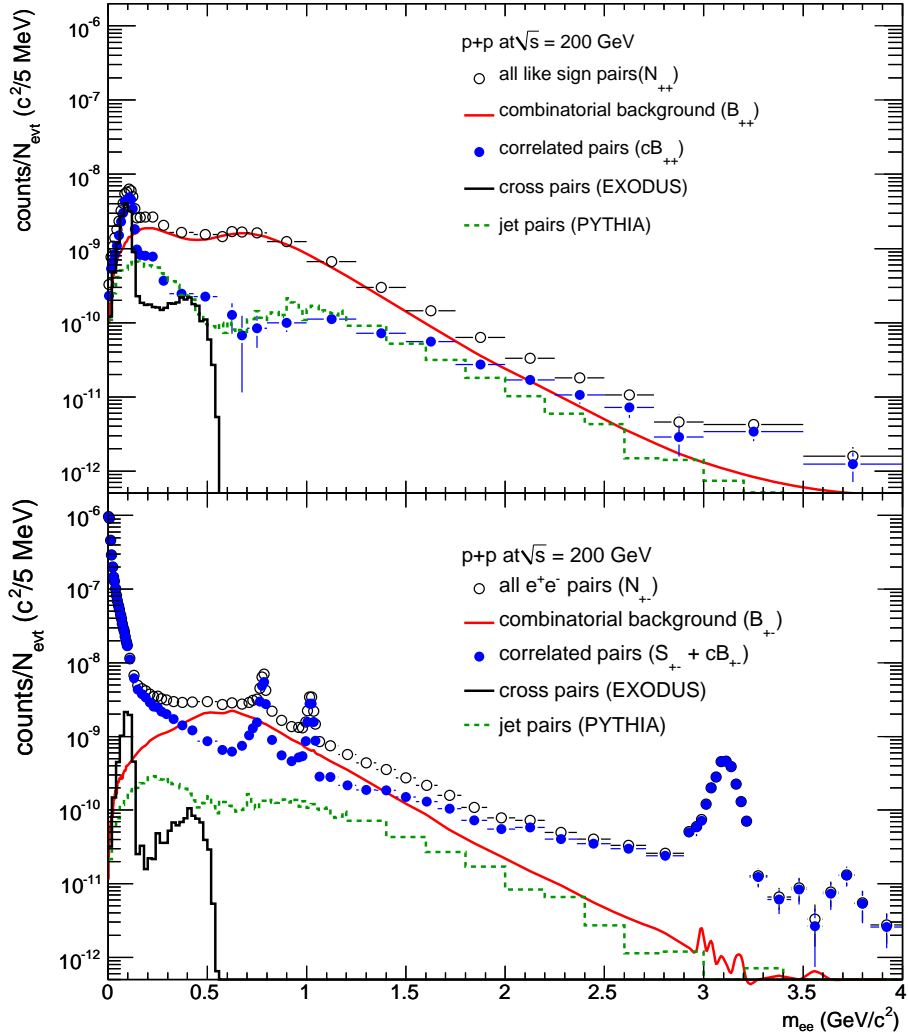}
  \caption[Invariant mass spectra of like- and unlike-sign pair in \pp.]{Raw
    dielectron spectra. The top panel shows like-sign pairs as
    measured in the experiment, the combinatorial background from
    mixed events, the correlated pair background obtained by
    subtracting the combinatorial background, and the individual
    contributions from cross and jet pairs to the correlated
    background (see text). The bottom panel shows the same
    distributions for unlike-sign pairs.  The correlated background in
    both panels is normalized to the measured like-sign pairs
    remaining after subtracting the combinatorial background.}
  \label{fig:mass_like_unlike}
\end{figure}

\fig{fig:mass_like_unlike} shows invariant mass spectra of like sign
pairs (top) and unlike-sign pairs (bottom) in real events, the mixed
event distribution normalized according to \eq{eq:regionA} and the
distributions after mixed event subtraction. The remaining like-sign
correlated pairs are fitted with the distribution of cross and jet
pairs in the two mass regions $0.06 < \mee < 0.12$~\gevcc and $0.80 <
\mee < 2.00$~\gevcc:
\begin{subequations}\label{eq:corrbg_norm}
  \begin{align}
    \int_{0.06}^{0.12} \left(\frac{dN}{dm}\right)_{\rm like} dm &= A \int_{0.06}^{0.12} \left(\frac{dN}{dm}\right)_{\rm cross} dm
    + B \int_{0.06}^{0.12} \left(\frac{dN}{dm}\right)_{\rm jet} dm\\
    \int_{0.8}^{2} \left(\frac{dN}{dm}\right)_{\rm like} dm &=  B \int_{0.8}^{2} \left(\frac{dN}{dm}\right)_{\rm jet} dm
  \end{align}
\end{subequations}
The resulting normalization factors are applied to the unlike-sign
distributions of cross and jet pairs which are shown in
\fig{fig:mass_like_unlike}. The unlike-sign signal results from
subtracting the correlated background from all correlated pairs.

\subsection{Like-sign Subtraction}
\label{sec:like_subtraction}
As a systematic check of the subtraction of combinatorial and
correlated background an independent, completely data driven method
has been developed, that does not make any assumption about the
background contributions. As stated at the beginning of the chapter,
the like-sign spectrum is free of physics signal and therefore a
measure of the background, whether correlated or uncorrelated. But the
different acceptance of unlike- and like-sign pairs in PHENIX, needs
to be considered before they can be subtracted from the real event
spectrum of \ee pairs.

The acceptance difference between unlike- and like-sign pairs is
preserved in mixed events, therefore the ratio of $BG_{+-}/(BG_{++} +
BG_{--})$ is the proper correction to convert the distribution of
like-sign pairs in real events into the background distribution of \ee
pairs. The signal of \ee pairs is calculated in \mee and \pt as
follows:
\begin{equation}
  S_{+-} = FG_{+-} - (FG_{++} + FG_{--}) \frac{BG_{+-}}{(BG_{++} BG_{--})}.
\end{equation}
In the special case that the real event like-sign spectrum is
perfectly described by the mixed events, this expression reduces to $S
= FG_{+-} - BG_{+-}$.

\begin{figure*}[h]
  \centering
  \subfloat[$BG_{+-}$]{\label{fig:unlikeacc}\includegraphics[width=0.44\textwidth]{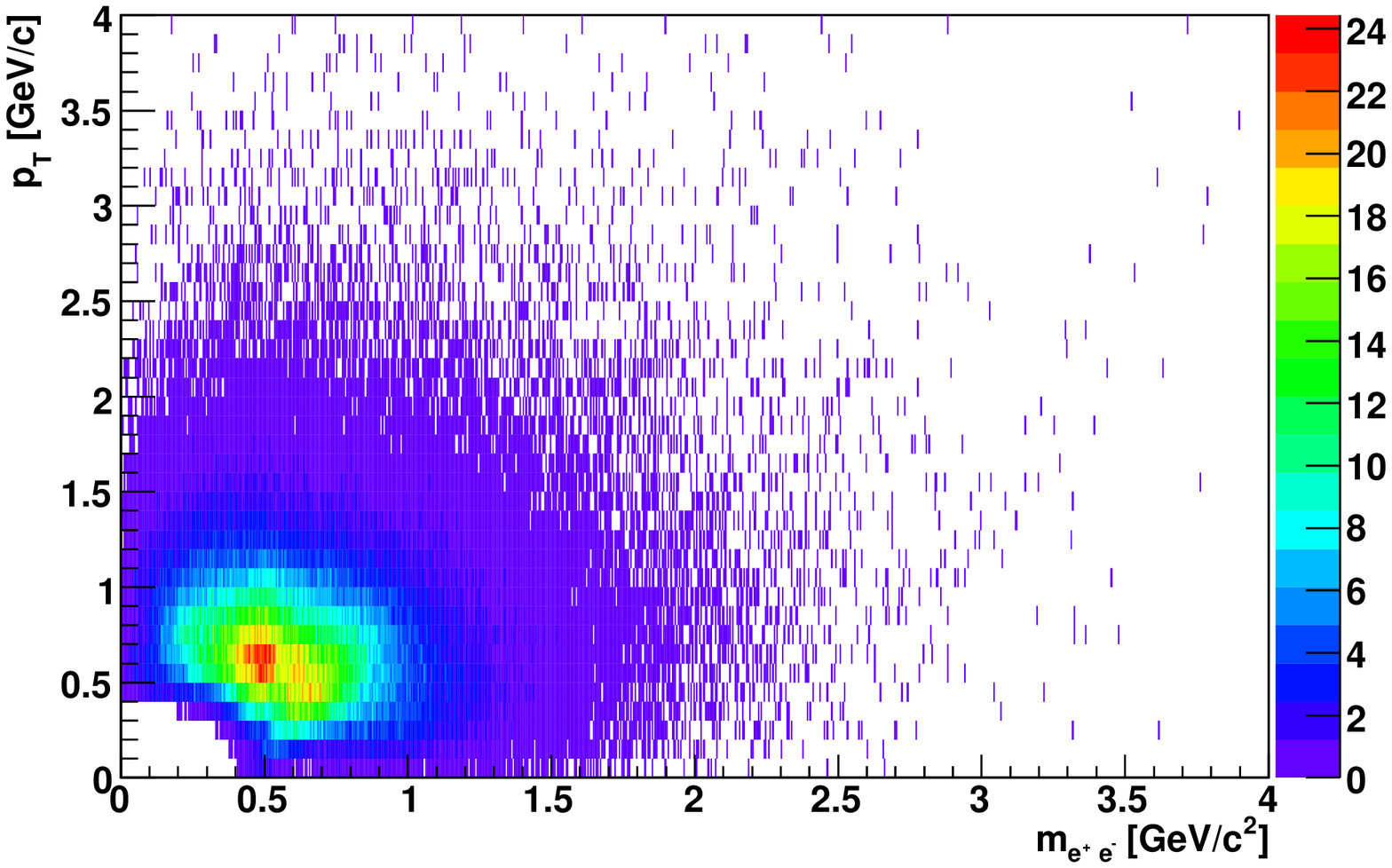}}
  \subfloat[$BG_{++} + BG_{--}$]{\label{fig:likeacc}\includegraphics[width=0.44\textwidth]{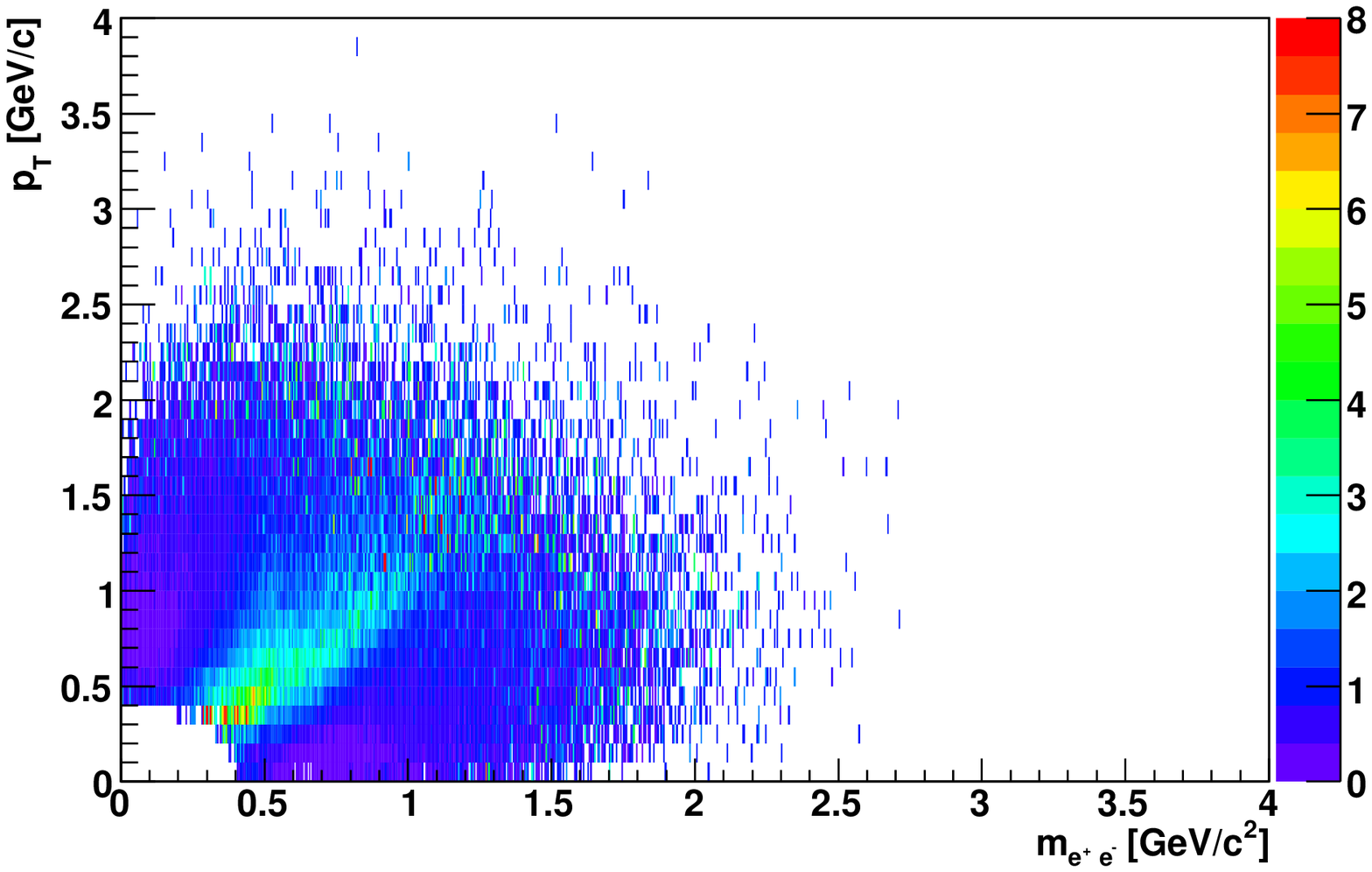}}\\
  \subfloat[Relative Acceptance]{\label{fig:relacc}\includegraphics[width=0.44\textwidth]{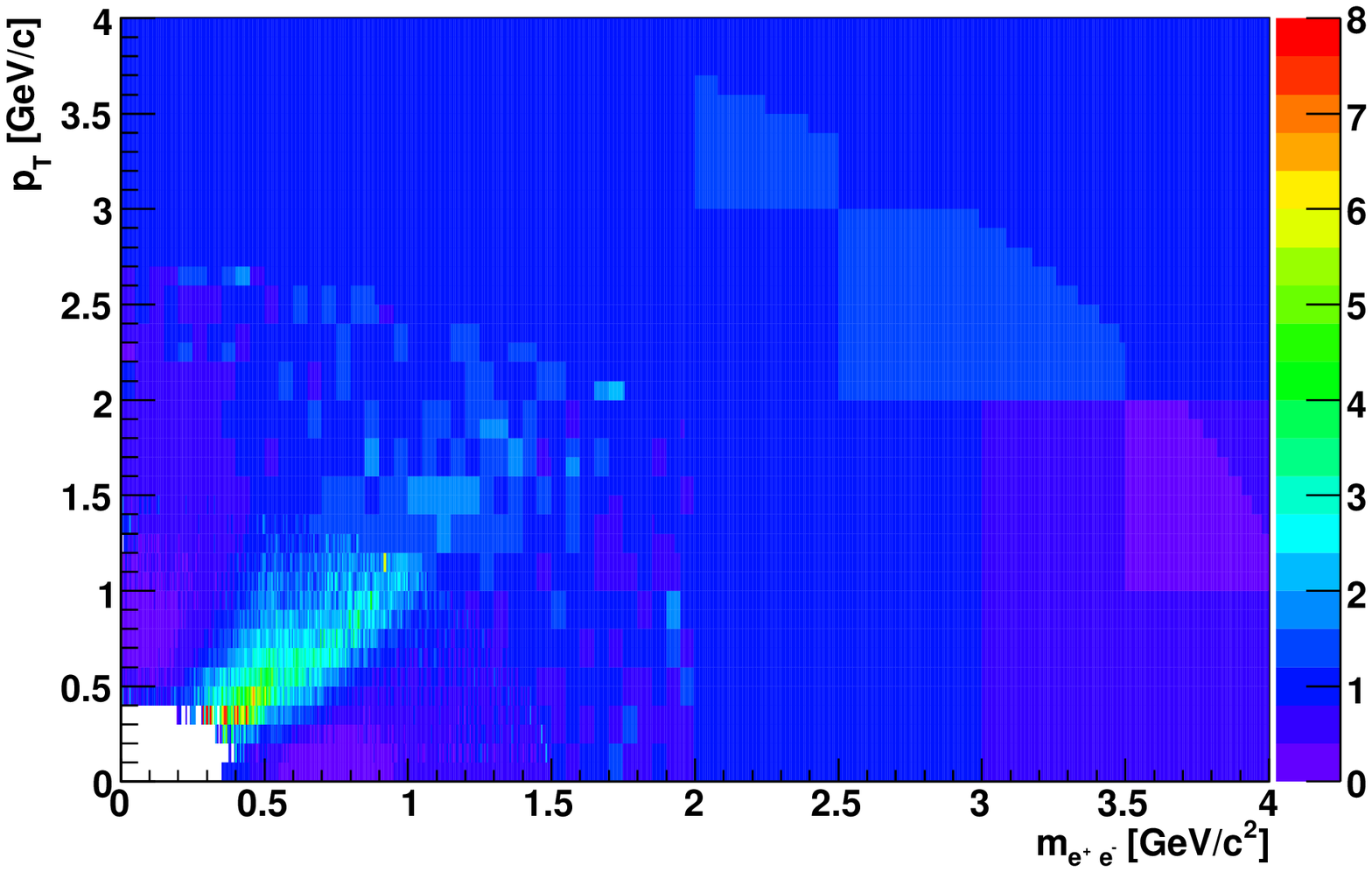}}
  \label{fig:acceptance}
\end{figure*}

The background distributions of like- and unlike-sign pairs are shown
in \fig{fig:acceptance} together with the relative acceptance.

\begin{figure}
  \centering
  \subfloat[LMR]{\label{fig:bgsubcomparison_lmr}\includegraphics[width=0.44\textwidth]{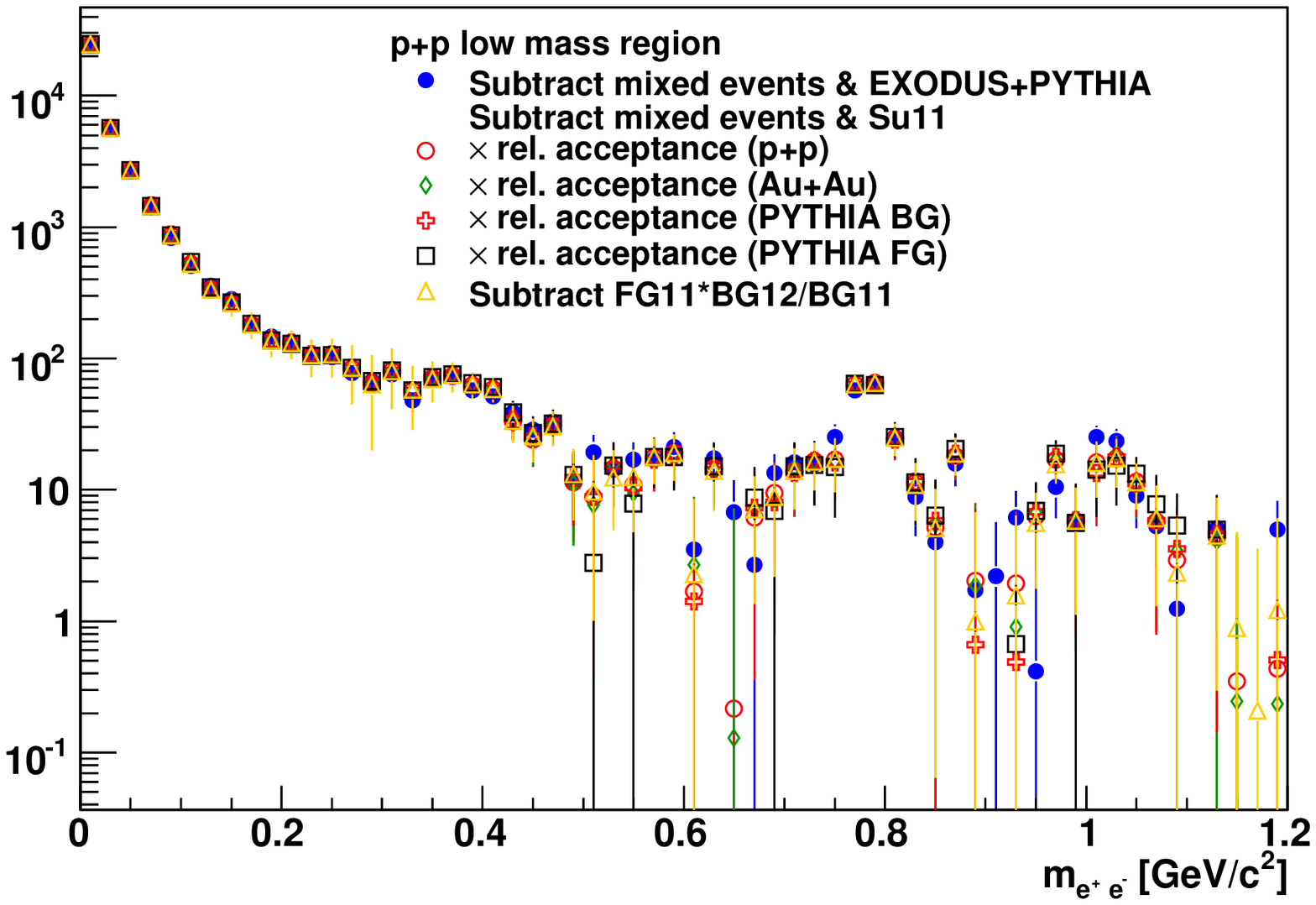}}
  \subfloat[IMR]{\label{fig:bgsubcomparison_imr}\includegraphics[width=0.44\textwidth]{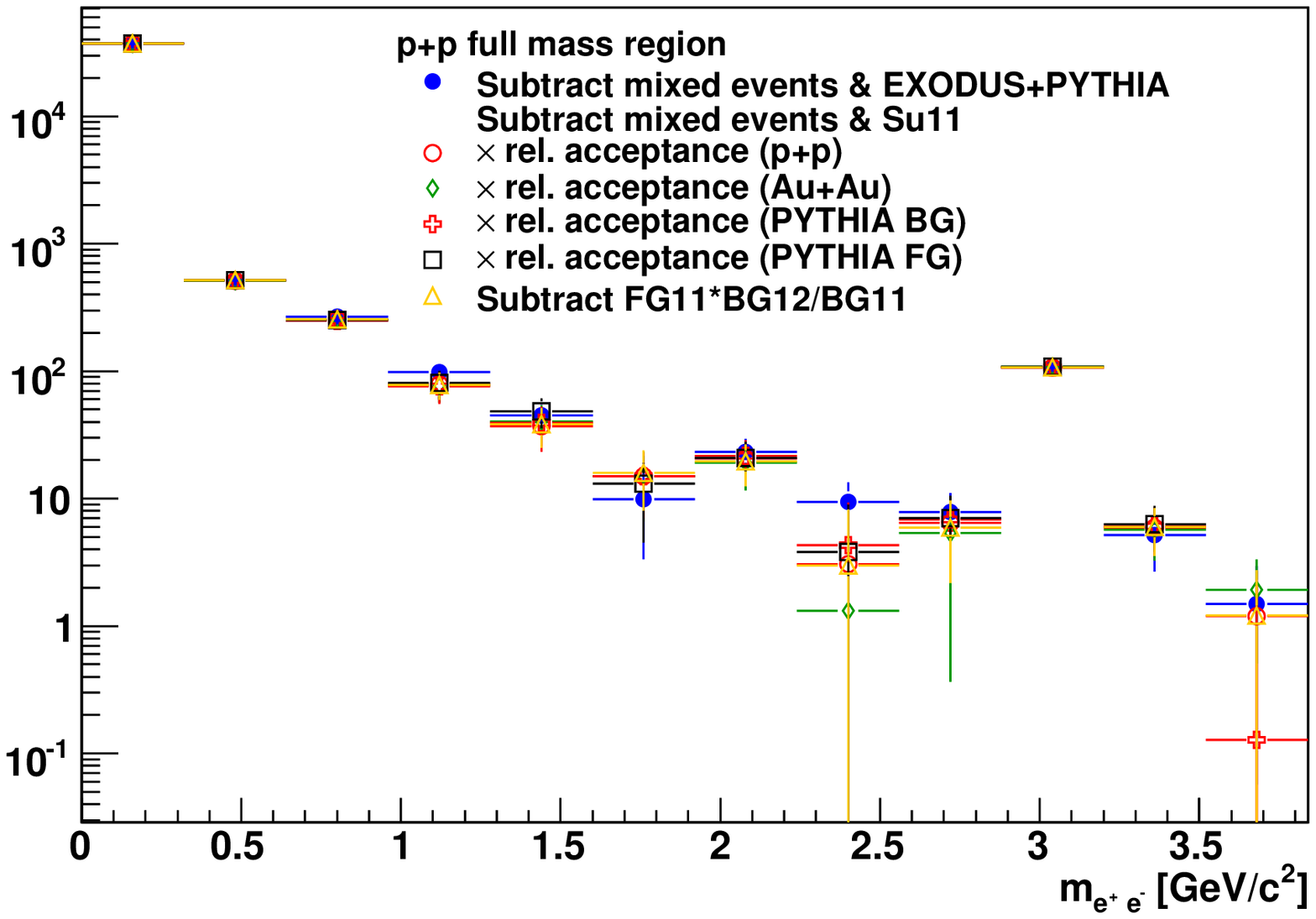}}
  \caption{Comparison of background subtraction methods.}
  \label{fig:bgsubcomparison}
\end{figure}

Alternatively to using the like sign foreground, one can first
subtract the mixed events (normalized as discussed in
Section~\ref{sec:background_norm}) and then subtract the mixed event
subtracted like sign signal, corrected for the relative acceptance,
from the mixed event subtracted unlike-sign spectrum everywhere but in
the normalization region\footnote{Note that a subtraction including
  this region would be arithmetically identical to the previous
  method, which skips the mixed event subtraction.}. To increase
statistics at high \pt and large invariant mass, the Au + Au mixed
events have been used as an alternative description of the relative
acceptance. The relative acceptance for correlated and uncorrelated
pairs may differ. This effect has been studied its effect by using
\pythia foreground spectra to determine the relative acceptance, \ie,
$FG_{+-} /(FG_{++} + FG_{--})$. In addition the mixed events from
\pythia have been used to describe the relative
acceptance. \fig{fig:bgsubcomparison} shows a comparison of the
subtracted unlike-sign spectra with the presented methods.

There are two systematic uncertainties in the background subtraction,
one due to the statistics available in the normalization region for
the various background sources:
\begin{itemize}
\item mixed events: 2.43\%
\item cross pairs: 2.49\%
\item jet pairs: 6.86\%
\end{itemize}

\begin{figure}
  \centering
  \includegraphics[width=0.9\textwidth]{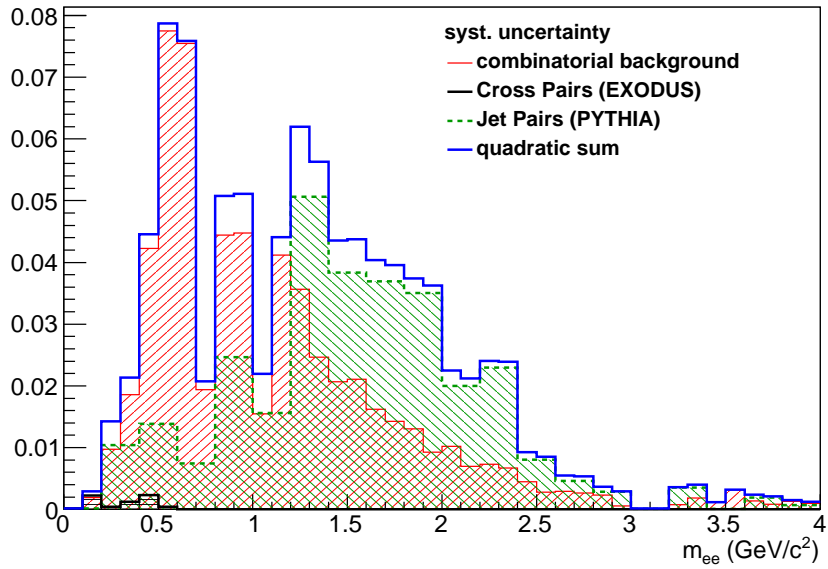}
  \caption{Comparison of background subtraction methods.}
  \label{fig:bgsys}
\end{figure}

\begin{figure}
  \centering
  \subfloat[LMR]{\label{fig:bgrms_lmr}\includegraphics[width=0.44\textwidth]{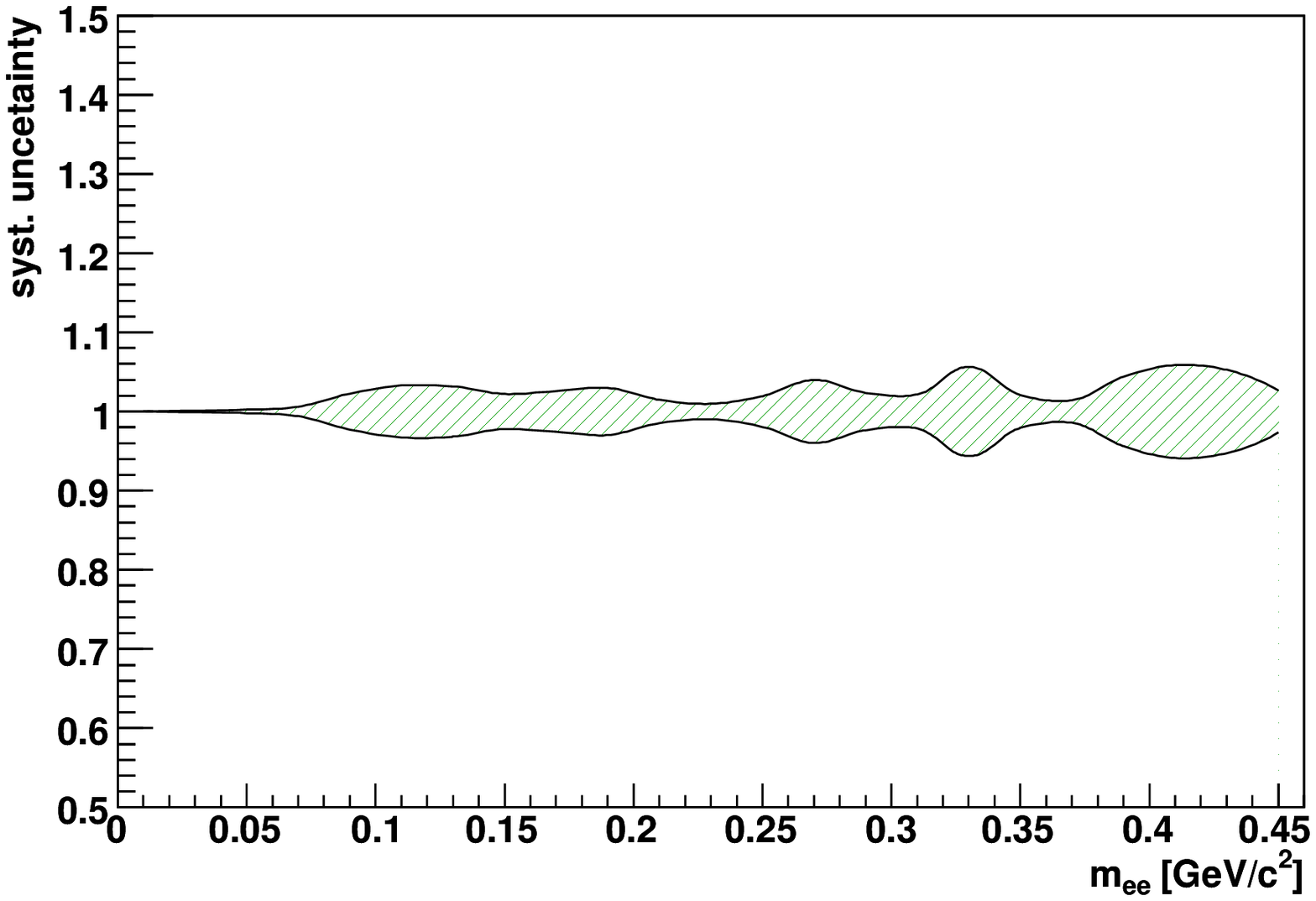}}
  \subfloat[IMR]{\label{fig:bgrms_imr}\includegraphics[width=0.44\textwidth]{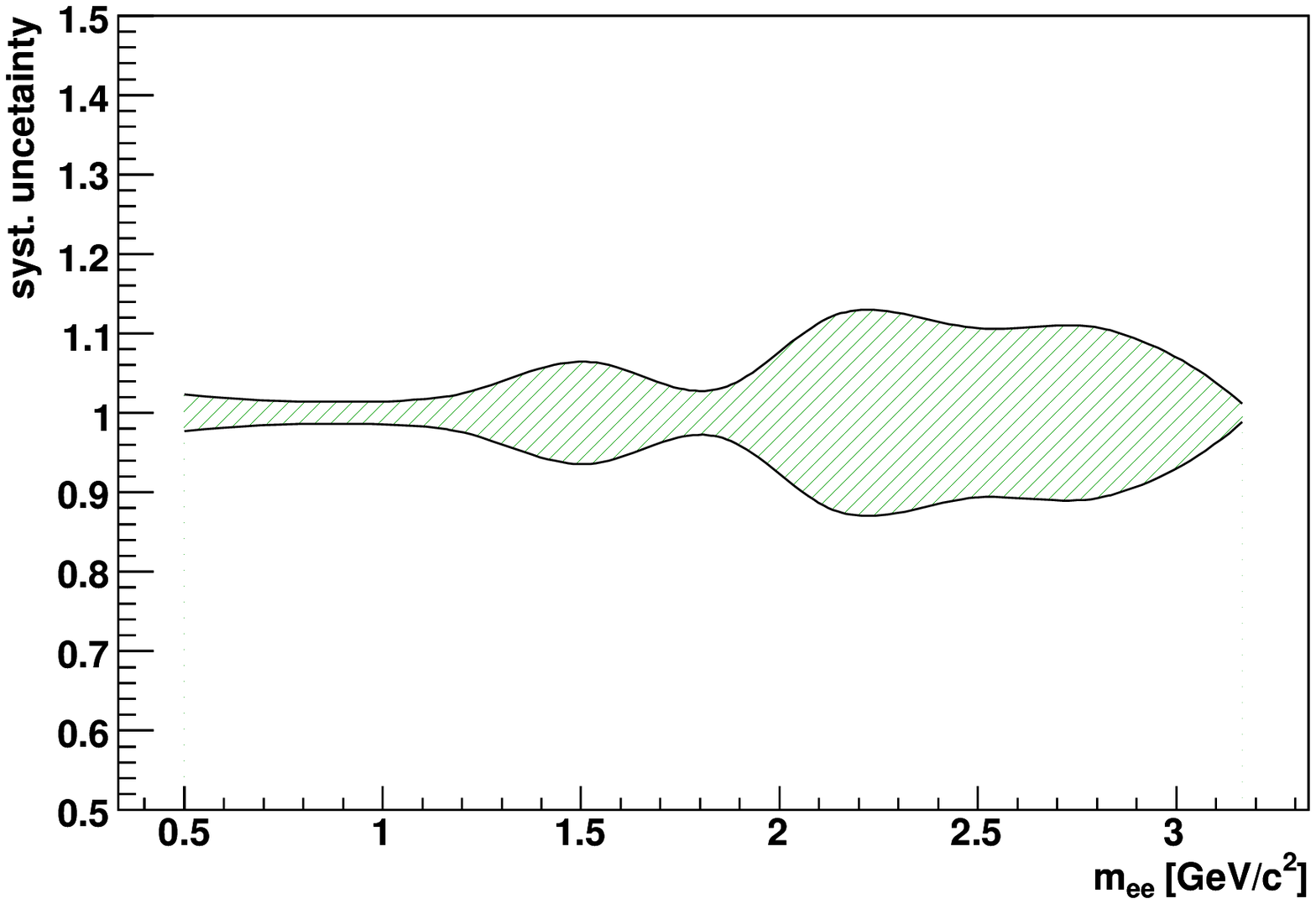}}
  \caption{RMS of all methods as function of \mee.}
  \label{fig:bgrms}
\end{figure}

The systematic uncertainty on the measured signal is shown in
\fig{fig:bgsys}. The other systematic uncertainty is in the shape of
the subtracted correlated background, which is estimated from the RMS
of the various subtraction methods described above, which is shown in
\fig{fig:bgrms}. For $m_{ee} < 2$ GeV a systematic uncertainty of 5\%
(10\% above) has been assigned.

\subsection[Background Subtraction in \AuAu]{Background Subtraction in $\boldsymbol{\rm Au+Au}$}
\label{sec:bgsub_au}
\begin{figure}
  \centering
  \includegraphics[width=0.9\textwidth]{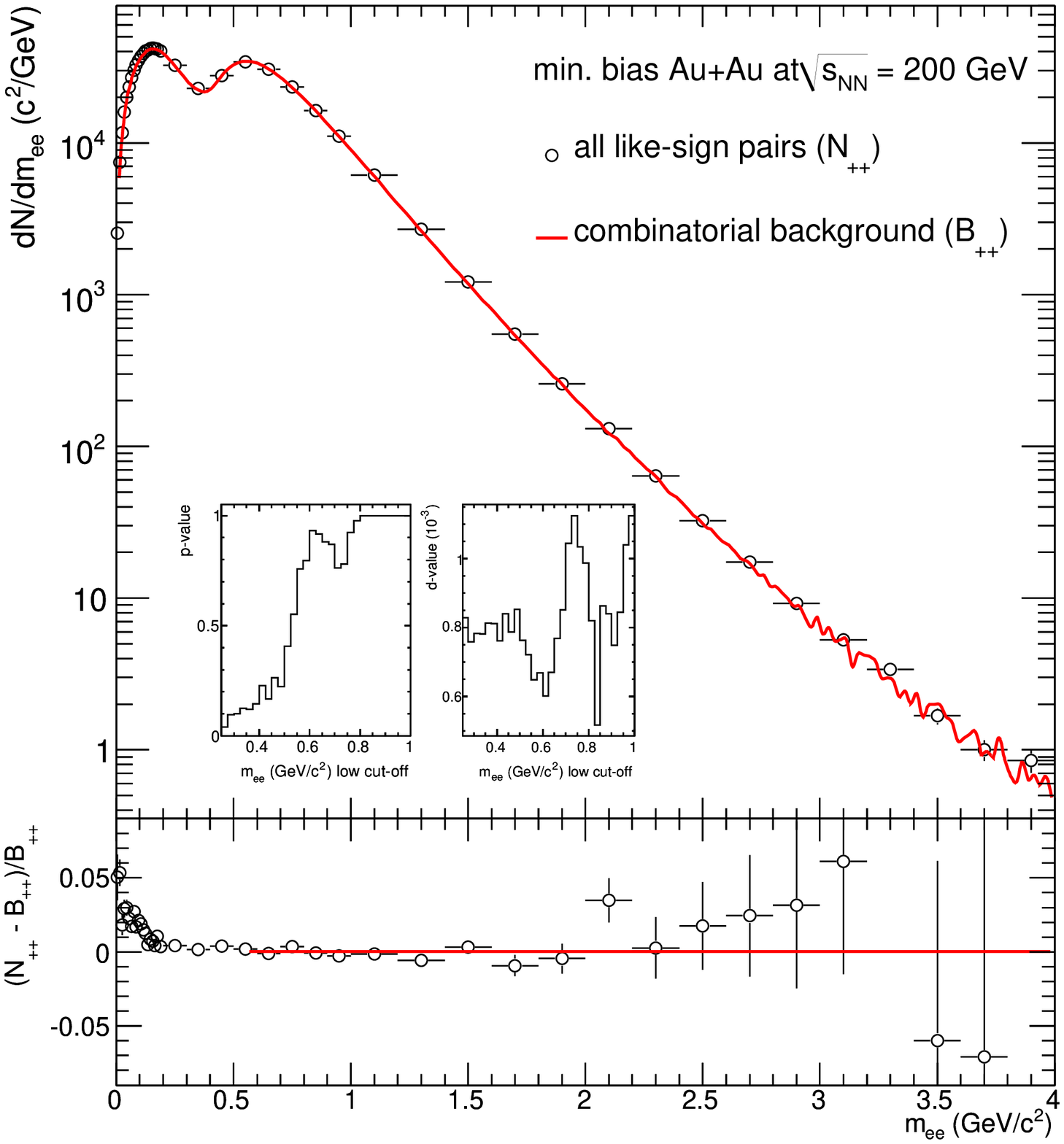}
  \caption[Invariant mass spectra of like-sign pairs in real and mixed
  events in MB \AuAu.]{Invariant mass distribution of like sign pairs
    in real events ({\em open circles}) and mixed events ({\em red
      line}). The mixed events to real events are normalized above
    700~\mevcc. The bottom panel shows the ratio of the mixed event
    subtracted spectrum and the mixed events.}
  \label{fig:au_mb_like}
\end{figure}

\begin{figure}
  \centering
  \includegraphics[width=0.9\textwidth]{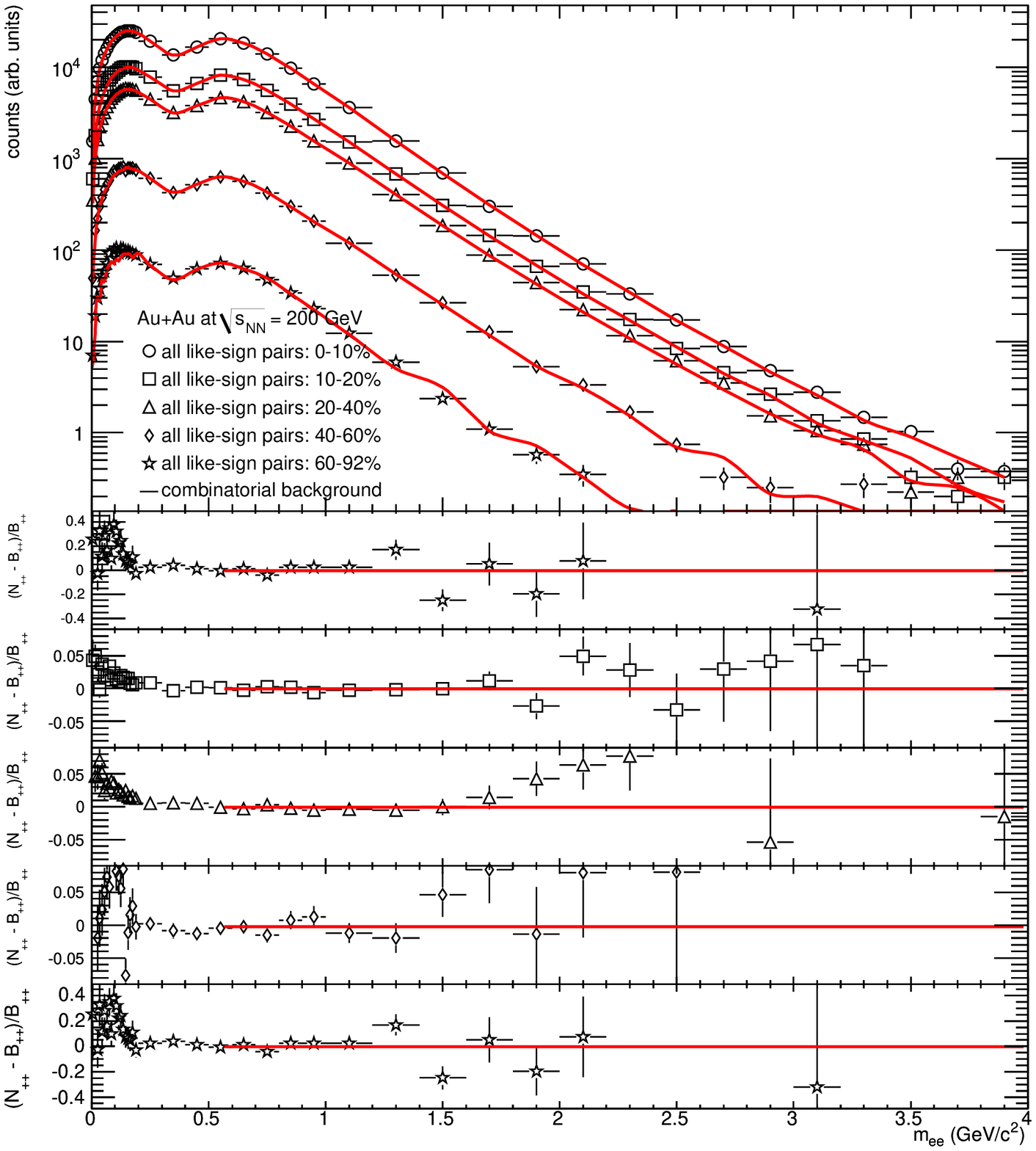}
  \caption[Like-sign pairs in real and mixed events for various \AuAu
  centrality bins.]{Invariant mass distribution of like sign pairs in
    real events ({\em open circles}) and mixed events ({\em red line})
    for different centrality bins. The mixed events to real events are
    normalized above 700~\mevcc. The bottom panel shows the ratio of
    the mixed event subtracted spectrum and the mixed events.}
  \label{fig:au_cent_like}
\end{figure}

The like-sign distribution of real and mixed events in minimum bias
\AuAu collisions are compared in \fig{fig:au_mb_like}. Their shapes
agree not only in the region defined in \eq{eq:regionA}, but for all
masses above 550~\mevcc. Qualitatively, this can be explained with the
suppression of away-side jets observed in \AuAu
collisions~\cite{adare:014901}. At small masses, below 200~\mevcc a
like-signal from cross pairs is observed.
The ratio of mixed event subtracted to mixed events is fitted above
550~\mevcc to a constant, which gives a result of $-0.000259 \pm
0.000633$ well in agreement with zero and a $\chi^2/DOF = 1.45$.
\fig{fig:au_cent_like} shows the like-sign mass spectra of real and
mixed events divided in centrality bins of 0--10\%, 10--20\%,
20--40\%, 40--60\% and 60--90\%. The same fits as for the MB spectra
are performed and summarized in \tab{tab:au_like_fit}, which also
includes the results of a statistical $\chi^2$ test and the maximum
deviation of a Kolmogorov-Smirnov test (also see inserts to
\fig{fig:au_mb_like}). The maximum deviations of Kolmogorov-Smirnov
test are small compared to the uncertainties of the absolute
normalization of the mixed event background. The $p$-values
corresponding the the $\chi^2$ test, which are greater than 0.999 for
all centrality classes, confirm the hypothesis that the two
distributions are compatible and can be accepted for any commonly used
significance level.
\begin{table}
  \centering
  \caption[Fit to ratios of like-sign real and mixed events in MB \AuAu]{\label{tab:au_like_fit}Fit parameters to ratios shown in \fig{fig:au_mb_like} and \fig{fig:au_cent_like}.\\}
  \begin{tabular}{rr@{.}l@{ $\pm$ }r@{.}lr@{.}lr@{.}lr@{.}l}\toprule
    \multicolumn{1}{c}{Centrality}   &  \multicolumn{4}{c}{Constant}   & \multicolumn{2}{c}{$\chi^2$/DOF} & \multicolumn{2}{c}{$\chi^2$ test} & \multicolumn{2}{c}{max. dist}\\\midrule
    0--10\%  & $ 0$&00063 & 0&00088 & 1&25 & 0&58 & 0&0013 \\
    10--20\% & $-0$&0009  & 0&0014  & 1&42 & 0&54 & 0&0018 \\
    20--40\% & $-0$&0024  & 0&0018  & 1&12 & 0&56 & 0&0037 \\
    40--60\% & $-0$&0085  & 0&0050  & 1&42 & 0&69 & 0&0099 \\
    60--92\% & $-0$&018   & 0&016   & 1&56 & 1&02 & 0&042  \\
    00--92\% & $-0$&00026 & 0&00063 & 1&45 & 0&51 & 0&0011 \\\bottomrule
  \end{tabular}
\end{table}

\begin{figure}
  \centering
  \includegraphics[width=0.9\textwidth]{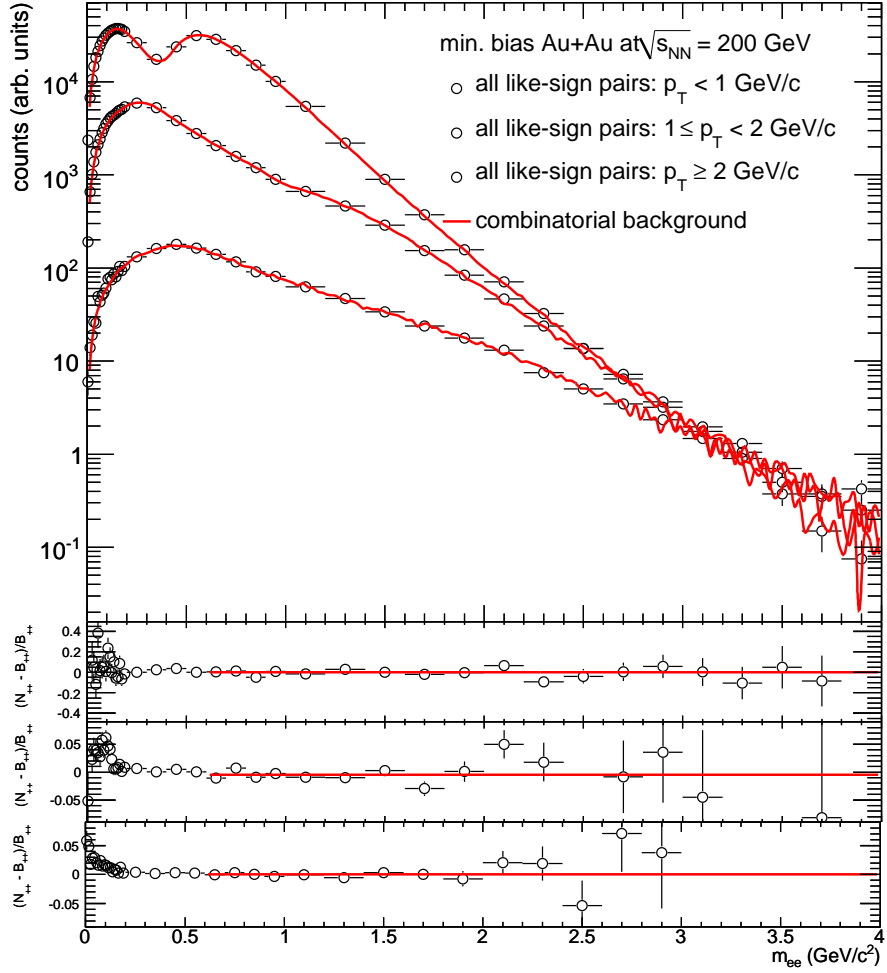}
  \caption[Invariant mass spectra of like-sign pairs in real and mixed
  events in MB \AuAu for various \pt bins.]{Invariant mass
    distribution of like sign pairs in real events ({\em open
      circles}) and mixed events ({\em red line}) for different \pt
    bins. The mixed events to real events are normalized above
    700~\mevcc. The bottom panel shows the ratio of the mixed event
    subtracted spectrum and the mixed events.}
  \label{fig:au_pt_like}
\end{figure}

Finally, \fig{fig:au_pt_like} shows the invariant mass spectra of
like-sign pairs in real and mixed event for different \pt ranges: $\pt
< 1$~\gevc, $1 \leq \pt < 2$\gevc, and $\pt \geq 2$~\gevc. Again no
deviation of the mixed event shape from the real event distribution is
observed for $\mee > 550$~\mevcc.

The mixed event are normalized to the like-sign pairs in real events
in the mass range $0.7 < \mee < 4$~\gevcc:
\begin{equation}
  \frac{\int_{0.7}^{4.0}BG_{++} + BG_{--}}{\int_{0.7}^{4.0}FG_{++} +
    FG_{--}} \equiv 1
\end{equation}
The lower limit of 0.7~\gevcc is chosen conservatively to avoid any
contamination from correlated background observed at low mass and
provides at the same time sufficient statistical accuracy of
0.12\%. The absolute normalization of the unlike-sign background is
given by the geometrical mean of the normalized like-sign integrals:
$BG_{+-} =\sqrt{BG_{++} + BG_{--}}$. An Additional 0.2\% uncertainty,
due to a correction of that normalization by $1.004\pm0.002$ for the
fact that the applied pair cuts removes a different fraction of like-
than unlike-sign pairs, yields to a total systematic uncertainty of
0.25\% on the background normalization, which is a conservative upper
limit of the uncertainty.

The correlated background, remaining after mixed event subtraction,
due to cross pairs and a contribution of jet-pairs at small masses is
subtracted analog to the \pp analysis. But, in contrast to \pp, there
is no room for a correlated signal due to away-side jet
correlations. Due to the known modifications of the away-side jet, the
\AuAu data are indeed expected to differ from the \pythia
simulation. Therefore, the near-side correlations, \ie, pairs with
$\Delta\phi < \pi/2$ are separated from the away-side contribution
with $\Delta\phi > \pi/2$ and the three components cross, near-, and
aways-side pairs are fitted to the like-sign correlated background
distribution:
\begin{equation}
  \left(\frac{dN}{dm}\right)_{\rm like} = A
  \left(\frac{dN}{dm}\right)_{\rm cross} + B
  \left(\frac{dN}{dm}\right)_{\rm near} + C \left(\frac{dN}{dm}\right)_{\rm away}
\end{equation}

\begin{figure}
  \centering
  \includegraphics[width=0.9\textwidth]{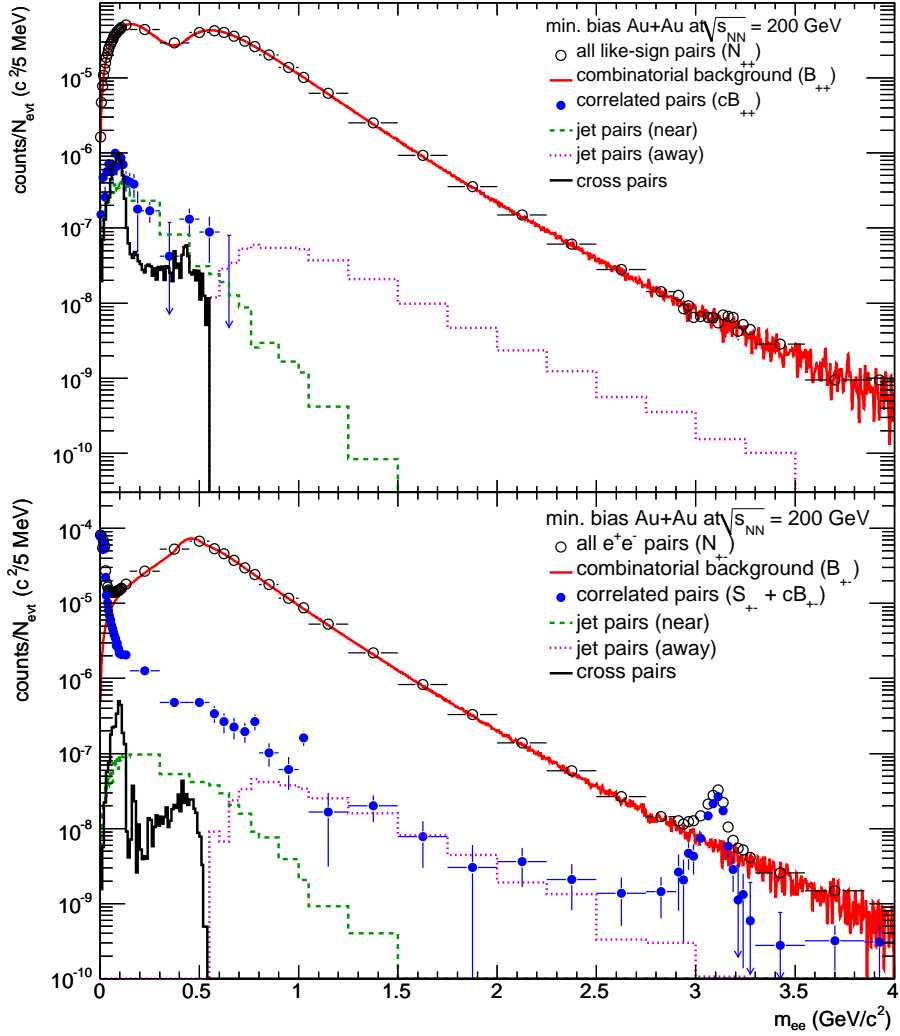}
  \caption[Invariant mass spectra of like- and unlike-sign pair in
  \AuAu.]{Raw dielectron spectra in min. bias \AuAu. The top panel
    shows like-sign pairs as measured in the experiment, the
    combinatorial background from mixed events, the correlated pair
    background obtained by subtracting the combinatorial background,
    and the individual contributions from cross and near- and
    away-side jet pairs to the correlated background (see text). The
    bottom panel shows the same distributions for unlike-sign pairs.
    The correlated background in both panels is normalized to the
    measured like-sign pairs remaining after subtracting the
    combinatorial background.}
  \label{fig:rawspectra_au}
\end{figure}

As the mixed events have been normalized at $\mee > 0.7$~\gevcc it
follows that $C=0$. The different background contributions are
illustrated in \fig{fig:rawspectra_au}. Here the away-side curve is
shown for illustration only and is normalized as the near-side
curve. The overall contribution of correlated pairs relative to the
unlike-sign signal is small, and comparing different subtraction
strategies leads to a systematic uncertainty of 10\% on correlated
background subtraction in the LMR.

\section{Efficiency Correction}
\label{sec:eff_corr}

The distribution of \ee pairs after subtraction of all background
sources needs to be corrected for the pair efficiencies of track
reconstruction ($\varepsilon_{\rm reco}^{\rm pair}$), electron identification
($\varepsilon_{\rm eID}^{\rm pair}$) and (in case of \pp) triggering by the ERT
($\varepsilon_{\rm ERT}^{\rm pair}$). The pair efficiencies for track
reconstruction and electron identification are the product of the
single efficiencies, \ie, $\varepsilon_{\rm reco}^{\rm pair} = \varepsilon_{\rm
  reco}^+ \varepsilon_{\rm reco}^-$ and $\varepsilon_{\rm eID}^{\rm pair} =
\varepsilon_{\rm eID}^+\varepsilon_{\rm eID}^-$, while the trigger
efficiency for a pair can be expressed in terms of the single trigger
efficiency as: $\varepsilon_{\rm ERT}^{\rm pair} = (1-\varepsilon_{\rm
  ERT}^+)\varepsilon_{\rm ERT}^- + (1-\varepsilon_{\rm
  ERT}^-)\varepsilon_{\rm ERT}^+ + \varepsilon_{\rm
  ERT}^+\varepsilon_{\rm ERT}^-$. Tracking and electron identification
efficiency include effects due to dead areas within the detector
acceptance. In addition the efficiency of the pair cuts
($\varepsilon_{\rm pair~cuts}$) needs to be taken into account. The
idea is that the invariant mass spectra are corrected ``into the
PHENIX acceptance'' as:
\begin{equation}\label{eq:yield_invmass}
  \frac{dN}{d\mee} = \frac{1}{N_{\rm evt}}\frac{1}{\Delta\mee}\frac{N_{ee}}{\varepsilon_{\rm pair}}\frac{\varepsilon_{\rm
        BBC}}{\varepsilon_{\rm bias}}
\end{equation}
with $\Delta\mee$ the bin width in \gevcc, $\varepsilon_{\rm pair} =
\varepsilon_{\rm reco}^{\rm pair}\varepsilon_{\rm eID}^{\rm
  pair}\varepsilon_{\rm ERT}^{\rm pair}\varepsilon_{\rm pair~cuts}$ is
the pair efficiency, $\varepsilon_{\rm BBC} = 0.545 \pm 0.06$ is the
BBC efficiency and $\varepsilon_{\rm bias} = 0.79 \pm 0.02$ is the BBC
trigger bias.

On the other hand, \pt distributions of \ee pairs are corrected for
the limited geometric acceptance ($\varepsilon_{\rm geo}^{\rm pair} =
\varepsilon_{\rm geo}^+ \varepsilon_{\rm geo}^-$) into full azimuth
($0 < \phi \leq 2\pi$) and into one unit of rapidity ($|y|<0.5$). The
spectra are shown as yield invariant under Lorentz transformation:
\begin{equation}\label{eq:invyield}
  \frac{1}{2\pi\, \pt}\frac{d^2N}{d\pt\, dy} = \frac{1}{2\pi\, \pt} \frac{1}{N_{\rm
      evt}\, \Delta\pt}\frac{N_{ee}}{\varepsilon_{\rm
      pair}\, \varepsilon_{\rm geo}^{\rm pair}}\frac{\varepsilon_{\rm
      BBC}}{\varepsilon_{\rm bias}}.
\end{equation}
This invariant yield can be converted into an invariant cross section
by multiplying with the inelastic \pp cross section $\sigma_{pp}^{\rm
  inel} = 42.2$~nb.
\begin{equation}
  E \frac{d^3\sigma}{dp^3} = \frac{1}{2\pi\, \pt}\frac{d^2N}{d\pt\, dy} \sigma_{pp}^{\rm inel}
\end{equation}

\subsection{Tracking and Electron Identification Efficiency}
\label{sec:eid_eff}

The tracking and electron identification efficiency is calculated for
single electrons based on a full Monte Carlo simulation of \pion
decays. 450M events are generated containing \pion's flat in rapidity
($|y| < 0.5$), azimuth ($0 < \phi \leq 2\pi$), vertex distribution
($|z_{\rm vertex}| < 30$~cm), and in \pt ($0 < \pt < 25$~\gevc). To
enhance statistics the Dalitz decay branching ratio is set to 100\%.

\begin{figure}
  \centering
  \includegraphics[width=0.9\textwidth]{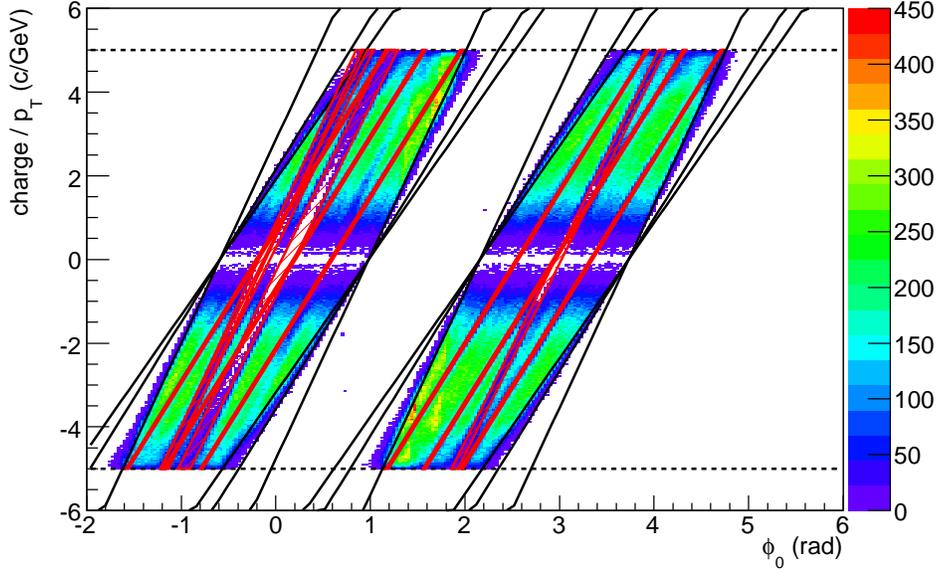}
  \caption[Single electron acceptance parameterization]{Single
    electron acceptance. The diagonal line represent the acceptance
    limits due to RICH and DC as defined in \eq{eq:track_acc}. The
    dashed lines indicate the low \pt cut off at 200~\mevc. Dead areas
    are removed with fiducial cuts shown as red shaded areas.}
  \label{fig:fid_acc}
\end{figure}

\begin{figure}
  \centering
  \subfloat[$e^-$]{\label{fig:pp_e_eid}\includegraphics[width=0.44\textwidth]{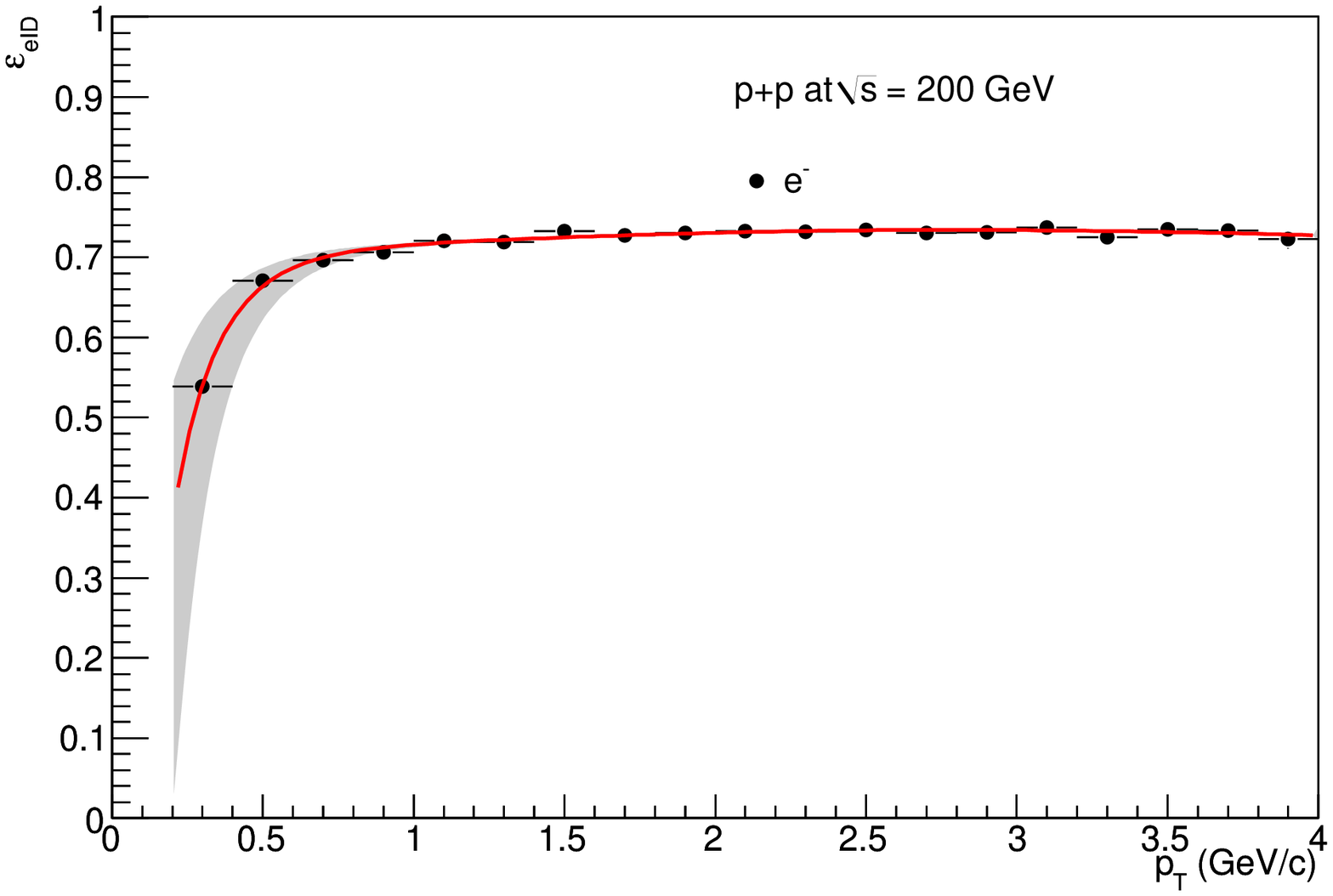}}
  \subfloat[$e^+$]{\label{fig:pp_p_eid}\includegraphics[width=0.44\textwidth]{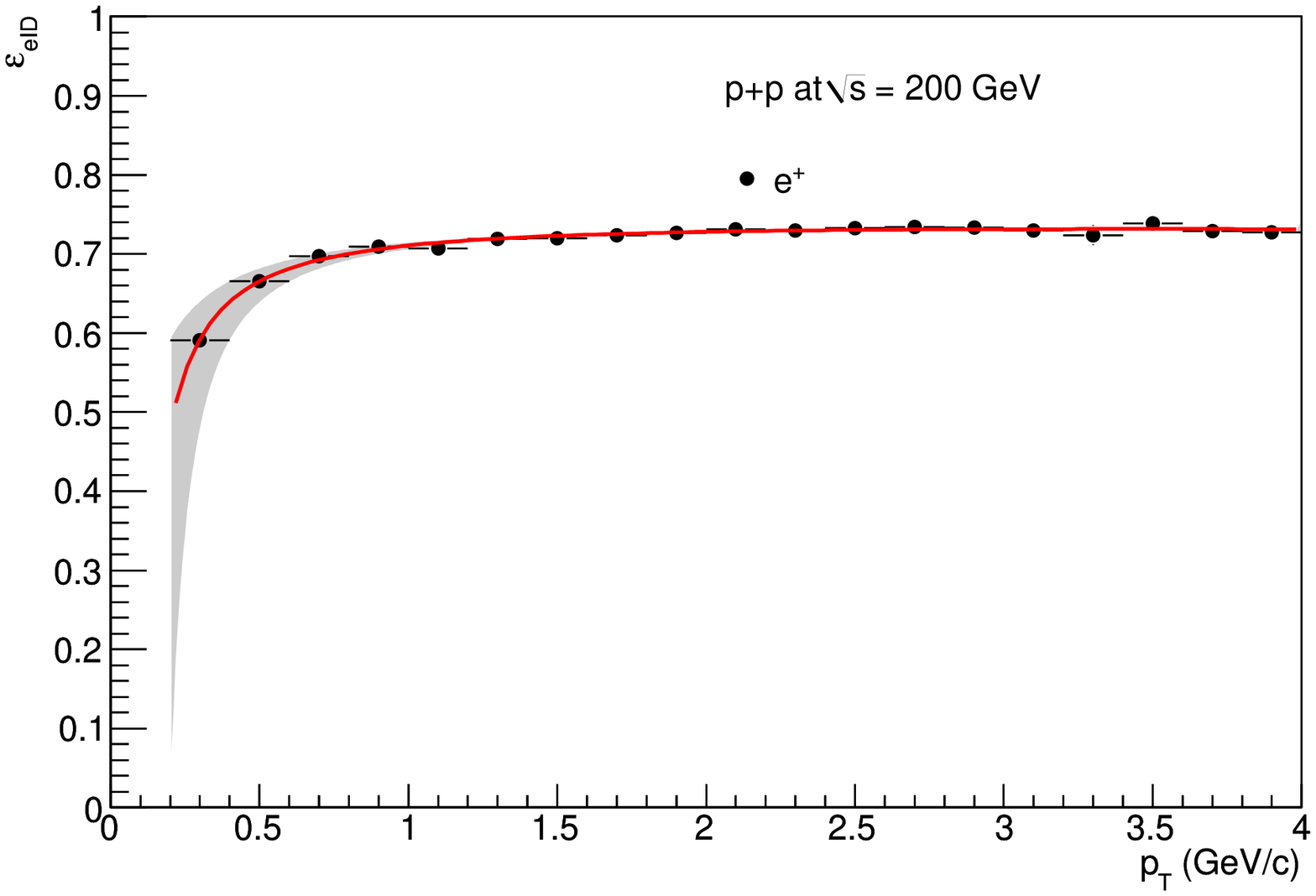}}\\
  \subfloat[$e^-$]{\label{fig:au_e_eid}\includegraphics[width=0.44\textwidth]{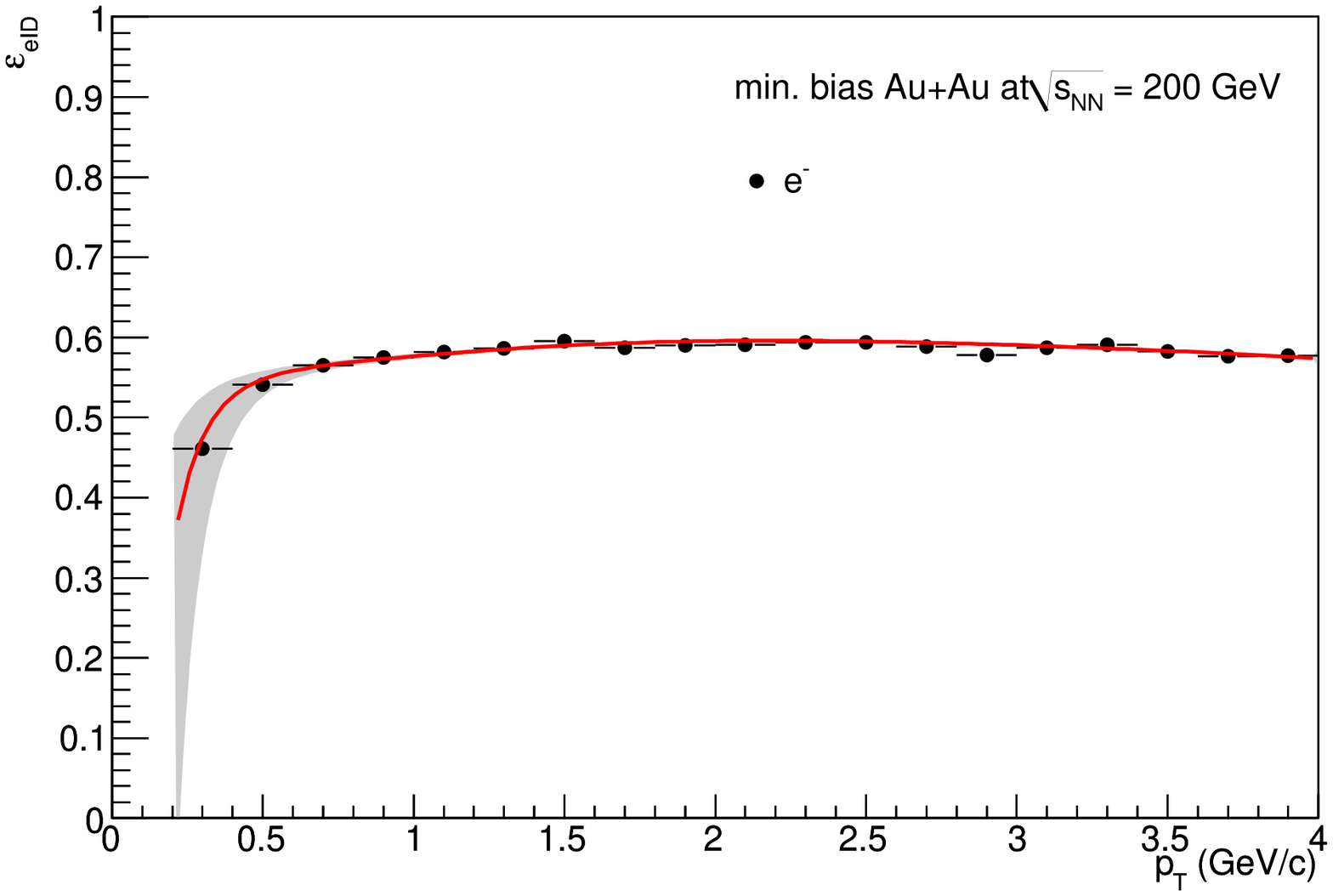}}
  \subfloat[$e^+$]{\label{fig:au_p_eid}\includegraphics[width=0.44\textwidth]{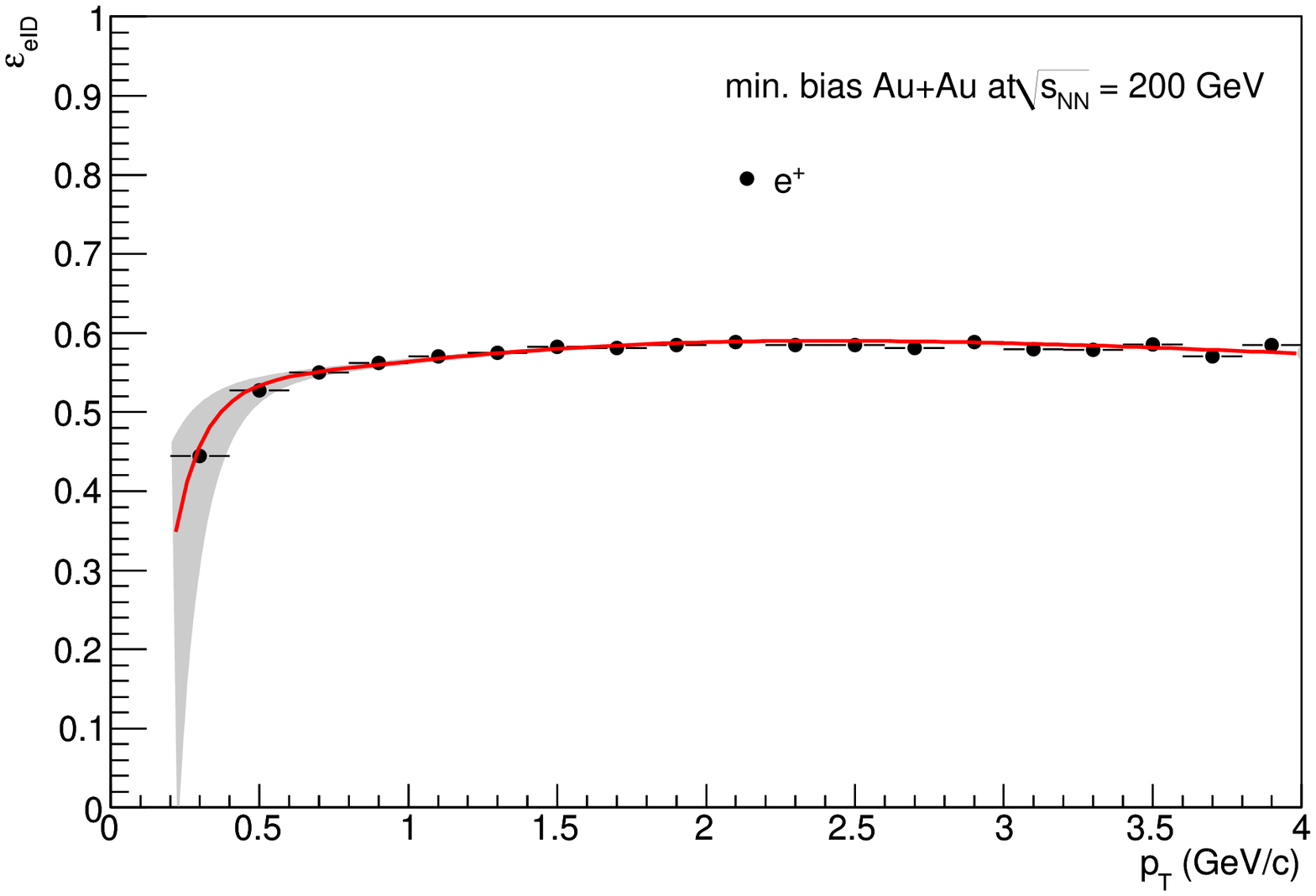}}
  \caption[Single electron identification efficiency in \pp an
  \AuAu]{Track reconstruction and electron identification efficiency
    in \pp collisions~\subref{fig:pp_e_eid}--\subref{fig:pp_p_eid} and
    in \AuAu~\subref{fig:au_e_eid}--\subref{fig:au_p_eid} as function
    of the single electron \pt.}
  \label{fig:single_eid}
\end{figure}

These events have been processed through a full GEANT simulation
program of the PHENIX detector~\cite{geant} which includes the details
of the detector response. The output is analyzed by the PHENIX event
reconstruction chain. The same single electron cuts are applied in
simulation as in real data. Every electron is weighted according to a
realistic \pt distribution of the parent \pion. The electron
identification efficiency is determined as function of the single
electron \pt:
\begin{equation}
  \varepsilon_{\rm eID}^{\pm} = \frac{dN_{\rm out}^{\pm}/d\pt^{\pm}}{dN_{\rm in}^{\pm}/d\pt^{\pm}}
\end{equation}
$dN_{\rm in}^{\pm}/d\pt^{\pm}$ is the \pt distribution of generated
electrons that fall into active area within the PHENIX acceptance as
parameterized in \eq{eq:track_acc}. The active area is parameterized
from data as function of the single particle charge sign $q$, \pt and
azimuthal angle $\phi_0$ as shown in \fig{fig:fid_acc}. Dead areas are
removed with fiducial cuts which are listed in \tab{tab:fidcuts} in
the simulations. $dN_{\rm out}^{\pm}/d\pt^{\pm}$ is the number of
generated electrons in the same fiducial area that pass the electron
identification cuts. \fig{fig:single_eid} shows the single electron
(positron) efficiency as function of \pt in \pp collisions.

\begin{table}
\centering
\caption[Fiducial Cuts]{\label{tab:fidcuts}
  Fiducial cuts in the PHENIX acceptance.\\}
\begin{tabular}{lr@{$<$}l@{~\&\&~}r@{$>$}r}
  \toprule
 ~ & \multicolumn{4}{c}{Dead Area}\\\midrule
 A & $-0.88 + 4.56 \phi_0$ & $q/\pt$ &  $q/\pt$ & $-0.28+4.56 \phi_0$ \\
 B & $0.28+4.56 \phi_0$    & $q/\pt$ & $q/\pt$  & $0.58+4.56 \phi_0$  \\
 C & $0.52+4.86 \phi_0$    & $q/\pt$ & $q/\pt$  & $0.88+4.86 \phi_0$  \\
 D & $-2.68+4.56 \phi_0$   & $q/\pt$ & $q/\pt$  & $-2.38+4.56 \phi_0$ \\
 E & $-13.78+4.56 \phi_0$  & $q/\pt$ & $q/\pt$  & $-13.48+4.56 \phi_0$\\
\bottomrule
\end{tabular}
\end{table}
The correction for \ee pairs is determined with the hadron decay
generator \exodus. The single electron (positron) efficiency is
parameterized by a fit of the following functional form:
\begin{equation}
  f(\pt) = a + b/\pt + c/\pt^2 + d \pt + \eexp{(e + f \pt)}
\end{equation}
The fit values for electrons and positrons in \pp and \AuAu
collisions, respectively, are tabulated in \tab{tab:eidfit}. This
parameterization has been implemented in \exodus and used as a weight
to each electron. In addition \ee pairs are rejected if at least one
of the electrons falls in one of the dead areas defined in
\tab{tab:fidcuts}. Then the efficiency to reconstruct and identify an
\ee pair is determined as function of the pair \pt and the invariant
mass.
\begin{table}
\centering
\caption[Electron identification efficiency]{\label{tab:eidfit}
  Parameterization of the single electron efficiency.\\}
\begin{tabular}{lr@{.}lr@{.}lr@{.}lr@{.}l}
  \toprule
  ~ & \multicolumn{4}{c}{\pp}&\multicolumn{4}{c}{\AuAu}\\
  ~ &  \multicolumn{2}{c}{electrons} & \multicolumn{2}{c}{positrons} & \multicolumn{2}{c}{electrons} & \multicolumn{2}{c}{positrons}\\\midrule
  $a$                &  0&991     & $-0$&662   &  0&824                 &  0&818   \\
  $b$ (\gevc)        & $-0$&244   & $-0$&0400  & $-0$&168               & $-0$&182 \\
  $c$ (\gevc)$^2$    &  0&00963   & $-0$&00431 &  8&12$\times 10^{-4}$  &  8&88$\times 10^{-4}$\\
  $d$ (\gevc)$^{-1}$ & $-0$&0260  & $-0$&00349 & $-0$&0324              & $-0$&0301 \\
  $e$                & $-0$&478   &  0&417     & $-0$&340               & $-0$&236  \\
  $f$ (\gevc)$^{-1}$ & $-2$&01    &  5&60$\times 10^{-9}$ & $-3$&14     & $-3$&15   \\     
\bottomrule
\end{tabular}
\end{table}

This method has been cross checked with a full Monte Carlo simulation
of \ee pairs. About 1M \ee pairs have been generated, one half of them
flat in mass (0--4~\gevcc) and \pt (0--4~\gevc), the other half
linearly falling in mass (0--1~\gevcc) and \pt (0--1~\gevc) to enhance
the statistics at low mass and \pt where the efficiency varies the
most. The pairs have been simulated flat in azimuth ($0 < \phi \leq
2\pi$) and rapidity ($|y| < 0.5$). Only pairs with both electron and
positron in the ideal PHENIX acceptance were processed by GEANT,
reconstructed and analysed with the same cuts as the data. As with the
previous method, the efficiency is calculated as function of the \ee
pair \pt and mass.

\begin{figure}
  \centering
  \subfloat[\pp]{\label{fig:pp_pair_eid}\includegraphics[width=0.44\textwidth]{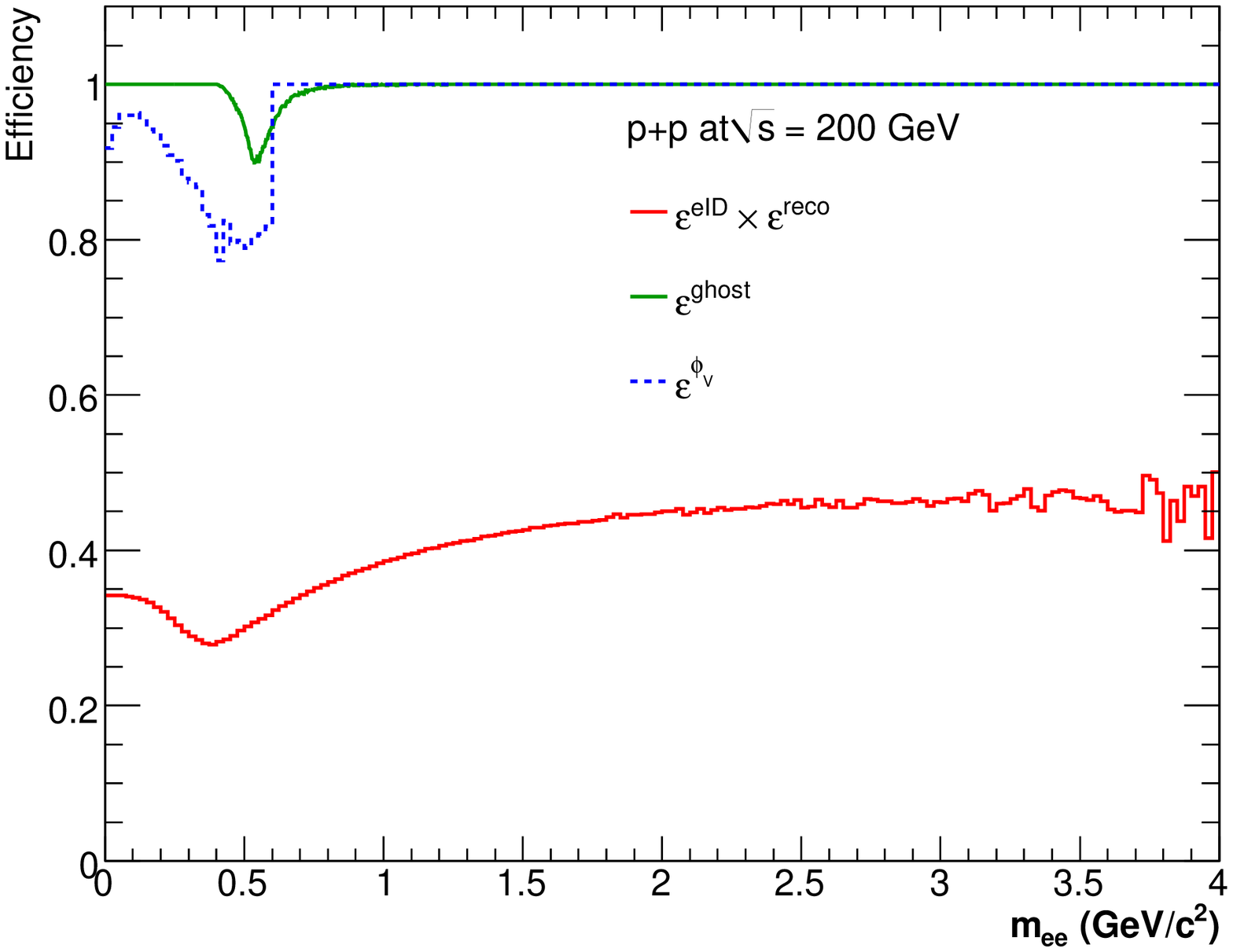}}
  \subfloat[\AuAu]{\label{fig:au_pair_eid}\includegraphics[width=0.44\textwidth]{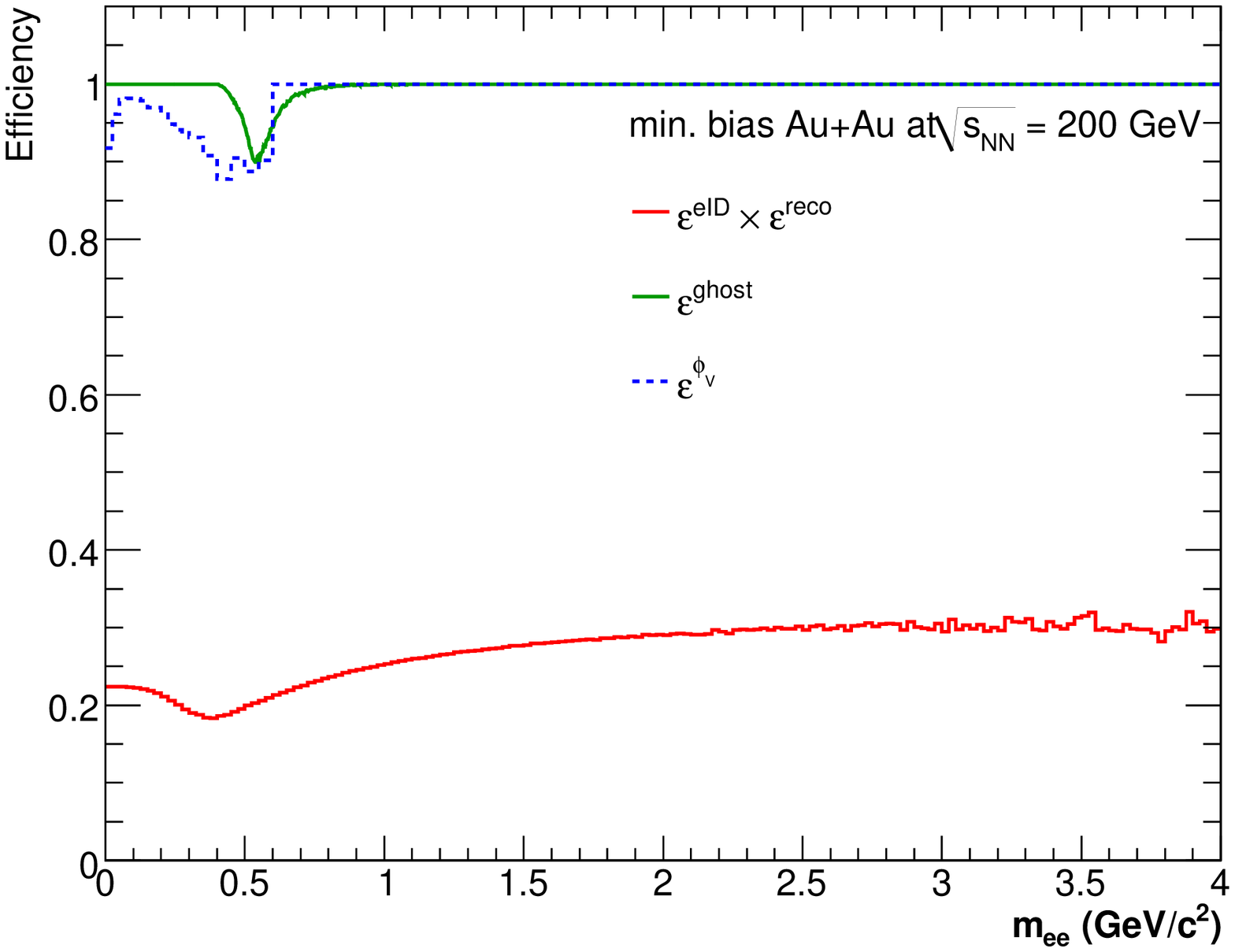}}
  \caption[Pair efficiency in \pp and \AuAu]{Track reconstruction and
    electron identification efficiency in \pp
    collisions~\subref{fig:pp_pair_eid} and in
    \AuAu~\subref{fig:au_pair_eid} as function of \ee pair invariant
    mass.}
  \label{fig:paireff_mass}
\end{figure}

The efficiency to reconstruct and identify an \ee pair is shown as
function of its invariant mass in \fig{fig:paireff_mass} averaged over
all \pt assuming the \pt shape from the hadron decay cocktail. At high
masses the pair efficiency of $\approx 45$\% is the square of the
plateau value of the single efficiency which consist out of the
efficiency of the track reconstruction (90\%), the electron
identification (75\%) and the dead areas (65\%) within the
acceptance. Towards smaller masses the shape of the pair efficiency is
the result of the convolution of the single electron efficiency which
steeply falls towards low \pt and the geometrical acceptance which
cuts the single electron spectrum at $\pt < 200$~\mevc. In the mass
range $0.4 < \mee \leq 0.8$~\gevcc the pair efficiency drops as a
result of the single efficiency decrease at low single electron
momenta. But for masses below 0.4~\gevcc the geometrical acceptance
removes all pairs with $\mt = \sqrt{\pt^2 + \mee^2} < 0.4$~\gevcc,
which leads to a higher average \pt for \ee pairs in this mass window
and therefore a rise of the efficiency.

The effect of the pair cuts on legitimate signal pairs is shown
separately in \fig{fig:paireff_mass}. The efficiency is due to
legitimate pairs that accidentally fulfill one of the pair cut
criteria, \eg, share the same ring in the RICH and are therefore
removed. This efficiency is studied with mixed events in which any
detector overlap is accidental by definition. The pair cut efficiency
is calculated as:
\begin{equation}
  \varepsilon_{\rm pair~cuts} = \frac{d^2N_{\rm mix}^{\rm pair~cuts}}{d\mee
    d\pt}\left/\frac{d^2N_{\rm mix}}{d\mee d\pt}\right.
\end{equation}
where $d^2N_{\rm mix}^{\rm pair~cuts}/(d\mee d\pt)$ is the mixed event
distribution of \ee pairs with pair cuts and $d^2N_{\rm mix}/(d\mee
d\pt)$ the one without pair cuts.

\begin{figure}
  \centering
  \includegraphics[width=0.9\textwidth]{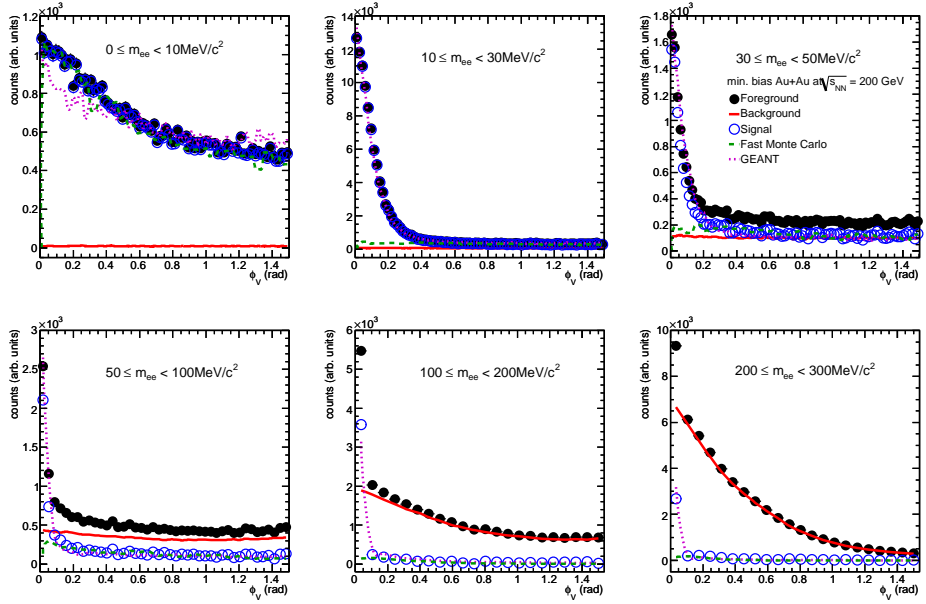}
  \caption[$\phi_V$ distributions in data and Monte Carlo]{Comparison
    of $\phi_V$ distributions for various mass regions in data
    (\AuAu), \exodus and the GEANT simulation.}
  \label{fig:phiv}
\end{figure}

The $\phi_V$ removes not only photon conversions but also signal pairs
which happen to be oriented in the magnetic field like
conversions. This effect is studied with the full GEANT Monte Carlo
simulation of \pion's which was already used for the electron
identification efficiency. Above the \pion mass this is extended with
the help of hadron decay simulations with \exodus. To simulate the
detector resolution, an empirical smearing of the magnitude and
direction of the 3-momentum vector of the electrons was implemented to
reproduce the $\phi_V$ distribution observed in data and the GEANT
simulation. The Gaussian smearing of the azimuthal and polar angle of
the 3-momentum vector $\vec{p}$ of a track with charge sign $q=\pm1$
was parameterized as follows:
\begin{subequations}\label{eq:angle_smearing}
  \begin{align}
    \mu_{\phi} &= q \cdot 0.002 \left(0.021 + \frac{0.16}{|\vec{p}|^2}\right) \\
    \sigma_{\phi} &= 0.0023 \sqrt{5.1+\frac{0.46}{|\vec{p}|^3}}\\
    \mu_{\theta} &= 0.0\\
    \sigma_{\theta} &= 0.001 \sqrt{0.54+\frac{0.36}{|\vec{p}|^3}}
  \end{align}
\end{subequations}
The resulting $\phi_V$ distribution is compared for different mass
ranges to data and the full GEANT simulation in \fig{fig:phiv}.
The efficiency is calculated from the ratio of \ee pairs with and
without applying the $\phi_V$ cut and is shown together with the other
efficiencies in \fig{fig:paireff_mass}.

\subsection{ERT Efficiency}
\label{sec:ert_eff}

\begin{figure}
  \centering
  \includegraphics[width=0.9\textwidth]{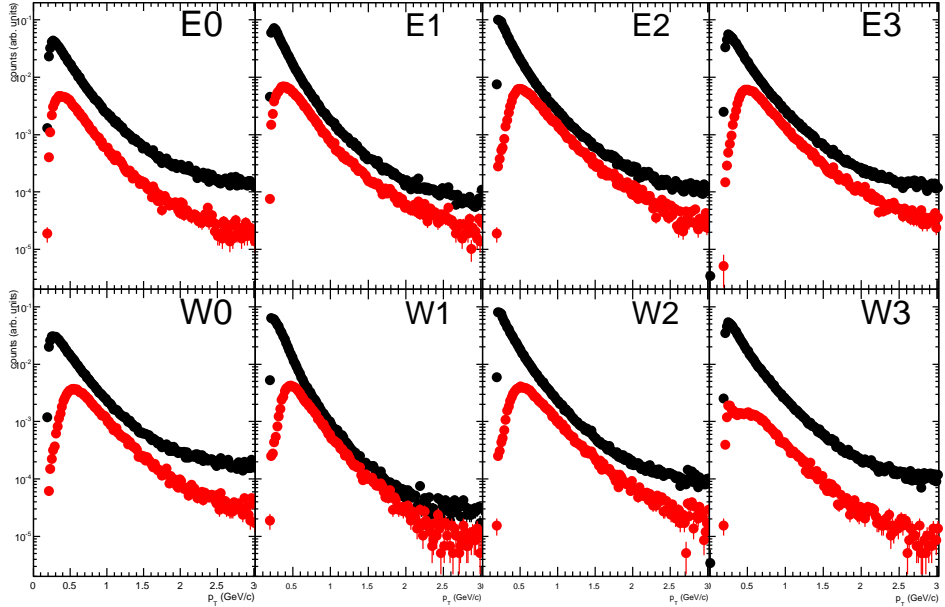}
  \caption[\pt distribution of ERT triggered electrons]{\pt
    distribution of single electrons ({\em black}) and single electrons
    which are associated with an ERT trigger hit ({\em red}).}
  \label{fig:ertpt_single}
\end{figure}

The dielectron analysis in \pp requires that at least one electron in
every event is associated with a hit in the ERT trigger. This biases
the distribution of \ee pairs and needs to be corrected to allow for a
meaningful comparison. The trigger efficiency of a single electron is
determined with the MB data sample. The Level-1 trigger decision is
recorded also in the MB data set. It is possible that an event is not
registered as ERT events, \ie, the Level-1 ERT trigger bit is not set,
but contains an electron associated to a hit in the ERT tiles. Such
events would lead to a miscounting of electrons which fired the
trigger, therefore, only MB events are considered in which the ERT
trigger bit is set. Within those the \pt spectrum of electron
candidates $dN_{\rm MB\&\&ERT}^{\pm}/d\pt^{\pm}$, \ie, tracks that
fulfill the electron identification cuts, that can be associated to a
fired ERT trigger tile is shown in \fig{fig:ertpt_single}. The
distribution is compared to the \pt spectrum of all electrons
candidates in all MB events $dN_{\rm MB}^{\pm}/d\pt^{\pm}$ (\ie, the
ERT trigger bit is not required).

\begin{figure}
  \centering
  \includegraphics[width=0.9\textwidth]{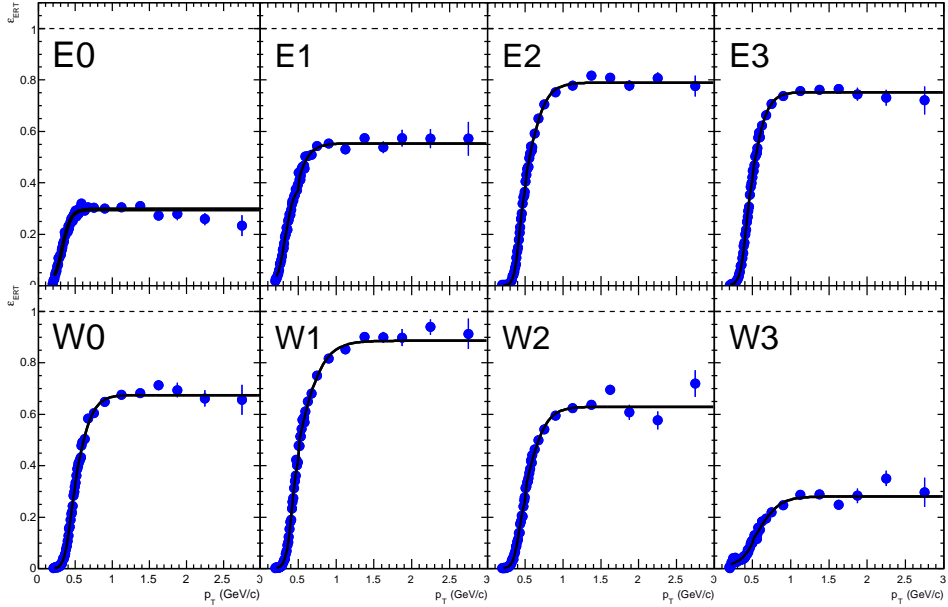}
  \caption[Single electron ERT trigger efficiency]{Single electron ERT
    trigger efficiency for all eight EMCal sectors. The fits are
    summarized in \tab{tab:erteff}.}
  \label{fig:erteff_single}
\end{figure}

The trigger efficiency is defined as the ratio of the two as in
\eq{eq:erteff}, which is shown in \fig{fig:erteff_single} for each
EMCal sector individually as function of the single electron \pt
\begin{equation}\label{eq:erteff}
  \varepsilon_{\rm ERT}^{\pm} = \frac{dN_{\rm MB\&\&ERT}^{\pm}/d\pt^{\pm}}{dN_{\rm MB}^{\pm}/d\pt^{\pm}}
\end{equation}

The trigger efficiency rises steeply above the trigger threshold of
400~MeV and reaches half of it's plateau value at $\pt \approx
600$~\mevcc. The plateau value itself strongly depends on the EMCal
sector, due to the variation in the number of active trigger
tiles. The trigger efficiencies are fit to two Fermi
functions\footnote{for EMCal sector E0 it was sufficient to fit the
  ERT efficiency with one Fermi function over the full \pt range}, one
below and the other above a single electron \pt of 0.5~\gevc:
\begin{equation}
  f(\pt) = \frac{\varepsilon_0}{\eexp{-(\pt - p_0)/k} + 1}\theta(\pt -
  0.5) + \frac{\varepsilon_0'}{\eexp{-(\pt - p_0')/k'} + 1}\theta(\pt +
  0.5)
\end{equation}
The fit results for every sector are tabulated in \tab{tab:erteff}

\begin{table}[tbh]
  \centering
  \caption[Single Electron ERT
  Efficiency]{\label{tab:erteff}Parameterization of the ERT efficiency
    for single electrons.\\}
  \begin{tabular}{ccccccc}
    \toprule
    Sector & $\varepsilon_0$ & $p_0$ (\gevc) & $k$ (\gevc) & $\varepsilon_0'$ & $p_0'$ (\gevc) & $k'$ (\gevc) \\\midrule
    E0 & 0.297 & 0.326 & 16.7 & ---   & ---   & ---\\
    E1 & 0.386 & 0.331 & 19.5 & 0.553 & 0.394 & 9.61\\
    E2 & 0.415 & 0.425 & 27.8 & 0.790 & 0.493 & 8.15\\
    E3 & 0.465 & 0.418 & 24.3 & 0.753 & 0.470 & 9.92\\
    \\
    W0 & 0.396 & 0.439  & 23.1 & 0.675 & 0.499 & 9.08\\
    W1 & 0.460 & 0.413  & 26.6 & 0.886 & 0.471 & 6.03\\ 
    W2 & 0.303 & 0.422  & 24.3 & 0.630 & 0.505 & 7.85\\
    W3 & 66.7  & 1.53   & 6.32 & 0.281 & 0.562 & 6.99\\
    \bottomrule
  \end{tabular}
\end{table}

\begin{figure}
  \centering
  \includegraphics[width=0.9\textwidth]{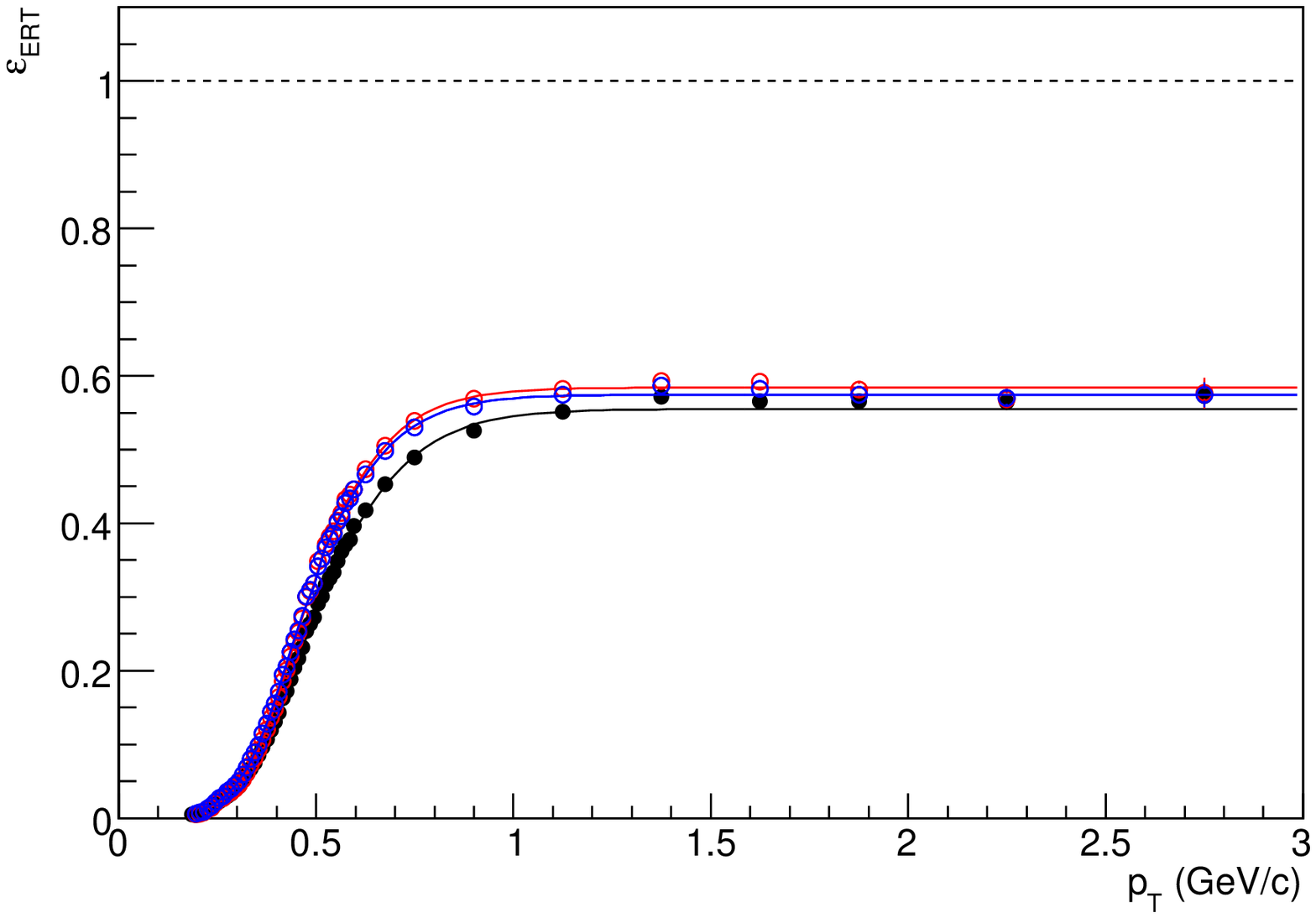}
  \caption[ERT trigger efficiency for different eID cuts]{ERT trigger
    efficiency for single electrons with different eID cuts. The
    trigger efficiency for strong eID cuts is shown in ({\em red}),
    for medium eID cuts ({\em blue}), and for lose eID cuts in ({\em
      black}).}
  \label{fig:erteff_eidcuts}
  \includegraphics[width=0.9\textwidth]{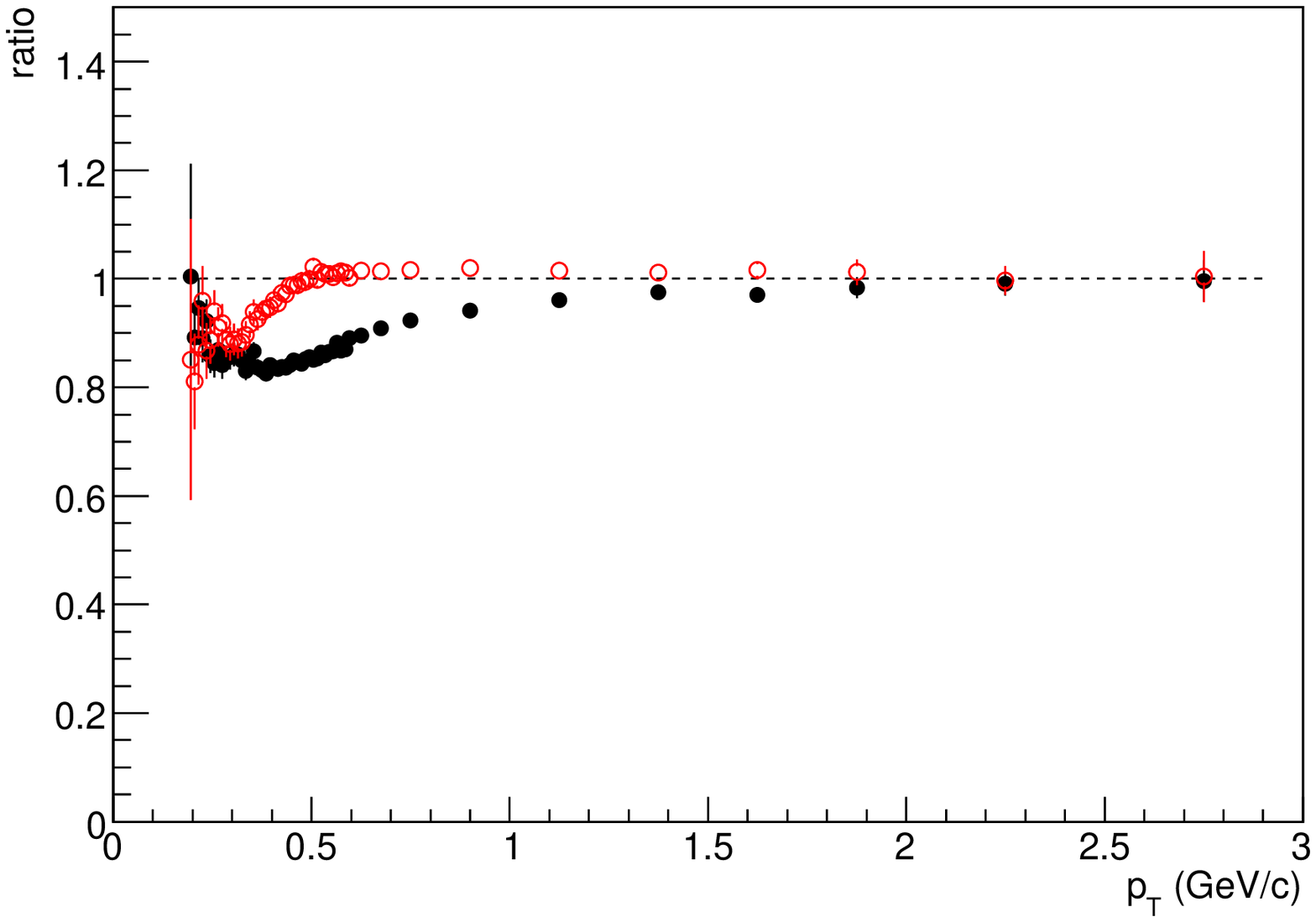}
  \caption[Comparison of ERT trigger efficiency with different eID
  cuts]{Shown are the ratios of ERT trigger efficiencies with
    different eID cuts. Strong divided by medium is shown in ({\em
      red}), lose divided by medium in ({\em black}).}
  \label{fig:erteff_eidcuts_ratio}
\end{figure}

It is important to note, that to determine the ERT efficiency more
stringent electron identification cuts need to be applied to reduce
the hadron contamination in the electron sample to a negligible
amount. This is necessary as any hadron in the electron sample lowers
the apparent single electron trigger efficiency. But the contribution
of hadrons to the \ee pair spectrum is subtracted with the
combinatorial background, therefore, the hadron contamination in the
pair spectrum is much reduced to a negligible fraction. It is the
trigger efficiency of the electrons in this background subtracted
sample, that are of interest. Therefore, the following eID cuts are
applied:
\begin{itemize}
\item $0.20$~GeV  $\leq$ \pt $\leq 20.0$~GeV
\item DC track quality $== 63~||~51~||~31$
\item ${\rm n}_0 \geq 3$
\item disp $< 5$
\item $\chi^2/{\rm npe}_0 < 10.0$
\item $0.8 < E/p < 1.2$
\item $\sqrt{{\rm emcsdphi\_e}^2 + {\rm emcsdz\_e}^2} < 2 \sigma$
\end{itemize}
On the other hand, this introduces a bias towards electrons which are
more likely to fire the ERT trigger, \eg, the harder ${\rm n}_0$ cut
prefers electrons that emit more Cherenkov light in the RICH. This
effect has been studied, by varying the electron identification
cuts. The trigger efficiency is compared for the strong eID cuts
listed above, the standard eID cuts used for the pair analysis and a
third set of cuts:
\begin{itemize}
\item $0.20$~GeV  $\leq$ \pt $\leq 20.0$~GeV
\item DC track quality $== 63~||~51~||~31$
\item ${\rm n}_0 \geq 2$
\item disp $< 5$
\item $\chi^2/{\rm npe}_0 < 10.0$
\item $E/p > 0.8$
\item $\sqrt{{\rm emcsdphi\_e}^2 + {\rm emcsdz\_e}^2} < 3 \sigma$
\end{itemize}
which in its strength lies in between the two other sets. The
resulting trigger efficiencies are shown in
\fig{fig:erteff_eidcuts_ratio}. The ratio with respect to the last set
of eID cuts shows only a small effect for $\pt>1.5$~\gevc, but for
$\pt < 1.5$~\gevc differences up to 20\% appear, which is assigned as
systematic uncertainty on the trigger efficiency.

The efficiency of an \ee pair is determined analog to the calculation
of the electron identification efficiency with \exodus. \ee pairs from
hadron decays are generated and the EMCal sector that an electron
would hit is determined as function of its \pt and azimuthal
emission. Then its probability to fire the ERT trigger is given by the
trigger efficiency in this EMCal sector which has been parameterized
from the fit results in \tab{tab:erteff}. For an \ee pair to survive,
at least one of the two electrons must have fired the ERT trigger.

\begin{figure}
  \centering
  \includegraphics[width=0.9\textwidth]{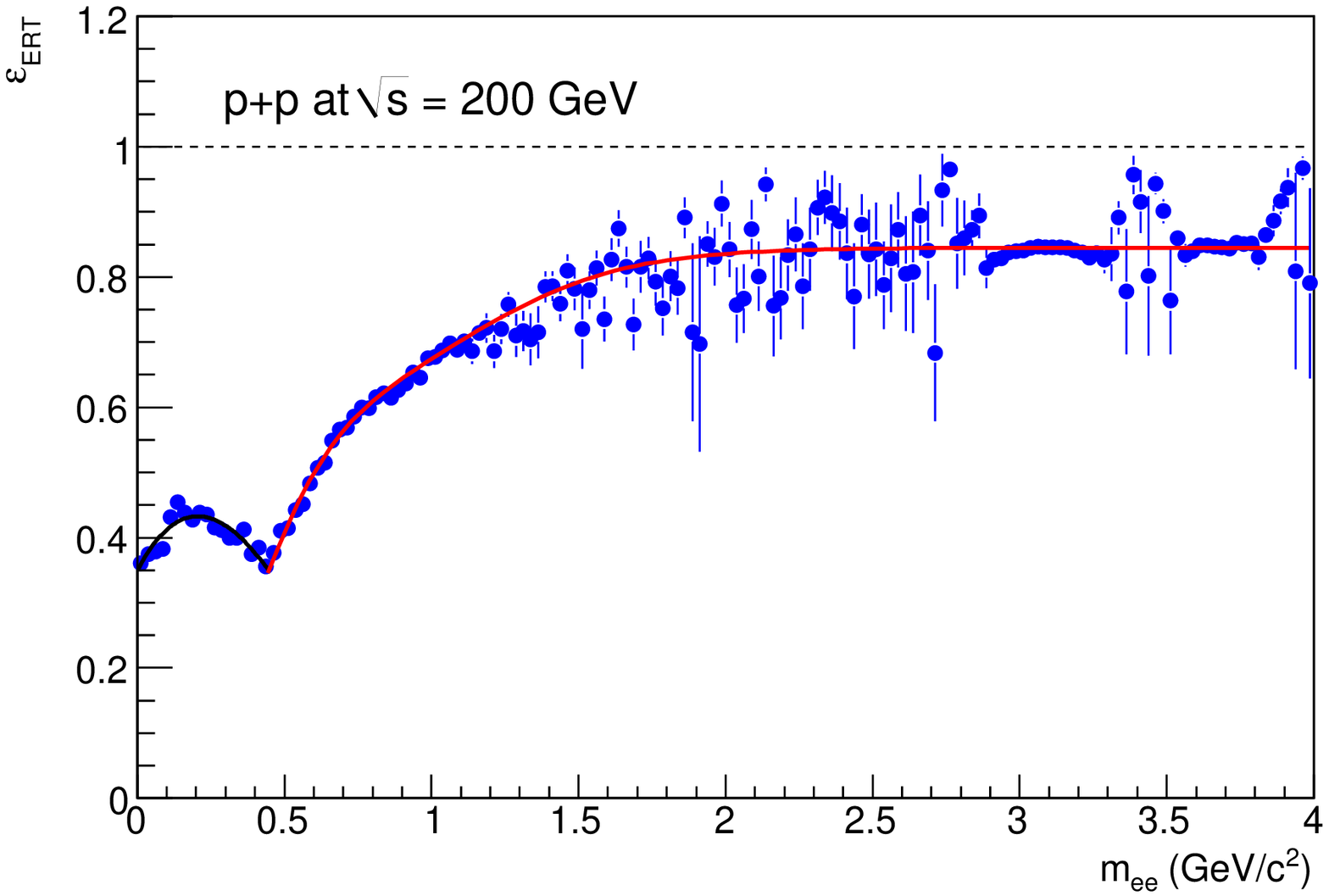}
  \caption[ERT trigger efficiency for \ee pairs]{ERT trigger
    efficiency for \ee pairs as function of \mee. The fits are
    summarized in \tab{tab:erteff_mass}.}
  \label{fig:erteff_mass}
\end{figure}

The ratio of \ee pairs surviving the ERT trigger condition and all
pairs is determined in mass and \pt of the pair. For the \pt inclusive
mass spectrum a trigger efficiency is applied in mass only as shown in
\fig{fig:erteff_mass}. The projection onto the mass axis is weighted by
the \pt distribution of \ee pairs from \exodus.  The plateau at large
invariant mass corresponds to the average single electron trigger
efficiency at high \pt. The shape is similar to the eID efficiency,
but has a more pronounced minimum, as the trigger condition needs to
be matched by only one of the electrons.

\begin{table}[tbh]
  \centering
  \caption[ERT efficiency as function of mass]{\label{tab:erteff_mass}Parameterization of trigger
    efficiency as function of mass.\\}
  \begin{tabular}{ccccccc}
    \toprule
    ~&$a$ ($c$/GeV) & $b$ ($c^2$/GeV$^2$) & $c$ ($c^3$/GeV$^3$) &  $d$ ($c^4$/GeV$^4$)&~\\\midrule
    ~&0.349 & 0.861 & $-2.47$ & 1.21 &~ \\\bottomrule
    \\\toprule
    $\varepsilon_0$ & $p_0$ (\gevc) & $k$ (\gevc) &  $\varepsilon_0'$ & $p_0'$ (\gevc) & $k'$ (\gevc)\\\midrule
    0.260     & 1.14 & 3.79 & 0.585 & 0.408 & 7.24\\\bottomrule
  \end{tabular}
\end{table}

The mass range $1 < \mee < 3$~\gevcc has large statistical
uncertainties due to the absence of hadron decays in this
region. Therefore, and to smoothen other statistical fluctuations the
result is parameterized by a fit. For $\mee < 400$~\mevcc a third
order polynomial is fit. Above, the sum of two Fermi functions is fit
to the ERT efficiency. The result is summarized in
\tab{tab:erteff_mass}.

\begin{figure}
  \centering
  \includegraphics[width=0.9\textwidth]{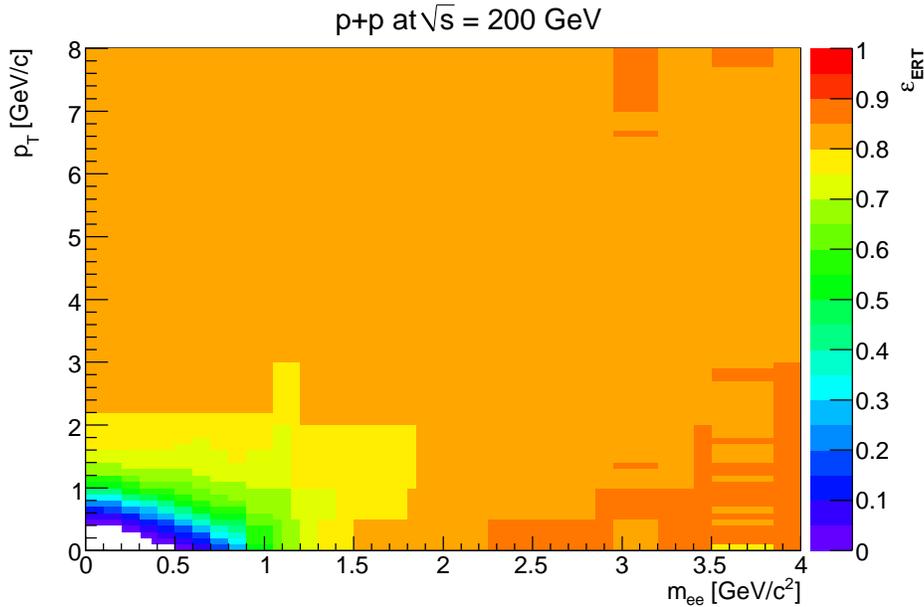}
  \caption[ERT trigger efficiency for \ee pairs in mass and \pt]{ERT
    trigger efficiency for \ee pairs as function of \mee and \pt.}
  \label{fig:erteff_mpt}
\end{figure}

For the differential analysis in mass and \pt, the two dimensional
trigger efficiency distribution is smoothened by fitting the \pt
distribution of \ee pairs in narrow mass slices in the LMR, the
$J/\psi$ and $\psi'$. The IMR is fitted in slices of \pt analog to the
\pt inclusive trigger efficiency in \fig{fig:erteff_mass}. The results
are combined into a smoothed parameterization of the ERT efficiency as
function of mass and \pt which is shown in \fig{fig:erteff_mpt}.

The uncertainty in the trigger efficiency due the electron
identification cuts of 20\% at low \pt affects \ee pairs with
$\mt<1$~\gevcc, for which this uncertainty is assigned. Elsewhere a
5\% systematic uncertainty is assigned. This includes uncertainties
due in the precise location of dead trigger tiles, which has been
studied by shuffling EMCal sectors in the simulation of the ERT pair
efficiency.

\begin{figure}
 \includegraphics[width=1.0\linewidth]{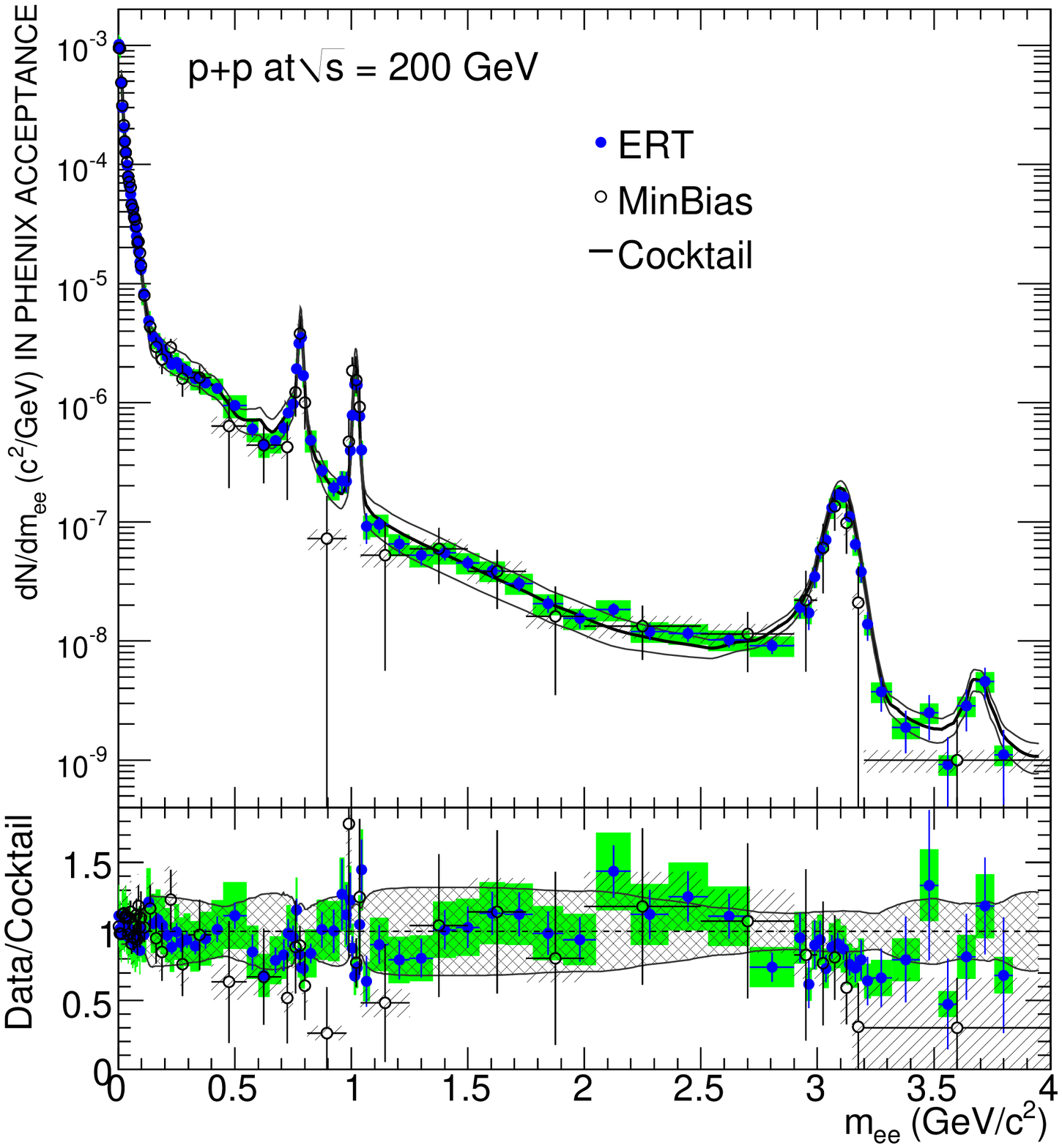}
 \caption[Comparison of dielectron spectra in \pp from MB and ERT
 triggered events]{Invariant mass spectra for $p+p$ data with the MB
   ({\em hollow}) and ERT ({\em solid}) data sets. The agreement
   between the two data sets is excellent.}
 \label{fig:ertmb}
\end{figure}

Figure~\ref{fig:ertmb} compares of the invariant mass spectra for the
$p+p$ data obtained with the MB and ERT data sets. The ERT data set has
been corrected by ERT trigger efficiency and eID efficiency, while the
MB is corrected only for the eID. The agreement between the two data
sets confirms the assessed accuracy of the ERT correction.

\subsection[Occupancy Correction in \AuAu]{Occupancy Correction in $\boldsymbol{\rm Au+Au}$}
\label{sec:occupancy_corr}

In \AuAu collisions there is also a finite efficiency loss for
particle detection due to the presence of other particles nearby.  To
get a quantitative understanding of the multiplicity-dependent
efficiency loss, single electrons and positrons are generated with a
full GEANT Monte Carlo simulation of the PHENIX detector and then
embedded into data files containing detector hits from real \AuAu
events.  Next, these new files containing the embedded $e^\pm$ are run
through the entire reconstruction software to produce track candidates
containing the variables used for the electron identification cuts.

Since all the detectors used in the analysis are located after the
pair has been opened by the magnetic field, the pair embedding
efficiency is defined as the square of the single electron efficiency
\begin{align}
  \epsilon_{\rm embed}^{\rm pair} &= \left(\varepsilon_{\rm embed}^{\pm}\right)^2 \\
  &= \left(\frac{\# \ \textrm{reconstructed} \ e^\pm \ \textrm{from
        embedded data}}{\# \ \textrm{reconstructed} \ e^\pm \
      \textrm{from single track data}}\right)^2
\end{align}
where a reconstructed particle from embedded data has most of its
detector hits associated with hits from the simulated particle.

\begin{table}
  \centering
  \caption[Embedding Efficiency]{\label{tab:embedding}Embedding efficiency.\\}
  \begin{tabular}{rr@{.}lr@{.}l} \toprule
    \multicolumn{1}{c}{Centrality}  &
    \multicolumn{2}{c}{$\varepsilon_{\rm embed}^{\pm}$} &
    \multicolumn{2}{c}{$\varepsilon_{\rm embed}^{\rm pair}$} \\\midrule
    0--10\%  &  0&86 & 0&74\\
    10--20\% &  0&91 & 0&83\\
    20--40\% &  0&93 & 0&87\\
    40--60\% &  0&97 & 0&95\\
    60--92\% &  0&99 & 0&98\\
    0--92\%  &  0&96 & 0&81\\
    \bottomrule
  \end{tabular}
\end{table}
\tab{tab:embedding} lists the embedding efficiencies for the
centrality classes used in the analysis. For the minimum bias the
occupancy correction is weighted by the effective pair signal in each
centrality class; since most of the yield is concentrated in the most
central classes, the resulting pair efficiency is 0.81 instead of
$0.92=0.96^2$.  A systematic uncertainty of $\approx 3\%$ has been
assigned by calculating the occupancy efficiency with a data-driven
method~\cite{Lebedev:2002AN137}.

\subsection{Acceptance}
\label{sec:acc_corr}
In addition to corrections for the track reconstruction, electron
identification and ERT trigger efficiency, the \pt spectra of \ee
pairs are also corrected for the geometrical acceptance of the two
central arm detectors $\varepsilon_{\rm geo}^{\rm pair}$. It accounts
for the fraction of \ee pairs over full azimuth $0 < \phi \leq 2\pi$
and one unit of rapidity ($|y|<0.5$) of which at least one electron
was not within the detector acceptance.

The acceptance correction has been calculated based on a Monte Carlo
simulation of Dalitz decays of pseudo-scalar mesons (\pion, $\eta$,
$\eta'$) and direct decays of vector mesons ($\rho$, $\omega$, $\phi$,
$J/\psi$, $\psi'$). The rapidity distribution of all mesons is assumed
to be flat, which is well justified by measurements of
BRAHMS~\cite{PhysRevLett.90.102301}. The \pt distribution of these
mesons is taken from PHENIX measurements (see
Section~\ref{sec:exodus}), their polarization from the PDG
book~\cite{pdg}. Vector mesons are assumed to be unpolarized.

\begin{figure}
  \centering
  \includegraphics[width=0.9\textwidth]{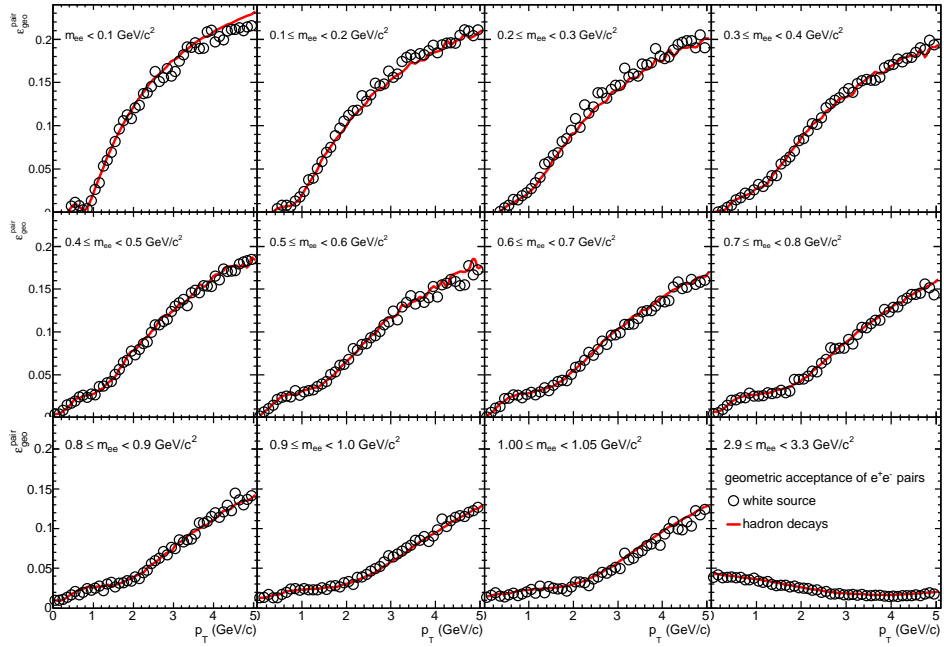}
  \caption[Geometric acceptance of \ee pairs]{Acceptance of \ee pairs
    as function of \pt for various mass ranges: 0--0.1, 0.1--0.2,
    0.2--0.3, 0.3--0.4, 0.4--0.5, 0.5--0.6, 0.6--0.7, 0.7--0.8,
    0.8--0.9, 0.9--1.0, 1.0--1.05, and 2.9--3.3~\gevcc (from left to
    right, top to bottom). The acceptance based on the hadronic
    cocktail is shown in black, the one based on white pairs in blue.}
  \label{fig:acc_pair}
\end{figure}

The acceptance correction is calculated as all other corrections in
mass and \pt by dividing the number of \ee pairs within the detector
acceptance, \ie, both tracks fulfill \eq{eq:track_acc}, by all pairs
in $0 < \phi \leq 2\pi$ and $|y| < 0.5$. The systematic uncertainty on
the polarization and due to the \pt parameterization is studied with a
simulation of unpolarized pairs with a flat mass and \pt distribution,
\footnote{which has already been used as cross check for the eID
  efficiency}. The resulting acceptance corrections are compared in
\fig{fig:acc_pair} for a number of mass ranges as function of \pt. The
shapes of the two acceptance corrections are very similar and agrees
within 5\% for the lowest mass bin, and better at higher masses.

In the intermediate mass region (1--3~\gevcc) charmed meson decays
dominate the \ee invariant mass spectrum. Their contribution has been
simulated with \pythia and normalized to the measured $c\overline{c}$
cross section measured in Ref.~\cite{adare:252002}. These \ee pairs do
not share the same parent, both are produced by semi-leptonic decays
of charmed mesons, \eg $D$, but their parents are correlated as they
contain each one of the $c$ quarks which are always produced in
pairs. $c\overline{c} \rightarrow D \overline{D} \rightarrow \ee$. For
masses $\mee < 0.5$~\gevcc the charm contribution is negligible, above
a systematic uncertainty of 5\% has been added due to the uncertainty
in the charm cross section of 30\%.

\subsection{Bin Shift Correction}
\label{sec:binshift}

For the \pt spectra of \ee pairs a further correction is applied to
take into account the difference between the average \pt of \ee pairs
within a bin and the bin center due to the finite bin width.

\begin{figure*}[t]
  \centering
  \includegraphics[width=0.9\textwidth]{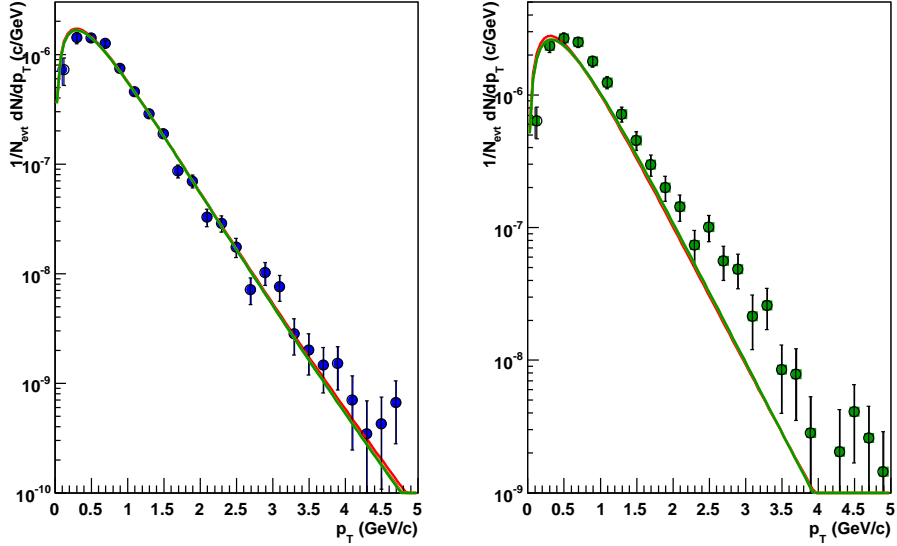}
  \caption[Bin shift correction for $\omega$ and $\phi$
  yields]{Yield of $\omega$ ({\em left}) and $\phi$
    ({\em right}) before bin shift correction. The data are fit to a
    power-law to determine the bin sift correction. The first
    iteration is shown in red, the third iteration in green. Open
    points show the bin shift corrected bin centers.}
  \label{fig:binshift}
\end{figure*}

The bin shift correction of the extracted yield is determined by
iteratively fitting the $dN/d\pt$ spectra with a power law and shifting
the yield in every \pt bin the fits by the relative difference between
the yield at the bin center ($\pt^{\rm cent}$) and at the average \pt
of the bin $\langle\pt\rangle$. Three iterations are performed, but
already after the first iteration the bin shift correction changes
insignificantly. As example, the fit results are shown in
\fig{fig:binshift} for the $\omega$ and $\phi$ mesons yields (also see
Section~\ref{sec:omega_phi}). Instead of correcting the bin center for
the bin shift, the yield is adjusted by a factor:
\begin{equation}
  \delta(\pt^{\rm cent}) = \frac{\int\limits_{\pt^{\rm cent}-\Delta/2}^{\pt^{\rm cent}+\Delta/2} \frac{dN}{d\pt}\,d\pt}{\frac{dN}{d\pt}\bigr\rvert_{\pt = \pt^{\rm cent}}}
\end{equation}
where $\Delta$ is the bin width. The bin shift correction is applied
on the final binning and only after this correction is the yield
$dN/d\pt$ converted into an invariant yield or cross section
$Ed^3\sigma/dp^3$.

The correction for the $\omega$ yield in \pt bins of 200 MeV/$c$ is
$\sim 3.28\%$ below 3~\gevc and $\sim 1.8\%$ above. The bin shift
correction for the $\phi$ cross section is $\sim 2.2\%$ in the full
range of 0--5~\gevc.

\section{Systematic Uncertainties}
\label{sec:syst_uncertainties}
The various contribution to the total systematic uncertainty, which
are summarized in \tab{tab:syst_uncertainties}, are described in the
following. (i) The uncertainty on the \ee pair reconstruction, which
is twice the uncertainty on the single electron reconstruction
efficiency\cite{adare:252002,adare:172301}, includes uncertainties on
the efficiency of electron identification cuts, geometrical
acceptance, and run-by-run fluctuations; the uncertainty on (ii) the
conversion rejection cut, (iii) the ERT and minimum bias trigger
efficiencies. These uncertainties are included on the yield of \ee
pairs in the invariant mass spectra. The uncertainties on the track
reconstruction do not have a strong dependence on the \ee pair \pt,
therefore they are assigned independent of \pt. Pair and conversion
cuts are localized in mass ($\mee < 600$~\mevcc) and also rather \pt
independent.

\begin{figure}
  \centering
  \subfloat[]{\label{fig:signal_bg_pt}\includegraphics[width=0.44\textwidth]{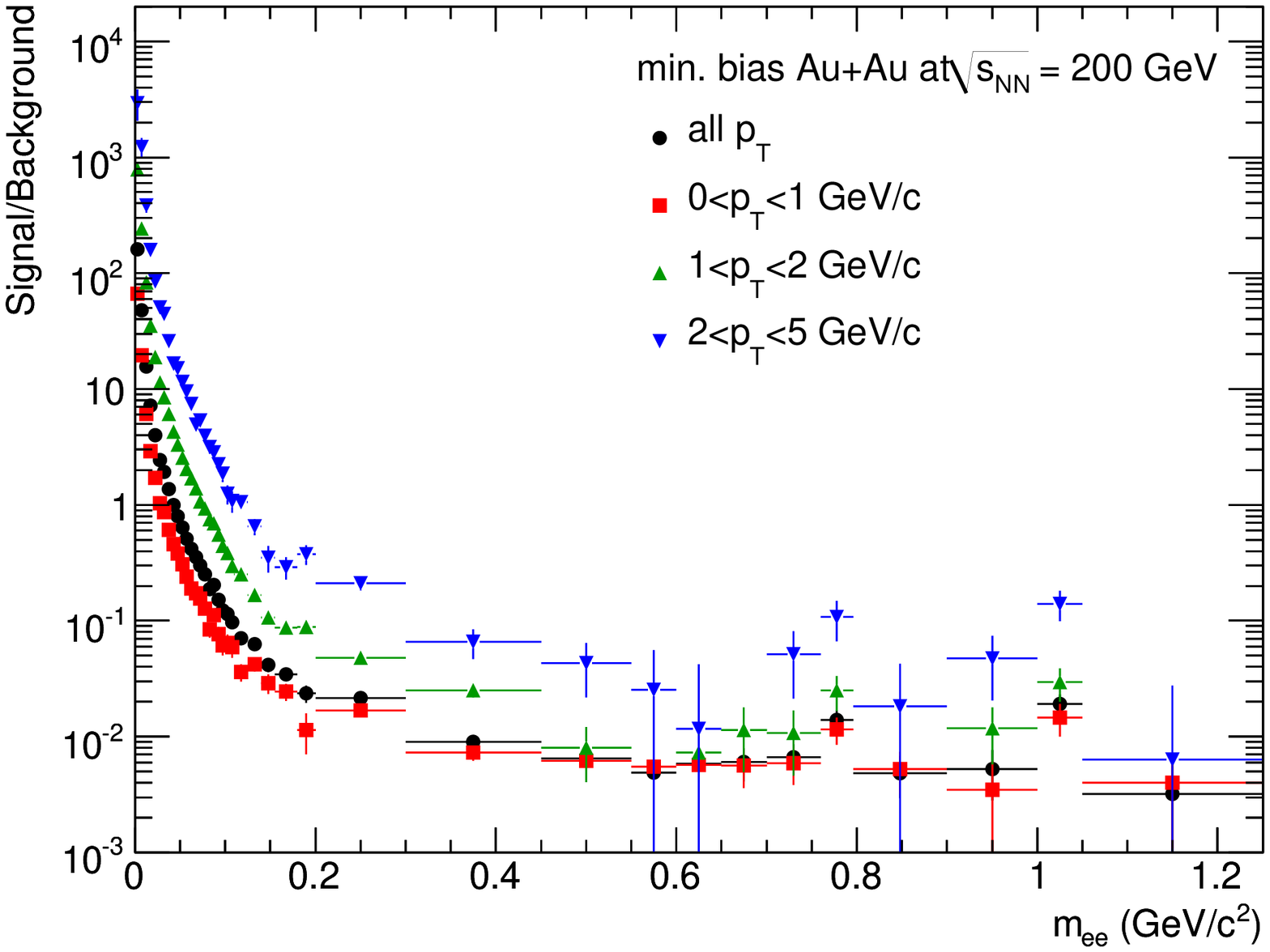}}
  \subfloat[]{\label{fig:signal_bg_cent}\includegraphics[width=0.4\textwidth]{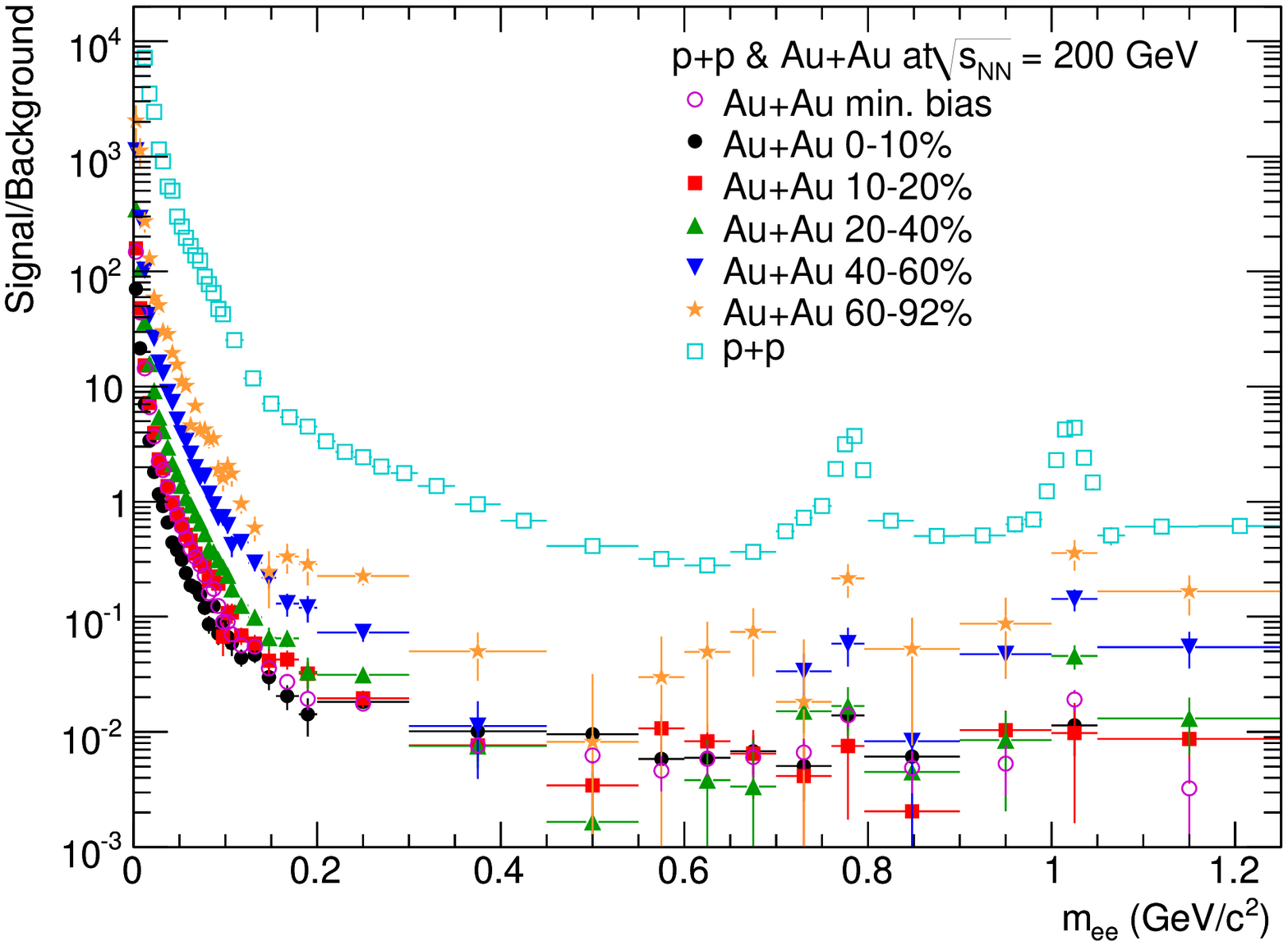}}
  \caption[Signal-to-Background Ratio in \pp and \AuAu
  collisions]{Signal-to-Background ratio for \ee pairs in min. bias
    \AuAu collisions with different \pt ranges~\subref{fig:signal_bg_pt}
    and for different centralities~\subref{fig:signal_bg_cent} which also
    includes \ee pairs in \pp collisions.}
  \label{fig:signal_bg}
\end{figure}

The largest contribution is the systematic uncertainty on the
background subtraction. The mixed event normalization enters with $3\%
\times B/S$ in \pp and in \AuAu with $0.25\times B/S$. But the
background-to-signal ratio, which is strongly mass and \pt dependent,
is two orders of magnitude worse for \AuAu collisions than in \pp as
shown in \fig{fig:signal_bg}. As the signal-to-background ratio shows
a \pt dependence, it is assigned for each \pt bin individually.

For the \pt spectra the uncertainty on the acceptance correction of
10\%, and for $\mee > 0.5$~\gevcc an additional 5\% due to the charm
contribution, are included in the systematic uncertainty.
\begin{table}[th]
  \caption[Systematic Uncertainties]{\label{tab:syst_uncertainties}Systematic uncertainties of the dilepton yield due to different sources and indication of mass range where the error is applied.\\}
  \centering
  \begin{tabular}{lr@{.}lr@{.}lc}\toprule
    Syst. Uncert. component    & \multicolumn{2}{c}{\pp}   & \multicolumn{2}{c}{\AuAu} & Mass Range (\gevcc)\\ \midrule
    pair reconstruction        & 14&4\%                    &  13&4\%                   & 0--4\\
    conversion rejection       &  6&0\%                    &   6&0\%                   & 0--0.6\\
    pair cuts                  &  5&0\%                    &   5&0\%                   & 0.4--0.6\\
    occupancy efficiency       & \multicolumn{2}{c}{---}   &   3&0\%                   & 0--4\\
    ERT efficiency             &  5&0\%                    & \multicolumn{2}{c}{---}   & $\mt > 1$\\
                               & 20&0\%                    & \multicolumn{2}{c}{---}   & $\mt \leq 1$\\
    combinatorial background   &  3&0\%$\cdot B/S$         &   0&25\%$\cdot B/S$       & 0--4\\
    correlated background (``near-side'')     & 3&0\%      &  10&0\%                   & 0--0.6\\
    correlated background (``away-side'')     & 11&0\%     &  \multicolumn{2}{c}{---}  & $>1$\\
    centrality                 & \multicolumn{2}{c}{---}   &  10&0\%                   & 0--4\\
    acceptance correction      & 10&0\%                    &  10&0\%                   & 0--4\\
    charm acceptance           &  5&0\%                    &   5&0\%                   & $>$0.5\\ \bottomrule
  \end{tabular}
\end{table}

\section[\exodus Simulation of Hadron Decays]{EXODUS Simulation of Hadron Decays}
\label{sec:exodus}

\exodus is a fast Monte Carlo simulation of hadron decays. Key input
is the rapidity density $dN/dy$ of neutral and charged pions which is
determined by a fit to PHENIX data~\cite{adare:051106} with a modified
Hagedorn function:
\begin{equation}\label{eq:hagedorn}
  E \frac{d^3\sigma}{dp^3} = A \left(\mathrm{e}^{-\left(a \pt + b \pt^2\right)} + \pt/p_0\right)^{-n}
\end{equation}
with $A = 377 \pm 60$~mb~GeV$^{-2}$, $a = 0.356 \pm
0.014$~(\gevc)$^{-1}$, $b = 0.068 \pm 0.019$~(\gevc)$^{-2}$, $p_0 =
0.7 \pm 0.02$~\gevc, and $n = 8.25 \pm 0.04$.

\begin{figure}[p]
  \centering
  \includegraphics[width=0.9\textwidth]{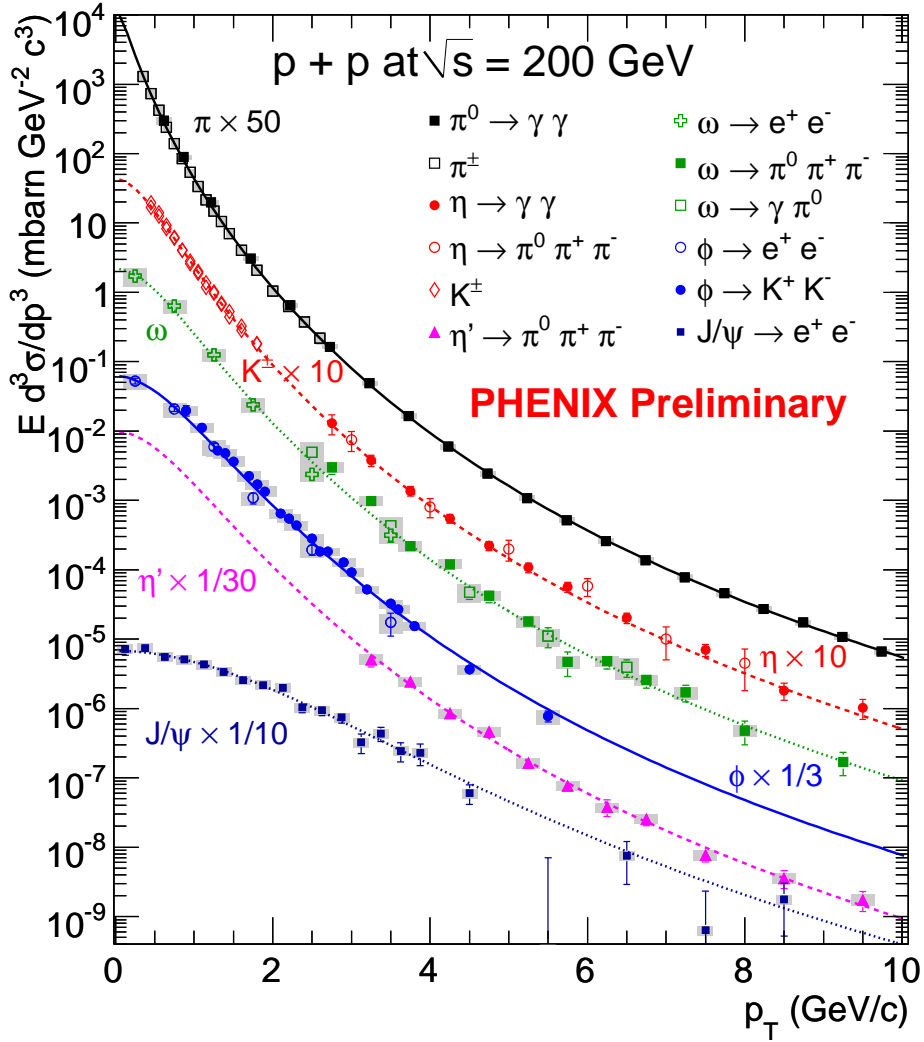}
  \caption[Meson cross sections in \pp collisions at \sqrts =
  200~GeV]{Compilation of meson production cross sections in \pp
    collisions at \sqrts = 200~GeV. Shown are data for
    neutral~\cite{adare:051106} and charged pions~\cite{adler:024904},
    $\eta$~\cite{adler:024909}, kaons~\cite{adler:024904},
    $\omega$~\cite{Dahms:2007AN614,Kijima:2007AN610,Sharma:2007AN618,adler:051902,0954-3899-34-8-S127,Ivanishchev:2006AN535},
    $\phi$~\cite{Dahms:2007AN614,Kijima:2007AN610,Sharma:2007AN618,0954-3899-34-8-S127,Ivanishchev:2007AN600},
    $\eta'$~\cite{Ivanishchev:2008AN692}, and
    $J/\psi$~\cite{adare:232002}. The data are compared to the
    parameterization based on \mt scaling used in our hadron decay
    generator.}
  \label{fig:cocktail_pp}
\end{figure}

PHENIX cross section measurements of $\pion$, $\pi^{\pm}$, $\eta$,
$\eta'$, $\omega$, $\phi$, and $J/\psi$ are shown in
\fig{fig:cocktail_pp}. The fit of the charged and neutral pion data
with \eq{eq:hagedorn} is also shown.

To parameterize the other hadron spectra, the modified Hagedorn fit of
the pion data is \mt scaled, replacing \pt in \eq{eq:hagedorn} by
$\sqrt{\left(\pt/c\right)^2 - m_{\pion}^2 + m_h^2}$, where $m_h$ is
the mass of the hadron and fit with a free normalization factor to the
hadron spectra. The good description of the data by the fits indicates
that meson production in \pp collisions follows \mt
scaling. Accordingly, this observation is extended to other mesons for
which no data are available.

In order to extract the meson yield per inelastic \pp collision the
fits are integrated over all \pt. Results, systematic uncertainties,
and references to data are given in \tab{tab:exinput_pp}. For the
$\rho$ meson $\sigma_{\rho}/\sigma_{\omega} = 1.15 \pm 0.15$ is
assumed, consistent with values found in jet
fragmentation~\cite{pdg}. The $\eta'$ yield is scaled to be consistent
with jet fragmentation $\sigma_{\eta'}/\sigma_{\eta} = 0.15 \pm
0.15$~\cite{pdg} which is consistent with preliminary measurements by
PHENIX~\cite{Ivanishchev:2008AN692}. The $\psi'$ is adjusted to the
value of $\sigma_{\psi'}/\sigma_{J/\psi} = 0.14 \pm
0.03$~\cite{Gavai:1994in} in agreement with preliminary PHENIX
results~\cite{Donadelli:2008AN722}. For the $\eta$, $\omega$, $\phi$,
and $J/\psi$ the quoted uncertainties include those on the data as
well as those using different shapes of the \pt
distributions. Specifically their spectra have been fitted with the
functional form given in \eq{eq:hagedorn} with all parameters free and
also an exponential distribution in \mt. For the $\rho$, $\eta'$, and
$\psi'$ the uncertainty is given by the uncertainty assumed for the
cross section ratios. It is important to note that the dilepton
spectra from meson decays are rather insensitive to the exact shape of
the \pt distribution.

\begin{figure}[p]
  \centering
  \includegraphics[width=0.9\textwidth]{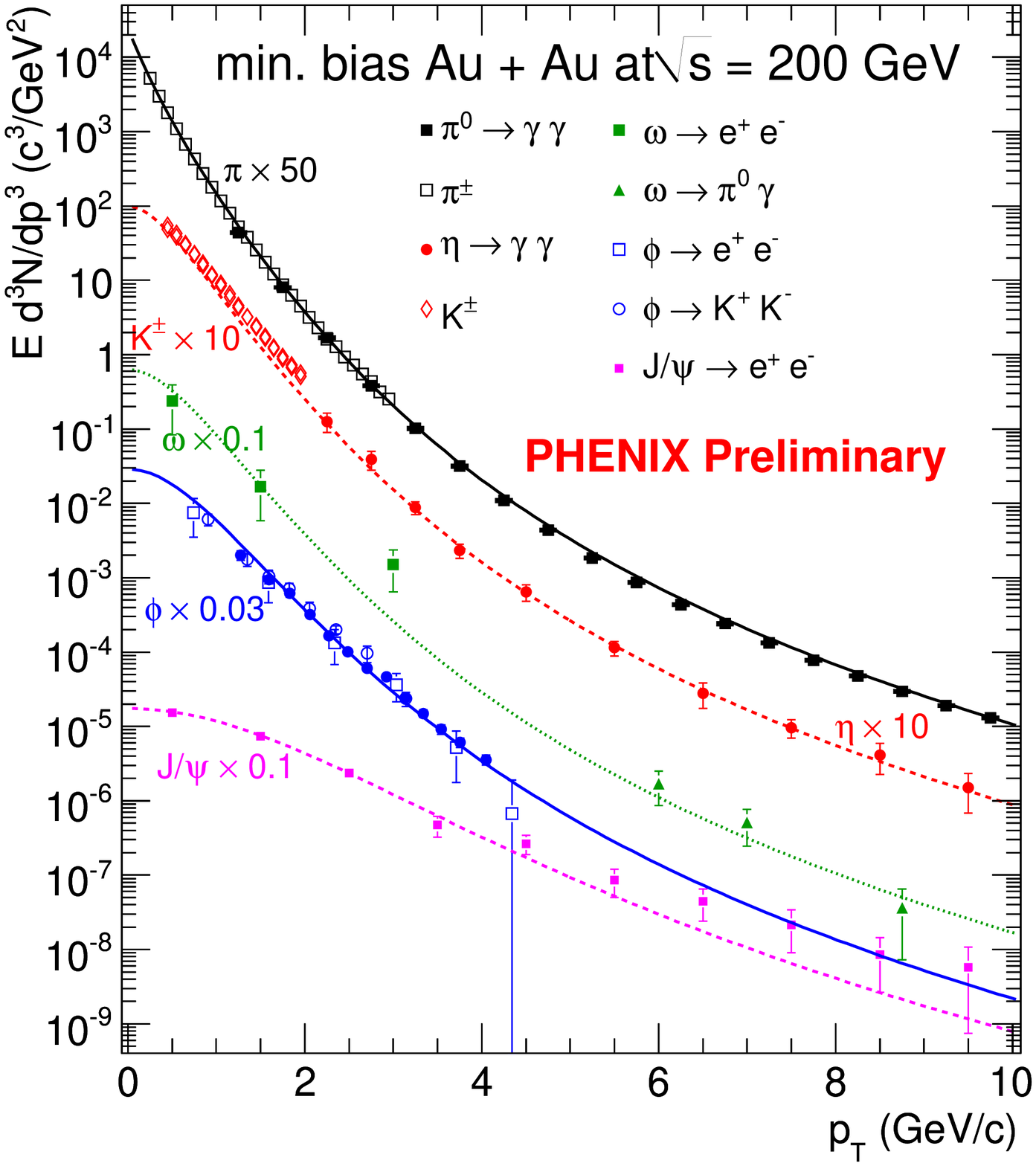}
  \caption[Meson cross sections in \AuAu collisions at \sqrtsnn =
  200~GeV]{Compilation of meson production cross sections in \AuAu
    collisions at \sqrtsnn = 200~GeV. Shown are data for
    neutral~\cite{adler:072301} and charged pions~\cite{adler:034909},
    $\eta$~\cite{adler:024909}, kaons~\cite{adler:034909},
    $\omega$~\cite{0954-3899-34-8-S127},
    $\phi$~\cite{adler:014903,0954-3899-34-8-S127} and
    $J/\psi$~\cite{adare:232301}. The data are compared to the
    parameterization based on \mt scaling used in our hadron decay
    generator.}
  \label{fig:cocktail_au}
\end{figure}

\begin{table}[tbh]
  \centering
  \caption[Hadron rapidity densities in \pp]{\label{tab:exinput_pp}Hadron rapidity
    densities in \pp collisions at \sqrts = 200 GeV used in \exodus.\\}
  \begin{tabular}{cr@{.}l@{ $\pm$ }r@{.}lcr@{.}ll} \toprule
    ~ & \multicolumn{4}{c}{$\frac{dN}{dy}\bigr\rvert_{y=0}$} & rel. uncert. & \multicolumn{2}{c}{meson/\pion} & data used \\ \midrule
    \pion    & 1&065 & 0&11  & 10\% & 1&0    & PHENIX~\cite{adare:051106,adler:024904}\\
    $\eta$   & 0&11  & 0&03  & 30\% & 0&1032 & PHENIX~\cite{adler:024909}\\
    $\rho$   & 0&089 & 0&025 & 28\% & 0&0834 & jet fragmentation~\cite{pdg}\\
    $\omega$ & 0&078 & 0&018 & 23\% & 0&0732 & PHENIX~\cite{Kijima:2007AN610,Dahms:2007AN614,Sharma:2007AN618,adler:051902,0954-3899-34-8-S127,Ivanishchev:2006AN535}\\
    $\phi$   & 0&009 & 0&002 & 24\% & 0&0084 & PHENIX~\cite{Kijima:2007AN610,Dahms:2007AN614,Sharma:2007AN618,0954-3899-34-8-S127,Ivanishchev:2007AN600}\\
    $\eta'$  & 0&016 & 0&005 & 40\% & 0&0127 & PHENIX~\cite{Ivanishchev:2008AN692} and~\cite{pdg}\\
    $J/\psi$ & (1&77 & 0&27)$\times10^{-5}$ & 15\% & 0&0000166 & PHENIX~\cite{adare:232002}\\
    $\psi'$  & (2&5  & 0&7)$\times10^{-6}$  & 27\% & 0&0000023 & world average~\cite{Gavai:1994in}\\ \bottomrule
  \end{tabular}
\end{table}
\begin{table}[tbh]
  \centering
  \caption[Hadron rapidity densities in \AuAu]{\label{tab:exinput_au}Hadron rapidity densities in \AuAu
    collisions at \sqrtsnn = 200 GeV used in \exodus.\\}
  \begin{tabular}{cr@{.}l@{ $\pm$ }r@{.}lcr@{.}ll} \toprule
    ~ & \multicolumn{4}{c}{$\frac{dN}{dy}\bigr\rvert_{y=0}$} & rel. uncert. & \multicolumn{2}{c}{meson/\pion} & data used \\ \midrule
    \pion    & 97&72 & 9&5  & 10\%  & 1&0     & PHENIX~\cite{adler:034909,adler:072301}\\
    $\eta$   & 10&77 & 3&2  & 30\%  & 0&112   & PHENIX~\cite{adler:024909}\\
    $\rho$   &  8&60 & 2&6  & 30\%  & 0&08981 & jet fragmentation~\cite{pdg}\\
    $\omega$ &  9&88 & 3&0  & 30\%  & 0&1032  & PHENIX~\cite{0954-3899-34-8-S127}\\
    $\phi$   &  2&05 & 0&6  & 30\%  & 0&0214  & PHENIX~\cite{adler:014903,0954-3899-34-8-S127}\\
    $\eta'$  &  2&05 & 2&05 & 100\% & 0&02146 & jet fragmentation~\cite{pdg}\\
    $J/\psi$ & (1&79 & 0&26)$\times10^{-5}$ & 15\% & 0&0000182 & PHENIX~\cite{adare:232301}\\
    $\psi'$  & (2&6  & 0&7)$\times10^{-6}$  & 27\% & 0&0000027 & PHENIX~\cite{Donadelli:2008AN722} and~\cite{Gavai:1994in}\\ \bottomrule
  \end{tabular}
\end{table}

Once the meson yields and \pt spectra are known the dilepton spectrum
is given by decay kinematics and branching ratios, which are
implemented in our decay generator \exodus following earlier work
published in~\cite{adare:252002}. The branching ratios are taken from
the compilation of particle properties in~\cite{pdg}.  For the Dalitz
decays \pion, $\eta$, $\eta' \rightarrow \gamma e^+ e^-$ and the decay
$\omega \rightarrow \pion e^+ e^-$ the Kroll-Wada
expression~\cite{PhysRev.98.1355} is used with electromagnetic
transition form factors measured by the Lepton-G
collaboration~\cite{lepton-g,landsberg}. For the decays of the vector
mesons $\rho$, $\omega$, $\phi \rightarrow e^+ e^-$ the expression
derived by Gounaris and Sakurai~\cite{PhysRevLett.21.244} is used,
extending it to 2~\gevcc, slightly beyond its validity range. For the
$J/\psi$ and $\psi' \rightarrow e^+ e^-$ the same expression modified
to include radiative corrections is used as discussed
in~\cite{adare:232002}. All vector mesons are assumed to be
unpolarized. For the Dalitz decays of which the third body is a
photon, the angular distribution is sampled according to $1 + \lambda
\cos^2\theta_{\rm CS}$ distribution, where $\theta_{\rm CS}$ is the
polar angle of the electrons in the Collins-Soper frame.

The resulting systematic uncertainties depend on mass and range from
10 to 25\%. They are combined of the uncertainty of the measured \pion
yield and the meson-to-pion ratios which is the dominant
uncertainty. They also include uncertainties on the measured
electromagnetic transition form factors, in particular for the $\omega
\rightarrow \pion \ee$ decay whose contribution, however, is only
visible in the mass range around $500 < \mee < 600$~\mevcc.

\begin{figure}
  \centering
  \includegraphics[height=0.44\textheight]{figs/pp_only_cocktail.eps}
  \caption[Cocktail of \ee pairs in \pp collisions at \sqrts =
  200~GeV]{Expected invariant mass spectrum of \ee pairs from hadron decays
    (cocktail) in \AuAu collisions at \sqrts = 200 GeV from \exodus.}
  \label{fig:pp_exodus_mass}
  \vspace{5\unitlength}
  \includegraphics[height=0.44\textheight]{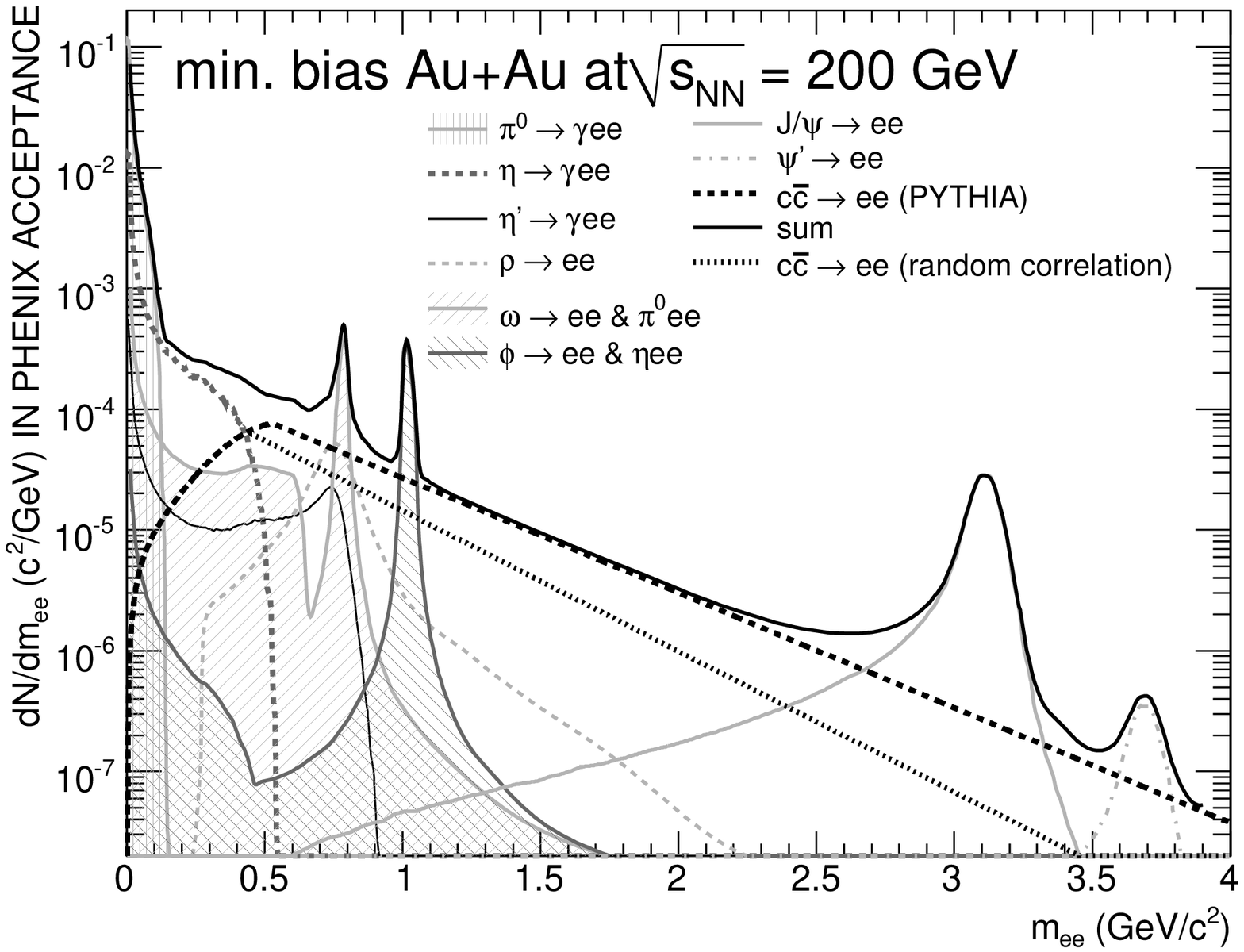}
  \caption[Cocktail of \ee pairs in \AuAu collisions at \sqrtsnn =
  200~GeV]{Expected invariant mass spectrum of \ee pairs from hadron decays
    (cocktail) in \AuAu collisions at \sqrtsnn = 200 GeV from \exodus.}
  \label{fig:au_exodus_mass}
\end{figure}

The \ee pairs from hadron decays are filtered into the PHENIX
acceptance as defined in \eq{eq:track_acc} and their momenta and
angular distributions are smeared to take into account the detector
resolution, as defined in \eq{eq:dc_res} and
\eq{eq:angle_smearing}. The resulting invariant mass spectra are shown
in \fig{fig:pp_exodus_mass} for \pp collisions and in
\fig{fig:au_exodus_mass} for \AuAu collisions, respectively. The
contribution of the various sources are shown individually and include
a contribution from open charm decays which is discussed in
Section~\ref{sec:charmxsec}.

\chapter{Results}
\label{cha:results}

This chapter summarizes the various results of the dielectron
continuum analysis. At first the invariant mass spectrum of \ee pairs
in \pp collisions at \sqrts = 200 GeV is shown in
Section~\ref{sec:known_sources} followed by the measurement of the
charm cross section via the \ee pair yield in the intermediate mass
region in Section~\ref{sec:charmxsec}. Section~\ref{sec:omega_phi}
presents the \pt spectra of the $\omega$ and $\phi$ mesons measured
via their resonance decays into \ee pairs.  The results of the
dielectron continuum analysis of \pp collisions are concluded with \pt
spectra of the low mass continuum and the contributions of virtual
direct photons in Section~\ref{sec:pp_dir_photons}. While the analysis
of \pp collisions is certainly the major achievement of this thesis, a
detailed comparison to the dielectron continuum measured in \AuAu
collisions is given. The results are presented in
Section~\ref{sec:auau_comparison} and compared to theoretical models
in Section~\ref{sec:model-comparisons}.

\section[The Dielectron Continuum in \pp Collisions]{The Dielectron Continuum in $\boldsymbol{p+p}$ Collisions}
\label{sec:known_sources}

\begin{figure}[p]
  \centering
  \includegraphics[width=0.9\textwidth]{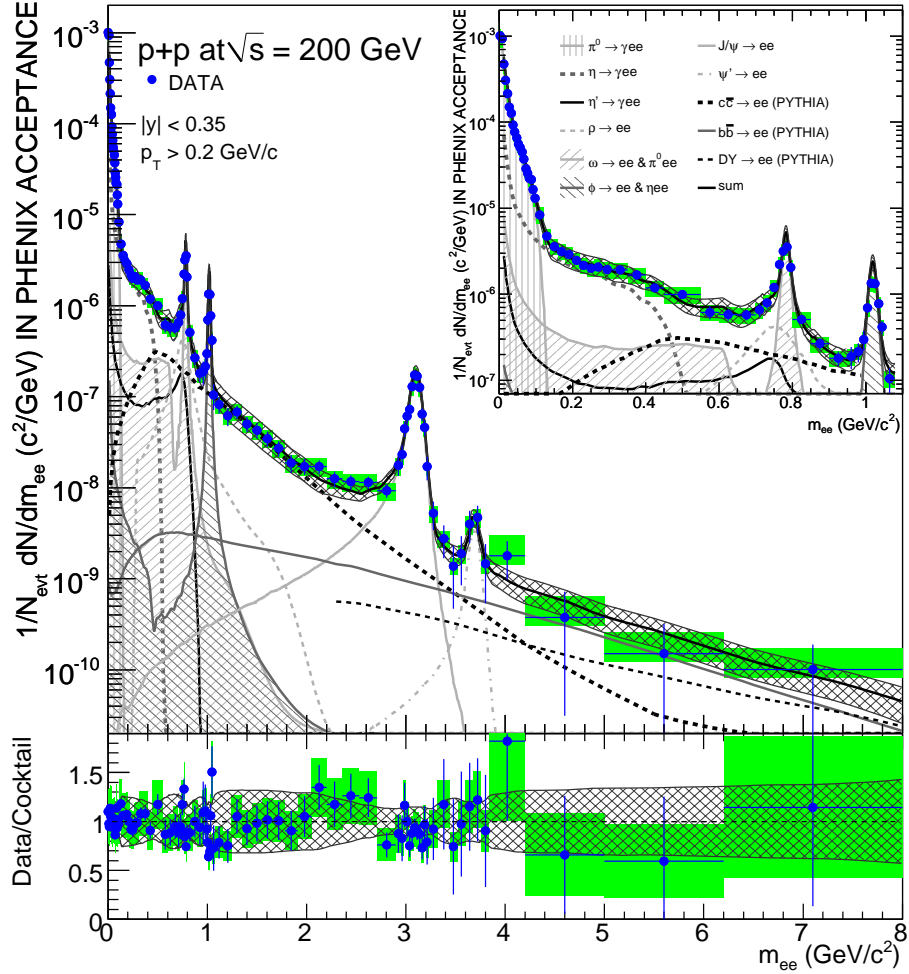}
  \caption[Invariant mass spectrum of \ee pairs in \pp
  collisions]{Electron-positron pair yield per inelastic \pp
    collision as function of pair mass. Data show statistical (bars)
    and systematic (shades) errors separately. The yield per event can
    be converted to a cross section by multiplying with the inelastic
    \pp cross section of 42.2~mb. The data are compared to a cocktail
    of known sources. The contribution from hadron decays is
    independently normalized based on meson measurements in PHENIX,
    the systematic uncertainties are given by the error band. The
    contribution from open charm production is scaled to match the
    data ($\sigma_{c\overline{c}}$ = 544 $\pm$ 39(stat.) $\pm$ 142(syst.)
    $\pm$ 200(model) $\mu$b). The inset shows the same data but
    focuses on the low mass region. The bottom panel shows the ratio
    of data to the cocktail of known sources. The systematic
    uncertainties of the data are shown as boxes, while the
    uncertainty on the cocktail is shown as band around 1.}
  \label{fig:pp_cocktailcomp}
\end{figure}

The invariant mass distribution of \ee pairs calculated according to
\eq{eq:yield_invmass} is compared to the hadronic cocktail as shown in
\fig{fig:pp_cocktailcomp}. The contribution from semileptonic heavy
quark decays is discussed in the Sec.~\ref{sec:charmxsec}. The data
are well described by the expected contributions from hadron decays
and semi-leptonic decays of heavy flavored mesons over the entire mass
range. The resonance peaks of $\omega$, $\phi$, $J/\psi$, and the
$\psi'$ are well reproduced and their width in good agreement with the
expected mass resolution. The ratio of data to cocktail is unity
within the quoted uncertainties as shown in the bottom panel
of~\fig{fig:pp_cocktailcomp}.

\subsection{Charm Cross Section}
\label{sec:charmxsec}

Except for the vector meson peaks, the dilepton yield in the mass
range above 1.1~\gevcc is dominated by semileptonic decays of $D$ and
$B$ mesons correlated through flavor conservation.

In order to extract the charm cross section, first the measured \ee
pair yield is integrated in the mass range $1.1 < \mee < 2.5$~\gevcc
and compared to the cocktail yield and the yield from semileptonic
decays of open heavy quarks (charm and bottom) as well as Drell-Yan
calculated with
\pythia~\cite{Sjostrand:1993yb,Sjostrand:2000wi}. \pythia 6.205 with
{\sc cteq5l} parton distribution function~\cite{Lai:1999wy} has been
used. Following earlier
analyses~\cite{Butsyk:2005qx,PhysRevLett.88.192303} a number of
modifications to \pythia parameters have been made:
\begin{itemize}
\item {\sc parp}(91) = 1.5 (\kt), 
\item {\sc mstp}(33) = 1 (use common K factor)
\item {\sc parp}(31) = 3.5 (K factor),
\item {\sc mstp}(32) = 4 ($Q^2$ scale)
\item[] in addition for charm production:
\item {\sc msel} = 4 ($c\overline{c}$ production)
\item {\sc pmas}(4,1) = 1.25 ($m_c$),
\item[] for bottom:
\item {\sc msel} = 5 ($b\overline{b}$ production)
\item {\sc pmas}(5,1) = 4.1 ($m_b$),
\item[] and for Drell-Yan:
\item {\sc msel} = 11 ($Z$ or $\gamma^{\ast}$ production)
\item {\sc parp}(31) = 1.8 (K factor)
\item {\sc ckin}(3) = 2.0 (min. parton \pt).
\end{itemize}

The following cross sections are assumed:
\begin{itemize}
\item $\sigma_{c\overline{c}} = 567 \pm 57 ({\rm stat.}) \pm 193 ({\rm syst.})~\mu$b~\cite{adare:252002}
\item $\sigma_{b\overline{b}} = 3.7 \pm 3.7~\mu$b~\cite{jaroschek}
\item $\sigma_{\rm DY} = 42 \pm 42~$nb~\cite{Gavin:1995}
\end{itemize}

The integrated yield of \ee pairs in $1.1 < \mee < 2.5$~\gevcc in the
PHENIX acceptance are:
\begin{itemize}
\item measured: $Y_{\rm data} = (4.53 \pm 0.28 ({\rm stat.}) \pm 0.97 ({\rm syst.})) \times 10^{-8}$
\item \pythia charm: $Y_{c\overline{c}} = 4.08 \times 10^{-8}$
\item \pythia bottom: $Y_{b\overline{b}} = 2.62 \times 10^{-9}$
\item \pythia Drell-Yan: $Y_{\rm DY} = 2.66 \times 10^{-10}$
\item \exodus cocktail: $Y_{\rm hadr}= 3.20 \times 10^{-9}$
\end{itemize}

The contribution from the cocktail is essentially only due to the tail
of the $\rho$, which is not known at this high mass. Therefore, an
uncertainty of 100\% is assigned. However, since the cocktail
contributes only 7\% to the data in this mass range, the systematic
uncertainty on the measured yield due to the $\rho$ contribution is
only 7\%. Also the bottom cross section from FONLL has a 100\%
uncertainty, which for the same reason translates into only a 6\%
systematic uncertainty on the measured yield.

The ratio of $(Y_{\rm data} - Y_{\rm hadr})/(Y_{c\overline{c}} +
Y_{b\overline{b}} + Y_{\rm DY}) = 0.96$ can be used to convert the charm
cross section measured in Ref.~\cite{adare:252002} which is used as
input for \pythia:
\begin{align*}
  \frac{d\sigma_{c\overline{c}}}{dy} &= 123 \pm 12 ({\rm stat.})\pm 37 ({\rm syst.})~\mu{\rm b}\\
  \sigma_{c\overline{c}} &= 567 \pm 57 ({\rm stat.}) \pm 193 ({\rm syst.})~\mu{\rm b}
\end{align*}
into what is measured in this analysis:
\begin{align*}
  Y_{\rm data} - Y_{\rm hadr} &= (4.21 \pm 0.28 ({\rm stat.}) \pm 1.02 ({\rm syst.})) \times 10^{-9}\\
  \frac{d\sigma_{c\overline{c}}}{dy} &= 118.1 \pm 8.38 ({\rm stat.})\pm 30.73 ({\rm syst.}) \pm 39.5 ({\rm model})~\mu{\rm b}\\
  \sigma_{c\overline{c}} &= 544 \pm 39 ({\rm stat.}) \pm 142 ({\rm syst.}) \pm 200 ({\rm model})~\mu{\rm b}
\end{align*}
This measurement is an important confirmation of the result reported
in Ref.~\cite{adare:252002}, in particular, as the STAR Collaboration
has reported a charm cross section in \pp
collisions~\cite{abelev:192301,adams:062301}, which is a factor of two
larger than observed by PHENIX. Currently, the source for this
discrepancy is unknown, but if PHENIX was to overestimate its single
electron identification efficiency by a factor of two, this effect
would enter quadratically into the \ee pair measurement and therefore
lead to a significant difference between the dielectron and single
electron results.

\begin{figure}
  \centering
  \includegraphics[width=0.9\textwidth]{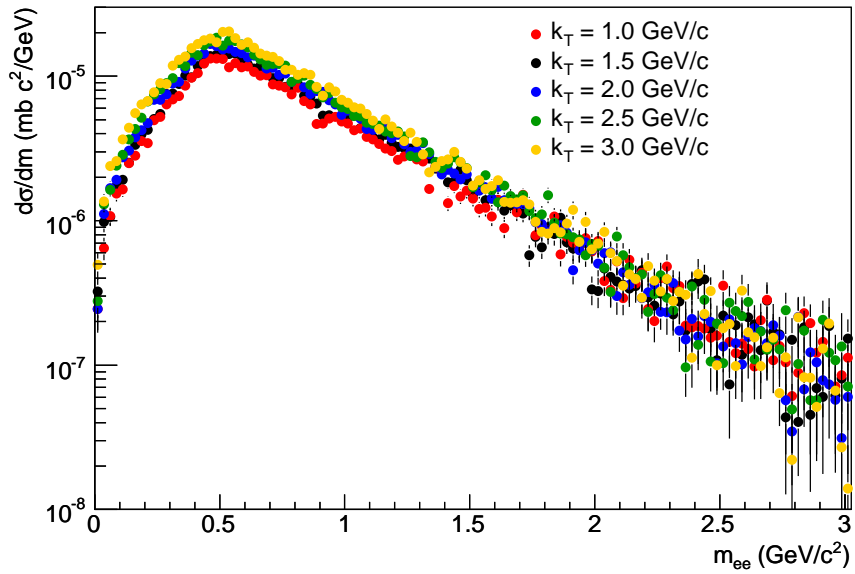}
  \caption[\pythia \ee pairs from open charm for different \kt values]{\ee pairs from semileptonic open charm decays for different
    values of \kt: 1.0~\gevc ({\em blue}), 1.5~\gevc ({\em black}), 2.0~\gevc
    ({\em yellow}), 2.5~\gevc ({\em magenta}), and 3.0~\gevc({\em cyan}).}
  \label{fig:pythia_kt}
\end{figure}

The systematic and model uncertainties on this measurement are due to
a number of sources. As the fraction of \ee pairs from correlated
heavy quark decays at mid-rapidity depends on the dynamical
correlation between the quarks, additional systematic uncertainties
beyond the parameterization of the PHENIX acceptance (10\%). The
momentum of a parton inside the colliding hadrons are expected to have
a finite ``primordial'' component \kt which is perpendicular to the
beam direction due to Fermi motion. In \pythia the `primordial \kt'
parameter is used to define the width of the Gaussian distribution of
the primordial parton \kt.  The value of \kt affects the azimuthal
correlation between $c$ and $\overline{c}$. While a nominal value of
1.5~\gevc is used, \kt has been varied between 1 and 3~\gevc. The
resulting \epair spectra are shown in \fig{fig:pythia_kt} and the
integrated yields are:
\begin{align*}
  Y_{c\overline{c}}(\kt=1.0~{\rm GeV}/c) &= 3.74 \times 10^{-4}\\
  Y_{c\overline{c}}(\kt=1.5~{\rm GeV}/c) &= 4.12 \times 10^{-4}\\
  Y_{c\overline{c}}(\kt=2.0~{\rm GeV}/c) &= 4.64 \times 10^{-4}\\
  Y_{c\overline{c}}(\kt=2.5~{\rm GeV}/c) &= 5.10 \times 10^{-4}\\
  Y_{c\overline{c}}(\kt=3.0~{\rm GeV}/c) &= 5.53 \times 10^{-4}
\end{align*}
Due to the variation in the yield an uncertainty of 20\% is assigned.

The uncertainty of the relative abundance of charm quarks and the
branching ratios of semileptonic decays is 21\%~\cite{adare:252002}. A
total branching ratio of $c \rightarrow e$ of $9.5\% \pm 1\%$ was
calculated from the the particle ratios $D^+/D^0 = 0.45 \pm 0.1$,
$D_s/D^0 = 0.25 \pm 0.1$, and $\Lambda_c/D^0 = 0.1 \pm 0.05$ and the
branching ratios according to~\cite{pdg}. The change in the
distributions leads to an uncertainty of 15\%~\cite{adare:252002}.

Furthermore the systematic uncertainty on the longitudinal correlation
of the charm quarks (rapidity gap) has been studied by changing the
parton distribution function (PDF). A PDF $f_i^p(x,Q^2)$ describes
probability to find a parton with flavor $i$ inside a beam particle
$p$ participating in a hard-scattering process with momentum transfer
$Q^2$. As the PDF can not be calculated with pQCD, PDFs are
constructed by various groups by fits to experimental data for a fixed
momentum transfer $Q_0^2$ and are then extrapolated in the ($x,\,
Q^2$) plane.

\begin{figure}
  \centering
  \includegraphics[width=0.9\textwidth]{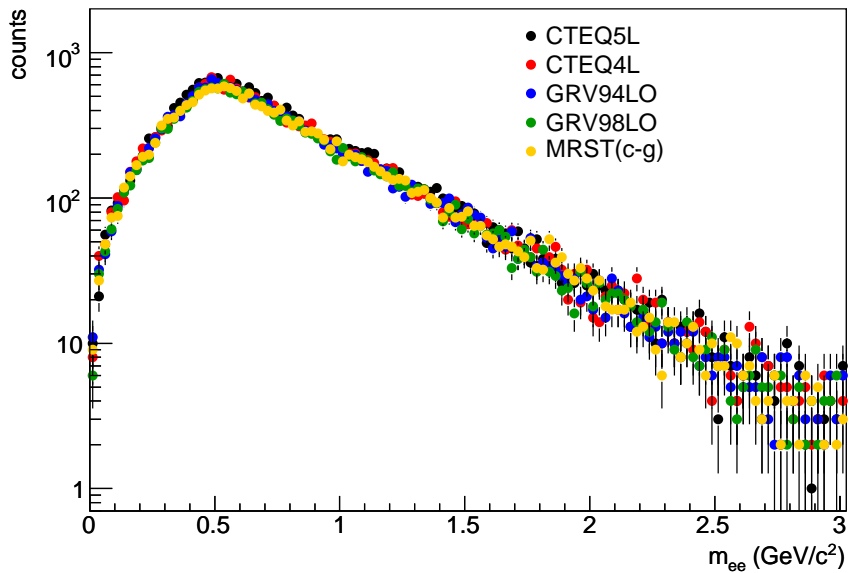}
  \caption[\pythia \ee pairs from open charm for different PDFs]{\ee
    pairs from semileptonic open charm decays for different PDFs: {\sc
      cteq5l} ({\em black}), {\sc cteq4l} ({\em blue}), {\sc grv94lo}
    ({\em red}), {\sc grv98lo} ({\em green}), and {\sc mrst}(c-g)
    ({\em yellow}).}
  \label{fig:pythia_pdf}
\end{figure}

The resulting spectra for five different PDFs ({\sc
  cteq5l}~\cite{Lai:1999wy}, {\sc cteq4l}~\cite{PhysRevD.55.1280},
{\sc grv94lo}~\cite{Gluck:1994uf}, {\sc grv98lo}~\cite{Gluck:1998xa},
and {\sc mrst}(c-g)\cite{Martin:1998sq}) are shown in
\fig{fig:pythia_pdf}. The variation in the \epair yield in the PHENIX
acceptance is 11\%.

\subsubsection{Drell-Yan}
\label{sec:drell-yan}

\begin{figure}
  \centering
  \includegraphics[width=0.9\textwidth]{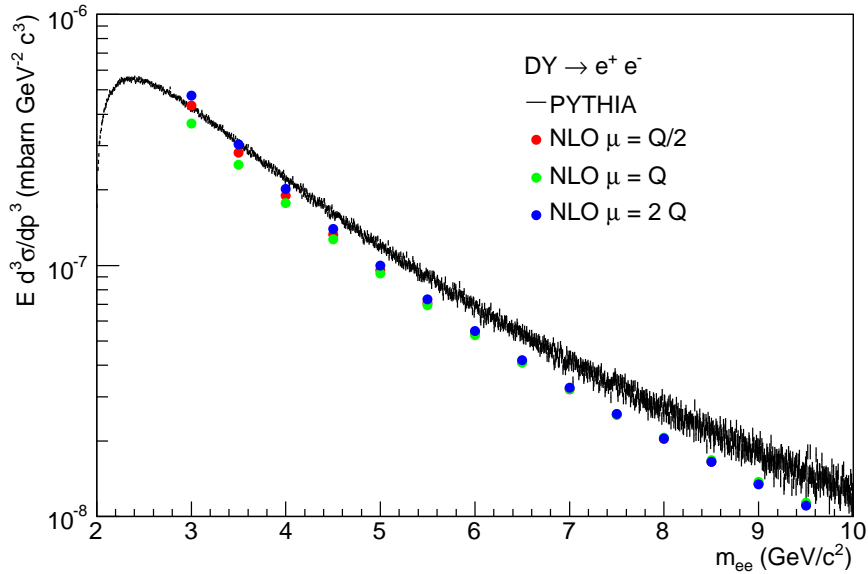}
  \caption[\ee pairs from Drell-Yan from \pythia and NLO pQCD]{\ee
    pairs from Drell-Yan in $|y_{ee}| < 0.5$ from \pythia (black) and
    a NLO pQCD calculation~\cite{vogelsang} for three scales: $\mu =
    q/2$ ({\em red}), $\mu = q$ ({\em green}), and $\mu = 2 q$ ({\em
      blue}).}
  \label{fig:nlo_dy}
\end{figure}

As estimate of the Drell-Yan contribution to the \epair spectrum a NLO
pQCD calculation by Werner Vogelsang has been
used~\cite{vogelsang}. The invariant mass spectrum of \ee pairs in
$|y_{ee}| < 0.5$ is shown in \fig{fig:nlo_dy} using {\sc cteq6m}
PDF~\cite{Pumplin:2002vw} with the scales $\mu = q/2$, $q$, and $2 q$.

The result is compared to a \pythia calculation which is shown in the
same figure. While above 4~\gevcc the two calculations disagree in
shape, the integrated yields agree within 20\% in the region of
interest.

\subsubsection{Fitting}
\label{sec:fitcharm}

\begin{figure}
  \centering
  \includegraphics[width=0.9\textwidth]{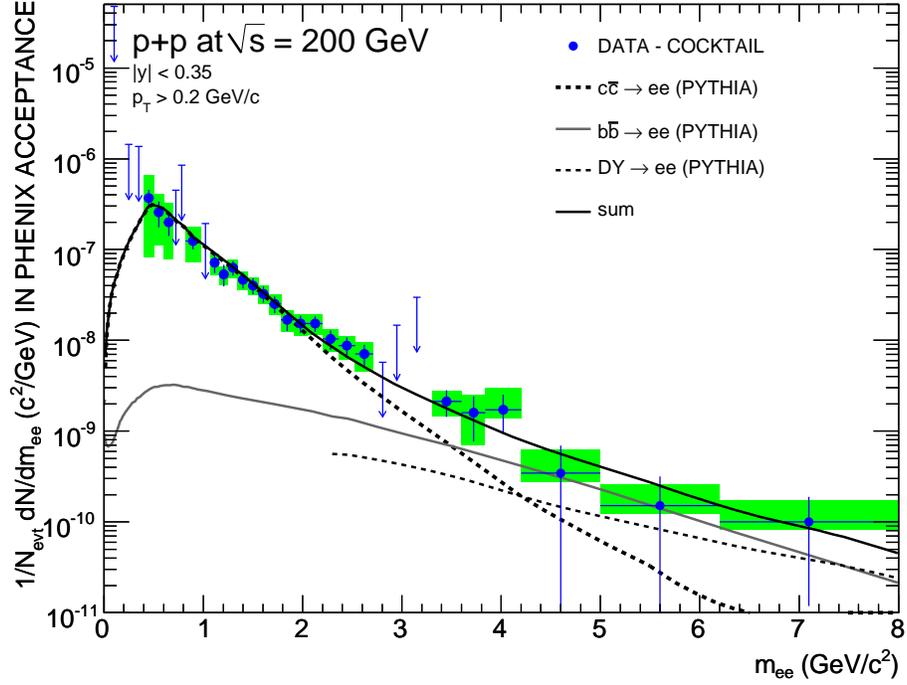}
  \caption[\ee pairs from semi-leptonic heavy flavor
  decays]{Electron-positron mass distributions from semileptonic
    decays of heavy flavor, obtained by subtracting the contribution
    from \pion, $\eta$, $\omega$, $\rho$, $\phi$, $\eta'$, $J/\psi$
    and $\psi'$ mesons from the inclusive \ee pair yield.  The arrows
    indicate upper limits (95\% CL) in the mass regions where the
    charm contribution is smaller or comparable to the systematic
    uncertainties. For all data points statistical error bars and
    systematic uncertainty boxes, including data and model
    contributions, are shown. Also shown are expected contributions
    from charm, scaled to data, and bottom as well as Drell-Yan.}
  \label{fig:cocktailsub}
\end{figure}

In a second approach the charm and bottom cross section have been
extracted with a simultaneous fit of the mass shapes from \pythia to
the cocktail subtracted data shown in \fig{fig:cocktailsub}.

The fit is performed in a mass range $1.1 < \mee < 7.0$~\gevcc, but
excludes the mass range $2.5 < \mee < 3.9$~\gevcc to avoid influences
from imperfections in the subtractions of $J/\psi$ and $\psi'$, with
the mass spectra from \pythia for \ee pairs from charm, bottom and
Drell-Yan:
\begin{equation}
  dN/d\mee = p_0 \cdot dN_{c\overline{c}}/d\mee + p_1 \cdot dN_{b\overline{b}}/d\mee + dN_{\rm DY}/d\mee
\end{equation}
where $p_0$ and $p_1$ denote the two fit parameters and
$dN_{c\overline{c}}/d\mee$, $dN_{b\overline{b}}/d\mee$, and $dN_{\rm
  DY}/d\mee$ are the mass distributions of simulated \ee pairs in the
PHENIX acceptance from charm, bottom and Drell-Yan with the cross
sections $\sigma_{c\overline{c}} = 567~\mu$b~\cite{adare:252002},
$\sigma_{b\overline{b}} = 3.7~\mu$b~\cite{jaroschek} and $\sigma_{\rm
  DY} = 42$nb~\cite{Gavin:1995}, respectively. The fit result is $p_0
= 0.914 \pm 0.082$ and $p_1 = 1.064 \pm 0.714$. When scaling the
\pythia Drell-Yan spectrum above 4\gevcc to the NLO pQCD calculation
by W. Vogelsang~\cite{vogelsang} the results are summarized in
\tab{tab:fits}.

\begin{table}[tbh]
  \centering
  \caption[\pythia Drell-Yan fits to NLO pQCD]{\label{tab:fits}Fit results for scaling the \pythia Drell-Yan
    spectrum to the NLO pQCD calculation~\cite{vogelsang} above 4~\gevcc.\\}
  \begin{tabular}{lr@{.}l@{ $\pm$ }r@{.}lr@{.}l@{ $\pm$ }r@{.}l} \toprule
    ~ & \multicolumn{4}{c}{$p_0$} & \multicolumn{4}{c}{$p_1$}\\ \midrule
    $\mu = q/2$&0&909&0&083&1&15&0&72\\
    $\mu = q$  &0&897&0&083&1&25&0&72\\
    $\mu = 2 q$&0&910&0&083&1&14&0&72\\ \bottomrule
  \end{tabular}
\end{table}

The resulting charm and bottom cross sections are therefore:
\begin{align*}
  \sigma_{c\overline{c}} &= 518 \pm 47 ({\rm stat.}) \pm 135 ({\rm syst.}) \pm 190 ({\rm model})~\mu{\rm b}\\
  \sigma_{b\overline{b}} &= 3.9 \pm 2.4 ({\rm stat.}) ^{+3}_{-2} ({\rm syst.})~\mu{\rm b}
\end{align*}
The fit result for the charm cross section is in good agreement with
yield integration method. Although the significance of the bottom
cross section is limited by large uncertainties, it is the first
measurement of bottom production at RHIC energies. A preliminary
result of the bottom to charm ratio measured via electron-hadron
correlations~\cite{Morino:2008nc} leads to a bottom cross section of $
\sigma_{b\overline{b}} = 4.61 \pm 1.31 ({\rm stat.}) ^{+2.57}_{-2.22}
({\rm syst.})~\mu{\rm b}$ which is in good agreement with the
dielectron result.

\subsubsection[Dependence on Collision Energy]{Dependence on Collision Energy}
\label{sec:charm_roots}

\begin{figure}
  \centering
  \includegraphics[width=0.9\textwidth]{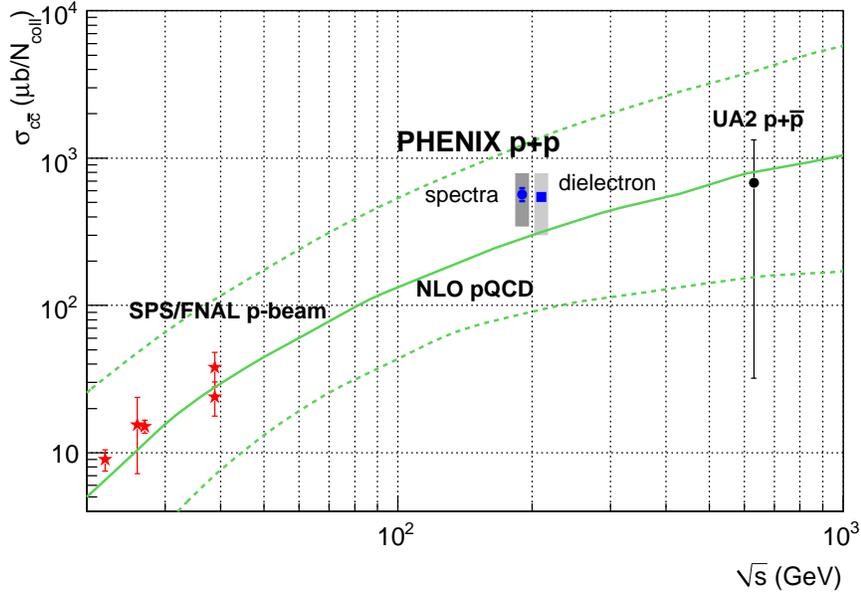}
  \caption[Total Charm Cross Section vs. \sqrts]{Shown is a NLO pQCD
    calculation of the total charm cross section as function of the
    collisions energy \sqrts. The central value is shown as {\em
      solid} line, the upper and lower ends of the uncertainty band
    are shown as {\em dashed} lines. The calculation is compared to
    experimental measurements of the charm cross section including the
    result of this thesis labeled {\em dielectron}.}
  \label{fig:charm_xsec_roots}
\end{figure}

The \sqrts dependence of charm production in \pp collisions as
predicted by a NLO pQCD calculations~\cite{Vogt:2007aw} is shown in
\fig{fig:charm_xsec_roots}. It is compared to experimental data from
SPS and FNAL experiments at $\sqrts \approx 20$~GeV, data from PHENIX
at \sqrts = 200 GeV and results from the UA2 Collaboration at \sqrts =
630 GeV. The two data points from PHENIX are the cross sections
measured with single electrons from semi-leptonic heavy flavor
decays~\cite{adare:252002} (labeled {\em spectra}) and the result
presented in this thesis (labeled {\em dielectron}). Within the large
theoretical uncertainties, NLO calculations describes the experimental
data over the full range of \sqrts.

\begin{figure}
  \centering
  \includegraphics[width=0.9\textwidth]{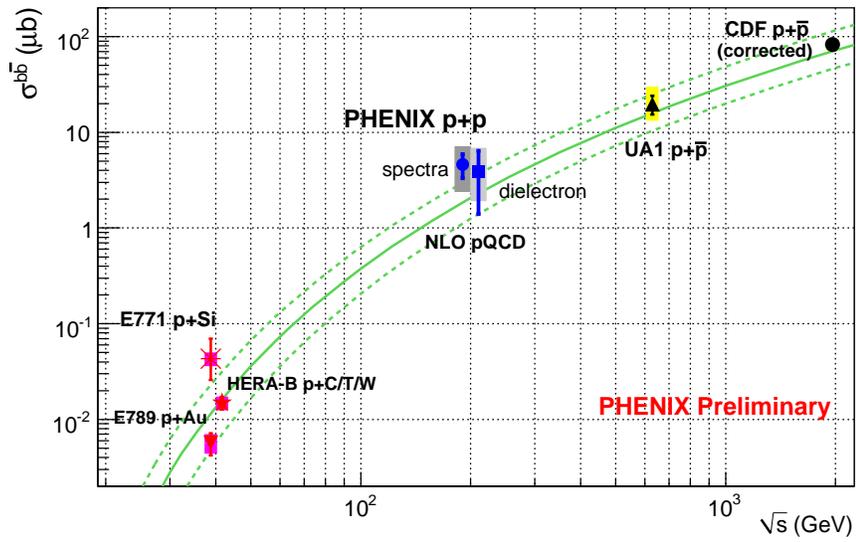}
  \caption[Total Bottom Cross Section vs. \sqrts]{Shown is a NLO pQCD
    calculation of the total bottom cross section as function of the
    collisions energy \sqrts. The central value is shown as {\em
      solid} line, the upper and lower ends of the uncertainty band
    are shown as {\em dashed} lines. The calculation is compared to
    experimental measurements of the charm cross section including the
    result of this thesis labeled {\em dielectron}.}
  \label{fig:bottom_xsec_roots}
\end{figure}

For the same NLO calculation, the bottom cross section is compared to
measurements as function of \sqrts in \fig{fig:bottom_xsec_roots}. The
PHENIX data point labeled {\em spectra} is the result of the
electron-hadron analysis~\cite{Morino:2008nc} and the point labeled
{\em dielectron} is the result of this thesis. NLO calculation.


\subsubsection[\pt Dependence]{Transverse Momentum Dependence}
\label{sec:charm_pt}

\begin{figure}
  \centering
  \includegraphics[width=0.9\textwidth]{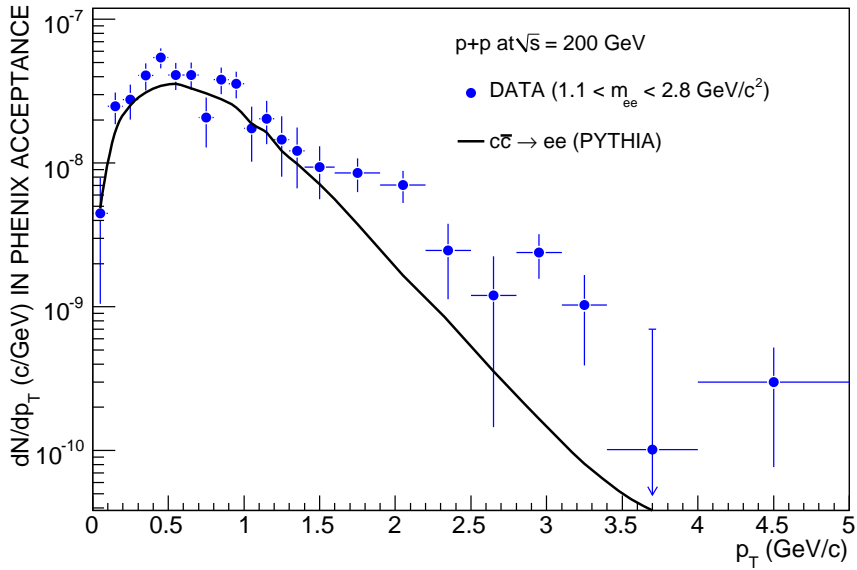}
  \caption[\pt spectrum of \ee pairs in the IMR]{Measured \pt
    distributions of \ee pairs in the mass region $1.1 < \mee <
    2.8$~\gevcc in. The data are corrected for eID and ERT efficiency
    but not for the geometric acceptance. The yield is compared to the
    expected yield from open charm calculated with \pythia filtered
    into the PHENIX acceptance. No systematic uncertainties are
    shown.}
  \label{fig:pp_imr_pt}
\end{figure}

While it is known that \pythia simulations as used in this analysis
fail to describe the single electron spectra at high
\pt~\cite{adare:252002}, it is of little concern here, as the IMR is
dominated by \ee pairs from $c\overline{c}$ decays with small \pt but large
opening angle. \fig{fig:pp_imr_pt} shows the measured yield of \ee
pairs in the mass range $1.1 < \mee < 2.5$~\gevcc as function of
\pt. The data are corrected for electron identification and ERT
trigger efficiency, but not for the geometric acceptance.  The
measured distribution of \ee pairs is compared to a \pythia calculation
filtered into the PHENIX acceptance. While at high \pt \pythia fails to
describe the \pt spectrum of \ee pairs, the low \pt region which
contains the dominant fraction of the yield is sufficiently well
described. Very much analog to the single electron \pt. It will be
interesting to study this \pt dependency in more detail and compare to
NLO calculations. To leading order heavy quark pairs are produced
back-to-back, thus their pair \pt sums to zero. Deviations from this
are due to the intrinsic \kt carried by the incoming partons. In
contrast, as shown in \fig{fig:nlo}, next-to-leading order processes
have three particles in the final state and therefore the distribution
of heavy quark pairs should be more isotropic and contribute more at
high \pt.

\subsubsection{Leading Order vs. Next-to-Leading Order Calculations}

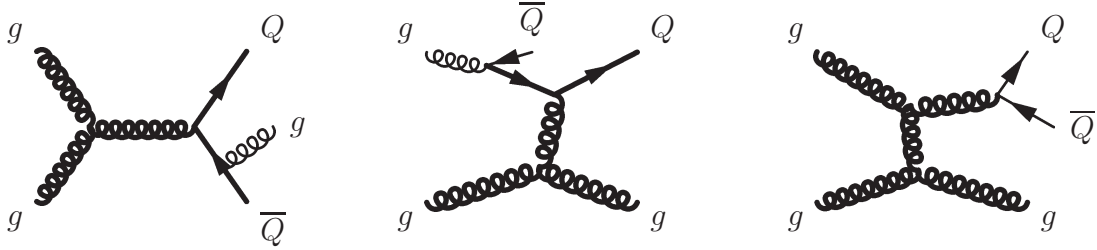
\begin{figure}
  \centering
  \begin{fmffile}{cch}
    \begin{fmfgraph*}(35,20)
      \fmfset{curly_len}{2mm}
      \fmfpen{thick}
      \fmfleft{i1,i2}
      \fmfright{o1,o2,o3}
      \fmf{gluon}{i1,v1,i2}
      \fmf{gluon}{v1,v2}
      \fmf{phantom_arrow}{o1,v2,o3}
      \fmflabel{$g$}{i1}
      \fmflabel{$g$}{i2}
      \fmflabel{$\overline{Q}$}{o1}
      \fmflabel{$Q$}{o3}
      \fmffreeze
      \fmf{plain}{o1,v3,v2,o3}
      \fmf{gluon, width=thin, tension=0}{v3,o2}
      \fmflabel{$g$}{o2}
    \end{fmfgraph*}
  \end{fmffile}
  \hspace{10\unitlength}
  \begin{fmffile}{cci}
    \begin{fmfgraph*}(35,20)
      \fmfset{curly_len}{2mm}
      \fmfpen{thick}
      \fmfleft{i1,i2}
      \fmfright{o1,o2}
      \fmftop{t1}
      \fmf{gluon}{i1,v1,o1}
      \fmf{gluon, width=thin, tension=1.6}{i2,v2}
      \fmf{gluon}{v1,v3}
      \fmf{phantom, tension=0.6}{t1,v2}
      \fmf{phantom}{v2,v3,o2}
      \fmffreeze
      \fmf{fermion}{v2,v3,o2}
      \fmf{fermion, width=thin}{t1,v2}
      \fmflabel{$g$}{i1}
      \fmflabel{$g$}{i2}
      \fmflabel{$g$}{o1}
      \fmflabel{$\overline{Q}$}{t1}
      \fmflabel{$Q$}{o2}
    \end{fmfgraph*}
  \end{fmffile}
  \hspace{10\unitlength}
  \begin{fmffile}{ccj}
    \begin{fmfgraph*}(35,20)
      \fmfset{curly_len}{2mm}
      \fmfpen{thick}
      \fmfleft{i1,i2}
      \fmfright{o1,o2,o3}
      \fmf{gluon}{i1,v1,o1}
      \fmf{gluon}{i2,v2,v1}
      \fmf{gluon}{v2,v3}
      \fmf{fermion, width=thin}{o2,v3,o3}
      \fmflabel{$g$}{i1}
      \fmflabel{$g$}{i2}
      \fmflabel{$g$}{o1}
      \fmflabel{$\overline{Q}$}{o2}
      \fmflabel{$Q$}{o3}
    \end{fmfgraph*}
  \end{fmffile}
  \\\vspace{10\unitlength}
  \caption[\pythia parton showers]{Examples of pair creation, flavor
    excitation, and gluon splitting in \pythia. The thick lines
    correspond to hard scattering processes, the thin ones to parton
    showers~\cite{Bedjidian:2004gd}.}
  \label{fig:lo++}
\end{figure}

In Ref.~\cite{Bedjidian:2004gd} parton showers in \pythia have been
used to approximate some of the real corrections to leading order hard
scattering processes. For example, a final state heavy quark can
radiate a gluon or an initial state gluon can split into a
quark-antiquark pair. According to the number of how many heavy quarks
are produced in the LO hard scattering process the following processes
are defined:
\begin{description}
\item[pair creation:] Two heavy quarks are created in the final
  state by quark-antiquark or gluon annihilation. These are the LO
  processes shown in \fig{fig:lo}
\item[flavor excitation:] One heavy quark, which is assumed to
  come from a gluon splitting process, is put on mass shell by hard
  scattering off a parton from the opposite hadron.
\item[gluon splitting:] No heavy quark is involved in the hard
  scattering, but a heavy quark pair is created in initial or final
  state showers from a gluon.
\end{description}
The corresponding Feynman graphs are shown in \fig{fig:lo++} in which
thick lines correspond to the hard scattering process and thin lines
to the parton shower.

\begin{figure}
  \centering
  \includegraphics[width=0.9\textwidth]{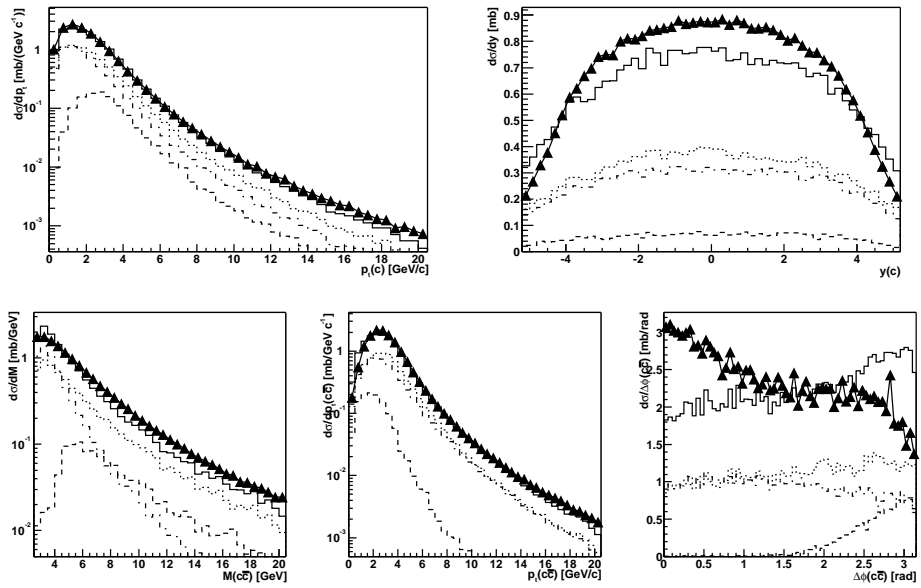}
  \caption[\pt spectrum of \ee pairs in the IMR]{Charm production in
    \PbPb collisions at \sqrts = 5.5~TeV calculated by \pythia including
    parton showers ({\em solid line}). The individual components are
    pair production ({\em dashed}), flavor excitation ({\em dotted}),
    and gluon splitting ({\em dot-dashed}). The \pythia result is
    compared to a NLO calculation ({\em
      triangles})\cite{Bedjidian:2004gd}.}
  \label{fig:charm_pythia++}
\end{figure}

The results of a \pythia calculation including parton showers tuned for
charm production in \PbPb collisions at \sqrts = 5.5~TeV are shown in
\fig{fig:charm_pythia++}. The total charm cross section has been
scaled to a NLO calculation which is shown in comparison. Transverse
momentum and rapidity distributions of single charm quarks, as well as
the invariant mass, \pt and $\Delta\phi$ of $c\overline{c}$ pair are
shown. Besides the fundamental differences in the two calculations,
the overall agreement is rather good. However,especially $\Delta\phi$
distributions show significant differences. It is very interesting to
note that the \pt distributions of the pair creation, which as the LO
processes has two hard scattered quarks in the final state, shows a
significantly softer \pt distribution than flavor excitation and gluon
splitting which have both a more isotropic distribution of the heavy
quark pair. This underlines the importance of higher order processes
at high \pt.

\subsection[$\omega$ and $\phi$ Cross Sections]{$\boldsymbol{\omega}$ and $\boldsymbol{\phi}$ Cross Sections}
\label{sec:omega_phi}

As discussed in Section~\ref{sec:medium_modifications} the in-medium
resonance decays of the vector mesons $\omega$ and $\phi$ provide
information about the hadronic phase of the medium created in heavy
ion collisions at the time of decay. To measure potential
medium-modifications to the line shape of the resonances, a baseline
measurement of the line shapes in \pp collisions is
crucial. Furthermore, the study of hadronic versus electromagnetic
decays of the $\phi$ are particularly interesting. As the $\phi$ mass
($m_{\phi} = 1019$~\mevcc) is very close to twice the kaon mass
($m_{K}=493.7$~\mevcc), the decay $\phi \rightarrow K^+ K^-$ is very
sensitive to any changes of the $\phi$ mass due to, \eg, the
restoration of chiral symmetry.

\begin{sidewaysfigure*}[p]
  \centering
  \includegraphics[width=\textwidth]{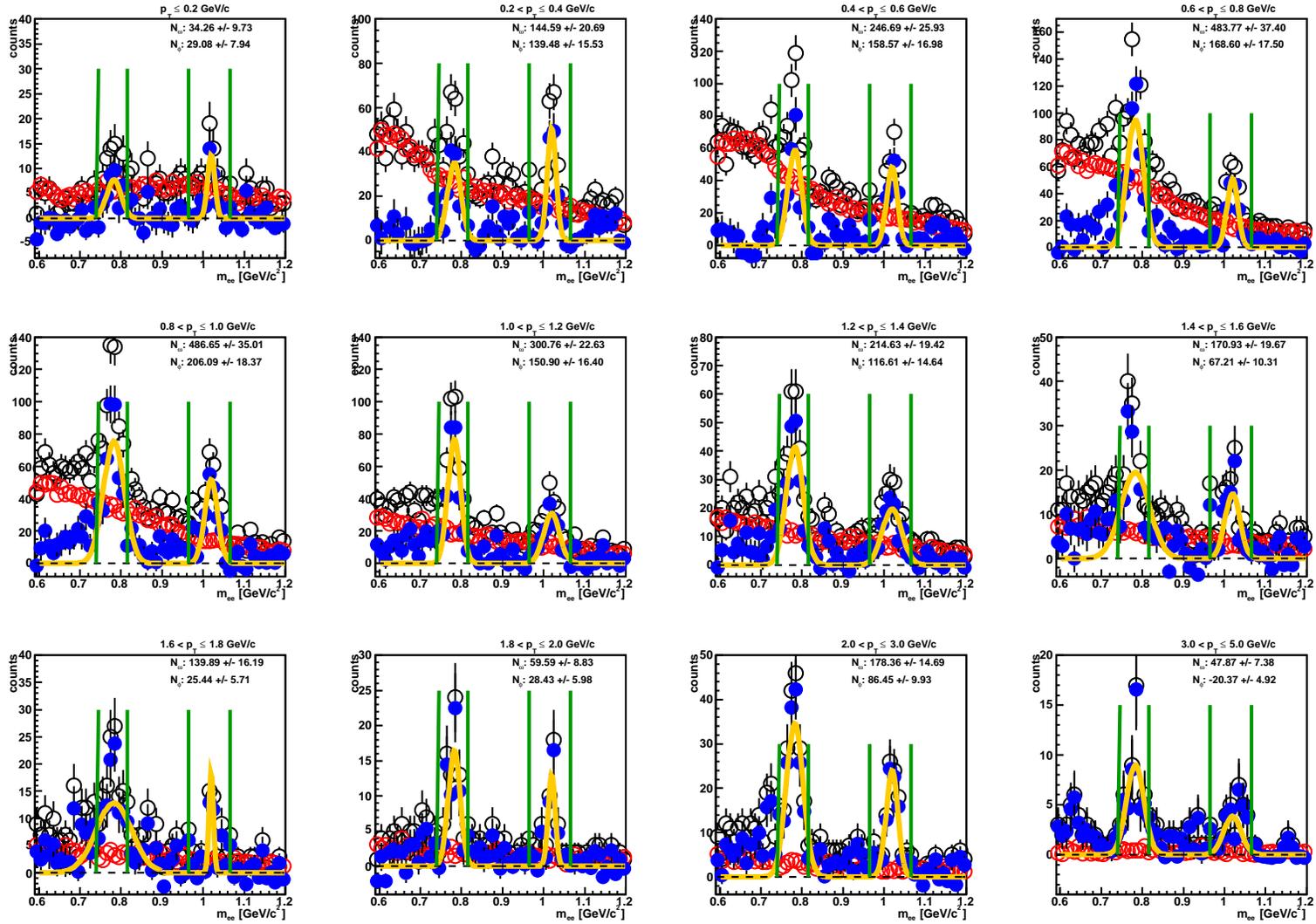}
  \caption[Invariant. mass spectrum of \ee pairs around the $\omega$
  and $\phi$]{Invariant mass spectrum of \ee pairs. Foreground ({\em
      black}), combinatorial background ({\em red}) and Signal ({\em
      blue}). Also shown are two Gaussian fits ({\em orange}) to
    $\omega$ and $\phi$, respectively.}
  \label{fig:om_phi_peaks}
\end{sidewaysfigure*}

To provide this baseline, the cross sections of $\omega$ and $\phi$
have been measured in \pp collisions via their resonance decays into
\ee pairs. \fig{fig:om_phi_peaks} shows the raw invariant mass spectra
of \ee pairs for various slices in \pt of 200~\mevc up to 2~\gevc
and beyond in two bins from 2--3~\gevc and 3--5~\gevc. They are
shown together with the the sum of correlated and combinatorial
background and the signal after background subtraction. Already before
background subtraction the $\omega$ and $\phi$ peaks are very
pronounced. The figure also shows a fit to a Gaussian with the mean
being fixed to the PDG value of the $\omega$ and $\phi$ mass
($m_{\omega} = 782.65$~\mevcc, $m_{\phi} =
1019.460$~\mevcc~\cite{pdg}), respectively.

The $\omega$ and $\phi$ yields (listed in Tab.~\ref{tab:counting}) are
extracted by counting \ee pairs in a mass region which corresponds to
a $3\sigma$ interval, \ie, 0.740--0.815~\gevcc and
0.965--1.065~\gevcc, respectively.

Alternatively, to determine the systematic uncertainty on the peak
extraction, the peaks are fitted with a Gaussian + $n^{\mathrm{th}}$
order polynomial ($n = 0,\,1,\,2$) to allow for contributions due to
the underlying continuum. The fits for the $\omega$ have been
performed in the mass range 0.6--0.9~\gevcc while for the $\phi$ the
range 0.9--1.2~\gevcc was chosen. The systematic uncertainties on the
peak extraction have been determined out of the variation of the yield
with the different fits. For the $\omega$ this uncertainty is 14\%,
while for the $\phi$ it is 20\% for $\pt > 1$~\gevc and 12\%
below. This uncertainty is added in quadrature to the ones listed in
\tab{tab:syst_uncertainties}.

Analog to \eq{eq:invyield} the invariant cross section of $\omega$ and
$\phi$ mesons is calculated as
\begin{equation}\label{eq:invxsecmeson}
  E\frac{d^3\sigma_{\omega,\phi}}{dp^3} = \frac{1}{2\pi\, \pt} \frac{1}{N_{\rm
      evt}\, \Delta\pt}\frac{N_{\omega,\phi}}{B_{ee}^{\omega}\, \varepsilon_{\rm
      pair}\,\varepsilon_{\rm geo}^{\rm pair}}\frac{\varepsilon_{\rm
      BBC}}{\varepsilon_{\rm bias}} \sigma_{pp}^{\rm inel}
\end{equation}
with the branching ratios for the decays $\omega \rightarrow \ee$:
$B_{ee}^{\omega} = 7.18 \pm 0.12\%$ and $\phi \rightarrow \ee$:
$B_{ee}^{\phi} = 2.97 \pm 0.04\%$, respectively~\cite{pdg}.

\begin{figure}
  \centering
  \includegraphics[width=\textwidth]{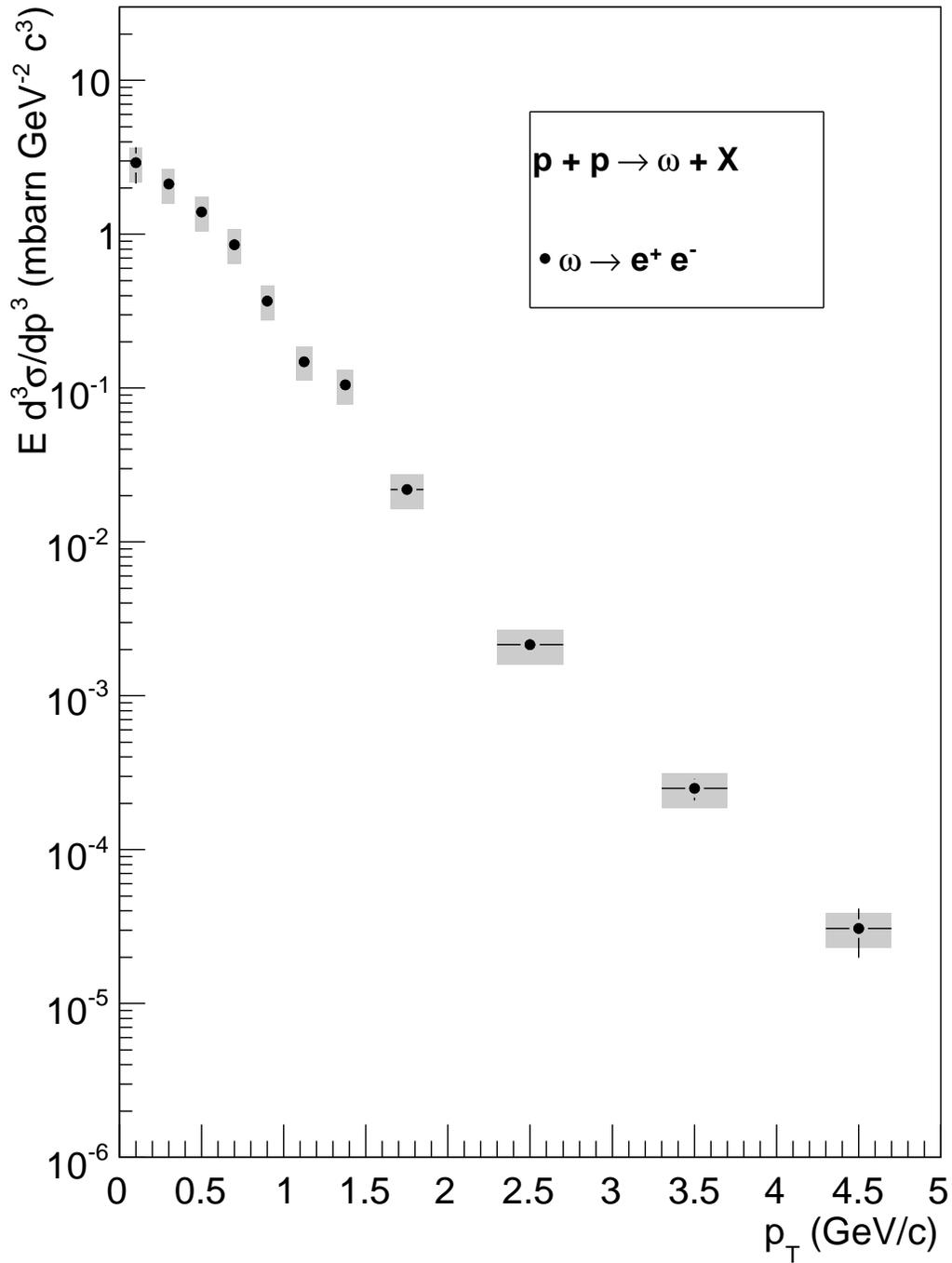}
  \caption[Invariant $\omega$ cross section]{Invariant cross section of $\omega$ in \pp collisions as
    function of \pt. The stat. errors are drawn as bars, the
    syst. uncertainties are represented with the gray boxes.}
  \label{fig:omegaxsec}
\end{figure}

\begin{figure}
  \centering
  \includegraphics[height=0.4\textheight]{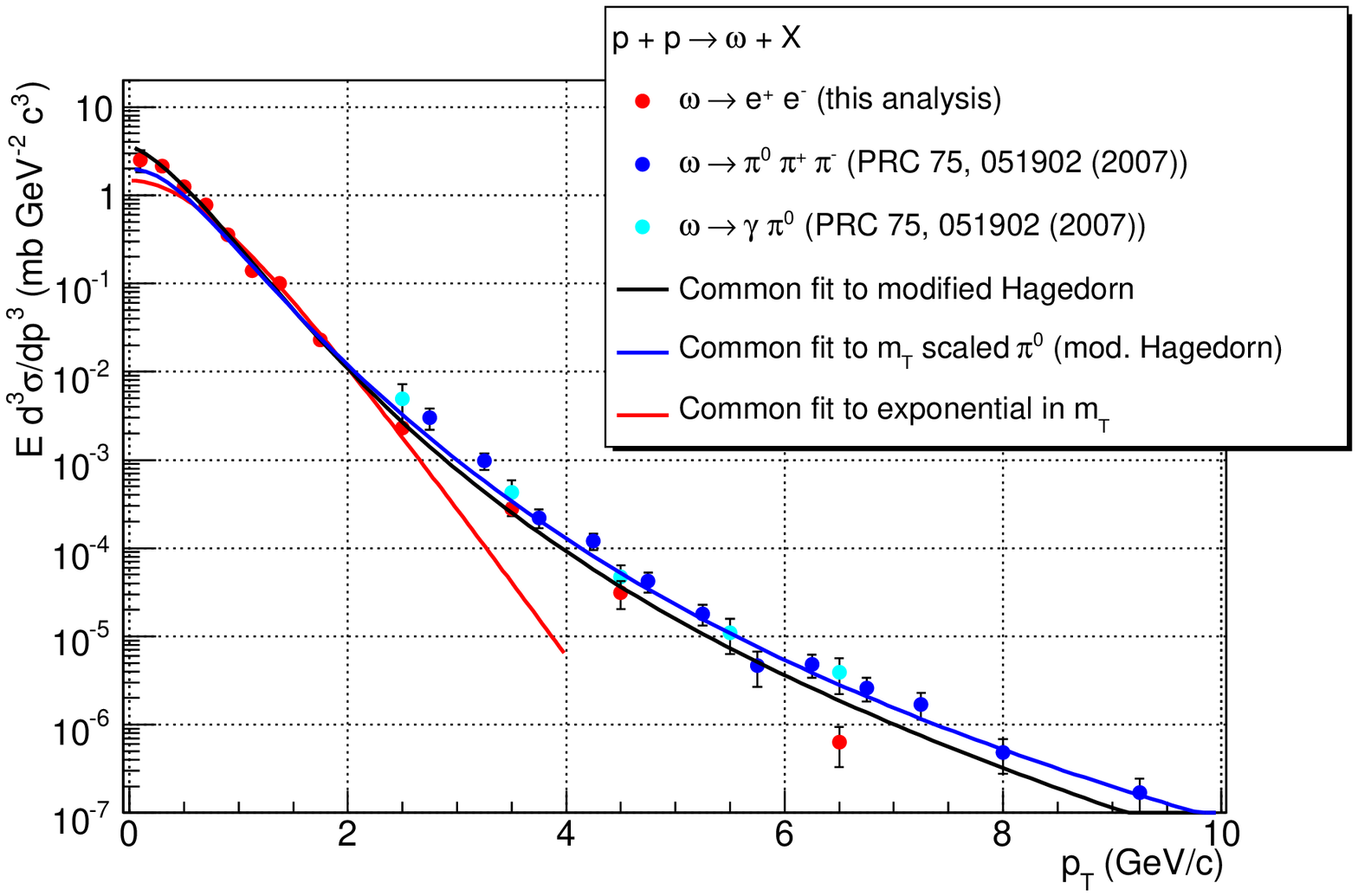}
  \caption[Invariant $\omega$ cross section in comparison to
  measurement of hadronic decay channels]{Invariant cross section of
    $\omega$ in \pp collisions as function of \pt measured in
    electromagnetic and hadronic decay channels. The results are
    fitted to the pion parameterization ({\em blue}), a modified
    Hagedorn ({\em black}) and an exponential in \mt ({\em red}).}
  \label{fig:omegaxsec_fit}
  \includegraphics[height=0.4\textheight]{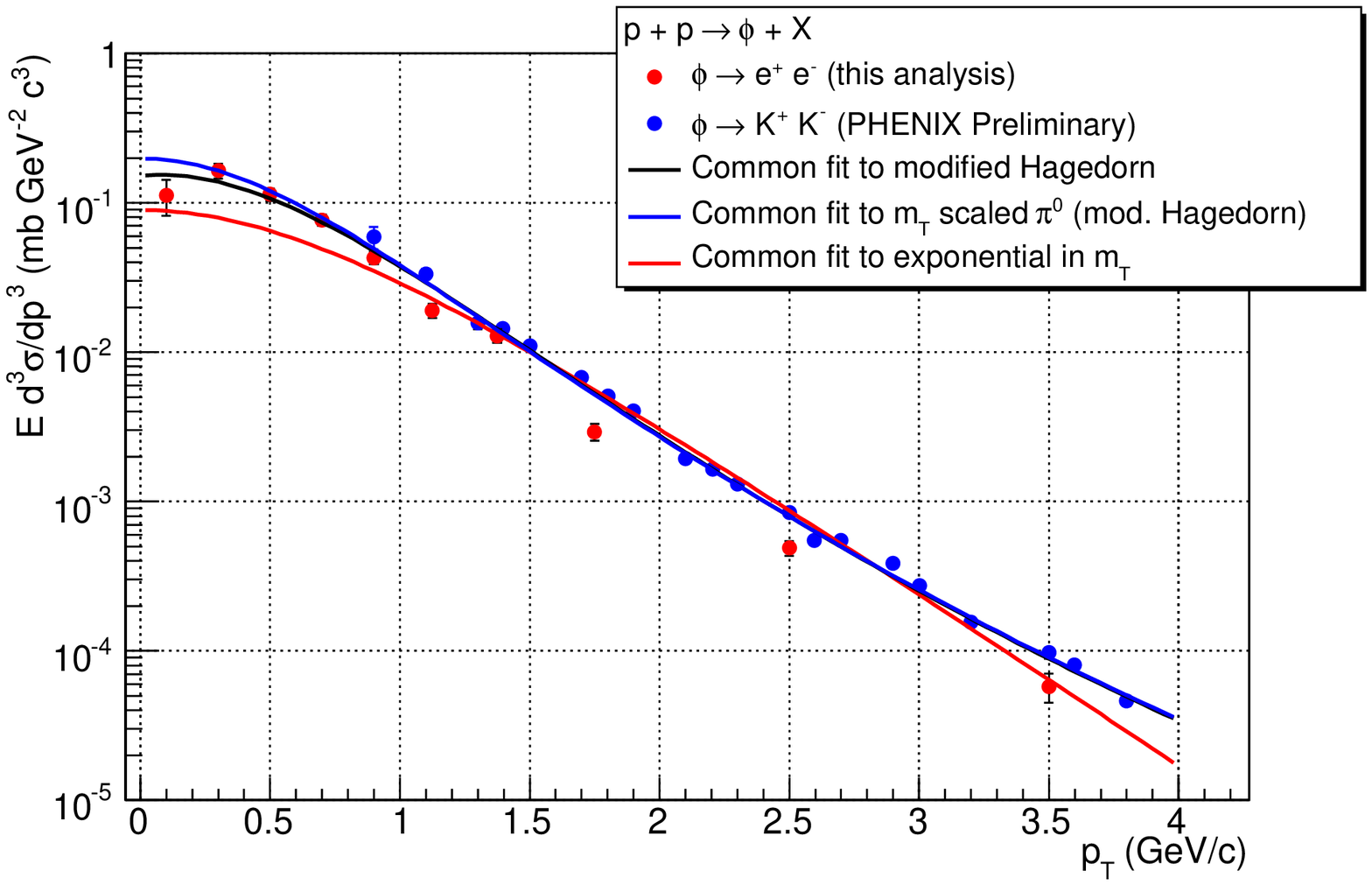}
  \caption[Invariant $\phi$ cross section in comparison to measurement
  of hadronic decay channels]{Invariant cross section of $\phi$ in \pp
    collisions as function of \pt measured in electromagnetic and
    hadronic decay channels. The results are fitted to the pion
    parametrization ({\em blue}), a modified Hagedorn ({\em black})
    and an exponential in \mt ({\em red}).}
  \label{fig:phixsec_fit}
\end{figure}

The resulting $\omega$ cross sections as function of \pt are shown in
\fig{fig:omegaxsec}. In \fig{fig:omegaxsec_fit} it is compared to
existing measurements at higher \pt via hadronic decays, \ie, $\omega
\rightarrow \pi^0 \pi^+\pi^-$ and $\omega \rightarrow \pi^0
\gamma$~\cite{Ivanishchev:2006AN535,0954-3899-34-8-S127,adler:051902}. In
the \pt region in which both measurements exist, they are in good
agreement. Also shown is a common fit to a \mt scaled modified
Hagedorn parameterization of charged and neutral pions as defined in
\eq{eq:hagedorn} with only the normalization factor as free parameter,
which has been used as parameterization of the $\omega$ in the \exodus
cocktail calculation as described in Section~\ref{sec:exodus}. In
addition, alternative fits with a modified Hagedorn and all parameters
free and a simple exponential in \mt ($E \frac{d^3\sigma}{dp^3} = A
\mathrm{e}^{-\left(\mt/T_{\rm eff}\right)}$) are shown.

\begin{figure}
  \centering
  \includegraphics[width=\textwidth]{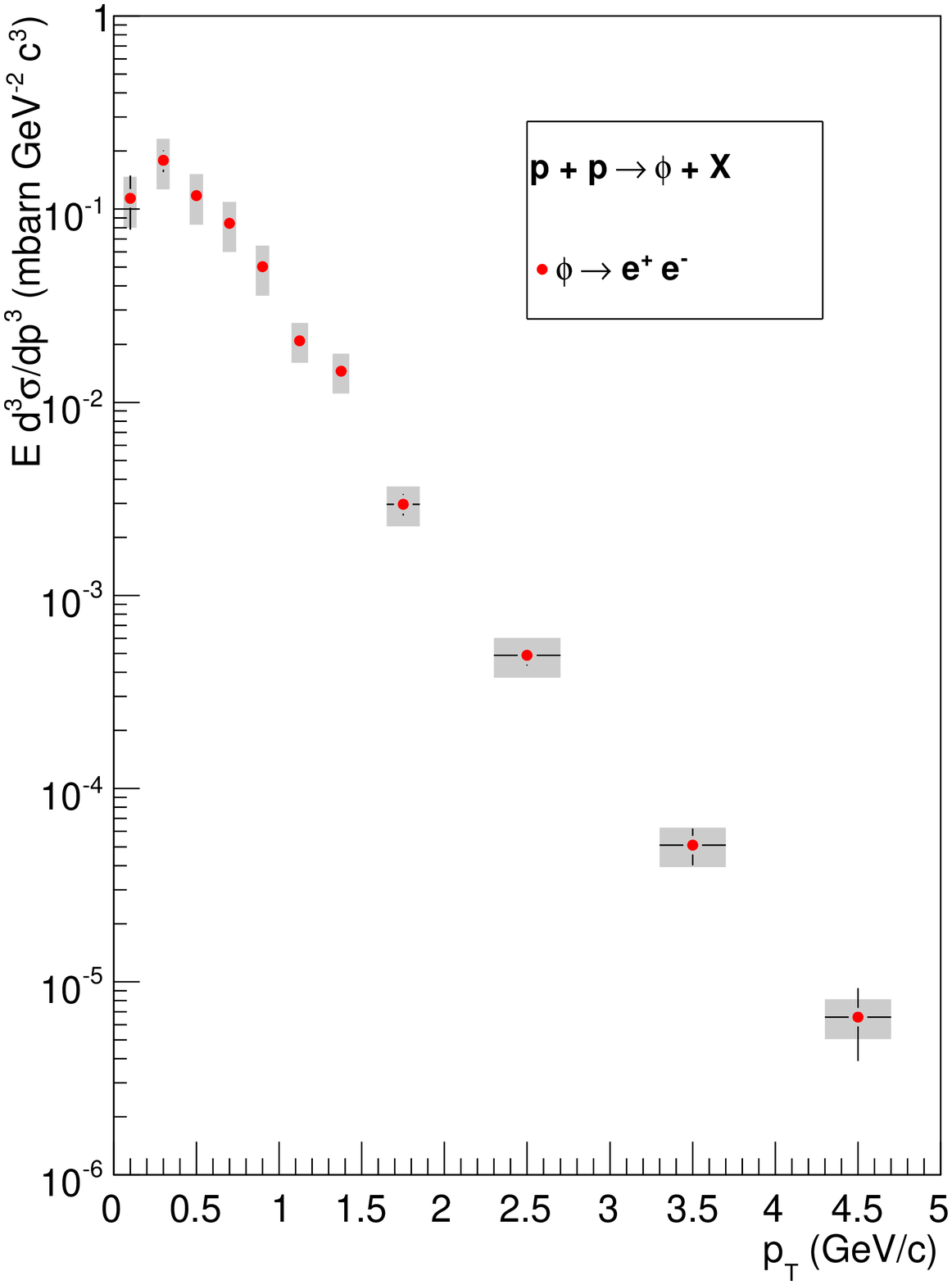}
  \caption[Invariant $\phi$ cross section]{Invariant cross section of
    $\phi$ in \pp collisions as function of \pt. The stat. errors are
    drawn as bars, the syst. uncertainties are represented with the
    gray boxes.}
  \label{fig:phixsec}
\end{figure}

\fig{fig:phixsec} shows the measured $\phi$ cross section, which in
\fig{fig:phixsec_fit} is compared to the result of the hadronic decay
into two kaons~\cite{Ivanishchev:2007AN600,0954-3899-34-8-S127}. The
agreement in the region where both measurements are available is not
very good and currently not understood. A further investigation of the
problem is ongoing and both results. As for the $\omega$, the data are
fit to the \mt scaled modified Hagedorn parameterization of charged
and neutral pions, a modified Hagedorn with all parameters left free
and an exponential in \mt.

For both, $\omega$ and $\phi$ cross sections, the fits with a
modified Hagedorn and all parameters free results in a shape very
comparable to the fit of charged and neutral pions. The exponential
fit in \mt offers only a good description at low \pt but fails for
$\pt > 2$~\gevc.

\subsection[Low Mass \pt Spectra]{Low Mass $\boldsymbol{p_T}$ Spectra}
\label{sec:pp_lowmass_pt}

\begin{figure}
  \centering
  \includegraphics[width=0.9\textwidth]{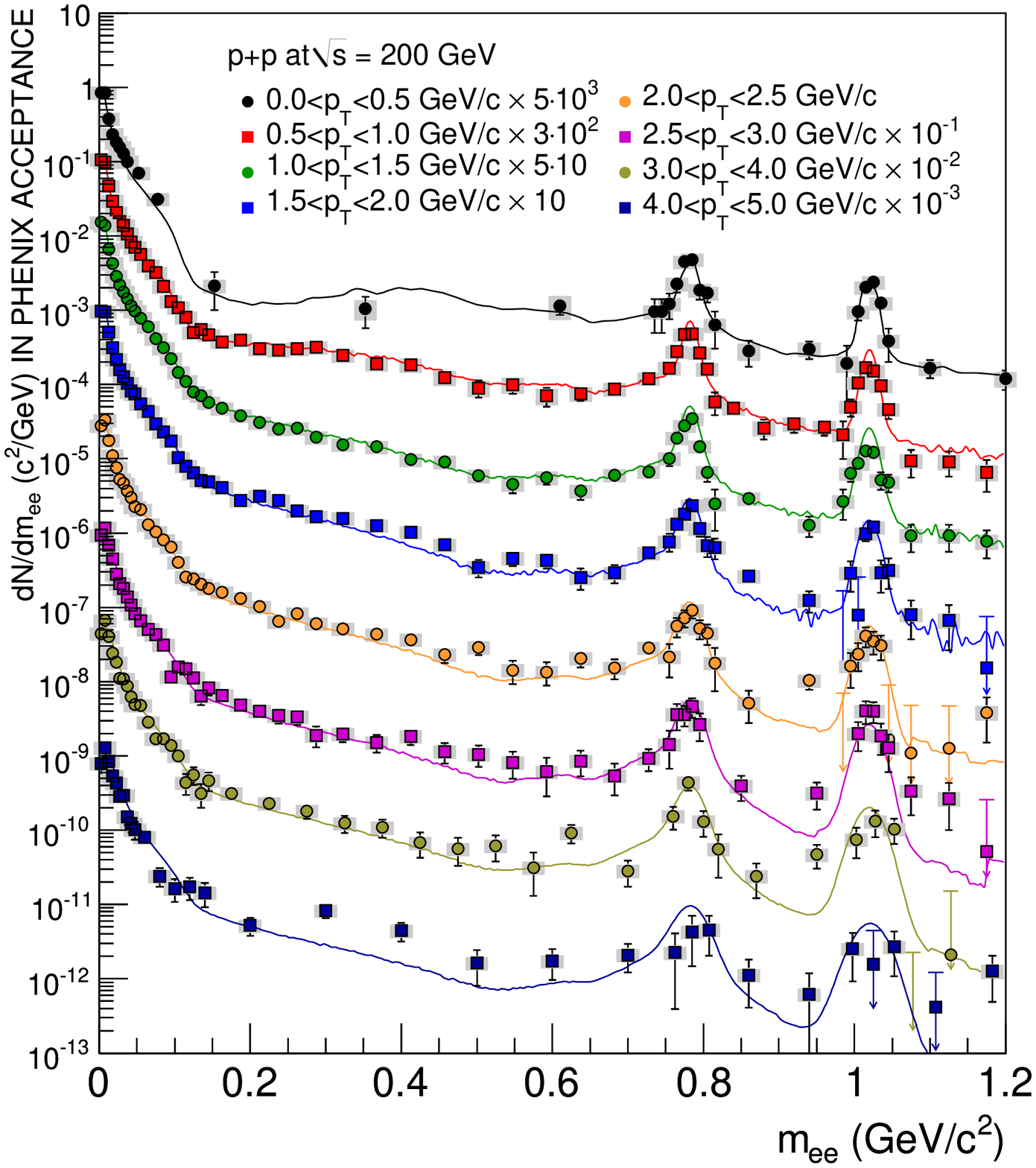}
  \caption[Invariant mass for different \pt ranges in \pp]{The \ee
    pair invariant mass distributions in \pp collisions. The \pt
    ranges are shown in the legend.  The solid curves represent an
    estimate of hadronic sources.}
  \label{fig:pp_mass_varpt}
\end{figure}

To study the \pt dependency of the dielectron continuum the invariant
mass spectra for \ee pairs in various \pt ranges is shown in
\fig{fig:pp_mass_varpt}. The results are again compared to the
cocktail, which has been normalized to the measured \ee pair yield in
$\mee < 30$~\mevcc. A cocktail with absolute normalization agrees with
the data in this region within the systematic uncertainty of
20\%. Again, the \pp data are in good agreement with the cocktail over
the full mass range and for all \pt bins, except for some small
deviations at high \pt which are discussed in
Section~\ref{sec:pp_dir_photons}.

\begin{figure}
  \centering
  \includegraphics[width=0.9\textwidth]{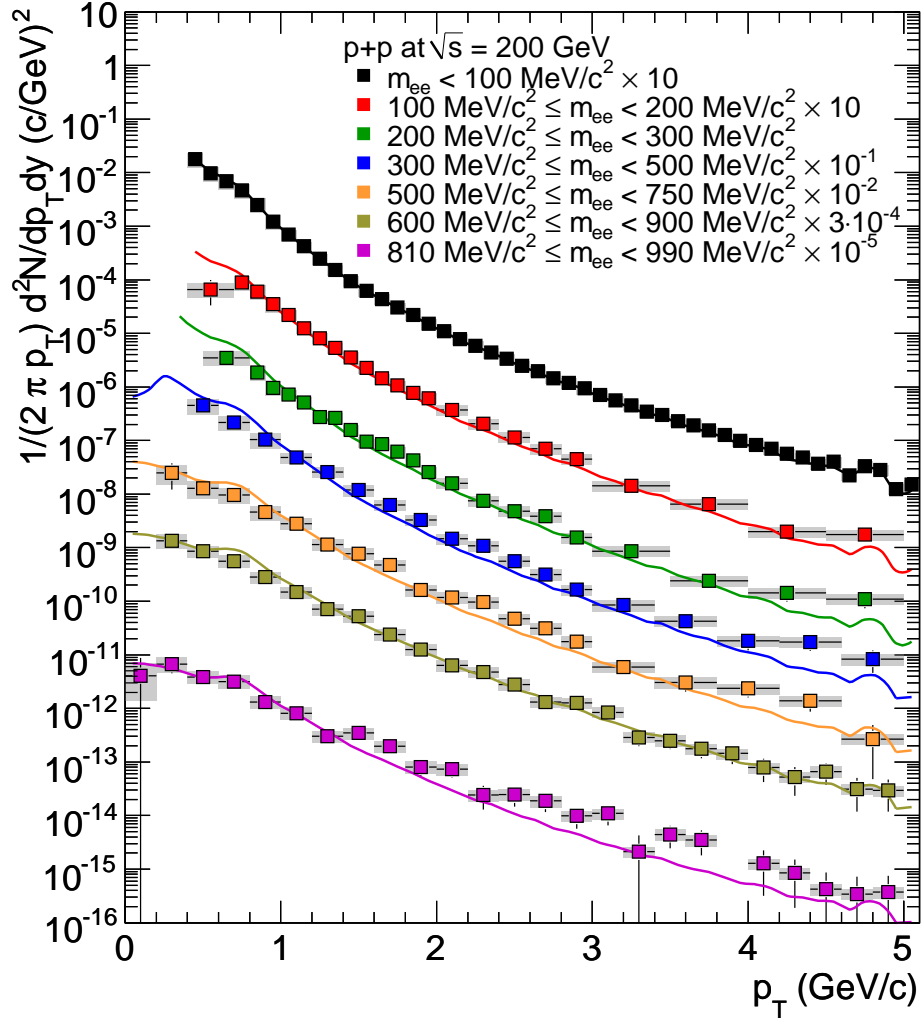}
  \caption[Invariant Yield as function of \pt for \ee pairs in
  different mass ranges in \pp collisions]{Shown is the invariant
    yield of \ee pairs in different mass ranges in \pp collisions. The
    mass ranges are defined in the legend.}
  \label{fig:pp_pt_varmass}
\end{figure}

In a next step the yield of \ee pairs in small mass ranges is analyzed
as function of \pt. The yield is now corrected for the geometric
acceptance of the PHENIX central arm detectors. The invariant yield as
function of the \ee pair \pt as defined in \eq{eq:invyield} is shown
in \fig{fig:pp_pt_varmass} for \ee pairs in different mass ranges and
compared to the cocktail. The \pt spectrum of \ee pairs with $\mee <
0.1$~\gevcc, dominated by \pion Dalitz decays, is in excellent
agreement with the expectation of hadron decays. Also the \pt spectra
of the other mass windows agree within the systematic uncertainties
well with the cocktail for $\pt < 2$~\gevc. Above a small excess,
which was also visible in the high \pt mass spectra in
\fig{fig:pp_mass_varpt}, is observed. This excess is analyzed in the
next Chapter.

\subsection[High \pt Direct Photons]{High $\boldsymbol{p_T}$ Direct Photons}
\label{sec:pp_dir_photons}

As discussed in Section~\ref{sec:direct_photons} any source of real
direct photons can also produce a virtual photon which subsequently
converts internally into an \ee pair. The mass distribution is given
by \eq{eq:kroll-wada} and is proportional to $1/m$ for $\pt \gg \mee$.

\begin{figure}
  \centering
  \includegraphics[width=0.9\textwidth]{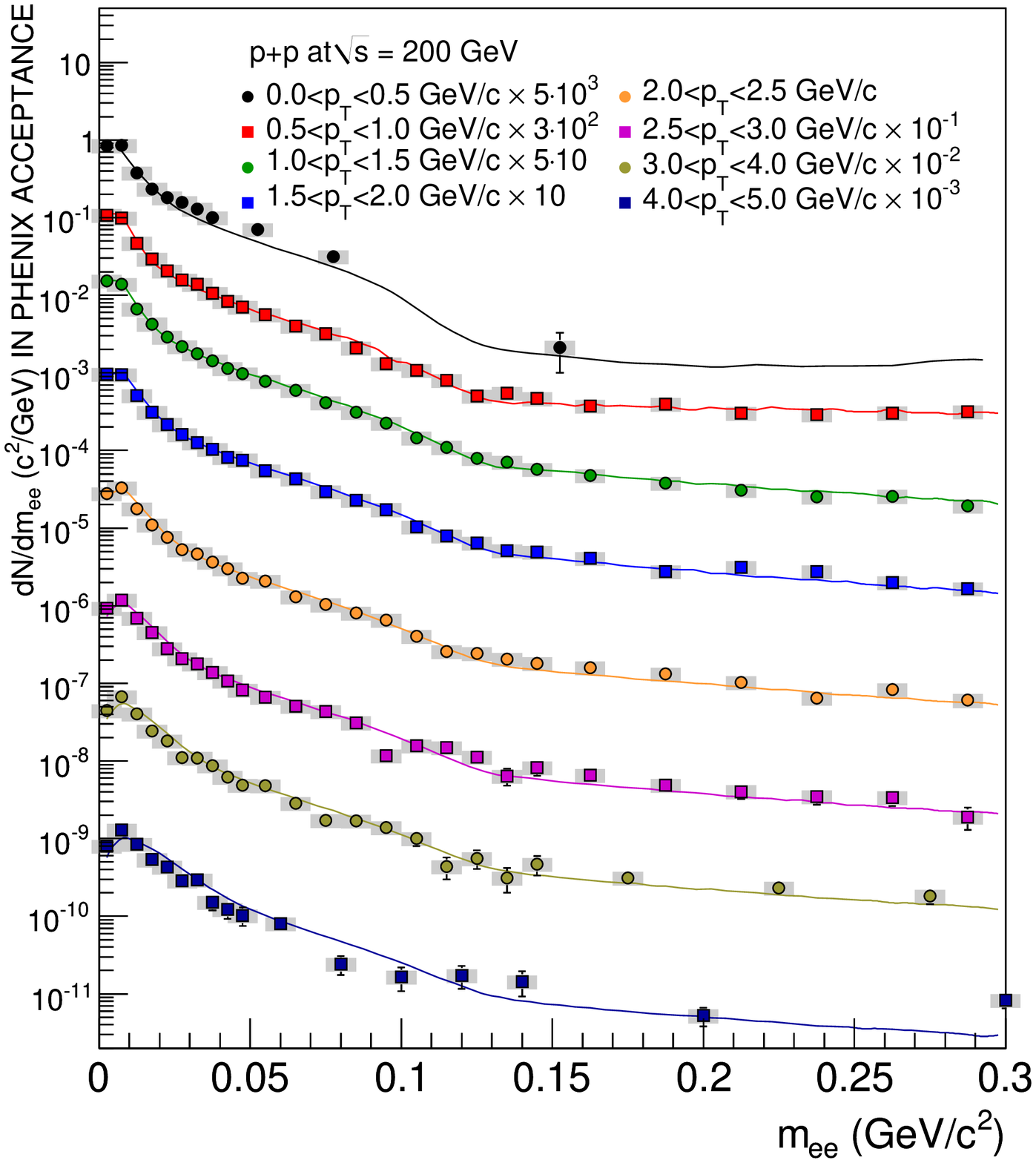}
  \caption[Invariant mass for different \pt ranges in \pp]{The \ee
    pair invariant mass distributions in \pp collisions. The \pt
    ranges are shown in the legend.  The solid curves represent an
    estimate of hadronic sources. This is a zoomed version of
    \fig{fig:pp_mass_varpt}.}
  \label{fig:pp_mass_varpt_zoom}
\end{figure}

In this Chapter an attempt is made to analyze the small excess
observed at high \pt in the low mass region of the dielectron
continuum, which is shown more detailed again in
\fig{fig:pp_mass_varpt_zoom}, under the assumption that it is solely
due to internal conversions of direct virtual photons. While at zero
mass \ee pairs from hadron decays have the same shape in mass as
internal conversions of direct photons, the suppression due to $S$
when approaching the hadron mass changes the shape of \ee pairs from
hadron decays. Therefore, one can fit, after an initial normalization
to the mass range 0--30~\mevcc, the two expected shapes for \ee pairs
from hadronic decays $f_{\rm cocktail}$ (as shown
in~\fig{fig:pp_pt_varmass}) and from direct photons $f_{\rm direct}$
in the mass range 80--300~\mevcc in which the \pion contribution is
severely suppressed. The only fit parameter is the relative fraction
of direct photons $r$:
\begin{equation}\label{eq:rdirect}
  f(\mee) = (1-r) f_{\rm cocktail}(\mee) + r f_{\rm direct}(\mee).
\end{equation}
\begin{figure}
  \centering
  \includegraphics[width=0.9\textwidth]{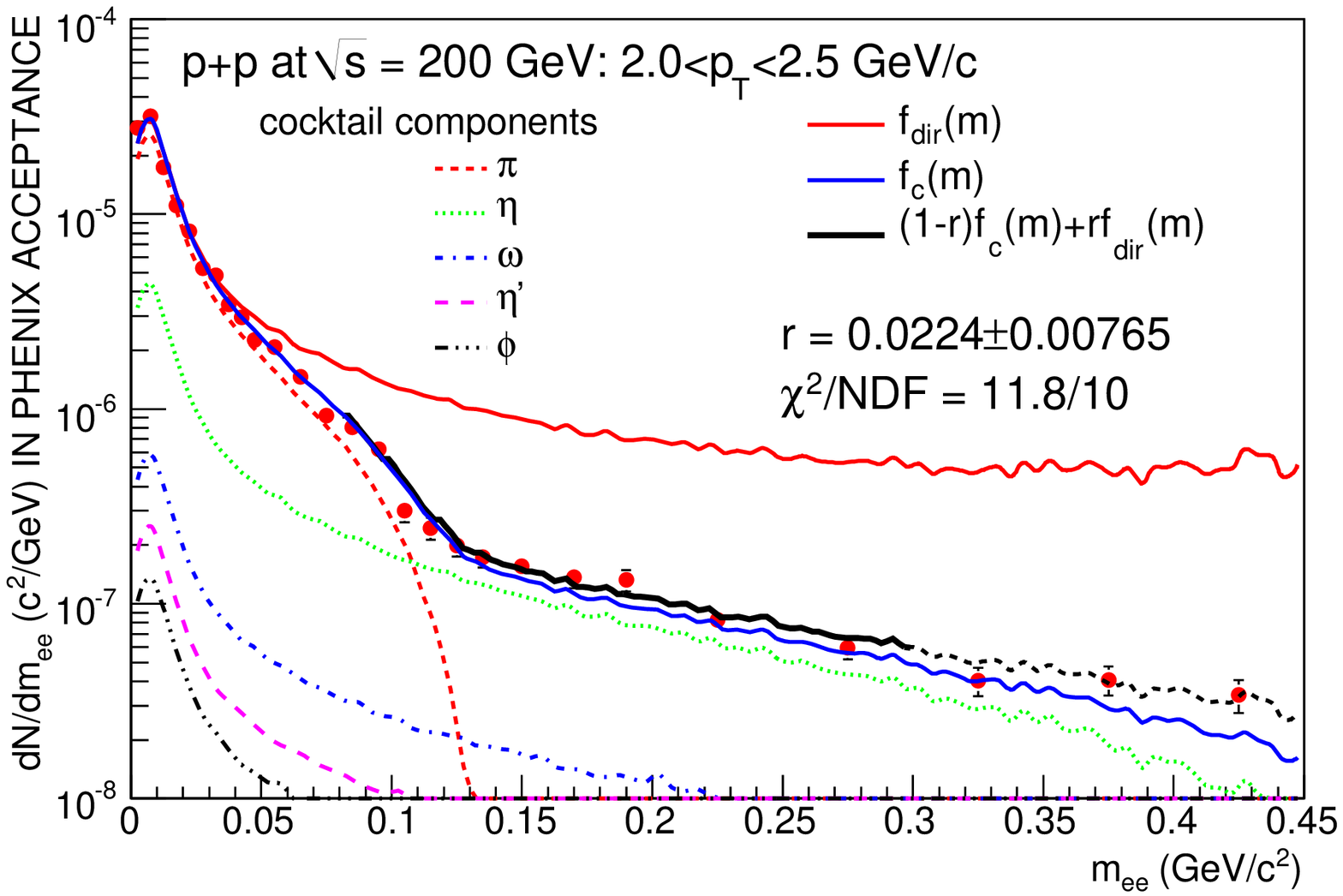}
  \caption[Fit of internal conversion shape to data in \pp]{Invariant
    mass distribution of \ee pairs in \pp collisions with $2.0 < \pt <
    2.5$~\gevc. The fit with \eq{eq:rdirect}in the range $80 < \mee <
    300$~\mevcc is explained in the text. The {\em black dashed} curve
    shows $f(\mee)$ outside of the fit range.}
  \label{fig:pp_mass_fit}
\end{figure}

As an example, the fit result is shown in~\fig{fig:pp_mass_fit} for
\ee pairs with $2.0 < \pt < 2.5$~\gevc in \pp collisions. The quality
of the fit result (\eg, $\chi^2/NDF = 11.8/10$ for the lowest \pt bin)
indeed justifies the assumption that the observed excess is due to
internal conversion of virtual photons. For each \pt bin, the fit is
performed for several mass ranges $m_{\rm low} < \mee <
300$~\mevcc. It is assumed that the form factor $F(\mee^2)$ for direct
photons is 1 independent of mass. This may not be the case for direct
photons from parton fragmentation or, in case of heavy ion collisions,
direct photon radiation from a hadron gas. If the form factor of
direct photons is arbitrarily set to the one of the $\eta$ meson, the
fraction of direct photons $r$ would decrease by $\simeq 10\%$.  The
mass region above 300~\mevcc is not used in the fits to avoid possible
effects due to the form factor and to stay well within the region of
$\mee \ll \pt$, in which the approximation $S=1$ is valid.

The largest uncertainty in the fit is the particle composition of the
cocktail, in particular the $\eta/\pion$ ratio, which is $0.48 \pm
0.03$ at high \pt, based on PHENIX
measurements~\cite{adler:202301,adler:024909}. This corresponds to a
7\% uncertainty in the \pp cocktail for the mass range $100 < \mee <
300$~\mevcc. Other sources cause only a few percent uncertainty in
this mass range, once the normalization of the cocktail is fixed to
the mass range $\mee < 30$~\mevcc. As the $\eta$ meson is the dominant
hadronic source in this mass range the data has been fit with three
components $f(\mee) = (1-r-r_{\eta}) f_{\rm cocktail}(\mee) + r f_{\rm
  direct}(\mee) + r_{\eta} f_{\eta}(\mee)$, with a constrain on
$r_{\eta}$ such that the $\eta/\pion$ range is not changed beyond
$0.48 \pm 0.03$. This alternative method results in values for $r$
which are consistent within the statistical uncertainties.

The fraction of direct photons $r$ is shown in the left panel of
\fig{fig:rdirect} for and compared to a NLO pQCD
prediction~\cite{PhysRevD.48.3136,vogelsang}. The curve shows the
predicted direct photon cross section $d\sigma_{\gamma}^{\rm
  NLO}(\pt)$ divided by the inclusive photon cross section
$d\sigma_{\gamma}^{\rm incl}$. The three curves which are shown
correspond to three different choices of the momentum scale $\mu =
0.5\pt,\, 1.0\pt,\, 2.0\pt$. For a perturbative treatment, the covered
\pt range of $1 < \pt \leq 5$~\gevc is very low, and the theoretical
uncertainties are quite large. Nevertheless, the NLO calculation seems
in quite good agreement with the data.

The fraction of direct photons $r$ can be converted into a cross
section of direct photons by multiplying with the inclusive photon
spectrum. The inclusive photon cross section is determined for each
\pt bin by $d\sigma_{\gamma}^{\rm incl} = \sigma_{pp}^{\rm inel}\,
dN_{ee}^{\rm data} \times dN_{\gamma}^{\rm cocktail}/dN_{ee}^{\rm
  cocktail}$, where $dN_{ee}^{\rm data}$ and $dN_{ee}^{\rm cocktail}$
are the yields of \epairs in $m < 30$~\mevcc for data and cocktail,
respectively, and $dN_{\gamma}^{\rm cocktail}$ is the yield of photons
from the cocktail. The resulting direct photon spectrum for \pp is
shown in \fig{fig:directxsec}. The pQCD calculation is consistent
with the \pp data for $\pt > 2$~\gevc within theoretical
uncertainties. A similar good agreement is observed for \pion
production in \pp~\cite{adare:051106}. The \pp data can be well
described by a modified power-law function:
\begin{equation}\label{eq:pp_directg_fit}
  E \frac{d\sigma^2}{dp^3} = A_{pp}\, \left(1 + \frac{\pt^2}{p_0}\right)^{-n}
\end{equation}
as shown by the dashed line in \fig{fig:directxsec}.

In summary the dielectron continuum in \pp collisions at \sqrts = 200
GeV can be fully understood by known hadronic sources, correlated
semi-leptonic decays of $D$ and $B$ mesons and direct photons produced
in hard-scattering processes of the incoming partons. The measurement
of ``quasi-real'' direct photons in a mass region in which \pion
Dalitz decays are suppressed has allowed the first measurement of
direct photons at low \pt. These results will be crucial for the
understanding and interpretation of the observed dielectron yields in
\AuAu collisions.

\section[Comparison to \AuAu Collisions]{Comparison to $\boldsymbol{\rm Au + Au}$ Collisions}
\label{sec:auau_comparison}

\begin{figure}[p]
  \centering
  \includegraphics[width=0.9\textwidth]{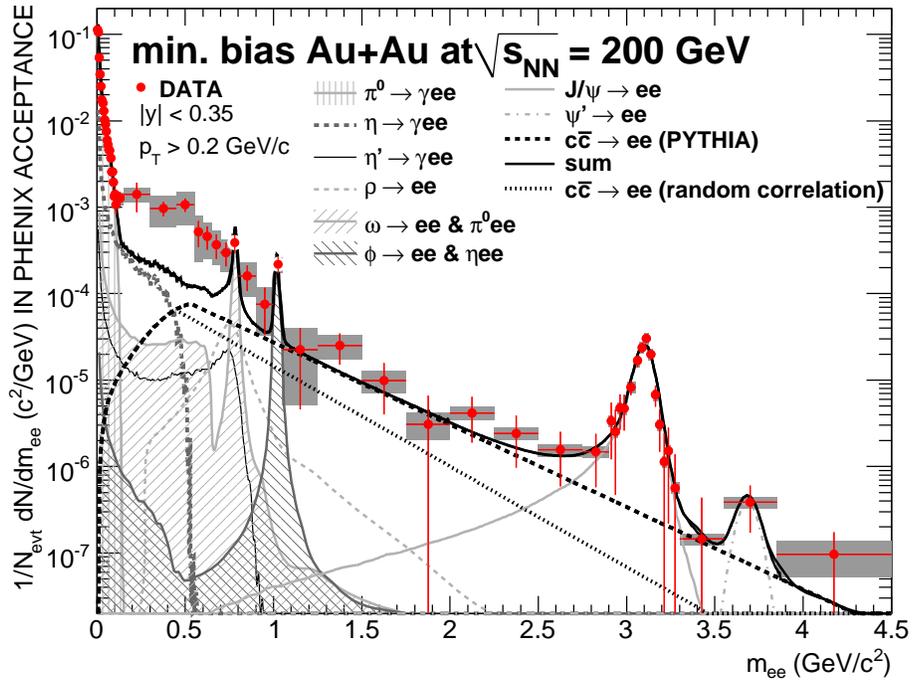}
  \caption[Invariant mass spectrum of \ee pairs in min. bias \AuAu
  collisions]{Electron-positron pair yield as function of pair
    mass. Data show statistical (bars) and systematic (shades) errors
    separately. The data are compared to a expected sources from the
    decays of light hadron and correlated charm
    decays~\cite{afanasiev:2007xw}. The charm contribution expected if
    the dynamical correlation of the $c\overline{c}$ pair is broken, is
    shown separately. The systematic uncertainty on the cocktail is
    not shown.}
  \label{fig:au_cocktailcomp}
\end{figure}

The yield \ee pairs in the PHENIX acceptance measured in min. bias
\AuAu collisions at \sqrtsnn = 200 GeV is shown in
\fig{fig:au_cocktailcomp} as function of invariant mass. The yield is
compared to the expected sources from hadron decays from \exodus and
open charm from \pythia. The open charm contribution is scaled to the
measured charm cross section in \pp ($\sigma_{c\overline{c}} = 567 \pm
57 ({\rm stat.}) \pm 193 ({\rm syst.})~\mu{\rm
  b}$)~\cite{adare:252002} multiplied by the average number of binary
collisions: $\langle N_{\rm coll} \rangle = 257.8 \pm 25.4$ (see
\tab{tab:glauber}).

The data are well described by the cocktail of hadron decays in the
mass region below 150~\mevcc. Due to the larger background to signal
ratio the resonance peaks of $\omega$ and $\phi$ are not as well
measured as in \pp but within the systematic uncertainties in
agreement with the expected yield. However, the yield of \ee pairs in
the mass range $150 < \mee < 750$~\mevcc is significantly enhanced
above the expected cocktail yield. The enhancement factor in this mass
range, defined by the ratio of the measured yield to the expected
yield from hadronic sources, is $4.0 \pm 0.3 {\rm (stat.)} \pm 1.5
{\rm (syst.)} \pm 0.8 {\rm (model)}$. Here the first uncertainty is
the statistical error, the second the systematic uncertainty on the
data, and the last error the uncertainty on the cocktail. The last two
uncertainties are very conservative estimates and not equivalent to
one standard deviation ($1\sigma$).

\begin{figure}
  \centering
  \subfloat[]{\label{fig:pp_au_mass}\includegraphics[width=0.44\textwidth]{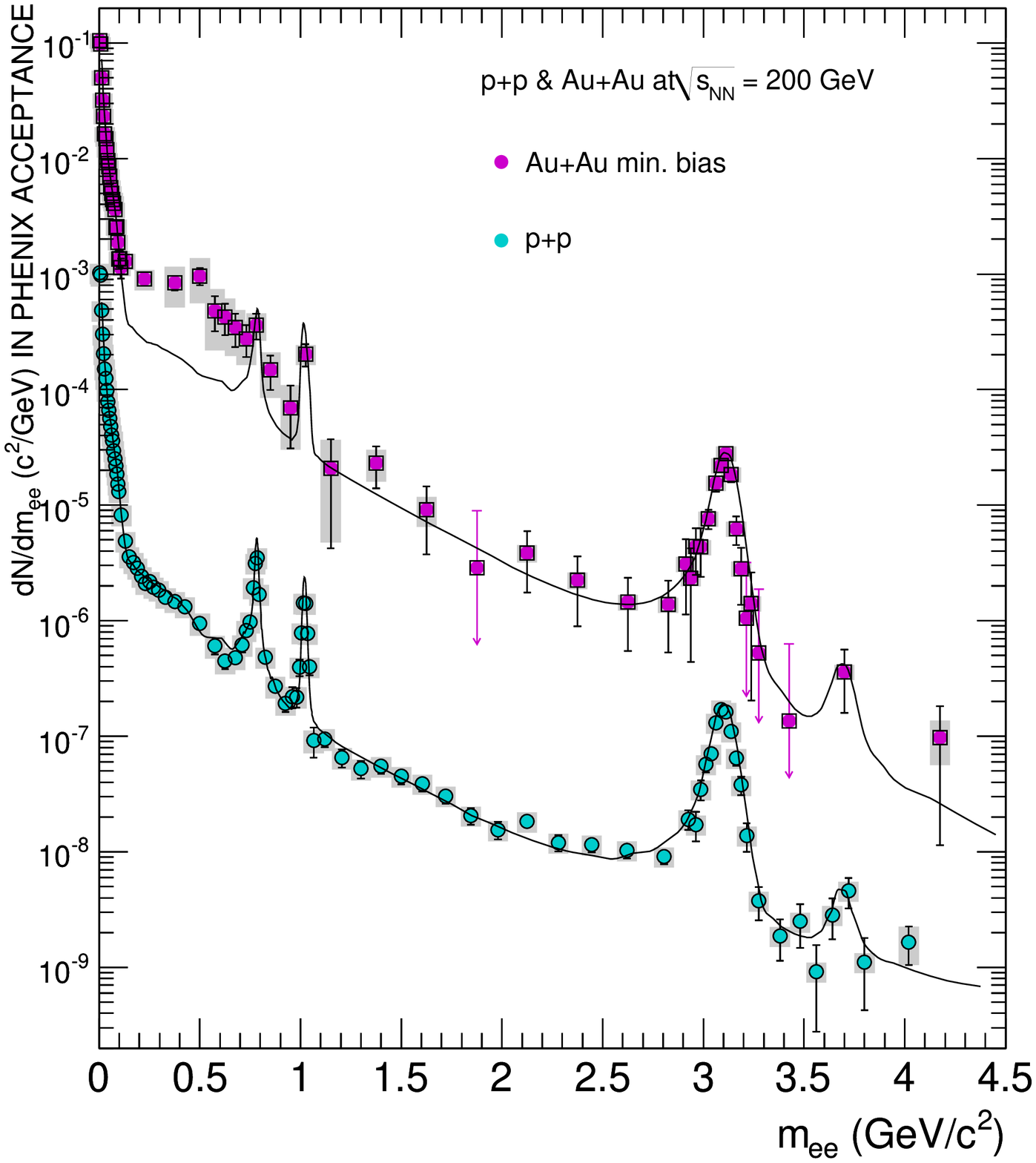}}
  \subfloat[]{\label{fig:au_mass_varcent}\includegraphics[width=0.44\textwidth]{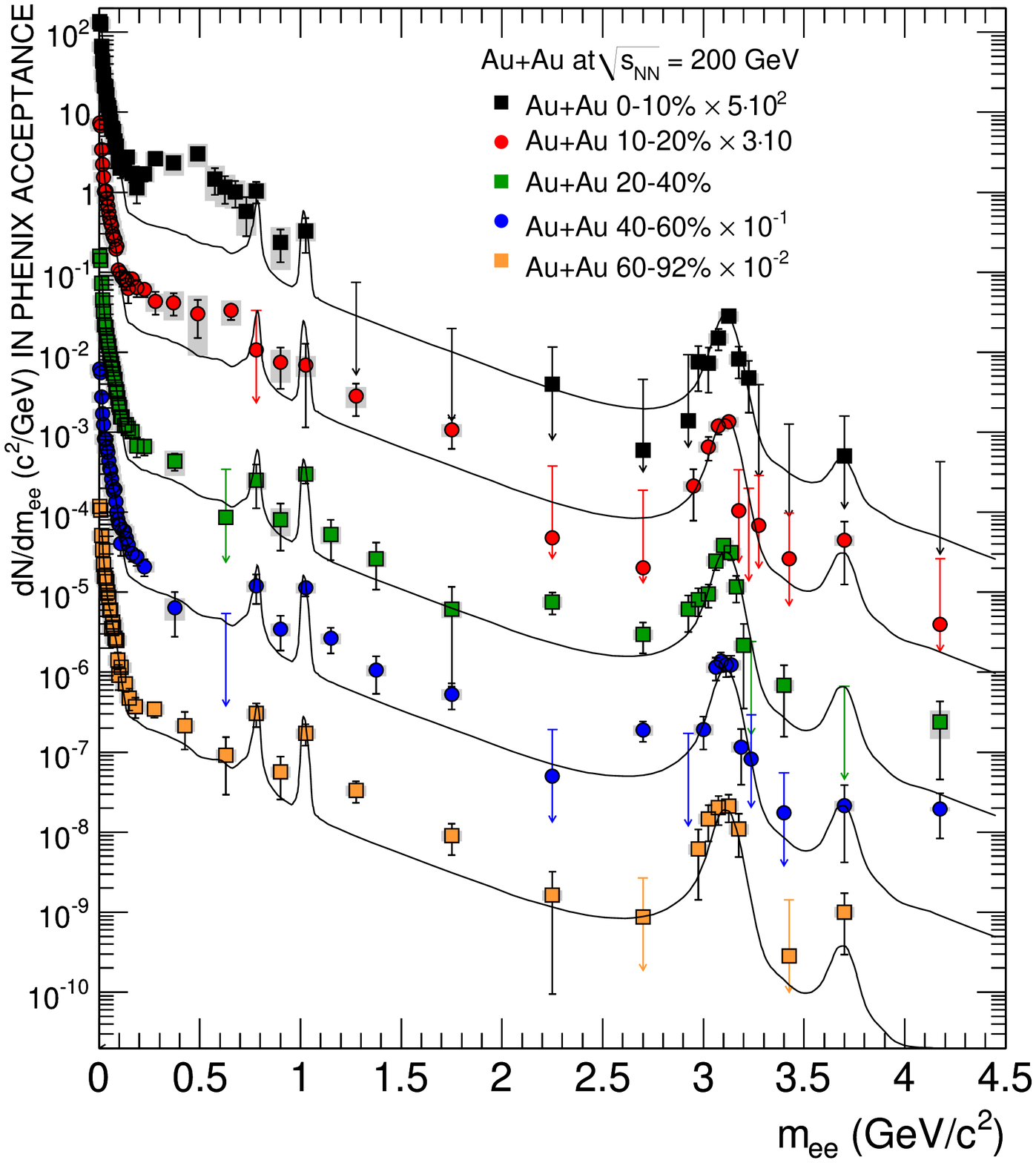}}
  \caption[Invariant mass spectrum of \ee pairs in \pp and various
  centrality bins of \AuAu collisions]{\subref{fig:pp_au_mass} shows
    the invariant mass spectrum in \pp and min. bias \AuAu
    collisions. \subref{fig:au_mass_varcent} shows the invariant mass
    spectra of five centrality classes in \AuAu collisions. The data
    are shown with statistical (bars) and systematic (shades) errors
    separately. The data are compared to the expected sources from the
    decays of light hadron calculated with \exodus and correlated
    charm decays based on \pythia.}
  \label{fig:pp_au_mass_varcent}
\end{figure}

In a direct comparison to the measured dielectron continuum in \pp the
result of min. bias \AuAu is shown in \fig{fig:pp_au_mass}. The
invariant mass spectra of \AuAu collisions in five centrality bins
(0--10\%, 10--20\%, 20--40\%, 40--60\% and 60--92\%, see
\tab{tab:glauber}) are shown in \fig{fig:au_mass_varcent} and compared
to the individual cocktail expectations. The low mass enhancement is
concentrated in the 20\% most central collisions and becomes less
significant in more peripheral ones. The yield in the IMR is
consistent with the \pythia calculation for all centrality bins. In
the following the centrality dependence of the IMR and LMR are further
investigated.

\subsection{The Intermediate Mass Region}
\label{sec:imr}

The intermediate mass region which served in the \pp analysis as a
window to extract the charm cross section is well described by the
\pythia calculation of open charm pairs. This result is somewhat
surprising as the measurement of single electron from semi-leptonic
charm decays shows a strong modification of charm, \ie, a suppression
of high \pt electrons and a significant elliptic
flow~\cite{adare:252002}. Therefore, one would expect this
modification to also influence the pair properties. An extreme
scenario is shown in addition to the \pythia curve as dotted line in
\fig{fig:au_cocktailcomp} assuming the initial correlation of the
$c\overline{c}$ pair is completely broken, \ee pairs are generated
with a single \pt distribution following the measured spectrum of
single electrons from semi-leptonic heavy-flavor
decays~\cite{adare:252002}, but a random azimuthal opening angle. This
distribution is much softer, than the \pythia curve, which would leave
room to other contributions such as thermal radiation via
$q\overline{q}$ annihilation.

\begin{figure}
  \centering
  \includegraphics[width=0.9\textwidth]{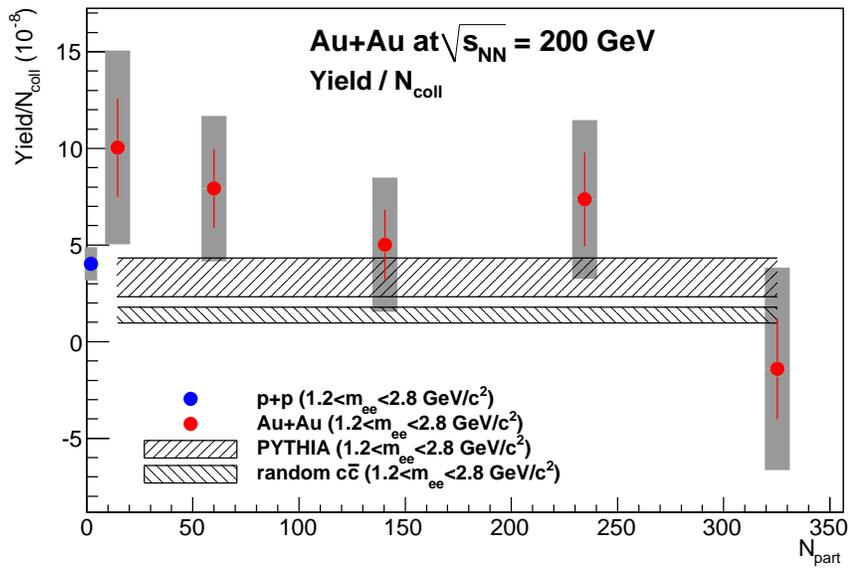}
  \caption[IMR yield per $N_{\rm coll}$ as function of $N_{\rm
    part}$]{Dielectron yield per binary collision as function of
    $N_{\rm part}$ for the mass range $1.2 < \mee <
    2.8$~\gevcc. Statistical and systematic uncertainties are shown
    separately. Also shown are two bands corresponding to two
    different estimates of the contribution from charmed meson
    decays. The width of the bands reflect the uncertainties on the
    charm cross section only.}
  \label{fig:imr_cent}
\end{figure}

If the yield in the IMR is dominated by open charm, it is expected to
increase proportional to the number of binary collisions. The yield
per number of binary collision $N_{\rm coll}$ in the mass range $1.2 <
\mee < 2.8$~\gevcc is shown as function of $N_{\rm part}$ in
\fig{fig:imr_cent}. The data show no significant centrality dependence
and are consistent with the expected yield calculated with
\pythia. But the apparent scaling with the number of binary collisions
may be a coincidence of two counteracting effects: (i) the suppression
of \ee pairs from open charm in the IMR due to modifications of charm
which increases with $N_{\rm part}$ and (ii) an additional
contribution due to thermal radiation from $q\overline{q}$
annihilation which is expected to increase faster than proportional to
$N_{part}$. As discussed in Section~\ref{sec:na60}, such a coincidence
may have been observed at the SPS~\cite{Abreu:2000nj}, where a prompt
component has been suggested by NA60~\cite{0954-3899-34-8-S149}.

\subsection{The Low Mass Enhancement}
\label{sec:low_mass_enhancement}

\begin{figure}
  \centering
  \includegraphics[width=0.9\textwidth]{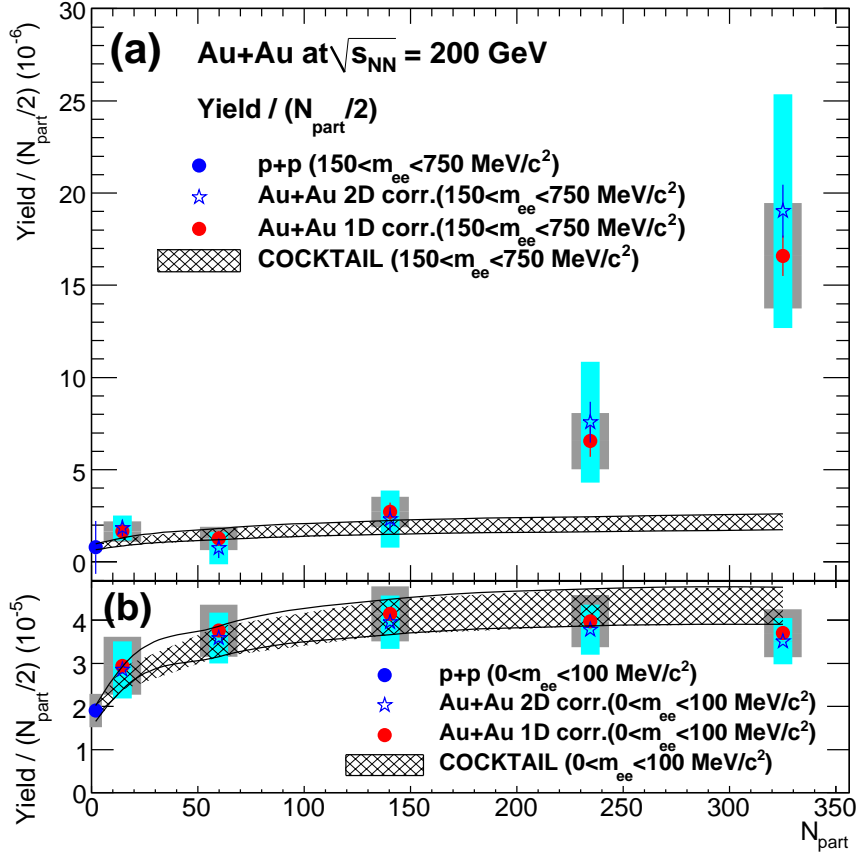}
  \caption[LMR yield per $N_{\rm part}/2$ as function of $N_{\rm
    part}$]{Dielectron yield per participating nucleon pair as
    function of $N_{\rm part}$ for the mass range $150 < \mee <
    750$~\mevcc (a) and $\mee < 100$~\mevcc (b). Statistical and
    systematic uncertainties are shown separately. Also shown are the
    expected cocktail yields. The band reflects the systematic
    uncertainty on the cocktail}
  \label{fig:lmr_cent}
\end{figure}

The centrality dependence of the yield in the low mass region has been
studied more quantitatively. The first range, $\mee < 100$~\mevcc, is
dominated by \pion Dalitz decays. As the \pion multiplicity roughly
scales with the number of participating nucleons, the yield in this
range should scale linearly with $N_{\rm part}$. \fig{fig:lmr_cent}
shows the yield in $\mee < 100$~\mevcc divided by the number of
participating nucleon pairs ($N_{\rm part}/2$) as function of $N_{\rm
  part}$. The yields in all centrality bins as well as \pp are in good
agreement with the cocktail, and show the expected scaling.

The top panel of \fig{fig:lmr_cent} shows the yield per participating
nucleon pair for the mass region $150 < \mee < 750$~\mevcc, in which
the enhancement is observed. While the agreement with the cocktail is
good for \pp and the peripheral collisions, an increase of the yield
is observed in the data above $N_{\rm part} \approx 200$ that is
stronger than linear. For the most central bin, the yield is a factor
$7.6 \pm 0.5 {\rm (stat.)} \pm 1.5 {\rm (syst.)} \pm 1.5 {\rm
  (model)}$ above the cocktail. This stronger than linear rise of the
yield is qualitative understood with the notion of $q\overline{q}$ or
$\pi\pi$ annihilation as source of the enhancement.

\subsection{Direct Photons}
\label{sec:auau_dir_photons}

\begin{figure}
  \centering
  \includegraphics[width=0.9\textwidth]{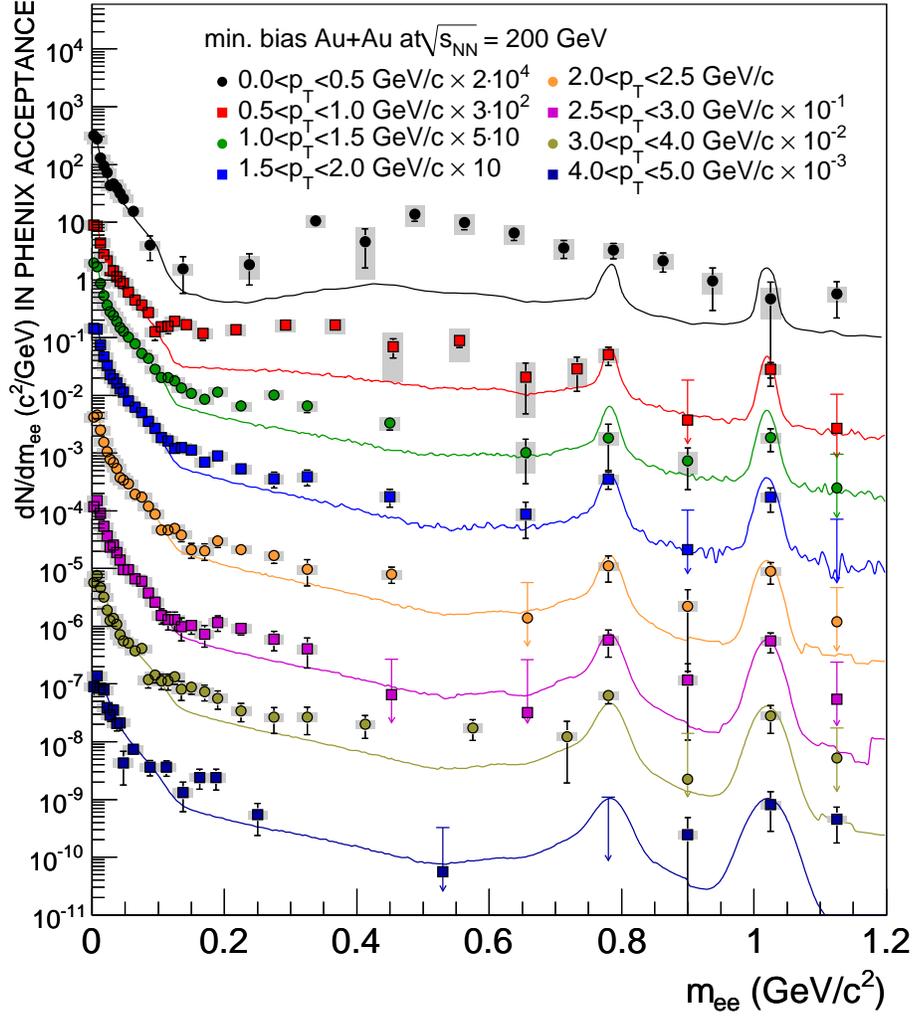}
  \caption[Invariant mass for different \pt ranges in \AuAu]{The \ee
    pair invariant mass distributions in \AuAu collisions. The \pt
    ranges are shown in the legend.  The solid curves represent an
    estimate of hadronic sources.}
  \label{fig:au_mass_varpt}
\end{figure}

\begin{figure}
  \centering
  \includegraphics[width=0.9\textwidth]{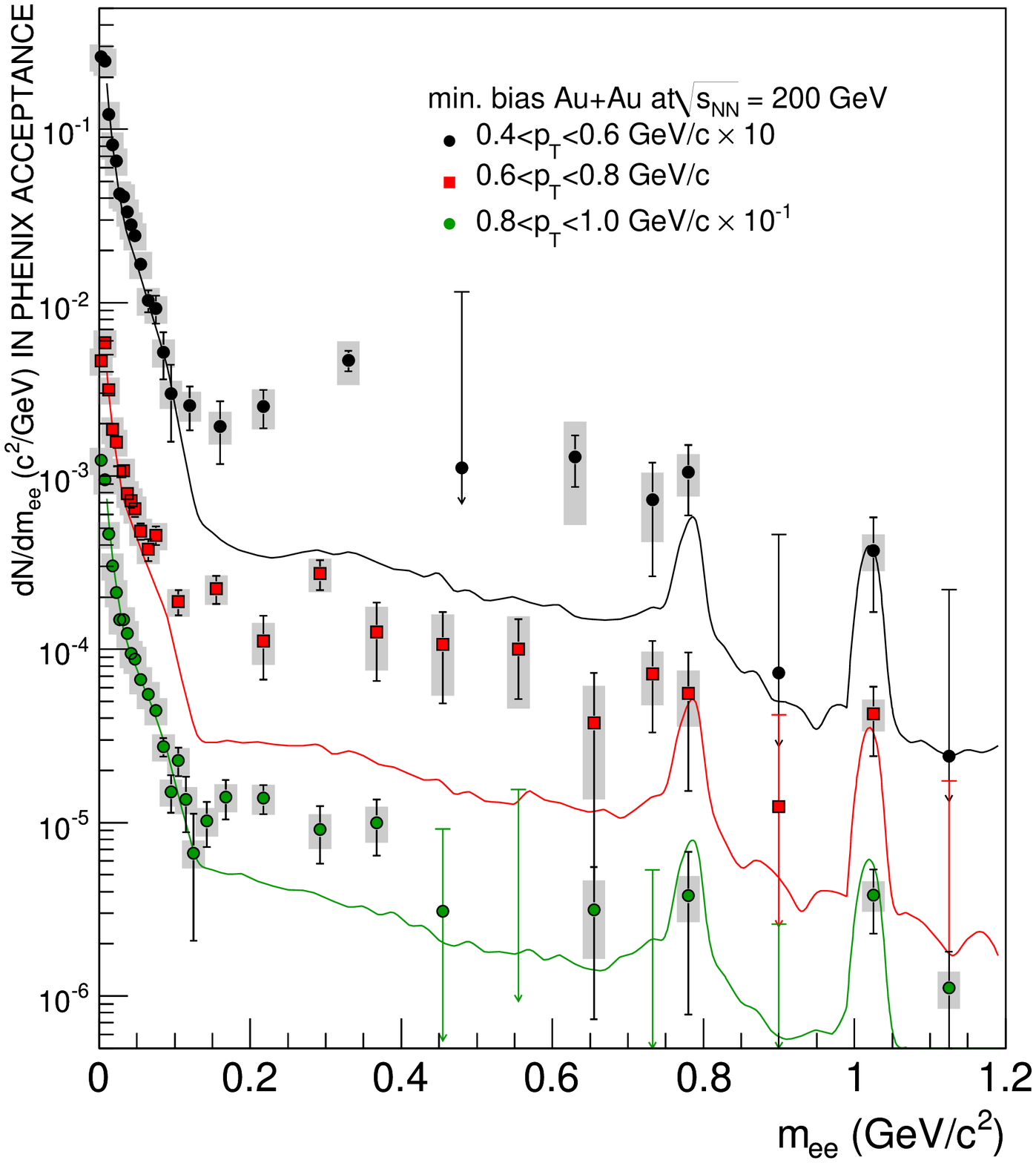}
  \caption[Invariant mass at low \pt in \AuAu]{The \ee pair invariant
    mass distributions in \AuAu collisions for the low \pt range. The
    solid curves represent an estimate of hadronic sources.}
  \label{fig:au_mass_lowpt}
\end{figure}

In this chapter it is tested whether the excess observed in the low
mass region is compatible with a source of virtual direct photons. The
analysis has been performed with the same method as for \pp which has
been discussed in
Section~\ref{sec:pp_dir_photons}. \fig{fig:au_mass_varpt} shows the
yield of \ee pairs as function mass below 1.2~\gevcc for various \pt
ranges in min. bias \AuAu collisions. The low mass enhancement is
concentrated at low transverse momenta: $\pt < 1$~\gevc. For $\pt >
1$~\gevc the enhancement, although reduced, is still quite significant
(a factor $\approx 2$ above the expected yield), but is quite
different in shape. The low \pt mass spectrum divided in finer \pt
bins as shown in \fig{fig:au_mass_lowpt}. While the low \pt
enhancement has a ``bump'' like structure, the shape at higher \pt is
consistent with the $1/\mee$ shape expected from internal conversions
of direct virtual photons.

\begin{figure}
  \centering
  \includegraphics[width=0.9\textwidth]{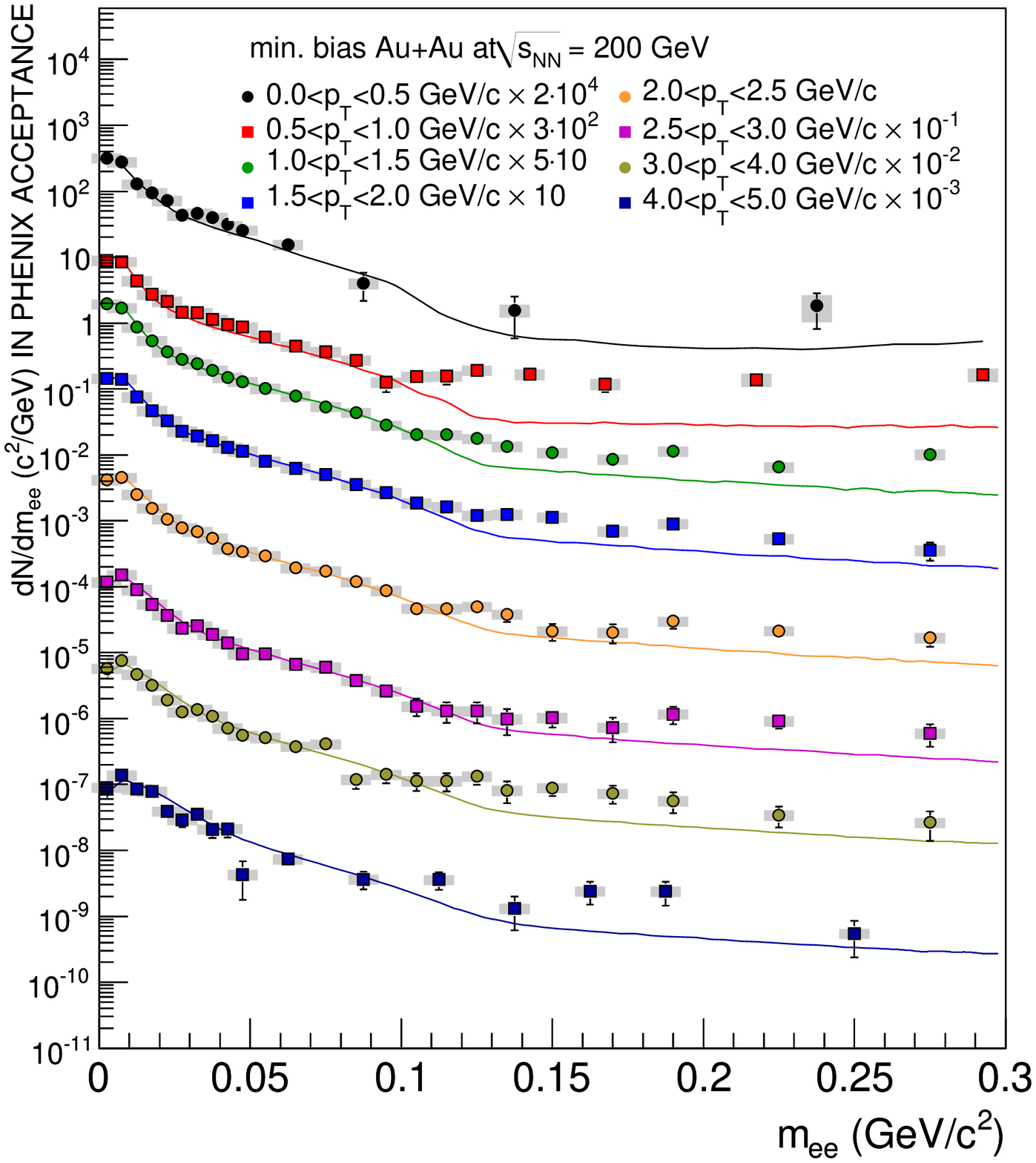}
  \caption[Invariant mass for different \pt ranges in \AuAu]{The \ee
    pair invariant mass distributions in \AuAu collisions. The \pt
    ranges are shown in the legend.  The solid curves represent an
    estimate of hadronic sources. This is a zoomed version of
    \fig{fig:au_mass_varpt}.}
  \label{fig:au_mass_varpt_zoom}
\end{figure}

\begin{figure}
  \centering
  \includegraphics[width=0.9\textwidth]{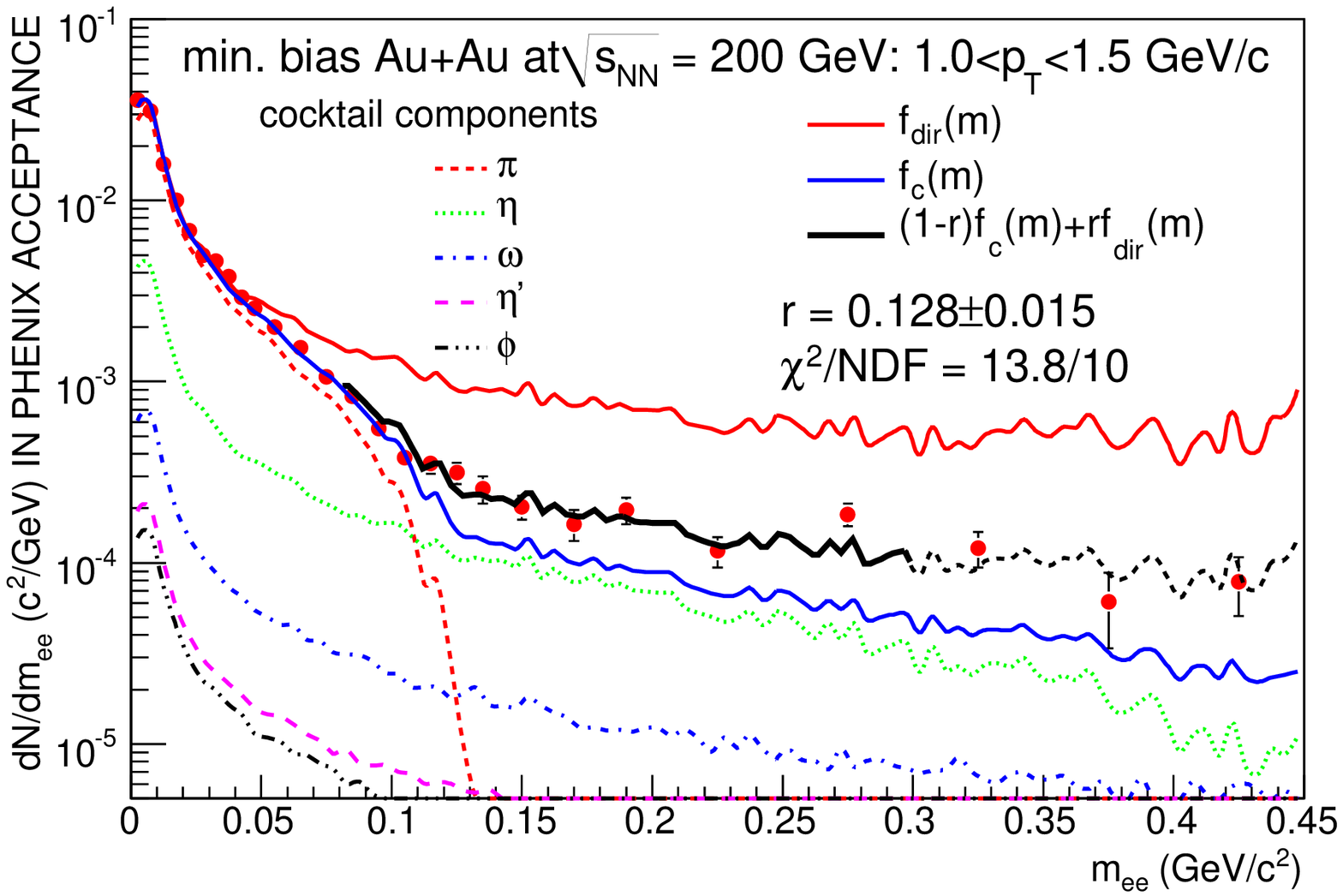}
  \caption[Fit of internal conversion shape to data in
  \AuAu]{Invariant mass distribution of \ee pairs in \AuAu collisions
    with $1.0 < \pt < 1.5$~\gevc. The fit with \eq{eq:rdirect}in the
    range $80 < \mee < 300$~\mevcc is explained in the text. The {\em
      black dashed} curve shows $f(\mee)$ outside of the fit range.}
  \label{fig:au_mass_fit}
\end{figure}

Thus the excess of \ee pairs for $\pt > 1$~\gevc is treated as
entirely due to internal conversions and as in \pp the mass spectra
are fitted with the sum of cocktail and a direct photon component with
a $1/\mee$ dependence. An example for $1.0 < \pt < 1.5$~\gevc is
shown in \fig{fig:au_mass_fit}. The uncertainty on the $\eta/\pion$
ratio is a bit larger than in \pp, but the same central value is
assumed: $0.48 \pm 0.08$, which is the more conservative approach,
since the actual PHENIX measurement of the ratio \AuAu is $\approx
0.4$ with an uncertainty of
11\%~\cite{adler:202301,adler:024909}. This leads to a 17\%
uncertainty due to the $\eta$ contribution in the mass range $80 <
\mee < 300$~\mevcc. As in \pp a fit with a variable $\eta$
contribution within a $\eta/\pion$ ratio of $0.48 \pm 0.08$ gives
consistent results within the statistical uncertainties.

\begin{figure}
  \centering
  \includegraphics[width=0.9\textwidth]{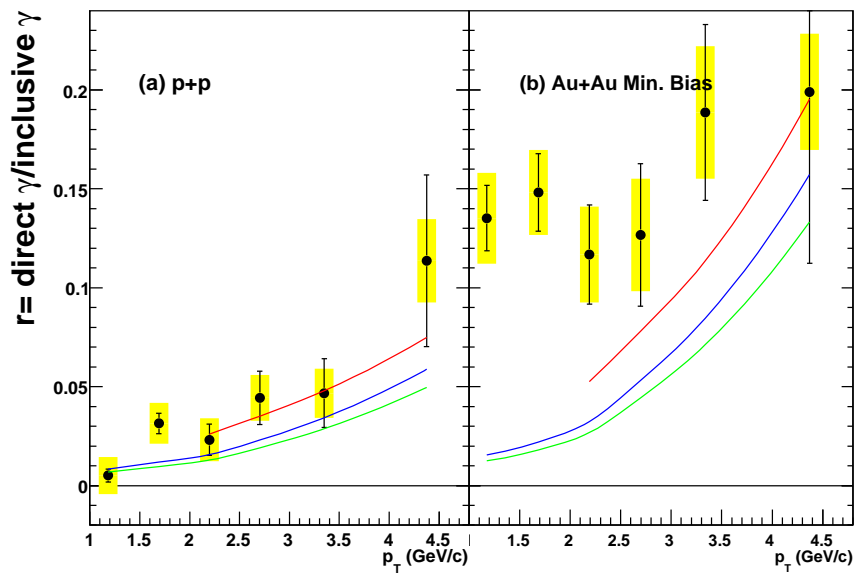}
  \caption[Fraction of direct photons $r$ in \pp and \AuAu]{The
    fraction of direct photons as function of \pt in (a) \pp and (b)
    \AuAu collisions. The error bars and the error band represent
    statistical and systematic uncertainties, respectively. The curves
    show the direct photon prediction of a NLO pQCD calculation.}
  \label{fig:rdirect}
\end{figure}

The fraction of direct photons $r$ in min. bias \AuAu is shown as
function of \pt in the right panel of \fig{fig:rdirect}.  It is
compared to the NLO pQCD prediction of direct photons for \pp divided
by the inclusive photon yield and scaled by the nuclear overlap factor
$T_{AA}$ ($T_{AA}\, d\sigma_{\gamma}^{\rm NLO}(\pt)/dN_{\gamma}^{\rm
  incl}(\pt)$), which is also shown for the three theory scales $\mu =
0.5\pt,\, 1.0\pt,\, 2.0\pt$. While the fraction of direct photons
measured in \pp was consistent with the NLO prediction, the fraction
measured in \AuAu is larger than than the calculation for $\pt <
4$~\gevc.

\begin{figure}
  \centering
  \includegraphics[width=0.9\textwidth]{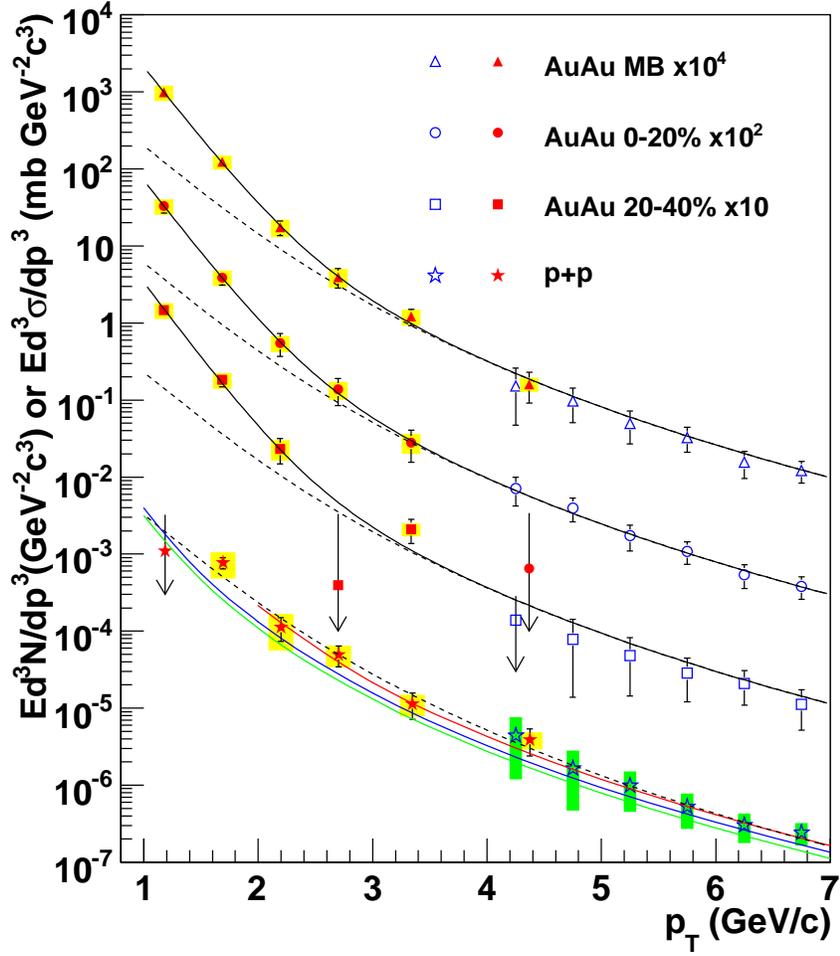}
  \caption[Direct Photon Spectrum in \AuAu and \pp
  collisions]{Invariant cross section (\pp) and invariant yield
    (\AuAu) of direct photons as function of \pt. The {\em filled}
    point are from this analysis and the {\em open} points are
    from~\cite{adler:232301,adler:012002}. The three curves on the \pp
    data represent the NLO pQCD calculation of direct photons, the
    {\em dashed} curves show a power-law fit to the \pp data, scaled
    by $T_{AA}$. The {\em solid black} curves are exponential plus
    $T_{AA}$ scaled \pp fit.}
  \label{fig:directxsec}
\end{figure}

As for \pp, the fraction of direct photons $r$ is converted into a
direct photon yield by multiplying with the inclusive photon
yield. The inclusive photon yield is determined for each \pt bin by
$dN_{\gamma}^{\rm incl} = dN_{ee}^{\rm data} \times dN_{\gamma}^{\rm
  cocktail}/dN_{ee}^{\rm cocktail}$, where $dN_{ee}^{\rm data}$ and
$dN_{ee}^{\rm cocktail}$ are the yields of \epairs in $m < 30$~\mevcc
for data and cocktail, respectively, and $dN_{\gamma}^{\rm cocktail}$
is the yield of photons from the cocktail. The resulting direct photon
spectrum for 0--20\%, 20--40\% and minimum bias \AuAu collisions is
shown in \fig{fig:directxsec} together with the direct photon cross
section measured in \pp collisions. At high \pt the direct photon
spectrum is extended by the EMCal measurement of direct
photons~\cite{adler:232301,adler:012002}, which show good agreement
with the NLO calculation. It is worth to point out the significant
improvement of the systematic uncertainties by the internal
conversion method with respect to the EMCal measurement

The measured direct photon yield in \AuAu is above the \pp fit to
\eq{eq:pp_directg_fit} scaled by $T_{AA}$ for $\pt < 2.5$~\gevc, which
indicates the the direct photon yield at low \pt increases faster than
the binary scaled \pp cross section. To quantify this excess, an
exponential plus a modified power-law are fit to the \AuAu data. The
power-law is fixed to the $T_{AA}$ scaled \pp fit:
\begin{equation}\label{eq:au_directg_fit}
  \frac{1}{2\pi \pt} \frac{d^2N}{d\pt\,dy} = A\, \eexp{-\pt/T} + T_{AA} \times A_{pp}\, \left(1 + \frac{\pt^2}{p_0}\right)^{-n}
\end{equation}
The inverse slope of the exponential $T$ and the $A$ are the only free
parameters. The systematic uncertainties of this fit are estimated
by changing the \pp fit parameters within the uncertainties. The
results are summarized in \tab{tab:au_directg_fit}, for which the fit
result of $A$ has been converted into a rapidity density
$dN/dy(\pt>1~{\rm GeV}/c)$. For the most central collision an inverse
slope of $T = 221 \pm 23 \pm 18$~MeV is found. If the direct photons
in \AuAu collisions are of thermal origin, the inverse slop $T$ is
related to the initial temperature $T_{\rm init}$ of the dense
matter. In thermal models, $T_{\rm init}$ is 1.5--3 times $T$ due to
the space-time evolution~\cite{d'Enterria:2005vz}. Comparisons to a
few models are shown in Section~\ref{sec:pt-model-comp}.

\begin{table}
  \centering
  \caption[Fits to direct photon spectra in \AuAu]{\label{tab:au_directg_fit}Summary of the fits with
    \eq{eq:au_directg_fit} to the direct photon spectra measured in
    \AuAu at \sqrtsnn = 200GeV. The first and second errors are
    statistical and systematical, respectively.\\}
  \begin{tabular}{lr@{.}l@{ $\pm$ }r@{.}l@{ $\pm$ }r@{.}lr@{ $\pm$ }r@{ $\pm$ }rr@{.}l@{/}l}\toprule
    Centrality & \multicolumn{6}{c}{$dN/dy(\pt>1~{\rm GeV}/c$)} & \multicolumn{3}{c}{$T$~(MeV)}  &\multicolumn{3}{c}{$\chi^2/{\rm DOF}$}\\\midrule
    0--20\% & 1&10 & 0&20 & 0&30 & 221 & 23 & 18 & 3&6 & 4\\
    20--40\%& 0&52 & 0&08 & 0&14 & 214 & 20 & 15 & 5&2 & 3\\
    MB      & 0&33 & 0&04 & 0&09 & 224 & 16 & 19 & 0&9 & 4\\\bottomrule
  \end{tabular}
\end{table}

\subsection[\pt of the Low Mass Enhancement]{$\boldsymbol{p_T}$ of the Low Mass Enhancement}
\label{sec:pt-spectra}

As shown in \fig{fig:au_mass_varpt}, the dominant fraction of the low
mass enhancement is localized at low \pt. Clearly, in this region the
approximation $\mee \ll \pt$ does not hold any longer, and
consequently $S \neq 1$. Therefore, the excess in this mass and \pt
range cannot be analyzed rigorously under the assumption of internal
conversions of direct photons as done at high \pt in the previous
chapter.

\begin{figure}
  \centering
  \includegraphics[width=0.9\textwidth]{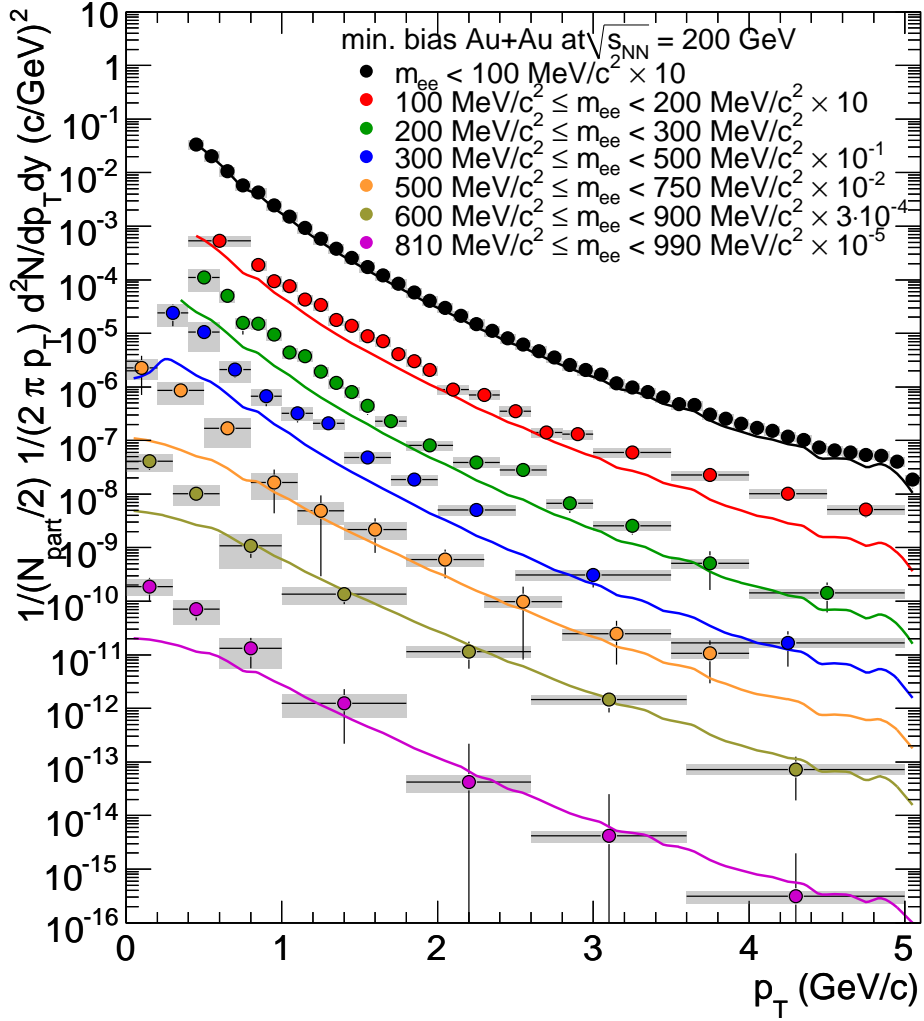}
  \caption[Invariant Yield as function of \pt for \ee pairs in
  different mass ranges in min. bias \AuAu collisions]{Shown is the
    invariant yield of \ee pairs in different mass ranges in min. bias
    \AuAu collisions. The mass ranges are defined in the legend.}
  \label{fig:au_pt_varmass}
\end{figure}

The invariant yield of \ee pairs as function of \pt is calculated as
defined in \eq{eq:invyield} and compared to the cocktail of hadronic
and charmed meson decays for various mass ranges in
\fig{fig:au_pt_varmass}. The systematic uncertainties are dominated by
the background subtraction which contributes as $\sigma_S/S = 0.25\%
B/S$. As the signal to background ratio improves towards high \pt, the
systematic uncertainties decrease. For the mass ranges below
400~\mevcc the data are cut at low \pt due to the single electron \pt
cut at 200~\mevcc.

In the mass region $\mee < 100$~\mevcc the \AuAu data are in agreement
with the cocktail, as expected. In the higher mass bins, the data show
a large excess both at low and high \pt. At high \pt this excess has
been interpreted as direct photons in the previous chapter. The \pt
spectra of direct photons have been successfully described with the
functional form given in \eq{eq:au_directg_fit} an the fit values in
\tab{tab:au_directg_fit}. The yield of direct photons can be converted
into a yield of \ee pairs in a mass range $m_1 < \mee < m_2$ according
to \eq{eq:kroll-wada}:
\begin{align}\label{eq:conv_directg}
  \frac{1}{2\pi \pt}\frac{d^2N_{ee}}{d\pt\,dy} &= \frac{1}{2\pi \pt} \frac{d^2N_{\gamma}}{d\pt\,dy}
  \int_{m_1}^{m_2} \frac{2\alpha}{3\pi}\frac{1}{\mee}
  \sqrt{1-\frac{4m_e^2}{\mee^2}}\left( 1+\frac{2m_e^2}{\mee^2}\right) S\, dm_{ee}\notag\\
  &= A\, \eexp{-\pt/T} + T_{AA} \times A_{pp}\, \left(1 + \frac{\pt^2}{p_0}\right)^{-n} \notag\\
  &\quad\cdot \int_{m_1}^{m_2}\frac{2\alpha}{3\pi}\frac{1}{\mee} \sqrt{1-\frac{4m_e^2}{\mee^2}}\left( 1+\frac{2m_e^2}{\mee^2}\right)
\end{align}

\begin{figure}
  \centering
  \includegraphics[width=0.9\textwidth]{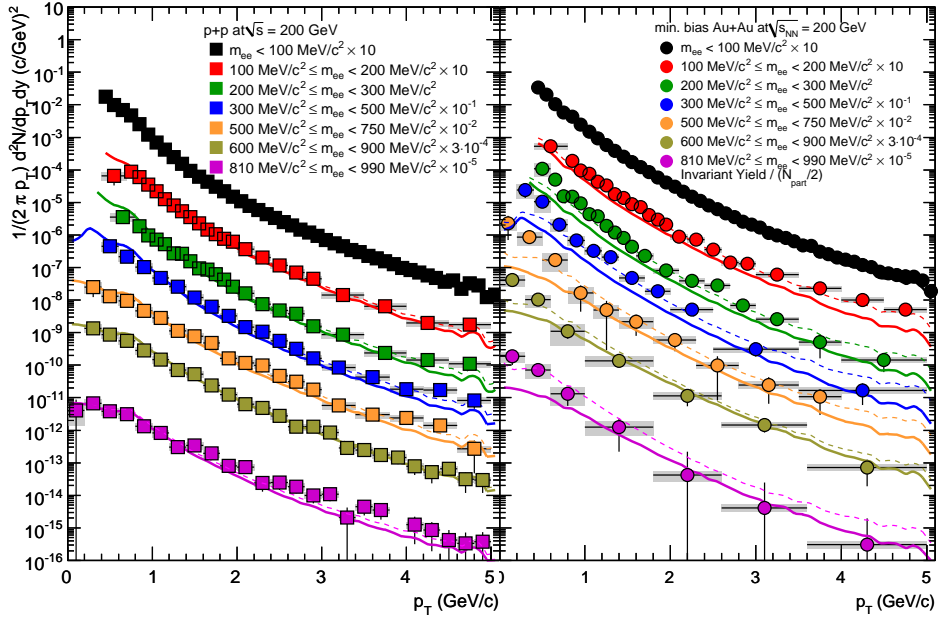}
  \caption[Invariant Yield as function of \pt for \ee pairs in
  different mass ranges for \pp and min. bias \AuAu collisions]{Shown
    is the invariant yield of \ee pairs in different mass ranges in
    \pp (left) min. bias \AuAu collisions (right). The mass ranges are
    defined in the legend. The data are compared to the cocktail from
    hadronic decays ({\em solid line}) and the sum of hadronic
    cocktail and direct photons ({\em dashed line}).}
  \label{fig:pp_au_pt_varmass}
\end{figure}

As previously, the distributions of direct photons and cocktail are
normalized to the data in the mass range $\mee < 30$~\mevcc, where
both have nearly identical shapes. The extension of the direct photon
contribution to lower \pt is done using the suppression factor $S$ for
quark-gluon Compton scattering defined in \eq{eq:qg_sfactor}:
\begin{align*}
  S &= 1 + \frac{2 u}{t^2 + s^2} \mee^2\\
  &= 1 - \frac{2x}{(x + \sqrt{1+x^2}) (3x^2+1+2x\,\sqrt{1+x^2})}\tag{\ref{eq:qg_sfactor}}
\end{align*}
with $x = \pt/\mee$. This contribution is added to the cocktail and
the sum is shown as dashed line in \fig{fig:pp_au_pt_varmass}, which
shows the \pt spectra of \ee pairs in both \pp and \AuAu
collisions. The excess at at $\pt \gtrsim 1$~\gevc with respect to the
hadronic cocktail agrees with the sum of cocktail and direct
photons. For the mass range 100--300~\mevcc this is of course
redundant, as the excess in exactly this region of phase space had
been used to measure this direct photon component, but also above
300~\mevcc the excess at $\pt > 1$~\gevc is in good agreement with a
direct photon contribution.

At low \pt, however, a significant excess beyond the extrapolated
direct photon contribution remains in the \AuAu data. This excess
increases in strength towards low \pt, despite the restriction on the
single electrons to have at least a transverse momentum of $\pt >
0.2$~\gevc to be fall within the PHENIX acceptance. This trend is
quite remarkable and opposite to the behavior of the cocktail
component. It is also reminiscent of the steepening of the \mt spectra
of the dimuon excess yield measured by NA60~\cite{arnaldi:022302}
(also see Section~\ref{sec:na60}). To further study the low \pt
excess, the spectra are combined to a one mass range of $300 < \mee <
750$~\mevcc.

In Ref.~\cite{adler:034909} PHENIX has measured that the differential
cross section of charged hadrons ($\pi^{\pm}$, $K^{\pm}$, $p$, and
$\overline{p}$) in \AuAu collisions at \sqrtsnn = 200 GeV can be
described by an exponential in $\mt - m_0$, where $m_0$ is the
particle mass, and $\mt = \sqrt{\pt^2 + m_0^2}$. This is also known
from lower beam energies for \pp, $p + A$, and $A + A$
collisions. Such an exponential shape of the invariant yield can be
parameterized as:
\begin{equation}\label{eq:mtexpo}
  \frac{1}{2\pi \mt} \frac{d^2N}{d\mt\, dy} = \frac{1}{2\pi T (T_{\rm eff} + m_0)}\,
  A\, \exp\left(-\frac{\mt-m_0}{T_{\rm eff}}\right)
\end{equation}
where $T_{\rm eff}$ denotes the inverse slope or effective
temperature, and $A$ is a normalization parameter which is related to
the rapidity density $dN/dy$. It has been observed that the inverse
slope increases with the particle mass, which is consistent with a
hydrodynamical description of particle emission from a source with
transverse
flow~\cite{PhysRevC.48.2462,PhysRevC.54.1390,PhysRevC.67.034904,PhysRevLett.86.4783,Teaney:2001av}. Under
such conditions the measured effective temperature $T_{\rm eff}$ is
blue shifted with respected to the freeze out temperature $T_{\rm
  fo}$:
\begin{equation}
  T_{\rm eff} = T_{\rm fo} + m \langle \beta_T \rangle^2.
\end{equation}
where $\langle \beta_T \rangle$ is a measure of the strength of the
transverse flow. The inverse slopes measured for kaons are larger then
200~MeV for all centralities. Together with the effective temperatures
measured for charged pions, protons, and anti-protons a centrality
independent freeze out temperature of $\approx 180$~MeV has been
extracted~\cite{adler:034909}.

Following this approach to characterize the low \pt excess, the \pt
spectrum of \ee pairs with $300 < \mee < 750$~\mevcc is fitted with
the sum of hadronic cocktail, direct photons and an \mt exponential as
defined in \eq{eq:mtexpo}. The direct photon contribution to the
cocktail at low \pt is 10\% at \mee = 0, which leads to equal
contribution of cocktail and direct photon in the fitted mass
range. The total contribution of cocktail and direct photons is fixed
by the normalization to the data in $\mee < 30$~\mevcc.

\begin{figure}
  \includegraphics[width=1.0\linewidth]{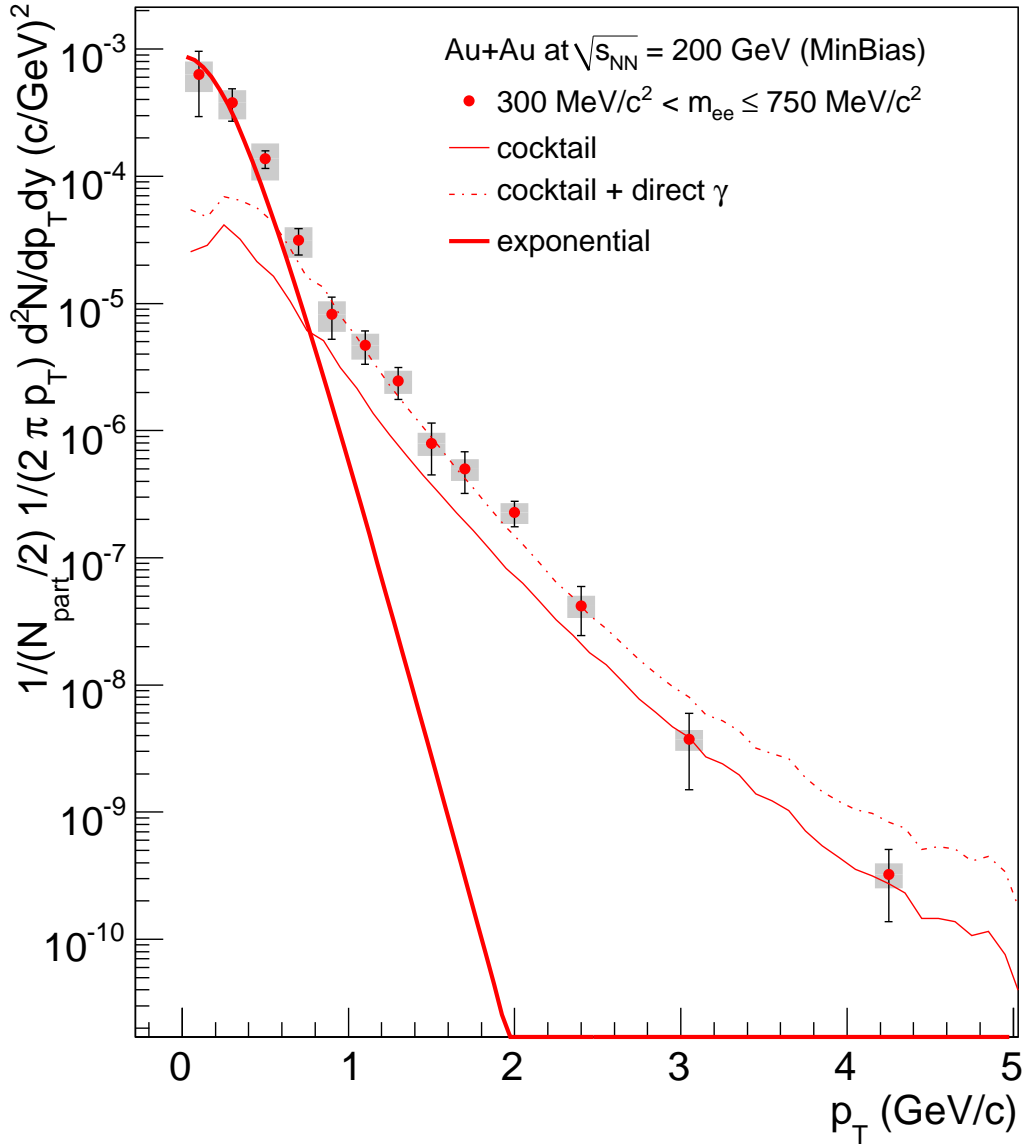}
  \caption[Fits of \mt spectra of \ee pairs in $0.3 < \mee <
  0.75$~\gevcc]{The figure shows \pt spectra for the mass range $300 <
    \mee < 750$~\mevcc. The spectrum is fitted to the sum of the
    cocktail expectation ({\em thin line}), direct photons and an
    exponential function in \mt ({\em thick line}). The cocktail
    include contributions from charmed mesons. The {\em dashed line}
    shows the expectation from the cocktail plus the direct photon
    contribution. The exponential function is also shown.}
  \label{fig:au_mtfit_300750}
\end{figure}

The \pt spectrum and the fit result are shown in
\fig{fig:au_mtfit_300750}, together with the individual
components. The data are well reproduced by the fit which gives an
inverse slope of $T_{\rm eff} = 88 \pm 11{\rm (stat.)} ^{+8}_{-13}{\rm
  (syst.)}$~MeV with a $\chi^2/DOF \approx 1$. The systematic
uncertainty accounts for the uncertainty on the data and the
uncertainty on the cocktail normalization (20\%). The yield attributed
to the exponential accounts for more than 50\% of the total \ee yield.

If the source of this excess is due to thermal radiation from the
fireball, which is dominated by $\pi \pi$ annihilation, the excess
yield in the LMR should exhibit a similar mass dependent effective
temperature due to transverse flow as the hadrons. The effective
temperature is significantly lower than for any of the identified
hadron spectra~\cite{adler:034909}. In particular the kaon, with a
mass comparable to the average invariant mass of the \ee pairs in the
fitted mass region, has an effective temperature more than twice as
large. This seems to advocate a source of dielectrons that does not
exhibit the rise with mass typical for a radially expanding
source. Also the temperature extracted from the direct photons at high
\pt ($T_{\rm eff} = 221 \pm 23 {\rm (stat.)} \pm 18{\rm (syst.)}$~MeV)
is significantly larger than the one of the excess yield at low \pt.

\section{Model Comparisons}
\label{sec:model-comparisons}

In this section the enhanced production of direct photons and the low
mass enhancement of the dielectron continuum observed in \AuAu
collisions are compared to theoretical models

\subsection{Thermal Photons}
\label{sec:pt-model-comp}

\begin{figure}
  \centering
  \includegraphics[width=0.9\textwidth]{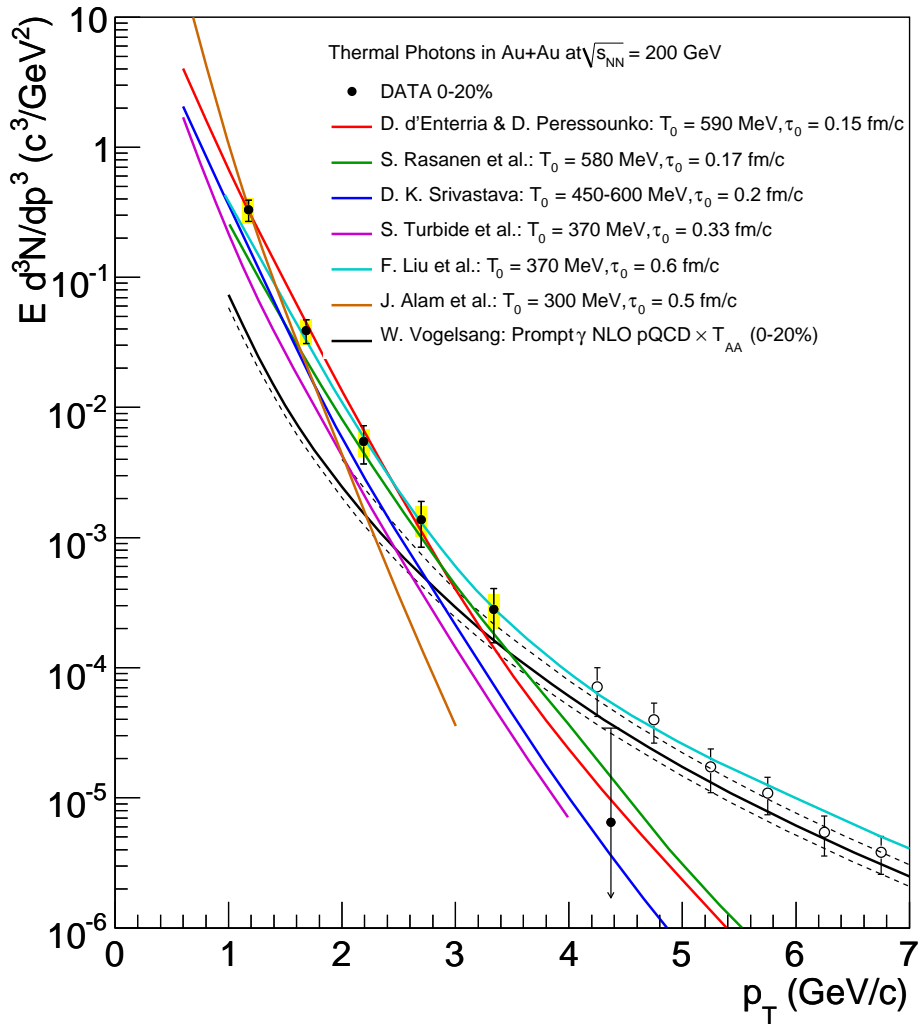}
  \caption[Comparison of direct photon yield \AuAu to theoretical
  predictions]{Comparison of direct photon yield \AuAu to theoretical
    predictions~\cite{PhysRevC.69.014903,d'Enterria:2005vz,Huovinen:2001wx,PhysRevC.64.034902,PhysRevC.63.021901,Liu:2008eh}. Note
    that the {\em cyan} curve by Liu {\em et al.} has a pQCD
    contribution included, while all other model curves only show the
    thermal contribution. For a comparison to the data above 3~\gevc a
    pQCD contribution, \eg the one shown by
    W. Vogelsang~\cite{PhysRevD.48.3136,vogelsang} needs to be added.}
  \label{fig:au_directg_theory_comp}
\end{figure}

\begin{figure}
  \centering
  \includegraphics[width=0.9\textwidth]{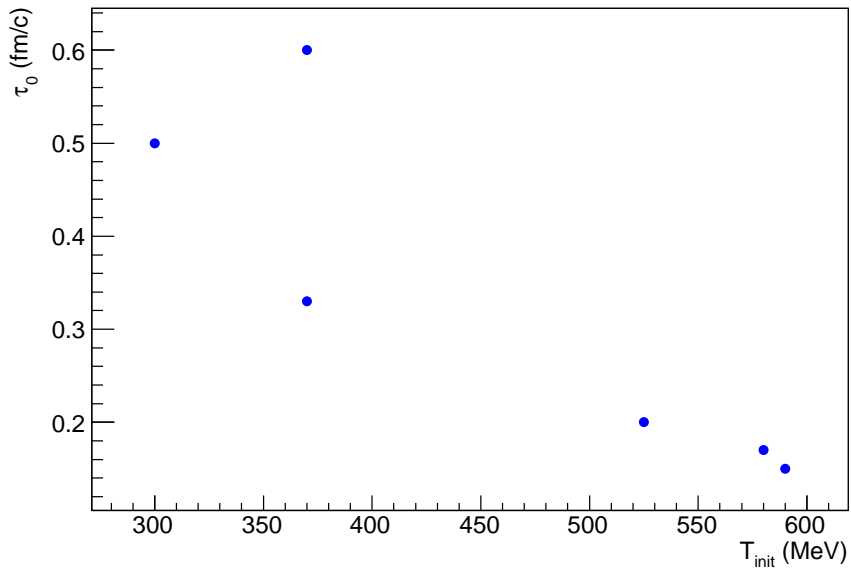}
  \caption[Comparison of direct photon yield \AuAu to theoretical
  predictions]{Comparison of formation time $\tau_0$ and initial
    temperature $T_{\rm init}$ of the thermal photon predictions.}
  \label{fig:tau_vs_T}
\end{figure}

In \fig{fig:au_directg_theory_comp} the direct photon spectrum
measured in 0--20\% \AuAu collisions is compared to a number of
hydrodynamical
models~\cite{PhysRevC.69.014903,d'Enterria:2005vz,Huovinen:2001wx,PhysRevC.64.034902,PhysRevC.63.021901,Liu:2008eh}.
All of these models are in good agreement with the data, and with each
other within a factor of two. This fact is quiet remarkable as the
initial temperature and formation times of these models vary in the
range of $T_{\rm init} = 300$~MeV, $\tau_0=0.6$~fm/$c$ to $T_{\rm
  init} = 600$, $\tau_0 = 0.6$~fm/$c$. \tab{tab:initial_cond} lists
the initial conditions of all the models mentioned here. They are
also shown as pairs of $(T_{\rm init},\tau_0)$ in \fig{fig:tau_vs_T},
which illustrates the apparent anti-correlation between the formation
time $\tau_0$ and the initial temperature $T_{\rm init}$ which
produces time averaged direct photon spectra all in agreement with the
data.

\begin{table}
  \centering
  \caption[Initial conditions of thermal photon
  predictions]{\label{tab:initial_cond} The initial Temperature
    $T_{\rm init}$ and formation time $\tau_0$ for various models
    predictions of thermal photons in \AuAu collisions at \sqrtsnn =
    200 GeV.\\}
  \begin{tabular}{lcr@{.}lc}\toprule
    & $T_{\rm init}$ (MeV) & \multicolumn{2}{c}{$\tau_0$ fm/$c$}& notes \\\midrule
    David d'Enterria {\em et al.} & 590 & 0&15 &  \cite{d'Enterria:2005vz}\\
    Rasanen {\em et al.}& 580 & 0&17 &  \cite{Huovinen:2001wx,Rasanen:2002qe}\\
    Srivastava          & 450--600 &0&2 &  \cite{PhysRevC.64.034902,Srivastava:2001hz}\\
    Turbide{\em et al.} & 370 & 0&33 &  \cite{PhysRevC.69.014903}\\
    L. Liu {\em et al.} & 370 & 0&6  &  \cite{Liu:2008eh}\\
    J. Alam {\em et al.} & 300 & 0&5  &  \cite{PhysRevC.63.021901}\\
    \bottomrule
  \end{tabular}
\end{table}
The general good agreement of the hydrodynamical models with the data
confirms that the \pt range of 1--3~\gevc is dominated by thermal
radiation. To disentangle the details of the initial condition and the
hydrodynamic evolution of the system more detailed studies will be
necessary.

\subsection{Low mass dileptons}
\label{sec:mass-model-comp}

While the excess of direct photons produced in \AuAu collision in
comparison to \pp seems to be qualitatively explained by thermal
radiation from a medium with temperatures of 300--600~MeV, the
situation for the low mass enhancement of dielectrons of the expected
cocktail of hadron decays is quiet different.

\begin{figure}
  \centering
  \includegraphics[width=0.9\textwidth]{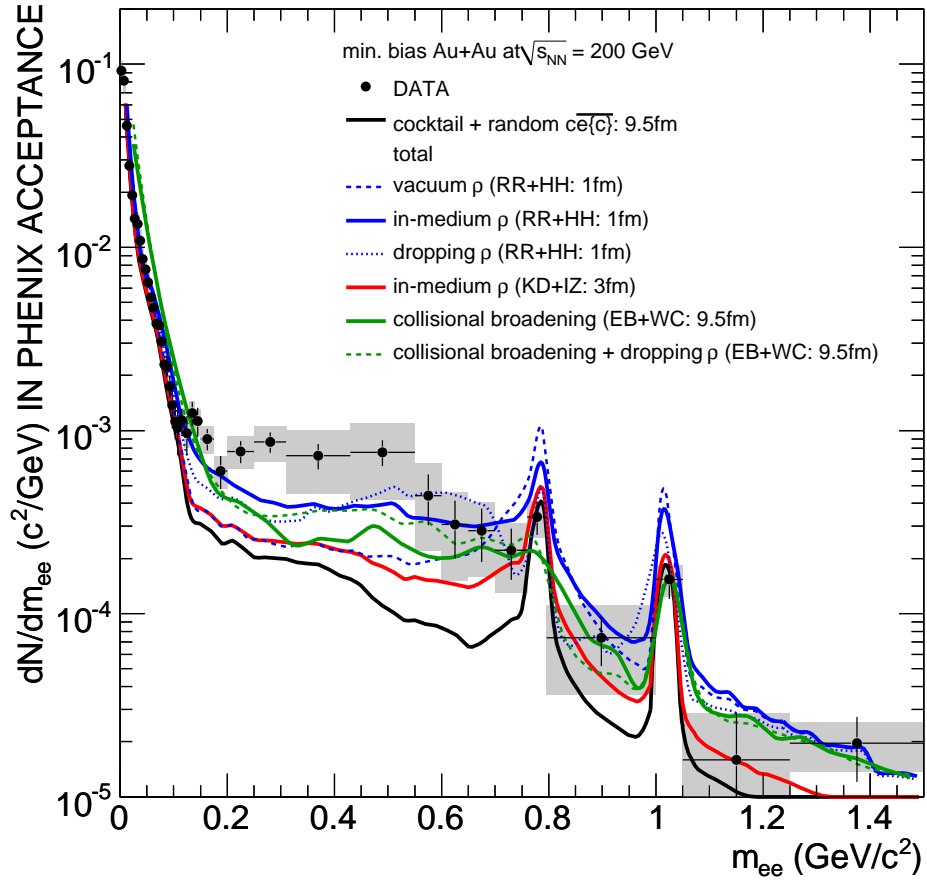}
  \caption[Comparison of low mass enhancement in \AuAu to theoretical
  predictions]{Invariant mass distribution of \ee pairs in min. bias
    \AuAu collisions compared to theoretical predictions including
    dropping mass and broadening scenarios by Rapp and
    Hees~\cite{Rapp:2002mm}, Dusling and Zahed~\cite{Dusling:2007su},
    Bratkovskaya and Cassing~\cite{Bratkovskaya:2008bf}.}
  \label{fig:au_mass_theory_comp}
\end{figure}

In \fig{fig:au_mass_theory_comp} the LMR of the dielectron continuum
in min. bias \AuAu collision at \sqrtsnn = 200 GeV is compared to a
number of predictions involving different scenarios for in-medium
modifications of the vector mesons by R. Rapp and H. van
Hees~\cite{Rapp:2002mm}, K. Dusling and
I. Zahed~\cite{Dusling:2007su}, and E. L. Bratkovskaya and W. Cassing
\cite{Bratkovskaya:2008bf}. As discussed in
Section~\ref{sec:exp-results}, an expanding fireball model with a
modified in-medium spectral function by R. Rapp successfully describes
the dilepton spectra measured by CERES (see \fig{fig:ceres_pbau}) and
NA60 (see \fig{fig:na60_lmr_theory})~\cite{hees:102301}. Also the
other authors find good agreement of their models with
NA60\cite{dusling:024908,Bratkovskaya:2008bf}. Both, Rapp and Dusling
include a contribution from $q\overline{q}$ annihilation during the
QGP phase in their predictions which, however, is important only in
the intermediate mass range.

Predictions for different $\rho$ spectral functions (vacuum, dropping
mass, and collisional broadening) are filtered into the PHENIX
acceptance and added to the hadronic cocktail which beforehand has its
vacuum $\rho$ contribution subtracted as this is yield is part of the
prediction. The results by R. Rapp~\cite{Rapp:2002mm} are shown as
blue lines in \fig{fig:au_mass_theory_comp}, the vacuum rho is shown
as {\em dashed}, the dropping $\rho$ as {\em dotted} and the
broadening scenario (for which the $\rho$ spectral function was shown
in \fig{fig:rapp_inmedium}) as {\em solid} line. The predictions are
in agreement with the data within the systematic uncertainties for
$\mee \gtrsim 0.5$~\mevcc, but clearly do not provide enough yield in
the mass range $150 \lesssim \mee \lesssim 0.5$~\mevcc to describe the
data. All three scenarios have in common a sharp drop of the $\rho$
spectral function below $\mee < 2m_{\pi}$, something that is not
observed in data. In data the enhancement stays at an almost constant
level down to $\mee \approx 100$~\mevcc.

Also the hydrodynamic calculation by K. Dusling is able to describe
the NA60 data~\cite{dusling:024908} with dilepton rates following from
a chiral virial expansion~\cite{PhysRevD.56.5605}, does not describe
the low mass enhancement below $\approx600$~\mevcc, as shown with the
{\em red solid} line in \fig{fig:au_mass_theory_comp}.

The third theoretical calculation by
E. L. Bratkovskaya~\cite{Bratkovskaya:2008bf} which is based on a
microscopic HSD (Hadron String Dynamics) transport
model~\cite{Ehehalt:1996uq} provides two scenarios, one is a
collisional broadening of the $\rho$ meson, the other includes in
addition a drop of the $\rho$ mass. They are shown in
\fig{fig:au_mass_theory_comp} as {\em solid} green and {\em dashed}
green line, respectively. As the other model calculations, also this
does not account for the enhancement observed below 600~\mevcc.

\chapter{Summary and Outlook}
\label{cha:summary}

This thesis has presented the analysis of the dielectron continuum in
\pp collisions at \sqrts = 200 GeV which allowed to extract a variety
of physics signals: The measurement of the total charm and bottom
cross section in the intermediate and high mass region
($\sigma_{c\overline{c}} = 544 \pm 39 ({\rm stat.}) \pm 142 ({\rm
  syst.})  \pm 200 ({\rm model})~\mu{\rm b}$ and
$\sigma_{b\overline{b}} = 3.9 \pm 2.4 ({\rm stat.}) ^{+3}_{-2} ({\rm
  syst.})~\mu{\rm b} $), the differential cross section measurement of
$\omega$ and $\phi$ mesons and the \pt dependence of the low mass
dielectron continuum. A detailed study of the spectral shape in the
mass region $100 < \mee < 300$~\mevcc resulted in the measurement of
direct photons via internal conversions in the \pt range of $1< \pt <
5$~\gevc which is consistent to NLO pQCD prediction.

These results provide an important baseline for any measurement of the
dielectron continuum in heavy ion collisions. As part of this thesis a
discussion of the dielectron continuum measured in \AuAu collisions
has been given. The results can be summarized briefly as follows: a
dielectron yield is observed in the mass range $150 < \mee < 750$
which is enhanced by a factor of $4.0 \pm 0.3 {\rm (stat.)} \pm 1.5
{\rm (syst.)} \pm 0.8 {\rm (model)}$ in min. bias \AuAu collisions
with respect to the expected contribution from hadron decays and
semi-leptonic charm decays. This enhancement is localized in central
collisions and is absent in \pp and peripheral collisions. It
increases faster than linear with $N_{\rm part}$ and is moreover
localized at low \pt. The \pt dependence of the enhancement has an
inverse slope of $T_{\rm eff} = 88 \pm 11{\rm (stat.)} ^{+8}_{-13}{\rm
  (syst.)}$~MeV and does not show any signs of radial flow. The
concentration of the enhancement at low \pt is qualitatively
consistent with results from SPS
experiments~\cite{Agakichiev:2005ai,arnaldi:022302}, but the missing
evidence for radial flow and the very small effective temperatures
seem to indicate some differences to the observations at SPS. This
becomes more evident when comparing the enhancement with current
models of in-medium modifications to the $\rho$ meson which are able
to explain the SPS results with collisional broadening of the
$\rho$~\cite{hees:102301,dusling:024908,Bratkovskaya:2008bf}, but fail
to give a complete description of the enhancement observed at
RHIC. The measurement of direct photons via internal conversions at
high \pt shows an excess compared to the \pp measurement. This excess
has an inverse slope of $T_{eff} = 221 \pm 23 {\rm (stat.)} \pm 18{\rm
  (syst.)}$~MeV in 0--20\% most central collisions and can be
described by hydrodynamical models with an initial temperature in the
range of $300 \leq T_{\rm init} \leq 600$~MeV at formation times of
$0.12 \leq \tau_0 \leq 0.6$~fm/$c$ as thermal radiation.

With upcoming data taking periods and the expected increases in
luminosity, one of the most interesting regions, besides a more
quantitative analysis of the low mass enhancement, will be the
intermediate mass region. Currently, the statistics available in \AuAu
does not allow to distinguish thermal from open charm contributions,
but with more statistics a study of the \pt dependence of the IMR may
provide new opportunities. This is equally true for the study of \pp
collisions. The comparison of the measured \pt spectrum of the IMR and
leading-order \pythia calculations suggests that higher order
processes will need to be considered at high \pt and invariant
masses. With more statistics to come the high mass region beyond the
$J/\psi$ will allow more precise measurements of the bottom cross
section. The Silicon Vertex Detector (VTX)
upgrade~\cite{Heuser:2003df} will accurately measure the collision
vertex as well as secondary decay vertices. This will tremendously
improve the open heavy flavour measurement, as it will allow to
differentiate between prompt decays and the off-vertex decays of
charmed and bottom meson.

Returning to the low mass region, it may be possible (at least in \pp)
to measure low momentum $\eta$ via their Dalitz decays by combining
\ee pairs with $100 < \mee < 500$ with photons measured in the
calorimeter. This would be very analog to the beam pipe conversion
analysis presented in Appendix~\ref{sec:bpanalysis}. A low momentum
$\eta$ measurement would significantly reduce the uncertainties on the
hadronic cocktail and may, together with a good understanding of the
open charm background, lead the way to a measurement of the $\omega$
form factor in \pp collisions at \sqrtsnn = 200 GeV.

The Hadron Blind Detector (HBD) upgrade~\cite{Fraenkel:2005wx} should
significantly improve the signal-to-background ratio of any low mass
dielectron analysis, by identifying \ee pairs with small opening angle
as originating from \pion Dalitz decays and photon conversion in a
magnetic field free region surrounding the beam pipe. The HBD is a
windowless proximity-focusing Cherenkov detector with a radiator
length of 50~cm operating in pure CF$_4$ as radiator and detector
gas. Cherenkov photons emitted by electrons create photoelectrons in a
CsI photocathode, which are avalanched by a triple stack of Gas
Electron Multipliers (GEM) and read out by PCB boards. The operation
in a reverse bias mode together with a shielding mesh above the
photocathode makes the HBD blind to minimum ionizing particles, \ie
hadron blind, as their charge is collected on the mesh. \ee pairs from
\pion Dalitz decays or photon conversions with small opening angle
will produce overlapping avalanches which could be identified by their
signal height compared to avalanches from single electrons.

% ...


\renewcommand{\baselinestretch}{1}
\normalsize
\clearpage
\phantomsection
\addcontentsline{toc}{chapter}{\numberline{}{Bibliography}}
\bibliography{mybibliography}


\appendix

\chapter{Background Normalization}
\label{cha:background_normalization}

\section{Pairing of electrons and positrons}
\label{sec:pairing}
In the following it is assumed that positrons and electrons are always
produced in pairs. Let $N$ pairs be produced in a particular event and
$N$ is given by a probability distribution $P(N)$. Of the $N$ pairs
only a fraction $p$ is reconstructed, and then the number of
reconstructed pairs $n_p$ is given by a binomial distribution $B$
sampling out of $N$ ``events'' with a probability $\varepsilon_p$.

\begin{itemize}
\item Probability to get $n_p$ pairs from $N$ true pairs: $\omega(n_p)
  = B(n_p, N, \varepsilon_p)$
\item with an average: $\langle n_p\rangle = \varepsilon_p N$
\item and variance: $\sigma_p^2 = \varepsilon_p N (1-\varepsilon_p)$
\end{itemize}

Of the remaining pairs one track is reconstructed with a probability
$\varepsilon_+$ or $\varepsilon_-$. For a given $N$ and $n_p$ the
number of additional single positive tracks $n_+$ and negative tracks
$n_-$ follow a multinomial distribution $M$ with possible three
possible outcomes for each of the $N-n_p$ unreconstructed pairs: no
track, one $+$ track or one $-$ track.

The probability to get $n_+$ and $n_-$ single tracks from $N$ true
pairs with $n_p$ reconstructed pairs, \ie, from $N' = (N-n_p)$ not
fully reconstructed pairs is: $\omega(n_+, n_-) = M(n_+, n_-; N',
\varepsilon_+, \varepsilon_-)$

\begin{itemize}
\item with average: $\langle n_{\pm}\rangle = \varepsilon_{\pm} N'$
\item variance: $\sigma_{\pm}^2 = \varepsilon_{\pm} N' (1-\varepsilon_{\pm})$
\item and covariance: $\rm{cov}(n_+, n_-) = -N' \varepsilon_+ \varepsilon_-$
\end{itemize}

In this case the number of unlike-sign pairs for a given $N$ and
$n_p$ is:
\begin{align}
  \langle n_{+-} \rangle &= n_p^2 + n_p \sum_+ n_+ \omega(n_+) + n_p
  \sum_- n_- \omega(n_-) + \sum_+^{N-n_p} \sum_-^{N-n_p} n_+ n_-
  \omega(n_+, n_-)\nonumber\\
  &= n_p^2 + n_p \varepsilon_+ (N-n_p) + n_p \varepsilon_- (N-n_p) +
  \langle n_+ n_- \rangle\nonumber\\
  &= n_p^2 + n_p \varepsilon_+ (N-n_p) + n_p \varepsilon_- (N-n_p) +
  \varepsilon_+ \varepsilon_- (N-n_p)^2\nonumber\\
  &\quad - \varepsilon_+ \varepsilon_- (N-n_p)\nonumber\\
  &= n_p^2 + \varepsilon_+ N n_p - \varepsilon_+ n_p^2 + \varepsilon_-
  N n_p - \varepsilon_- n_p^2 + \varepsilon_+ \varepsilon_- N^2 - 2
  \varepsilon_+ \varepsilon_- N n_p\nonumber\\
  &\quad + \varepsilon_+ \varepsilon_- n_p^2 - \varepsilon_+
  \varepsilon_- N + \varepsilon_+ \varepsilon_-
  n_p\nonumber\\
  &= \left(n_p + \varepsilon_+ \left(N-n_p\right)\right)\left(n_p +
    \varepsilon_- \left(N-n_p\right)\right) - \varepsilon_+
  \varepsilon_- (N-n_p).
\end{align}
Similarly one can calculate the number of like-sign pairs:
\begin{align}
  2\langle n_{++} \rangle &= \sum_+ (n_p + n_+)(n_p + n_+ -1)
  \omega(n_+)\nonumber\\
  &= n_p^2 - n_p + \langle n_+^2 \rangle - \langle n_+ \rangle + 2 n_p
  \langle n_+ \rangle\nonumber\\
  &= n_p^2 - n_p + \varepsilon_+^2 (N-n_p)^2 +
  \varepsilon_+(1-\varepsilon_+)(N-n_p) - \varepsilon_+ (N-n_p)\nonumber\\
  &\quad + 2 \varepsilon_+ n_p (N-n_p)\nonumber\\
  &= n_p^2 - n_p + \varepsilon_+^2 (N-n_p)^2 - \varepsilon_+^2(N-n_p)
  + 2 \varepsilon_+ n_p (N-n_p)\\
\intertext{and}
2\langle n_{--} \rangle &= n_p^2 - n_p + \varepsilon_-^2 (N-n_p)^2 - \varepsilon_-^2(N-n_p)
  + 2 \varepsilon_- n_p (N-n_p).
\end{align}

To obtain the expected number of like- and unlike-sign pairs for a
fixed number of real pairs $N$ it is averaged over all possible
reconstructed pairs $n_p$:
\begin{align}
  \langle N_{+-} \rangle &=  \sum_{n_p} \langle n_{+-} \rangle B(n_p) \nonumber\\
  &= (1 - \varepsilon_+ - \varepsilon_- + \varepsilon_+ \varepsilon_-)
  \langle n_p^2 \rangle + (\varepsilon_+ N + \varepsilon_- N - 2
  \varepsilon_+ \varepsilon_- N + \varepsilon_+ \varepsilon_-) \langle
  n_p \rangle \nonumber\\
  &\quad + \varepsilon_+ \varepsilon_- N^2 - \varepsilon_+
  \varepsilon_- N \nonumber\\
  &= (1 - \varepsilon_+ - \varepsilon_- + \varepsilon_+
  \varepsilon_-)(\varepsilon_p^2N^2 + \varepsilon_p(1-\varepsilon_p)N) \nonumber\\
  &\quad + (\varepsilon_+ N + \varepsilon_- N - 2 \varepsilon_+ \varepsilon_- N
  + \varepsilon_+ \varepsilon_-) \varepsilon_p N + \varepsilon_+ \varepsilon_- N^2 - \varepsilon_+ \varepsilon_-
  N \nonumber\\
  &= (\varepsilon_p^2 - \varepsilon_p^2 \varepsilon_+ -
  \varepsilon_p^2 \varepsilon_- + \varepsilon_p^2 \varepsilon_+
  \varepsilon_- + \varepsilon_p \varepsilon_+ + \varepsilon_p
  \varepsilon_- - 2 \varepsilon_p \varepsilon_+ \varepsilon_- +
  \varepsilon_+ \varepsilon_-)\nonumber\\
  &\quad \cdot (N^2-N) + \varepsilon_p N\nonumber\\
  &= (\varepsilon_p + \varepsilon_+ (1-\varepsilon_p))(\varepsilon_p
  + \varepsilon_-(1-\varepsilon_p))(N^2-N) + \varepsilon_p N.
\end{align}

Now the like-sign background is calculated:
\begin{align}
  2 \langle N_{++} \rangle &= \sum_{n_p} 2 \langle n_{++} \rangle
  B(n_p) \nonumber\\
  &= \varepsilon_p^2 N^2 + \varepsilon_p(1-\varepsilon_p)N -
  \varepsilon_p N + \varepsilon_p^2 \varepsilon_+^2 N^2 +
  \varepsilon_+^2 \varepsilon_p(1-\varepsilon_p) N  - 2 \varepsilon_+^2 \varepsilon_p N^2\nonumber\\
  &\quad + \varepsilon_+^2 N^2 - \varepsilon_+^2 N + \varepsilon_+^2
  \varepsilon_p N + 2 \varepsilon_+ \varepsilon_p N^2 - 2
  \varepsilon_+ \varepsilon_p^2
  N^2 - 2 \varepsilon_+ \varepsilon_p (1-\varepsilon_p) N\nonumber\\
  &= \varepsilon_p^2 (N^2-N) + \varepsilon_+^2 \varepsilon_p^2 (N^2-N)
  + \varepsilon_+^2 \varepsilon_p N - 2 \varepsilon_+^2
  \varepsilon_p N^2 + \varepsilon_+^2(N^2-N)\nonumber\\
  &\quad + \varepsilon_+^2 \varepsilon_p N + 2\varepsilon_+
  \varepsilon_p N^2 - 2 \varepsilon_+
  \varepsilon_P^2 N^2 - 2 \varepsilon_+ \varepsilon_p N + 2 \varepsilon_+ \varepsilon_p^2 N\nonumber\\
  &= (\varepsilon_p^2 + \varepsilon_+^2 +
  \varepsilon_+^2\varepsilon_p^2)(N^2-N) - 2 \varepsilon_+^2
  \varepsilon_p (N^2-N) + 2 \varepsilon_+ \varepsilon_p
  (N^2-N)\nonumber\\
  &\quad - 2 \varepsilon_+ \varepsilon_p^2 (N^2-N)\nonumber\\
  \langle N_{++} \rangle &= \frac{1}{2} (\varepsilon_p + \varepsilon_+ (1-\varepsilon_p))^2(N^2-N)\\
  \intertext{and}
  \langle N_{--} \rangle &= \frac{1}{2} (\varepsilon_p + \varepsilon_-
  (1-\varepsilon_p))^2(N^2-N).
\end{align}

Finally, it is averaged over all $N$ to get the foreground unlike-sign
pairs:
\begin{align}
  \langle FG_{+-} \rangle &= \sum_{N} \langle N_{+-} \rangle
  P(N)\nonumber\\
  &= (\varepsilon_p + \varepsilon_+(1-\varepsilon_p))(\varepsilon_p +
  \varepsilon_-(1-\varepsilon_p))(\langle N^2 \rangle - \langle N
  \rangle) + \varepsilon_p \langle N \rangle\nonumber\\
  &= \langle BG_{+-} \rangle + \langle S \rangle.\\
  \intertext{The unlike-sign foreground $FG_{+-}$ consists of the sum
    of the unlike-sign background $BG_{+-}$ and the signal $S =
    \varepsilon_p \langle N \rangle$. Similarly the like-sign
    foreground is calculated as:} \langle FG_{++} \rangle &= \sum_{N}
  \langle N_{++} \rangle
  P(N)\nonumber\\
  &= \frac{1}{2}(\varepsilon_p +
  \varepsilon_+(1-\varepsilon_p))^2(\langle N^2 \rangle - \langle N
  \rangle)\nonumber\\
  &= \langle BG_{++} \rangle\\
  \intertext{and}
  \langle FG_{--} \rangle &= \langle BG_{--} \rangle.
\end{align}
The like-sign foreground contains no signal.

So due to the fact that electrons and positrons are always created in
pairs, the unlike-sign background is the geometric mean of the
like-sign backgrounds, independent of the primary multiplicity
distribution.
\begin{equation}
  \langle BG_{+-} \rangle = 2 \sqrt{\langle BG_{++} \rangle \langle BG_{--} \rangle}
\end{equation}

Comparing the background to the product of the average track
multiplicities one gets for a fixed $n_p$:
\begin{align}
  \langle n_+ \rangle &= \sum_+ (n_p + n_+)\omega(n_+)\nonumber\\
  &= n_p + \langle n_+ \rangle\nonumber\\
  &= n_p + \varepsilon_+ (N-n_p)\\
  \intertext{averaged over all possible $n_p$:}
  \langle N_+ \rangle &= \sum_{n_p} \langle n_+ \rangle\omega(n_p)\nonumber\\
  &= \varepsilon_p N + \varepsilon_+ N - \varepsilon_+ \varepsilon_p N\nonumber\\
  &= (\varepsilon_p + \varepsilon_+ (1-\varepsilon_p)) N\\
  \intertext{or averaged over all possible $N$:}
  \langle FG_+ \rangle &= \sum_N \langle N_+ \rangle P(N)\nonumber\\
  &= (\varepsilon_p + \varepsilon_+ (1-\varepsilon_p)) \langle N \rangle\\
  \intertext{and thus:}
  \langle FG_+\rangle \langle FG_+\rangle &= (\varepsilon_p +
  \varepsilon_+ (1-\varepsilon_p))(\varepsilon_p +
  \varepsilon_- (1-\varepsilon_p))\langle N\rangle^2\\
  \intertext{or}
  \frac{\langle BG_{+-}\rangle}{\langle FG_+ \rangle \langle FG_- \rangle}
  &= 1 + \frac{\sigma^2 - \langle N \rangle}{\langle N \rangle^2}.
\end{align}

So in general $\langle BG_{+-} \rangle \neq \langle FG_+ \rangle
\langle FG_- \rangle$, except for the special case that $P(N)$ is a
Poisson distribution. Note this is the opposite conclusion one derives
in the case that the sources of $+$ and $-$ tracks are independent,
\ie, $+$ and $-$ tracks are produced as singles and not as pairs like
they are for muons. In that case $\langle FG_+ \rangle \langle
FG_-\rangle$ is the correct background normalization.

\chapter{Beam Pipe Conversions}
\label{cha:bpconv}

\section{Introduction}
\label{sec:bpintro}
In this Appendix another approach to measure direct photons at low \pt
is presented. It is the measurement of real photons which convert
externally in the beam pipe, surrounding the collision vertex at a
radius of 4~cm. These are the pairs which are removed with the
$\phi_V$ cut in the regular dielectron continuum and internal
conversion analysis. The basic idea is to identify \ee pairs from beam
pipe conversions and to combine these with photons in the EMCal to tag
conversion photons from \pion decays. It is certainly not as strong as
the virtual photon analysis but is nevertheless an interesting
approach.

\section{Analysis}
\label{sec:bpanalysis}
To identify \ee pairs from photon conversions, a single electron
identification cut is applied, which requires signals from at least two
phototubes in the Ring Imaging Cherenkov Detector (RICH) matching to a
reconstructed charged track in the Drift Chamber (DC). No further
electron identification cuts were applied since the pair cuts (see
Section~\ref{sec:bpconversions}) to separate conversion photons from other
\ee pairs are more efficient and powerful enough to provide a very clean
photon conversion sample.

The extracted photon conversions are tagged with photons reconstructed
in the EMC to determine the contribution from $\pi^0 \rightarrow
\gamma \gamma$ decays (see Section~\ref{sec:bptagging}).

All yields are measured as a function of \pt of the
e$^+$e$^-$--pair, which makes a direct comparison of the inclusive
photon yield, $N_{\gamma}^\mathrm{incl}$, and the tagged photon yield,
$N_{\gamma}^{\pi^0 \mathrm{tag}}$, possible:
\begin{align}\label{eq:inclusive}
N_{\gamma}^\mathrm{incl}\left(\pt\right) &=
\epsilon_{\ee}~a_{\ee}~\gamma^\mathrm{incl}\left(\pt\right) \\
\label{eq:piontag}
N_{\gamma}^{\pi^0 \mathrm{tag}}\left(\pt\right) &=
\epsilon_{\ee}~a_{\ee}~
\epsilon_{\gamma}\left(\pt\right)~f~\gamma^{\pi^0}\left(\pt\right)
\end{align}
The measured yield of inclusive photons depends on the reconstruction
efficiency $\epsilon_{\ee}$ and the PHENIX acceptance $a_{\ee}$
of the conversion \ee pair. The tagged photon yield depends in addition
on the efficiency to reconstruct the second photon in the EMC
$\epsilon_{\gamma}(\pt)$ and on the conditional probability $f$ to
find it in the EMC acceptance, given that the \ee pair has been
reconstructed already. Here, $\epsilon_{\gamma}(\pt)$ is weighted with
the \pt distribution of the \ee pair. In the ratio
$N_{\gamma}^{incl}/N_{\gamma}^{\pi^0 tag}$ the \ee pair reconstruction
efficiency and acceptance correction factor cancel.

A ratio of the hadronic decay photon yield,
$N_{\gamma}^{\mathrm{hadr}}$, and the tagged photon yield from $\pi^0$
decays, $N_{\gamma}^{\pi^0 \mathrm{tag}}$, is calculated with
simulations.
\begin{equation}\label{eq:pions}
N_{\gamma}^{\pi^0 \mathrm{tag}}\left(\pt\right) = f~N_{\gamma}^{\pi^0}\left(\pt\right)
\end{equation}

The comparison of the ratio in data and in simulations in a double
ratio leads to an expression that is equivalent to the ratio of
inclusive and decay photons as shown in \eq{eq:doubleratio}.
\begin{equation}\label{eq:doubleratio}
  \frac{\gamma^\mathrm{incl}\left(\pt\right)}{\gamma^\mathrm{hadr}\left(\pt\right)} =
  \frac{\epsilon_{\gamma}\left(\pt\right) \cdot
    \left(\frac{N_{\gamma}^\mathrm{incl}\left(\pt\right)}{N_{\gamma}^{\pi^0 \mathrm{tag}}\left(\pt\right)}\right)_\mathrm{Data}}
  {\left(\frac{N_{\gamma}^\mathrm{hadr}\left(\pt\right)}{f~N_{\gamma}^{\pi^0}\left(\pt\right)}\right)_\mathrm{Sim}}
\end{equation}
The only remaining factors are the reconstruction efficiency of the
photon in the EMC, $\epsilon_{\gamma}(\pt)$, and the conditional
acceptance $f$ in the simulation part of the double ratio, which have
both been determined with Monte Carlo simulations (see
Section~\ref{sec:bpsimulations}).

\subsection{Photon Conversions}
\label{sec:bpconversions}
Since the PHENIX tracking algorithm assumes the track to originate
from the collision vertex, off-vertex conversion pairs are
reconstructed with an artificial opening angle which leads to an
invariant mass that is proportional to the radius at which the
conversion occurs.

\begin{figure}
  \centering
  \includegraphics[width=0.9\textwidth]{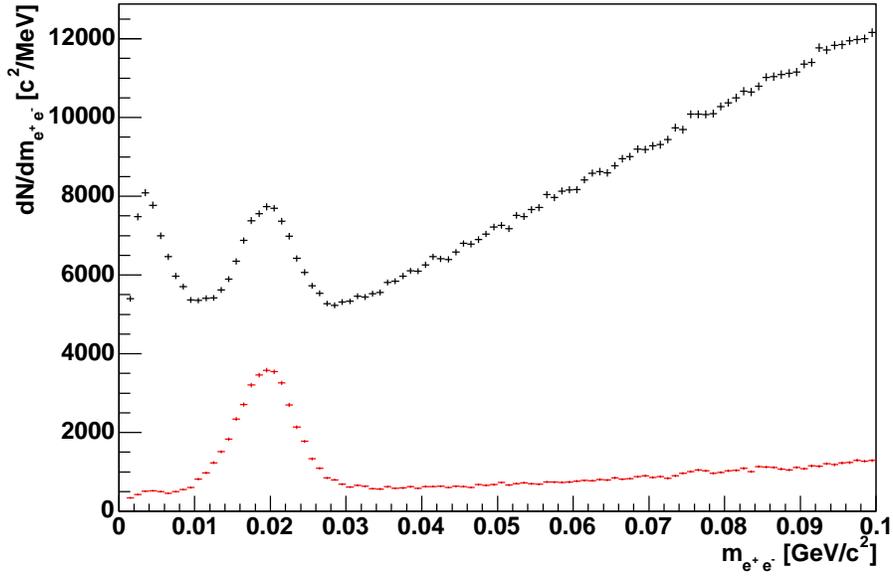}
  \caption[Invariant mass of \ee pairs from beam pipe
  conversion]{Invariant mass of \ee pairs before (black) and after
    (red) applying cuts on the orientation of the \ee pair in the
    magnetic field.}
  \label{fig:bp_allpairs}
\end{figure}

Therefore, photon conversions that occur in the beam pipe material
(Be, $0.3\,\%$ radiation length) at a radius of 4~cm are reconstructed
with an invariant mass of $\sim
20~\mathrm{MeV/c^2}$. \fig{fig:bp_allpairs} shows an invariant mass
spectrum of \ee pairs in the range 0--0.1~\gevcc. The peak from photon
conversions in the beam pipe at 20~\mevcc can be clearly separated
from Dalitz decays $\pi^0 \rightarrow \gamma \ee$, which dominate the
spectrum below 10~\mevcc, and combinatorial background pairs, whose
contribution increases toward higher invariant masses.

The photon conversion pairs, which have no intrinsic opening angle,
can be distinguished from Dalitz decays and purely combinatorial pairs
by cutting on the orientation of the \ee pair in the magnetic field.

\fig{fig:bp_allpairs} shows the invariant mass spectra of \ee pairs
before (black) and after (red) applying these pair cuts. The yield
from integrating the mass region $ < 35~\mathrm{MeV/c}^2$ of the
conversion peak is corrected for the remaining \pt dependent
contamination of $\sim 15.0 \pm 2.0~\mathrm{(syst)}\,\%$ due to
combinatorial \ee pairs which has been determined with mixed events.

\subsection{Tagging of Decay Photons}
\label{sec:bptagging}
\begin{figure}
  \centering
  \includegraphics[width=0.9\textwidth]{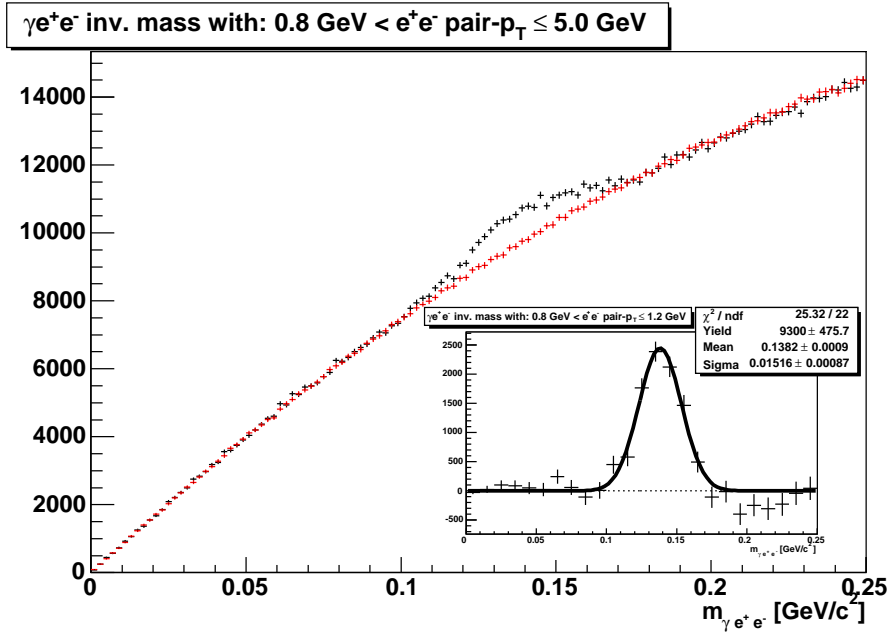}
  \caption[Invariant mass of $\gamma \ee$ triplets]{Invariant mass of $\gamma \ee$ triplets in same events
    (black) and normalized mixed events (red) for \ee pairs with
    $0.8~<~\pt~\leq~5.0$~\gevc. The insert shows the invariant mass of
    $\gamma \ee$ triplets after background subtraction for
    \ee pairs with $0.8 < \pt \leq 1.2$~\gevc. A fit with a Gaussian is
    drawn and the resulting parameters shown in the box in the upper
    right of the graph.}
  \label{fig:bp_invmasstriplets}
\end{figure}

To reveal which of these conversion photons come from $\pi^0
\rightarrow \gamma \gamma$ decays, the \ee pairs in the conversion
peak are combined with photons which have been measured in the EMC,
under loose cuts based on the time of flight and the shower profile
for photons with a minimum \pt of $0.3~\mathrm{\gevc}$, and their
invariant mass is calculated (see \fig{fig:bp_invmasstriplets}).

The reconstruction efficiency $\epsilon_{\gamma}(\pt)$ of the loose
photon has been estimated with a full GEANT simulation which embed
simulated photons into real EMC data, therefore providing a combined
information on the photon identification efficiency and occupancy
effects. The overall efficiency is determined to be $82 \pm 1\,\%$
independent of \pt beyond the minimum \pt cut off.

Conversion photons that are identified as decay products of $\pi^0$
can be tagged as $N_{\gamma}^{\pi^0 \mathrm{tag}}$. This signal has a
large combinatorial background due to the high photon multiplicity in
\AuAu collisions.

The combinatorial background is reproduced with an event mixing
method, which creates uncorrelated pairs of photons and \ee pairs from
different events. The mixed event spectrum is normalized to the same
event spectrum well outside the $\pi^0$ mass region (0--100~\mevcc,
170--250~\mevcc) and subtracted.

The statistical error on the normalization factor is on the order of
$0.2\,\%$ and depends only on the statistics in the same event
spectrum in the normalization region. As an example, the resulting
$\pi^0$ signal for \ee pairs with $0.8 < \pt \leq 1.2$~\gevc is shown as
insert in \fig{fig:bp_invmasstriplets}.

Mean and $\sigma$ are determined by a fit of the background subtracted
data with a Gaussian. The data are also fitted to the sum of a second
order polynomial and a Gaussian, to take into account the possibility
that the shape is not completely described by the mixed event
spectrum. The difference in the resulting mean and $\sigma$ is
negligible. The mean and $\sigma$ obtained by the fit are then used to
integrate the data in a region $\pm 1.5~\sigma$ around the mean,
chosen to optimize the signal to background ratio.

The statistical error on the extracted $\pi^0$ signal is given by:
\begin{equation}
  \sigma_S^2 = \sum_i FG(i) + \alpha \sum_i BG^{\prime}(i) +
  \left(\frac{\sigma_{\alpha}}{\alpha} \sum_i
    BG^{\prime}(i)\right)^2\label{eq:staterror}
\end{equation}
With $FG(i)$ and $BG^{\prime}(i)$ being the yields in bin $i$ of
invariant mass spectrum in same events and normalized mixed events,
respectively, the summations are performed over the integration
region. It is important to note that the last term in
\eq{eq:staterror}, is the square of the sum over the normalized
background, and therefore, depends on the integration region and is
not bin independent. Different integration regions have been
used. Variations in the resulting yield have been used to set a
systematic uncertainty on the yield extraction of $2.5\,\%$
independent of \pt.

The loss of $N_{\gamma}^{\pi^0 \mathrm{tag}}$ due to the external
conversion of the second photon is corrected by a factor $1-p_{conv} =
94 \pm 2\,\%$. In this factor $p_{conv}$ is the conversion probability
due to the material budget between the vertex and the Pad Chamber 3
(PC3) in front of the EMC.

\subsection{Simulations}
\label{sec:bpsimulations}
The contribution of hadronic decays has been determined with a fast
Monte Carlo simulation of $\pi^0$ and $\eta$ Dalitz decays. A
parameterization of the $\pi^0$ spectrum measured by
PHENIX~\cite{adler:072301} has been used as input. The $\eta$
distribution has been generated assuming $m_T$ scaling ($\pt
\rightarrow \sqrt{\pt^2 + m_{\eta}^2 - m_{\pi^0}^2}$) of the $\pi^0$
spectral shape and a normalization at high \pt to $\eta / \pi^0 = 0.45
\pm 0.04$, according to PHENIX data~\cite{adler:202301}. The relative
error of $9\,\%$ on the $\eta/\pi^0$ ratio is reduced by the branching
ratio of the two photon decay and results in a $3\,\%$ error in the
ratio $N_{\gamma}^{\mathrm{hadr}}/N_{\gamma}^{\pi^0}$.

The contamination due to neutral Kaons which decay before the beam
pipe has been found negligible ($\sim 1\,\%$) and has been folded into
the systematic error on the simulations.

The conditional probability $f$ that the photon is reconstructed in
the EMC once the \ee pair is reconstructed already was calculated with
a fast Monte Carlo simulation of $\pi^0 \rightarrow \gamma e^+
e^-$. The use of Dalitz decays is justified by the fact that the \pt
spectra of photons from $\pi^0 \rightarrow \gamma \gamma$ are
essentially identical to the \ee pair \pt spectrum from $\pi^0
\rightarrow \gamma \ee$.

After the \ee pair has been filtered in the detector acceptance, the
conditional acceptance $f$ for the second photon has been calculated
taking dead areas of the detector into account. Uncertainties in
calculating $f$ are found to be $5\,\%$, which is the largest source
of systematic errors. The limited energy resolution of the EMC of
$\sigma_E/E = 5\,\%~\oplus~9\,\%/\sqrt{E}$ introduced an additional
systematic error of $\sim 1\,\%$ due to the \pt cut at
$0.3~\mathrm{\gevc}$.

\section{Conclusions} 
\label{sec:bpconclusions}
\begin{figure}
  \centering
  \includegraphics[width=0.9\textwidth]{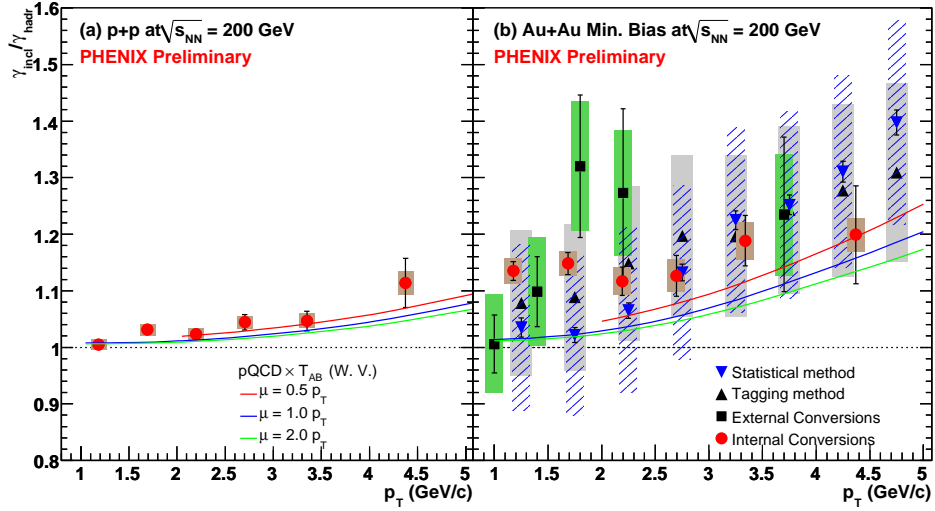}
  \caption[Comparison of direct photon excess in different
  methode]{Direct photon excess. (a) The fraction of the direct photon
    component as a function of \pt in \pp. (b) \AuAu (min. bias)
    compared to other measurements of direct photons as described in
    the legend. The curves are from a NLO pQCD
    calculation~\cite{PhysRevD.48.3136,vogelsang}.}
  \label{fig:bp_ratio}
\end{figure}

The right panel of \fig{fig:bp_ratio} shows the preliminary result for
the double ratio $\gamma^\mathrm{incl}\left(\pt\right) /
\gamma^\mathrm{hadr}\left(\pt\right)$ as in \eq{eq:doubleratio} for
Minimum Bias \AuAu collisions at $\sqrt{s_{NN}}$~=~200~GeV. The main
sources of systematic errors arise from the uncertainties in the
description of the detector active areas, in the peak extraction and
in the assumptions of the \pion shape and give a final systematic
error on the double ratio of $\sim 7\,\%$.
The result is compared to the conventional statistical subtraction of
hadronic decay photons~\cite{isobe:S1015}, and a
result~\cite{Gong:2007hr} which is based on the same tagging method,
but instead of photons coming from conversions in the beam pipe, the
clean photon sample is determined by selecting EMC clusters with very
strict photon identification cuts.

In addition the result of the internal conversion analysis is shown as
$1+r \approx \gamma_{\rm incl}/\gamma_{\rm hadr}$ in the right panel
for \AuAu and in the left of~\fig{fig:bp_ratio} for \pp
collisions. While all results agree within their uncertainties, the
improvement in statistical and systematic uncertainties with the
internal conversion analysis is quite significant. Also the NLO pQCD
predictions of W. Vogelsang~\cite{PhysRevD.48.3136,vogelsang} are
shown..

\chapter{Data Tables}
\label{cha:data_tables}

\begin{table}[tbh]
  \centering
  \caption[$\omega$ and $\phi$ yields]{\label{tab:counting}$\omega$ and $\phi$ yields.\\}
  \begin{tabular}{cr@{$~\pm~$}lr@{$~\pm~$}l}
    \toprule
    \pt (\gevc) & \multicolumn{2}{c}{$N_{\omega}$ (0.740--0.815~\gevcc)} & \multicolumn{2}{c}{$N_{\phi}$ (0.965--1.065~\gevcc)} \\\midrule
    0.10 & 33.6 & 9.0 & 30.4 & 9.5\\
    0.30 & 153  & 19  & 141  & 18\\
    0.50 & 372  & 28  & 185  & 19\\
    0.70 & 508  & 29  & 223  & 20\\
    0.90 & 358  & 24  & 193  & 18\\
    1.10 & 261  & 20  & 140  & 15\\
    1.30 & 182  & 16  & 93   & 12\\
    1.50 & 134  & 13  & 57.6 & 9.2\\
    1.70 & 68.3 & 9.3 & 40.5 & 7.4\\
    1.90 & 65.3 & 8.9 & 28.4 & 6.2\\
    2.50 & 142  & 13  & 91   & 10\\
    4.00 & 51.4 & 7.4 & 31.2 & 5.8\\
    5.50 & 3.8  & 2.0 & 0.9  & 1.0\\
    \bottomrule
  \end{tabular}
\end{table}

\newpage
\begin{center}
  \begin{longtable}{cr@{$~\pm~$}lr@{$~\pm~$}lr@{$~\pm~$}lr@{$~\pm~$}l}
    \caption[Fit results of $\omega$ yield extraction]{\label{tab:omegaextraction}Fit results of $\omega$ yield extraction.}\\
    \toprule
    \multicolumn{9}{c}{Yield}\\
    \pt (\gevc) & \multicolumn{2}{c}{Gauss} & \multicolumn{2}{c}{Gauss + Pol0} & \multicolumn{2}{c}{Gauss + Pol1} & \multicolumn{2}{c}{Gauss + Pol2} \\ \midrule \endfirsthead
    \multicolumn{9}{c}{{\bfseries \tablename\ \thetable{}:} (continued)} \\\addlinespace  \endhead
    \addlinespace\midrule \multicolumn{9}{r}{{Continued on next page}} \\ \midrule \endfoot
    \bottomrule \endlastfoot

    0.10 & 34.3 & 9.7 & 47   & 12  & 41   & 12  & 45   & 16\\
    0.30 & 145  & 21  & 126  & 23  & 125  & 23  & 116  & 26\\
    0.50 & 247  & 26  & 212  & 29  & 210  & 29  & 186  & 38\\
    0.70 & 484  & 37  & 359  & 33  & 363  & 33  & 299  & 34\\
    0.90 & 487  & 35  & 330  & 32  & 340  & 33  & 289  & 35\\
    1.10 & 301  & 23  & 230  & 22  & 234  & 22  & 215  & 24\\
    1.30 & 215  & 19  & 164  & 19  & 166  & 20  & 141  & 20\\
    1.50 & 171  & 20  & 10   & 14  & 104  & 15  & 85   & 15\\
    1.70 & 140  & 16  & 83   & 15  & 85   & 15  & 78   & 19\\
    1.90 & 59.6 & 8.8 & 51.2 & 9.0 & 49.6 & 9.1 & 43.4 & 9.2\\
    2.50 & 178  & 15  & 137  & 15  & 139  & 15  & 117  & 16\\
    4.00 & 47.9 & 7.4 & 50.6 & 7.7 & 43.0 & 7.6 & 52   & 10\\
    \\
    \multicolumn{9}{c}{Sigma (\mevcc)}\\
    \pt (\gevc) & \multicolumn{2}{c}{Gauss} & \multicolumn{2}{c}{Gauss + Pol0} & \multicolumn{2}{c}{Gauss + Pol1} & \multicolumn{2}{c}{Gauss + Pol2} \\\midrule
    0.10 & 17.6 & 5.4 & 22.5 & 6.2  & 20.2 & 6.0  & 21.6 & 6.9\\
    0.30 & 8.2  & 1.6 & 7.2  & 1.6  & 7.1  & 1.6  & 6.7  & 1.7\\
    0.50 & 16.7 & 2.4 & 14.3 & 2.7  & 14.1 & 2.7  & 12.2 & 3.6\\
    0.70 & 10.2 & 1.5 & 6.89 & 0.78 & 7.00 & 0.81 & 5.92 & 0.71\\
    0.90 & 25.7 & 3.3 & 16.1 & 1.8  & 16.6 & 1.9  & 14.5 & 1.9\\
    1.10 & 15.6 & 1.7 & 11.7 & 1.3  & 11.9 & 1.3  & 11.1 & 1.3\\
    1.30 & 20.6 & 3.0 & 14.7 & 2.1  & 15.0 & 2.2  & 13.0 & 2.0\\
    1.50 & 35.1 & 9.4 & 11.4 & 2.3  & 12.0 & 2.4  & 9.3  & 2.5\\
    1.70 & 43.5 & 9.7 & 21.2 & 4.2  & 21.8 & 4.5  & 20.4 & 4.9\\
    1.90 & 14.4 & 2.6 & 12.4 & 2.3  & 12.1 & 2.2  & 10.9 & 2.3\\
    2.50 & 20.6 & 2.9 & 14.7 & 1.7  & 15.0 & 1.7  & 13.1 & 1.6\\
    4.00 & 21.2 & 2.2 & 22.8 & 2.5  & 20.5 & 2.9  & 24.7 & 4.2\\
    \\
    \newpage
    \multicolumn{9}{c}{$\chi^2$/ndf}\\
    \pt (\gevc) & \multicolumn{2}{c}{Gauss} & \multicolumn{2}{c}{Gauss + Pol0} & \multicolumn{2}{c}{Gauss + Pol1} & \multicolumn{2}{c}{Gauss + Pol2} \\\midrule
    0.10 & \multicolumn{2}{c}{25.51/28} & \multicolumn{2}{c}{19.02/27} & \multicolumn{2}{c}{16.98/26} & \multicolumn{2}{c}{16.72/25}\\
    0.30 & \multicolumn{2}{c}{34.32/28} & \multicolumn{2}{c}{32.22/27} & \multicolumn{2}{c}{31.84/26} & \multicolumn{2}{c}{31.47/25}\\
    0.50 & \multicolumn{2}{c}{52.24/28} & \multicolumn{2}{c}{46.99/27} & \multicolumn{2}{c}{46.30/26} & \multicolumn{2}{c}{45.18/25}\\
    0.70 & \multicolumn{2}{c}{87.42/28} & \multicolumn{2}{c}{48.89/27} & \multicolumn{2}{c}{44.51/26} & \multicolumn{2}{c}{32.96/25}\\
    0.90 & \multicolumn{2}{c}{86.84/28} & \multicolumn{2}{c}{37.16/27} & \multicolumn{2}{c}{31.66/26} & \multicolumn{2}{c}{26.28/25}\\
    1.10 & \multicolumn{2}{c}{94.51/28} & \multicolumn{2}{c}{38.00/27} & \multicolumn{2}{c}{31.98/26} & \multicolumn{2}{c}{28.46/25}\\
    1.30 & \multicolumn{2}{c}{67.84/28} & \multicolumn{2}{c}{42.57/27} & \multicolumn{2}{c}{40.62/26} & \multicolumn{2}{c}{34.94/25}\\
    1.50 & \multicolumn{2}{c}{90.59/28} & \multicolumn{2}{c}{51.59/27} & \multicolumn{2}{c}{45.66/26} & \multicolumn{2}{c}{38.84/25}\\
    1.70 & \multicolumn{2}{c}{43.87/28} & \multicolumn{2}{c}{26.07/27} & \multicolumn{2}{c}{25.52/26} & \multicolumn{2}{c}{25.22/25}\\
    1.90 & \multicolumn{2}{c}{39.52/28} & \multicolumn{2}{c}{31.76/27} & \multicolumn{2}{c}{30.16/26} & \multicolumn{2}{c}{26.10/25}\\
    2.50 & \multicolumn{2}{c}{89.68/28} & \multicolumn{2}{c}{42.62/27} & \multicolumn{2}{c}{37.99/26} & \multicolumn{2}{c}{27.94/25}\\
    4.00 & \multicolumn{2}{c}{47.46/28} & \multicolumn{2}{c}{44.28/27} & \multicolumn{2}{c}{26.22/26} & \multicolumn{2}{c}{18.73/25}\\
  \end{longtable}
\end{center}

\newpage
\begin{center}
  \begin{longtable}{cr@{$~\pm~$}lr@{$~\pm~$}lr@{$~\pm~$}lr@{$~\pm~$}l}
    \caption[Fit results of $\phi$ yield extraction]{\label{tab:phiextraction}Fit results of $\phi$ yield extraction.}\\
    \toprule
    \multicolumn{9}{c}{Yield}\\
    \pt (\gevc) & \multicolumn{2}{c}{Gauss} & \multicolumn{2}{c}{Gauss + Pol0} & \multicolumn{2}{c}{Gauss + Pol1} & \multicolumn{2}{c}{Gauss + Pol2} \\ \midrule \endfirsthead
    \multicolumn{9}{c}{{\bfseries \tablename\ \thetable{}:} (continued)} \\\addlinespace \endhead
    \addlinespace\midrule \multicolumn{9}{r}{{Continued on next page}} \\ \midrule \endfoot
    \bottomrule \endlastfoot

    0.10 & 29.1 & 7.9 & 32.4 & 8.8 & 29.8 & 8.7 & 27.7 & 8.7\\
    0.30 & 139  & 16  & 120  & 16  & 119  & 16  & 124  & 17\\
    0.50 & 159  & 17  & 134  & 17  & 132  & 18  & 130  & 18\\
    0.70 & 169  & 18  & 150  & 18  & 139  & 18  & 134  & 18\\
    0.90 & 206  & 18  & 181  & 19  & 168  & 19  & 188  & 21\\
    1.10 & 151  & 16  & 139  & 18  & 138  & 18  & 142  & 21\\
    1.30 & 117  & 15  & 103  & 16  & 101  & 17  & 97   & 19\\
    1.50 & 67   & 10  & 71   & 12  & 75   & 13  & 100  & 31\\
    1.70 & 25.4 & 5.7 & 30.4 & 5.9 & 28.2 & 6.0 & 29   & 6.0\\
    1.90 & 28.4 & 6.0 & 35.1 & 6.5 & 27.3 & 6.1 & 27.3 & 6.2\\
    2.50 & 86.5 & 9.9 & 96   & 10  & 83   & 10  & 84   & 10\\
    4.00 & 20.4 & 4.9 & 25.8 & 5.1 & 23.2 & 5.5 & 21.4 & 5.6\\
    \\
    \multicolumn{9}{c}{Sigma (\mevcc)}\\
    \pt (\gevc) & \multicolumn{2}{c}{Gauss} & \multicolumn{2}{c}{Gauss + Pol0} & \multicolumn{2}{c}{Gauss + Pol1} & \multicolumn{2}{c}{Gauss + Pol2} \\\midrule
    0.10 & 9.1  & 3.8  & 10.4 & 4.2  & 9.4  & 4.1  & 8.6  & 5.2\\
    0.30 & 5.48 & 0.72 & 4.79 & 0.68 & 4.77 & 0.68 & 4.92 & 0.71\\
    0.50 & 13.1 & 1.3  & 11.8 & 1.3  & 11.7 & 1.3  & 11.6 & 1.3\\
    0.70 & 6.55 & 0.76 & 5.97 & 0.71 & 5.66 & 0.68 & 5.54 & 0.68\\
    0.90 & 15.9 & 1.4  & 14.4 & 1.4  & 13.7 & 1.4  & 14.8 & 1.6\\
    1.10 & 19.4 & 2.4  & 18.0 & 2.4  & 18.0 & 2.5  & 18.4 & 2.7\\
    1.30 & 23.7 & 3.6  & 21.2 & 3.5  & 20.8 & 3.6  & 20.3 & 3.8\\
    1.50 & 18.4 & 4.0  & 19.5 & 4.7  & 21.0 & 5.2  & 29.5 & 9.4\\
    1.70 & 5.5  & 1.9  & 7.0  & 1.4  & 6.4  & 1.7  & 6.5  & 1.6\\
    1.90 & 8.7  & 2.6  & 16.5 & 7.5  & 8.3  & 2.5  & 8.3  & 2.6\\
    2.50 & 14.3 & 2.0  & 16.9 & 3.2  & 13.6 & 1.9  & 13.8 & 1.9\\
    4.00 & 21.2 & 2.6  & 26.8 & 2.4  & 24.9 & 3.1  & 25.0 & 3.7\\
    \\
    \newpage
    \multicolumn{9}{c}{$\chi^2$/ndf}\\
    \pt (\gevc) & \multicolumn{2}{c}{Gauss} & \multicolumn{2}{c}{Gauss + Pol0} & \multicolumn{2}{c}{Gauss + Pol1} & \multicolumn{2}{c}{Gauss + Pol2} \\\midrule
    0.10 & \multicolumn{2}{c}{18.57/28} & \multicolumn{2}{c}{17.03/27} & \multicolumn{2}{c}{15.78/26} & \multicolumn{2}{c}{15.13/25}\\
    0.30 & \multicolumn{2}{c}{49.46/28} & \multicolumn{2}{c}{33.43/27} & \multicolumn{2}{c}{33.37/26} & \multicolumn{2}{c}{32.19/25}\\
    0.50 & \multicolumn{2}{c}{51.69/28} & \multicolumn{2}{c}{33.32/27} & \multicolumn{2}{c}{33.23/26} & \multicolumn{2}{c}{32.99/25}\\
    0.70 & \multicolumn{2}{c}{27.39/28} & \multicolumn{2}{c}{17.95/27} & \multicolumn{2}{c}{12.29/26} & \multicolumn{2}{c}{11.45/25}\\
    0.90 & \multicolumn{2}{c}{60.09/28} & \multicolumn{2}{c}{47.10/27} & \multicolumn{2}{c}{41.32/26} & \multicolumn{2}{c}{31.12/25}\\
    1.10 & \multicolumn{2}{c}{32.43/28} & \multicolumn{2}{c}{29.74/27} & \multicolumn{2}{c}{29.73/26} & \multicolumn{2}{c}{29.56/25}\\
    1.30 & \multicolumn{2}{c}{19.12/28} & \multicolumn{2}{c}{16.33/27} & \multicolumn{2}{c}{16.06/26} & \multicolumn{2}{c}{15.94/25}\\
    1.50 & \multicolumn{2}{c}{52.02/28} & \multicolumn{2}{c}{51.64/27} & \multicolumn{2}{c}{48.58/26} & \multicolumn{2}{c}{46.97/25}\\
    1.70 & \multicolumn{2}{c}{133.5/28} & \multicolumn{2}{c}{65.33/27} & \multicolumn{2}{c}{61.16/26} & \multicolumn{2}{c}{60.53/25}\\
    1.90 & \multicolumn{2}{c}{80.72/28} & \multicolumn{2}{c}{51.22/27} & \multicolumn{2}{c}{36.73/26} & \multicolumn{2}{c}{36.72/25}\\
    2.50 & \multicolumn{2}{c}{111.4/28} & \multicolumn{2}{c}{79.21/27} & \multicolumn{2}{c}{60.80/26} & \multicolumn{2}{c}{50.34/25}\\
    4.00 & \multicolumn{2}{c}{57.31/28} & \multicolumn{2}{c}{24.48/27} & \multicolumn{2}{c}{23.07/26} & \multicolumn{2}{c}{20.02/25}\\
  \end{longtable}
\end{center}

\begin{table}[p]
  \centering
  \caption[Invariant Cross Section of $\omega$]{\label{tab:omxsec}Invariant Cross Section of $\omega$.\\}
  \begin{tabular}{cr@{.}lr@{.}lr@{.}l}
    \toprule
    \pt (\gevc) & \multicolumn{2}{c}{$E d^3\sigma/dp^3$ (mbarn GeV$^{-2}$/$c^3$)}& \multicolumn{2}{c}{stat. error} & \multicolumn{2}{c}{syst. error} \\\midrule
    0.10 & 2&91 &  7&7$\times 10^{-1}$ & 7&4$\times 10^{-1}$\\
    0.30 & 2&12 &  2&7$\times 10^{-1}$ & 5&4$\times 10^{-1}$\\
    0.50 & 1&39 &  1&0$\times 10^{-1}$ & 3&5$\times 10^{-1}$\\
    0.70 & 8&56$\times 10^{-1}$  & 5&0$\times 10^{-2}$ & 2&2$\times 10^{-1}$\\
    0.90 & 3&67$\times 10^{-1}$  & 2&4$\times 10^{-2}$ & 9&3$\times 10^{-2}$\\
    1.12 & 1&49$\times 10^{-1}$  & 1&1$\times 10^{-2}$ & 3&8$\times 10^{-2}$\\
    1.38 & 1&049$\times 10^{-1}$ & 7&6$\times 10^{-3}$ & 2&6$\times 10^{-2}$\\
    1.75 & 2&20$\times 10^{-2}$  & 1&8$\times 10^{-3}$ & 5&5$\times 10^{-3}$\\
    2.50 & 2&15$\times 10^{-3}$  & 2&0$\times 10^{-4}$ & 5&4$\times 10^{-4}$\\
    3.50 & 2&50$\times 10^{-4}$  & 4&0$\times 10^{-5}$ & 6&3$\times 10^{-5}$\\
    4.50 & 3&1$\times 10^{-5}$   & 1&1$\times 10^{-5}$ & 7&7$\times 10^{-6}$\\
    6.50 & 3&7$\times 10^{-7}$   & 2&0$\times 10^{-7}$ & 2&0$\times 10^{-7}$\\
    \bottomrule
  \end{tabular}
\end{table}

\begin{table}[p]
  \centering
  \caption[Invariant Cross Section of $\phi$]{\label{tab:phixsec}Invariant Cross Section of $\phi$.\\}
  \begin{tabular}{cr@{.}lr@{.}lr@{.}l}
    \toprule
    \pt (\gevc) & \multicolumn{2}{c}{$E d^3\sigma/dp^3$ (mbarn GeV$^{-2}$/$c^3$)}& \multicolumn{2}{c}{stat. error} & \multicolumn{2}{c}{syst. error} \\\midrule
    0.10 & 1&14$\times 10^{-1}$ & 3&5$\times 10^{-2}$ & 3&4$\times 10^{-2}$\\
    0.30 & 1&79$\times 10^{-1}$ & 2&2$\times 10^{-2}$ & 5&2$\times 10^{-2}$\\
    0.50 & 1&17$\times 10^{-1}$ & 1&2$\times 10^{-2}$ & 3&4$\times 10^{-2}$\\
    0.70 & 8&47$\times 10^{-2}$ & 7&6$\times 10^{-3}$ & 2&5$\times 10^{-2}$\\
    0.90 & 5&03$\times 10^{-2}$ & 4&7$\times 10^{-3}$ & 1&5$\times 10^{-2}$\\
    1.12 & 2&09$\times 10^{-2}$ & 2&3$\times 10^{-3}$ & 4&9$\times 10^{-3}$\\
    1.38 & 1&45$\times 10^{-2}$ & 1&5$\times 10^{-3}$ & 3&4$\times 10^{-3}$\\
    1.75 & 2&97$\times 10^{-3}$ & 3&8$\times 10^{-4}$ & 6&9$\times 10^{-4}$\\
    2.50 & 4&90$\times 10^{-4}$ & 5&7$\times 10^{-5}$ & 1&1$\times 10^{-4}$\\
    3.50 & 5&1$\times 10^{-5}$  & 1&1$\times 10^{-5}$ & 1&2$\times 10^{-5}$\\
    4.50 & 6&6$\times 10^{-6}$  & 2&7$\times 10^{-6}$ & 1&5$\times 10^{-6}$\\
    6.50 & 1&9$\times 10^{-8}$  & 3&9$\times 10^{-8}$ & 4&0$\times 10^{-8}$\\
    \bottomrule
  \end{tabular}
\end{table}

% ...

\end{document}